\documentstyle[12pt]{article}
\normalsize
\begin{document}

\setcounter{page}{0}

\def\pp{\hskip 5mm}
\def\bq{{\bf q}}

%
%
\title{Fermi Systems with Strong Forward Scattering}
\author{Walter Metzner \\
{\em Sektion Physik, Universit\"at M\"unchen} \\
{\em Theresienstra{\ss}e 37, D-80333 M\"unchen, Germany} \\
\\
Claudio Castellani and Carlo Di Castro \\
{\em Dipartimento di Fisica, Universit\`a "La Sapienza",} \\
{\em P.le A. Moro 2, 00185 Roma, Italy}}
\maketitle
%
%
\begin{abstract}

\pp We review the theory of interacting Fermi systems whose 
low-energy physics is dominated by forward scattering, i.e.\  
scattering processes generated by effective interactions with small 
momentum transfers.
These systems include Fermi liquids as well as several important 
non-Fermi liquid phases: one-dimensional Luttinger liquids, systems 
with long-range interactions, and fermions coupled to a gauge field.
We report results for the critical dimensions separating different 
"universality classes", and discuss the behavior of physical quantities
such as the momentum distribution function, the single-particle 
propagator and low-energy response functions in each class.
\par
The renormalization group for Fermi systems will be reviewed and 
applied as a link between microscopic models and effective low-energy 
theories. 
Particular attention is payed to conservation laws, which constrain
any effective low-energy theory of interacting Fermi systems.
In scattering processes with small momentum transfers the velocity
of each scattering particle is (almost) conserved. This asymptotic
conservation law leads to non-trivial cancellations of Feynman diagrams 
and other simplifications, making thus possible a non-perturbative 
treatment of forward scattering via Ward identities or bosonization 
techniques.
\par
\medskip
PACS: 71.10.-w, 71.10.Ay, 71.10.Hf, 71.10.Pm                   

\end{abstract}

\vfill\eject


\pagenumbering{roman}
\setcounter{page}{1}

\def\df{\dotfill}

\def\bk{{\bf k}}
\def\bq{{\bf q}}
\def\ket{\rangle}

\centerline{\bf CONTENTS} 
\vskip 1cm

{\bf 1. Introduction                                       \hfill  1}
\begin{itemize}
\item[1.1.] Normal phases of interacting fermions             \df  1 
\item[1.2.] Theoretical routes to non-Fermi liquid behavior   \df  3 
\item[1.3.] Dominant forward scattering                       \df  7
\item[1.4.] Contents of this article                          \df 10
\end{itemize}
\medskip
{\bf 2. Renormalization group                             \hfill  12} 
\begin{itemize}
\item[2.1.] Functional integral representation                \df 13
\item[2.2.] Mode elimination and effective actions            \df 15
\item[2.3.] Low-energy coupling-space                         \df 21
\end{itemize}
\medskip
{\bf 3. Exact conservation laws                            \hfill 26}
\begin{itemize}
\item[3.1.] Charge conservation                               \df 26
\item[3.2.] Spin conservation                                 \df 29
\item[3.3.] Renormalization of response functions             \df 31
\end{itemize}
\medskip
{\bf 4. Fermi liquid                                       \hfill 34}
\begin{itemize}
\item[4.1.] Microscopic Fermi liquid theory                   \df 34
\item[4.2.] Effective action and fixed point                  \df 36
\item[4.3.] Response functions                                \df 39
\item[4.4.] Subleading corrections                            \df 40
\item[4.5.] Instabilities                                     \df 41
\end{itemize}
\medskip
{\bf 5. Forward scattering                                 \hfill 43}
\begin{itemize}
\item[5.1.] Global charge and spin conservation               \df 45
\item[5.2.] Velocity conservation                             \df 47
\begin{itemize}
\item[5.2.1.] Loop cancellation and response functions        \df 48
\item[5.2.2.] Density and current vertex                      \df 51
\item[5.2.3.] Single-particle propagator                      \df 52
\end{itemize}
\item[5.3.] Bosonization                                      \df 56
\end{itemize}
\medskip
{\bf 6. One-dimensional Luttinger liquid                   \hfill 63}
\begin{itemize}
\item[6.1.] The g-ology model                                 \df 64
\item[6.2.] Renormalization group and cutoffs                 \df 66
\item[6.3.] Charge/spin conservation and Ward identities      \df 68
 \begin{itemize}
 \item[6.3.1.] Total charge and spin conservation             \df 68
 \item[6.3.2.] Separate left/right conservation laws          \df 70
 \end{itemize}
\item[6.4.] Luttinger liquid fixed point                      \df 71
\item[6.5.] Instabilities                                     \df 77
\end{itemize}
\medskip
{\bf 7. Short-range interactions in d dimensions           \hfill 79}
\begin{itemize}
\item[7.1.] Continuation to non-integer dimensions            \df 80
\item[7.2.] Perturbative results                              \df 81
 \begin{itemize}
 \item[7.2.1.] Particle-hole bubble                           \df 81
 \item[7.2.2.] Particle-particle bubble                       \df 83
 \item[7.2.3.] Second order self-energy                       \df 84
 \end{itemize}
\item[7.3.] Random phase approximation                        \df 87
 \begin{itemize}
 \item[7.3.1.] Effective interaction                          \df 87
 \item[7.3.2.] Density-density response                       \df 88
 \item[7.3.3.] RPA self-energy                                \df 88
 \end{itemize}
\item[7.4.] Resummation of forward scattering                 \df 90
\item[7.5.] Critical dimensions and crossover                 \df 94
 \begin{itemize}
 \item[7.5.1.] Leading low-energy behavior                    \df 94
 \item[7.5.2.] Subleading corrections                         \df 96
 \end{itemize}
\end{itemize}
\medskip
{\bf 8. Long-range density-density interactions            \hfill 98}
\par
\medskip
{\bf 9. Fermions coupled to a gauge-field                 \hfill 103}
\begin{itemize}
\item[9.1.] Action                                           \df 103
\item[9.2.] Gauge-field propagator                           \df 105
\item[9.3.] Fermion propagator                               \df 108
\item[9.4.] Response functions                               \df 114
\end{itemize}
\medskip
{\bf 10. Conclusions                                      \hfill 116}
\par
\bigskip
{\bf Acknowledgements                                     \hfill 122}
\par
\bigskip
{\bf Appendices                                           \hfill 123}
\begin{itemize}
\item[A:] Loop cancellation for $N=3$                        \df 123
\item[B:] Spectral representation of $\bar D$                \df 123
\item[C:] Bubbles for a quadratic dispersion relation        \df 125
\item[D:] Perturbative self-energy in one dimension          \df 127
\item[E:] Limit $d \to 1$ for the propagator                 \df 127
\end{itemize}
\bigskip
{\bf References                                           \hfill 129}
\par
\bigskip
{\bf Figure captions                                      \hfill 137}

\vfill\eject

\pagenumbering{arabic}
\setcounter{page}{1}

\def\pp{\hskip 5mm}
\def\eps{\epsilon}
\def\bp{{\bf p}}
\def\bq{{\bf q}}
\def\bk{{\bf k}}
\def\bv{{\bf v}}
\def\xip{\xi_{\bp}}
\def\xik{\xi_{\bk}}
\def\xikq{\xi_{\bk+\bq}}
\def\tilk{\tilde k}
\def\tilth{\tilde\theta}
\def\tilxi{\tilde\xi}
\def\dph{\Delta^{ph}}
\def\dpp{\Delta^{pp}}
\def\Im{{\rm Im}}
\def\Re{{\rm Re}}
\def\bra{\langle}
\def\ket{\rangle}
\def\up{\uparrow}
\def\down{\downarrow}
\def\Dt{\tilde D}
\def\tht{\tilde\theta}
\def\omt{\tilde\omega}
\def\q0t{\tilde q_0}
\def\qrt{\tilde q_r}
\def\xit{\tilde \xi}
\def\tt{\tilde t}
\def\Lt{\tilde L}
\def\bv{{\bf v}}
\def\b0{{\bf 0}}
\def\bj{{\bf j}}
\def\bR{{\bf R}}
\def\br{{\bf r}}
\def\cdotr{\!\cdot\!}
\def\Lam{\Lambda}
\def\lam{\lambda}

\vspace*{1cm}
\centerline{\large 1. INTRODUCTION}
\vskip 1cm

{\bf 1.1. NORMAL PHASES OF INTERACTING FERMIONS} \par
\medskip
\pp Observed phases of interacting Fermi systems such as liquid $^3$He 
and electrons in metals fall into two broad categories: 
symmetry-broken and "{\em normal}\/".
The former contains superconductors or superfluids, magnetically 
ordered phases, charge density waves, and several others. 
While the variety of ordered phases has long been known to be very
rich, the low-energy behavior of normal Fermi systems seemed to be governed
by a remarkably small number of "universality classes". 
\par
\pp Actually most normal metals, as well as normal $^3$He, follow the
same type of low-energy behavior, the one described by {\em Fermi 
liquid theory} \cite{LAN,NOZ}. 
This originally purely phenomenological theory makes very specific 
predictions on the low-energy scaling (i.e.\ power-law) behavior of 
thermodynamic and 
transport properties as a function of temperature, frequency, and
other small energy or momentum scales. These predictions have been
experimentally verified to an extremely high precision in liquid
$^3$He, a rather exceptional substance in its high purity and its 
perfect isotropy \cite{VW}. In metals, the anisotropy imposed by the 
crystal structure and the unavoidable impurities blur some of the more
sensitive properties of pure and isotropic Fermi liquids, but many
others are robust with respect to such complications. 
Among these robust properties are the existence of a Fermi surface,
a specific heat proportional to temperature, a constant 
spin-susceptibility, a finite density of low-energy 
single-particle excitations and, of course, a metallic DC-conductivity
which increases for decreasing temperature, implying that charged
excitations are gapless. These properties behave qualitatively as
in a non-interacting Fermi gas. In Fermi liquid theory they are
phenomenologically, but quantitatively, described in terms of Landau's
"{\em quasi-particles}\/", i.e.\ fermionic elementary excitations which
have momenta near the Fermi surface and
obey a free-fermion-like energy-momentum relation. To leading order (in
energy scales), the quasi-particles are mutually independent, except
for a Hartree-type interaction. The {\em parameters}\/ characterizing 
these almost-free excitations vary drastically from one substance to 
another. In particular, the effective mass of quasi-particles in 
metals may be as much as a factor thousand bigger than the bare 
electron mass, as has been observed in the so-called heavy fermion 
systems \cite{HFS}.
In addition, the energy or temperature {\em scale} controlled
by Fermi liquid theory is strongly material dependent, too.
Usually at least three distinct scales must be distinguished:
the Fermi temperature $T_F$ given roughly by the average kinetic
energy, a coherence temperature $T^*$ below which Fermi liquid 
behavior sets in, and a transition temperature $T_c$ associated with
an instability of the normal Fermi liquid towards some sort of
symmetry breaking. For electrons in metals, $T_F$ is usually much
higher than room temperature. Phase transition temperatures may be
anything from zero to $1388\,K$, the Curie temperature of cobalt.
Both $T_c$ and $T^*$ are always much smaller than $T_F$. Usually $T_c$
is also much smaller than $T^*$, but sometimes they are comparable,
such that only very gross aspects of Fermi liquid behavior can be 
really observed.
\par
\pp Going to small but finite energy scales, there are subleading 
corrections due to quasi-particle scattering.
Among these are a quasi-particle decay rate proportional to the square
of its energy, a contribution proportional to $T^3 \log T$ to the
specific heat (in three dimensions) and a contribution of order $T^2$
to the electrical resistivity due to electron-electron interactions
in metals. In metals, subleading terms due to electron-electron
scattering cannot always be observed experimentally, because they are 
often superposed by other, much larger, contributions due to phonons 
or disorder. 
\par
\smallskip
\pp Up to recently, Fermi liquid theory seemed universally
applicable at least to all sufficiently pure interacting Fermi
systems, and its more gross features even to quite dirty ones, 
provided the normal phase is not destroyed
by symmetry breaking before the coherence temperature $T^*$ is 
reached.
This situation has changed during the last years with the discovery 
of new materials, where unusual scaling behavior is observed
above $T_c$, which differs in many respects from the predictions of 
Fermi liquid theory.
\par
\pp High temperature superconductors are certainly the most famous
among these new materials. Their unusual and unexpected properties
have stimulated a profound reinvestigation of basic concepts in the
theory of metals. The high superconducting transition temperature
is hard to obtain from the conventional phonon-induced pairing, and
the coherence length is too short for applying simple BCS mean field
theory, whatever the pairing mechanism may be. Even more striking
is the behavior of the normal metallic phase in these materials,
which does not fit in a simple Fermi liquid type description. 
Although photoemission experiments yield convincing evidence for the
existence of a large Fermi surface, many transport properties are 
incompatible with a Fermi liquid picture of low-lying 
excitations \cite{WYD,PKC}.
The most prominent among many other quantitities showing 
non-Fermi-liquid behavior is the electrical resistivity, whose 
temperature dependence deviates significantly from the $T^2$ behavior
at low $T$ 
predicted by Fermi liquid theory. The layer structure and the giant 
anisotropy observed in various properties like the resistivity
indicate that the normal phase is governed by electrons confined
essentially to two dimensions. The vicinity to an antiferromagnetic Mott
insulating phase reveals the importance of electron-electron interactions.
These observations have stimulated speculations on the existence of  
non-Fermi-liquid metallic phases in two-dimensional interacting electron 
systems \cite{AND90,VLS}.
\par
\pp Asymptotic low-energy behavior different from Fermi liquid 
predictions in a normal metallic phase has also been found recently
in various heavy-fermion alloys \cite{AT,AS,SML,LPP,OTT}.
A specific heat proportional to $T \log T$ and a $T$-linear 
contribution to the electrical resistivity was observed. 
The magnetic susceptibility as a function of temperature
has a cusp in one of these materials, and an algebraic divergence for
$T \to 0$ in another. 
\par

\bigskip

{\bf 1.2. THEORETICAL ROUTES TO NON-FERMI LIQUID BEHAVIOR} \par
\medskip
\pp Theoretically, the almost universal validity of Fermi liquid
theory in normal Fermi systems has been made plausible in various
ways. A justification within perturbation theory has already been
given by Landau and his coworkers \cite{AGD}. To push the
validity of the theory beyond the perturbatively accessible regime
in coupling space, only few basic assumptions on the asymptotic 
low-energy behavior of propagators and vertex functions need to be
made \cite{NOZ,AGD}.
\par
\pp A new point of view has recently been developed with the
adaption of Wilson's {\em renormalization group} to interacting 
Fermi systems \cite{FT,BG,FMRT,SHA91,SHA94}.
The idea is to integrate out high energy degrees of freedom in a 
path-integral representation of the full interacting theory. At
least in perturbation theory this corresponds to integrating out 
fermions with momenta far from the Fermi surface. Successive reduction
of the momentum cutoff generates a sequence of effective actions with 
renormalized kinetic terms and interactions. 
One may then classify different types of low-energy behavior in terms
of different asymptotic effective actions, defined on a thin shell 
around the Fermi surface of the interacting system.\footnote{
It is important to give a sufficiently general definition of a
{\em Fermi surface}.
In a Fermi liquid, the momentum distribution function $n_{\bk}$ is 
discontinuous across the Fermi surface, and this property is often used
to define the surface itself.
There are other systems where $n_{\bk}$ is continuous everywhere, but 
where one can still define a Fermi surface as a surface in momentum 
space where the excitations with lowest energy are located.}
Consistency of a certain hypothetical low-energy action can be
systematically checked by calculating the renormalization group
flow of its couplings.
The renormalization group thus provides effective actions representative
of the low-energy behavior of the system. These asymptotic low-energy
theories are prototype models of possible (metallic) phases, which can 
often be solved exactly by various techniques. 
\par
\pp An effective low-energy action with finite (not singular) 
renormalized interactions will usually lead to Fermi liquid behavior, 
with the possibility of a Cooper instability towards superconductivity. 
Different low-energy behavior may be found if the effective theory has
one of the following peculiar features: \par
\medskip
\begin{tabular}{rl}
a) & dimensionality $d=1$, \\
b) & special Fermi surfaces, \\
c) & vanishing Fermi velocities, \\
d) & singular interactions, \\
e) & coupling to gauge fields, \\
f) & coupling to other soft modes, \\
g) & local degrees of freedom. 
\end{tabular}
\par
\medskip
A {\em normal}\/ metallic phase of interacting fermions with a low-energy 
behavior that is not described by Fermi liquid theory is generally 
called a "{\em non-Fermi liquid}\/".
"Metallic" behavior implies that there are gapless charge fluctuations.
More generally, including neutral fermions such as $^3$He, one would 
require the existence of gapless density fluctuations.
Let us now address the various routes to non-Fermi liquid behavior one by
one.
\par
\smallskip
a) The breakdown of Fermi liquid theory in {\em one-dimensional}\/ 
interacting Fermi systems shows up already in second order perturbation 
theory: 
the perturbative contributions to the quasi-particle weight diverge 
logarithmically at the Fermi surface of the non-interacting system. 
The problem of treating these divergencies has been solved by the 
renormalization group, using an effective low-energy theory known as 
"g-ology" model \cite{SOL}. 
Assuming a scaling ansatz for the vertex
functions, one approaches the Fermi surface by rescaling the fields and the
coupling constants (a small number of "g's"). Depending on the values of the
bare couplings the renormalized couplings flow either to strong coupling, 
and hence out of the perturbatively controlled regime, or to the 
Luttinger model \cite{TL}, which is exactly solvable \cite{ML} .
In the latter case the system is a "{\em Luttinger liquid}\/", i.e.\
a normal (not symmetry-broken) metallic phase characterized by 
i) a continuous momentum distribution with a power-law singularity at the
Fermi surface, the exponent $\eta$ being non-universal;
ii) a single-particle density of states which vanishes as $\omega^{\eta}$
near the Fermi energy, implying the absence of fermionic quasi-particles;
iii) finite charge and spin density response for small wave vectors and
the existence of collective bosonic charge and spin density modes;
iv) power-law singularities in Cooper pair correlation functions and 
density correlation functions with large wave vectors;
v) separation of spin and charge degrees of freedom \cite{SOL,VOI95,SCH95}.
As introduced by Haldane \cite{HAL81}, the term "Luttinger liquid" denotes
the universality class of systems whose fixed point Hamiltonian is the
Luttinger model. 
\par
\smallskip
b) It is easy to see, in weak coupling analysis, that certain special
shapes of Fermi surfaces make a Fermi liquid unstable. 
In particular "{\em nesting}$\,$", i.e.\ the property that a finite 
fraction of the Fermi surface can be shifted to other points of the 
surface by adding a certain fixed momentum transfer, usually leads to
formation of density waves and thus breaking of translational
invariance. 
If a Fermi surface contains extended almost flat pieces, a whole
host of competing infrared singularities appears in perturbation
theory. The perturbative quasi-particle decay rate is then linear
in temperature (or frequency), signalling a breakdown of 
Fermi liquid theory. To obtain the real low-energy behavior in such a
case is a formidable task even at weak coupling, because the necessary
renormalization group calculation is complicated by the huge coupling
space one has to consider, and one must go at least to two-loop order 
to include singularities involving single-particle excitations. 
It has been suggested that the shape of Fermi surfaces in high-$T_c$ 
superconductors might be responsible at least for some of their 
anomalous properties \cite{VR}, but to our knowledge a controlled 
calculation that takes into account all the singular channels has not 
yet been performed. 
Hence it is not clear whether the Fermi surface shapes can
quantitatively explain the strange low-energy behavior in these 
materials over a rather wide range of different compositions.
\par
\smallskip
c) The Fermi velocity may vanish on or near parts of the Fermi surface.
Saddle points of the dispersion relation in momentum space lead to 
{\em van Hove singularities}\/ in the single-particle density of states.
In three
dimensions these singularities are only cusps, while in two dimensions 
logarithmic divergencies are obtained. If such a divergence comes close
to the Fermi energy, it will affect the low-energy behavior:
second order perturbation theory yields a quasi-particle decay rate
that depends linearly on energy \cite{PKN}.
It has been proposed that this linear behavior can explain the
anomalous transport in high-$T_c$ superconductors \cite{PKN,NTH}. 
On the other hand, it has been pointed out that transport properties 
are related to decay rates for currents rather than single-particle decay, 
and, within perturbation theory, these decay rates relevant for transport 
properties are not significantly affected by the van Hove singularities 
\cite{HR}.
In any case, the presence of singularities makes the application of 
standard perturbation theory problematic, and one has to wait for a 
more sophisticated analysis.
\par
\smallskip
d) Anderson \cite{AND90} has suggested that in a two-dimensional Fermi
system perturbation theory and Fermi liquid theory should break down 
even at {\em weak}\/ coupling, having in mind two intriguing signals: 
i) a finite phase shift for two interacting particles on the same point 
of the Fermi surface, and ii) an "antibound state" in the two-particle
spectrum, separated by an energy gap from the continuum of all other
states. He argued that conventional many body theory would miss these
effects, while they might be included by introducing a {\em singular
"pseudo-potential"} acting between electrons. The resulting state should
not be a Fermi liquid, but rather a two-dimensional "Luttinger 
liquid"\footnote{
According to Anderson's terminology, the term "Luttinger liquid"
is no more restricted to one-dimensional systems, but may also denote
higher-dimensional non-Fermi liquids, provided they have a Fermi surface
in the generalized sense described above.}
with properties similar to those known for one-dimensional systems. 
So far, nobody has succeeded to construct a sound theory based 
on these intuitive ideas.
Several authors have analyzed the phase shift in considerable detail,
but they all concluded that one cannot infer a breakdown of Fermi liquid
theory from that effect \cite{FHN,ER92,STA93,MC}.
Recent rigorous results confirm the general expectation that the Cooper
instability is the only weak coupling instability of generic 
two-dimensional Fermi liquids with short-range interactions \cite{FKL}.
It has also been shown that the mechanism leading to Luttinger liquid
behavior in one dimension does not destroy the Landau quasi-particle in 
any (possibly non-integer) dimension $d > 1$ \cite{CDM94}.
However, certain ad hoc {\em long-range}\/ interactions (with singular 
Fourier transforms in momentum space) can indeed destabilize the Fermi 
liquid in dimensions $d > 1$ \cite{STA92,BW,CD94}.
To overwhelm screening, which usually makes long-range interactions
in an electron system effectively short-ranged, these interactions must 
be even more singular than the Coulomb interaction.
\par
\smallskip
e) For lattice models of interacting fermions with a local constraint
(arising from very strong repulsion) an effective low-energy theory
involving fermions and bosons coupled to a ficticious {\em gauge field}\/
has been derived \cite{BA,IL,LN}. The coupling to the gauge field leads 
to infrared divergences in perturbation theory. In an approximate 
evaluation of the theory non-Fermi liquid behavior in several transport
properties has been obtained \cite{IL,LN}.
A controlled derivation of the low-energy behavior is however very 
difficult, and has not yet been fully achieved.
The real electromagnetic gauge field also leads to divergences when
coupled to a many fermion system \cite{HNP,REI}. The ensuing physical
effects are however only of order $v_F/c$ (where $v_F$ is the Fermi
velocity and $c$ the velocity of light) and are thus practically 
unobservable. 
Fermions coupled to a gauge field
may also describe interacting fermions in a strong magnetic field tuned
such that the highest occupied Landau level is precisely half-filled. 
Halperin, Lee and Read \cite{HLR} and Kalmeyer and Zhang \cite{KZ} have
indeed proposed that this system can be mapped on a system of spinless 
fermions in zero (average) magnetic field, but coupled to a fluctuating 
gauge field. 
\par
\smallskip 
f) Anomalous scattering mediated by soft modes appears to be a generic
consequence of proximity to a critical point. This mechanism has been
invoked to explain various non-Fermi liquid features in heavy fermion
systems close to a second-order phase transition \cite{LOE,STE}.
Proximity to a zero temperature antiferromagnetic and/or charge
instability has been also suggested to be responsible for the anomalous
metallic behavior in high-$T_c$ cuprates. 
\par
\pp Close to an antiferromagnetic instability, {\em antiferromagnetic spin 
fluctuations}\/ lead to a strongly enhanced quasi-particle decay rate,
pointing towards non-Fermi liquid behavior.
It has been argued that the anomalous quasi-particle decay carries over 
to anomalous transport properties,
such as those observed in high temperature superconductors \cite{MTU,MP}.
On the other hand, analyzing non-trivial solutions of the Boltzmann 
equation which take into account that spin fluctuations enhance scattering 
strongly only on special points of the Fermi surface ("hot spots"), 
Hlubina and Rice \cite{HR} have found that standard Fermi liquid behavior 
(for transport) holds up to a rather high crossover scale, which is
too high to explain the observed behavior in the high-$T_c$ materials. 
\par
\pp Charge fluctuations also lead to strongly singular scattering and 
non-Fermi liquid behavior \cite{EK,CDG,VAR}. In particular, in the 
proximity of phase separation the effective interactions are singular at 
small $\bq$ and lead to infrared singularities which share many 
similarities with those generated by coupling to a gauge field 
\cite{CDG}. When the long-range Coulomb force ensures macroscopic charge 
neutrality an incommensurate charge density wave may instead occur at a 
finite wave vector $\bq_{\rm CDW}$. Close to this instability the effective 
interaction becomes singular at $\bq_{\rm CDW}$ giving rise to non-Fermi 
liquid contributions in the quasi-particle decay rate and in transport 
properties \cite{CDG}. 
\par
\pp All the above results are mainly based on lowest order perturbation 
theory whose validity is questionable close to an instability. A 
consistent treatment of non-Fermi liquid properties produced by critical 
fluctuations in the proximity of an instability has not yet been achieved. 
It is however a formidable project which is worth carrying out.
\par
\smallskip
g) Non-Fermi liquid behavior has been well-established in special 
single-impurity models, where a sea of conduction electrons is coupled 
to a {\em local degree of freedom}. The first instance was given by a 
two-channel Kondo model, where a localized spin is coupled to two 
degenerate channels of conduction electrons \cite{NB,COX,AD,WT,AL}. 
Since then many other single impurity models with non-Fermi liquid
behavior have been discovered. A non-Fermi liquid phase has also been
found in the infinite-dimensional Falikov-Kimball model, a (translation
invariant) lattice model with two interacting species of electrons,
one itinerant, the other localized \cite{SKG}. In the Falikov-Kimball
model localized and itinerant electrons do not hybridize, but non-Fermi
liquid behavior has been reported also for a more general class of
infinite-dimensional two-band models, where localized and itinerant
degrees of freedom are connected by a hybridization term \cite{SK,SRK}.
\par

\bigskip

{\bf 1.3. DOMINANT FORWARD SCATTERING} \par
\medskip
\pp Forward scattering, i.e.\ scattering with a {\em small momentum 
transfer}\/ $\bq$, plays a prominent role in Fermi liquids,
as well as in several of the above-mentioned non-Fermi liquid systems:
one-dimensional Luttinger liquids, systems with singular interactions,
and fermions coupled to a gauge field. In all these systems forward
scattering governs the leading low-energy long-wavelength behavior of 
response functions, and in the non-Fermi liquid phases forward
scattering drives the instability of Landau's quasi-particles.
The breakdown of Fermi liquid theory in these systems is associated 
with divergences in perturbation theory, which require a suitable
resummation of all orders in the coupling constant. Sometimes the
divergencies can be treated by perturbative renormalization group
methods, but in general a non-perturbative construction is necessary.
\par
\pp Fortunately, in forward-scattering-dominated systems there is a
small parameter other than the coupling strength:
the dimensionless ratio $|\bq|/k_F$, where $\bq$ is a typical 
momentum transfer in a scattering process, and $k_F$ is the Fermi 
momentum (or, in anisotropic systems, an effective radius in momentum
space measuring the local curvature of the Fermi surface). 
The smallness of this parameter can be exploited in several
ways. To make things precise, it is useful to introduce a cutoff
$q_c \ll k_F$, and to allow only for scattering processes with a momentum
transfer $|\bq| < q_c$. This allows one to consider a formal 
expansion with respect to the small parameter $q_c/k_F$. 
Apart from certain finite renormalizations of parameters, the 
asymptotic low-energy behavior of forward-scattering-dominated systems 
turns out to be actually independent of $q_c$, showing that scattering 
processes with $|\bq| > q_c$ are ultimately irrelevant in the low-energy 
limit, however small $q_c$ may be.
In some cases even subleading corrections to the leading low-energy
behavior are completely determined by forward scattering, but in
general contributions with any generic $\bq$ contribute to these 
corrections.
For example, the decay rate of quasi-particles in systems with 
short-range interactions is dominated by forward scattering in dimensions
$d < 2$ \cite{CDM94}, while processes with any {\bf q} contribute to
the asymptotic decay rate in $d \geq 2$.
\par 
\pp To leading order in $q_c/k_F$, the {\em velocity}\/ of the particles
is {\em conserved}\/ in each scattering process. This leads to dramatic 
simplifications in diagrammatic perturbation theory, especially: 
\par
i) {\em "Loop cancellation"}: Fermionic loops containing more than
two insertions (density, current or interaction type) cancel each 
other.
\par
ii) {\em "Density-current relation"}: The irreducible current vertex
${\bf \Lam}(p;q)$ is related to the irreducible density vertex
$\Lam^0(p;q)$ via
$$ {\bf \Lam}(p;q) \> \sim \> \bv_{\bp} \> \Lam^0(p;q) $$
for small momenta $\bq$. Here $\bv_{\bp}$ is the velocity of a fermion 
with momentum $\bp$, and $p,q$ are energy-momentum variables, e.g.
$p = (p_0,\bp)$.
\par
\pp The loop-cancellation has been noticed long ago in the analysis of
the one-dimensional Luttinger model, where it is exact for arbitrary
$q_c$ \cite{DL}. In higher dimensions this cancellation has been more 
or less implicit in many works on forward-scattering-dominated problems. 
An explicit proof has been presented recently by Kopietz et al.\ 
\cite{KHS}; 
an alternative derivation obtained independently by us will be given
in Sec.\ 5. The loop-cancellation directly implies that self-energy
and vertex corrections cancel each other in polarization bubbles
with small momenta, i.e.\ the random phase approximation (summing
bare bubble-chains) yields the exact low-energy long-wavelength
response functions to leading order in $q_c/k_F$, even in cases
where the quasi-particle pole in the single-particle propagator is 
destroyed by small-$\bq$ scattering. Similarly, bare bubbles is all
one needs to construct effective interactions and dressed gauge-field 
propagators.
\par
\pp The density-current relation has first appeared (implicitly) in the 
work by Dzya\-loshinski and Larkin \cite{DL} on one-dimensional systems.
Its validity as an asymptotic relation in higher dimensions has been 
first pointed out by the present authors \cite{CDM94}, and also, in the 
context of gauge theories, by Ioffe et al. \cite{ILA}.
Combining the density-current relation with a Ward identity reflecting
global charge and spin conservation, one obtains a simple relation for 
the density and current vertex, expressed in terms of the single
particle propagator $G$. This latter relation can be viewed as an
{\em asymptotic Ward identity}\/ associated with conservation of charge 
and spin separately on each point of the Fermi surface. 
In the Luttinger model, where the Fermi "surface" is a discrete set 
consisting of two Fermi points, this identity is exact, while in higher 
dimensions it holds only asymptotically for $q_c \ll k_F$. 
Inserting the asymptotic Ward identity into a Dyson equation for the
self-energy, one obtains a complete system of equations for the single 
particle propagator $G$, valid to all orders in the coupling constant, 
to leading order in $q_c/k_F$ \cite{CDM94}.
\par
\pp Asymptotic Ward identities have been applied to several problems
where forward scattering is dominant: low-dimensional systems with
short-range interactions \cite{CDM94} or long-range interactions 
\cite{CD94}, and fermions coupled to gauge fields \cite{ILA}.
\par
\pp An alternative way of treating forward-scattering dominated problems
is {\em bosonization}\/, where fermionic creation and annihilation
operators are represented in terms of (bosonic) density fluctuation 
operators. This technique has been introduced by Mattis and Lieb 
\cite{ML} and by Luther and Peschel \cite{LP} to the analysis of 
one-dimensional systems, and has subsequently proved to be very useful 
\cite{SOL,HAL81,VOI95,SCH95}.
An early generalization of bosonization techniques to higher (than one)
dimensions by Luther \cite{LUT79} has met only with limited success, and
remained essentially unnoticed. Recently, however, Haldane \cite{HAL92} 
pioneered a different extension to $d>1$, which turned out to be
more successful. Haldane's scheme of d-dimensional bosonization has
been elaborated in detail by Houghton and Marston \cite{HM}, and 
others \cite{HKM,CF,KHVbos}. An alternative formulation of Haldane's 
bosonization, using functional integrals instead of operators and
Hamiltionians, has been proposed by Kopietz and Sch\"onhammer 
\cite{KS}, and further elaborated by Kopietz et al.\ \cite{KHS}.
This latter version allows one to derive at least formal
expressions for the corrections to the "non-interacting boson 
approximation", which yields only the leading terms in the small-$\bq$ 
limit correctly. The explicit evaluation
of corrections turned out to be quite difficult, however, and has so
far been achieved only to finite order in the coupling constant for 
a relatively simple quantity, i.e.\ the long-wavelength density-density 
correlation function \cite{KHS}.
\par
\pp To leading order in $q_c/k_F$, the results obtained from the
bosonization technique are identical to those from the asymptotic
Ward identity approach. Hence, both techniques can be regarded as 
different versions of an expansion in $q_c/k_F$.
In both techniques the calculation of corrections to the leading
order in $q_c/k_F$ is possible in principle but very difficult in 
practice.
\par
\pp A third approach that has been used in the analysis of singular
interactions \cite{STA92} and the gauge problem \cite{KSgau} is the
so-called {\em eikonal approximation}\/. This approximation can also
be motivated via an expansion in $q_c/k_F$, and yields, for small
$q_c$, essentially the same results as the other two methods 
\cite{KHVbos}.
\par

\bigskip

{\bf 1.4. CONTENTS OF THIS ARTICLE} \par
\medskip
\pp The main purpose of this review is to describe the structure of
conventional Fermi liquids and forward-scattering-dominated non-Fermi
liquids in one framework, making thus common aspects of at first
sight quite different systems obvious. 
Such a framework is provided by the combined power of the 
renormalization group, conservation laws and asymptotic properties
of forward scattering processes. The latter become crucial in
non-Fermi liquids and also in Fermi liquids with large subleading
corrections due to residual forward scattering.
\par
\pp The renormalization group, which provides a link between microscopic
systems and effective low-energy theories, will be introduced in 
Sec.\ 2.
The role of exact conservation laws, such as global charge and spin 
conservation is discussed in Sec.\ 3.
Special properties of and techniques for forward scattering are the
subject of Sec.\ 5. We derive the important loop-cancellation 
and present the asymptotic Ward identity approach in considerable 
detail, because no such presentation has so far been published. 
Bosonization techniques will also be reviewed, and the reader will be 
referred to the vast literature on the subject.
\par
\pp The remaining sections are devoted to a discussion of several 
distinct systems and universality classes.
In Sec.\ 4 the structure of Fermi liquids is reviewed from a 
renormalization group point of view. This section contains merely a
description of the most common universality class, not a detailed 
investigation of its stability.
In Sec.\ 6 we apply the general concepts from Secs.\ 2,3 and 5 to 
one-dimensional systems, obtaining thus in particular the well-known
low-energy structure of Luttinger liquids. 
Sec.\ 7 contains a quantitative analysis of 
residual forward scattering in d-dimensional systems with short-range
interactions, where d is continued to non-integer values. Such a
continuation is very instructive, since it connects otherwise isolated 
results. The critical dimension for the stability of quasi-particles
with respect to residual forward scattering and the crossover from
Luttinger liquid behavior in one dimension to Fermi liquid behavior
in higher-dimensional systems are determined.
Sec.\ 8 addresses similar issues as Sec.\ 7, but now for long-range
density-density interactions.
In Sec.\ 9 we discuss the low-energy physics of Fermi systems coupled 
to abelian gauge-fields.
\par
\pp As a byproduct of our analysis of systems with arbitrary 
dimensionality we have obtained several new explicit analytic results for 
various quantities such as particle-hole and particle-particle 
bubbles in d dimensions, which, except for results for the particle-hole
bubble in systems with a quadratic dispersion relation \cite{UR,BW}, can 
so far be found in the literature only separately in $d = 1,2,3$.
For a linearized dispersion relation this material, which may be useful
also in other contexts (e.g. $\epsilon$-expansions around a critical
dimension), is contained in Sec.\ 7, while results for a quadratic 
dispersion are listed in the Appendix.
\par
\pp We emphasize that this is {\em not}\/ a comprehensive review of 
{\em all}\/ known non-Fermi liquids, but only of those where the 
break-down of Fermi liquid theory is driven by {\em forward scattering}\/. 
In particular, models where non-Fermi liquid behavior is due to local 
degrees of freedom, special shapes of the Fermi surface, special 
band-structure, or strong scattering with large momentum transfers have 
been addressed only briefly in Sec.\ 1.2, but will not be discussed in 
the bulk of the paper. 
Note also that we deal only with {\em pure}\/ systems, i.e.\ disorder 
effects are not treated.
\par

\vfill\eject

\def\bk{{\bf k}}
\def\bQ{{\bf Q}}
\def\bq{{\bf q}}
\def\bP{{\bf P}}
\def\bp{{\bf p}}
\def\b0{{\bf 0}}
\def\bi{{\bf i}}
\def\bj{{\bf j}}
\def\bv{{\bf v}}
\def\eps{\epsilon}
\def\up{\uparrow}
\def\down{\downarrow}
\def\bra{\langle}
\def\ket{\rangle}
\def\FS{\partial{\cal F}}
\def\Re{{\rm Re}}
\def\Im{{\rm Im}}
\def\xik{\xi_{\bk}}
\def\cO{{\cal O}}
\def\cD{{\cal D}}
\def\cF{{\cal F}}
\def\cG{{\cal G}}
\def\cZ{{\cal Z}}
\def\Gam{\Gamma}
\def\Lam{\Lambda}
\def\dbm{\delta\bar\mu}
\def\scr{\scriptstyle}
\def\sg{\sigma}
\def\Sg{\Sigma}

\vspace*{1cm}
\centerline{\large 2. RENORMALIZATION GROUP}
\vskip 1cm
\pp To one-dimensional Fermi systems, renormalization group (RG) methods
have already been applied twenty years ago, driven by the necessity
to deal with a variety of competing infrared divergences obtained in 
weak coupling perturbation theory \cite{SOL}. Problems with singularities
in many different channels cannot be solved by common resummations
 of subsets of diagrams; rather a systematic loop expansion in a 
renormalization group framework must be done. For today's standards
this is quite simple in one-dimensional systems, where the Fermi
surface consists only of two discrete points, leading to a very
limited finite number of non-irrelevant coupling constants.
\par
\pp In higher dimensions, a general renormalization group for
Fermi systems has been formulated only much later. 
In $d > 1$ there is an infinite number of non-irrelevant couplings and
the scale $k_F$ does not simply drop out of the low-energy theory as
in one dimension. Nevertheless the reason for the delayed development
of a systematic RG approach for interacting Fermi systems in higher
dimensions was certainly not that it was too difficult to do; 
actually it might have been tackled immediately after Wilson's \cite{WIL} 
works. The reason was rather that nobody seemed to find it useful,
since no generic failure of perturbation theory was encountered. 
The few infrared singularities appearing in two- and three-dimensional
Fermi systems, especially in the Cooper channel, could usually be
handled by simpler means. A notable exception: disordered systems,
non-interacting or interacting, required special RG treatments which,
driven by necessity, have indeed been successfully developed \cite{DIS}.
In addition, {\em qualitative}\/ renormalization group ideas have been
underlying many phenomenological approaches, especially Fermi liquid
theory. Much of RG language has tacitly crept into the theory of
interacting Fermi systems \cite{SR}. 
Nevertheless, a quantitative and general formulation has been lacking 
up to recently. Motivated by the issue
of non-Fermi liquid behavior, and the related discussion on the
validity of perturbation theory, such a formulation has now been
worked out almost simultaneously by several groups, all of them
following Wilson's idea of integrating out fast modes.
\par
\pp Feldman and Trubowitz \cite{FT}, and independently Benfatto and
Gallavotti \cite{BG}, have provided a rigorous version of Wilson's 
RG, aiming at a non-perturba\-tive control of interacting 
Fermi systems at least within
a certain finite radius of convergence in coupling space. Rigorous
results have indeed been obtained in one-dimensional \cite{BG,BG1d,BGb} 
and two-dimensional \cite{FMRT,FKLT} systems. According to these studies,
there is a finite radius of convergence in coupling space
(at least for short-range interactions) inside 
which no hitherto unknown instabilities and/or
non-perturbative effects can occur.
Higher dimensions than two, where surprises are even less expected, 
seem to be technically more difficult, such that no significant 
rigorous results have so far been obtained.
An instructive review of the above developments has been given by Chen, 
Fr\"ohlich and Seifert \cite{CFS}.
\par
\pp A more intuitive formulation of Wilson's RG for interacting 
Fermi systems has been presented by Shankar \cite{SHA91,SHA94}. 
Basic renormalization group ideas, many technical details, and 
relations between RG and Fermi liquid theory are explained at length in 
Shankar's recent review article \cite{SHA94}. 
\par
\pp A brief heuristic description of the renormalization group for 
Fermi systems from an elementary particle physicists viewpoint has been 
provided by Polchinski \cite{POL}. Fermi liquid theory is viewed as
a natural effective (low-energy) field theory of a normal interacting
Fermi system. 
\par
\smallskip
\pp In this section we will present a concise introduction to the
renormalization group for Fermi systems, focussing mainly on those 
aspects that are important for the rest of the paper.
We shall use the RG as a link between microscopic systems and effective
low-energy theories, which are specified by renormalized effective
actions with running (i.e.\ cutoff-dependent) coupling functions.
The effective action is obtained (at least in principle) from the 
microscopic theory by integrating out high-energy states outside a
thin shell of width $\Lam$ around the Fermi surface.
In many cases of interest one reaches a {\em fixed point} in the
low-energy limit, i.e.\ an effective action $\bar S^*$ which is 
asymptotically invariant under further reduction of the cutoff $\Lam$. 
This fixed point action can often be 
solved exactly, i.e.\ the leading low-energy behavior can be expressed 
exactly in terms of fixed point couplings. Fermi liquids and 
one-dimensional Luttinger liquids provide two examples for this
favorable situation. In the former case the fixed point action yields
the quasi-particle velocity and the quasi-particle interactions. 
\par

\bigskip

{\bf 2.1. FUNCTIONAL INTEGRAL REPRESENTATION} \par
\medskip
\pp We consider a d-dimensional interacting Fermi system, which may be 
either a one-band lattice system or a continuum system.
The dynamics is specified by a Hamiltonian 
$$ H = \sum_{\bk,\sg} \eps_{\bk} \> a_{\bk\sg}^{\dag} a_{\bk\sg}
     + {\textstyle{1 \over 2V}} \sum_{\bk,\bk',\bq} \sum_{\sg\sg'}
       g_{\bk\bk';\bq}^{\sg\sg'} \>
       a_{\bk -\bq/2,\sg }^{\dag} a_{\bk +\bq/2,\sg }
       a_{\bk'+\bq/2,\sg'}^{\dag} a_{\bk'-\bq/2,\sg'}     \eqno(2.1) $$
where $a_{\bk\sg}^{\dag}$ and $a_{\bk\sg}$ are the usual creation 
and annihilation operators for spin-$1 \over2 $ fermions with 
momentum $\bk$ and spin projection $\sg$. 
Here $\eps_{\bk}$ is a general dispersion relation, while 
$g_{\bk\bk';\bq}^{\sg\sg'}$ parametrizes a two-particle interaction, 
and $V$ is the volume of the system. 
We are interested in the {\em low-energy}\/ thermodynamics, response 
and correlation functions. 
\par
\pp The thermodynamics of the system can be derived from the partition 
function $\cZ$, which can be written as \cite{NO,POP}
$$ \cZ = \int \prod_k d\psi_k d\psi^*_k \>
   e^{S_0[\psi,\psi^*] + S_I[\psi,\psi^*]}                 \eqno(2.2) $$ 
Here $\psi_k$, $\psi_k^*$ are Grassmann variables, where the index
$k = (k_0,\bk)$ includes fermionic Matsubara energies $k_0$ and 
momenta $\bk$. 
Spin variables have been suppressed here to avoid proliferation of 
indices; they are very easy to include whenever necessary.  
The action $S = S_0 + S_I$ contains a quadratic term
$$ S_0 = \int_k \psi^*_k (ik_0 - \eps_{\bk} + \mu) \psi_k 
                                                           \eqno(2.3) $$
which includes the kinetic energy $\eps_{\bk}$ and the chemical 
potential $\mu$, and a quartic term 
$$ S_I = - {\textstyle{1 \over 2}} \int_{k,k',q} g_{\bk\bk';\bq} \>
    \psi^*_{k-q/2} \psi^*_{k'\!+q/2} \psi_{k'\!-q/2} \psi_{k+q/2} 
                                                           \eqno(2.4) $$
describing two-body interactions. 
At zero temperature, $\int_k$ is a shorthand notation for 
$(2\pi)^{-(d+1)} \int dk_0 \int d^d\bk$. For finite temperatures $T$,
the energy integrals have to be replaced by Matsubara sums, i.e.\
$(2\pi)^{-1} \int dk_0 \mapsto T \sum_{k_0}$, where the summation
runs over discrete energies $(2n\!-\!1)\pi T$ with integer $n$.
The interaction term can also be written as
$$ S_I = - {\textstyle{1 \over 4}} \int_{k,k',q} \Gam_0(k,k';q) \>
    \psi^*_{k-q/2} \psi^*_{k'\!+q/2} \psi_{k'\!-q/2} \psi_{k+q/2}
                                                           \eqno(2.5) $$
where $\Gam_0$ is the bare vertex or antisymmetrized coupling, 
which is related to $g$ by
$$ \Gam_0(k,k';q) = g_{\bk\bk';\bq} - 
   g_{(\bk+\bk'\!-\bq)/2,(\bk+\bk'\!+\bq)/2;\bk-\bk'}      \eqno(2.6) $$
A diagrammatic representation of the bare interaction is shown in 
Fig.\ 2.1.
\par
\smallskip
\pp Correlation functions can be obtained from a generating functional
\cite{NO,POP}
$$ \cG[\chi,\chi^*] = 
   \log \Big\{ \, \cZ^{-1} \int \prod_k d\psi_k d\psi^*_k \> 
   e^{S_0[\psi,\psi^*] + S_I[\psi,\psi^*] 
      + \int_k (\psi_k^*\chi_k + \chi_k^*\psi_k)} 
   \Big\}                                                  \eqno(2.7) $$
where a source term with Grassmann variables $\chi_k$, $\chi_k^*$ as
source fields has been added to the action.
Functional derivatives with respect to the source fields generate 
connected Euclidean n-particle Green functions:
$$ \tilde G_n(k_1,..,k_n;k_1',..,k_n') = 
   (-)^n \bra \psi_{k_1} \dots \psi_{k_n} 
   \psi^*_{k_n'} \dots \psi^*_{k_1'} \ket_c =    \hskip 3cm           $$
\vskip -4mm
$$ \left. 
   {\delta^n \over \delta \chi_{k_1}^* \dots \delta \chi_{k_n}^*} 
   {\delta^n \over \delta \chi_{k_n'} \dots \delta \chi_{k_1'}}
   \cG[\chi,\chi^*] 
   \right|_{(\chi,\chi^*) = 0}                             \eqno(2.8) $$ 
where $\bra \dots \ket_c$ is the connected part of the average
$$ \bra \dots \ket = \cZ^{-1} \int \prod_k d\psi_k d\psi^*_k \>
   e^{S_0[\psi,\psi^*] + S_I[\psi,\psi^*]} \dots          \eqno(2.9) $$
Energy and momentum conservation implies that $\tilde G_n$ can be 
written as
$$ \tilde G_n(k_1,..,k_n;k_1',..,k_n') =
   (2\pi)^{d+1} \delta(k_1'\!+...+\!k_n'\!-\!k_1\!-...-\!k_n) \>
   G_n(k_1,..,k_n;k_1',..,k_n')                           \eqno(2.10) $$
and in particular
$$ \tilde G_1(k;k') = 
   (2\pi)^{d+1} \delta(k'\!-\!k) \> G(k)                  \eqno(2.11) $$
\pp Useful shorthand notations are
$ \cD[\psi,\psi^*] := \prod_k d\psi_k d\psi_k^*$
for the integration measure and
$ [\psi^*\chi + \chi^*\psi] :=
   \int_k (\psi_k^*\chi_k + \chi_k^*\psi_k)$
for the source term.
Thus, in compact notation, the partition function reads
$$ \cZ = \int \cD[\psi,\psi^*] \>
   e^{S_0[\psi,\psi^*] + S_I[\psi,\psi^*]}                \eqno(2.12) $$
and the generating functional becomes
$$ \cG[\chi,\chi^*] = 
   \log \Big\{ \> \cZ^{-1} \int \cD[\psi,\psi^*] \> 
   e^{S_0[\psi,\psi^*] + S_I[\psi,\psi^*] + 
     [\psi^*\chi + \chi^*\psi]}                         
   \Big\}                                                 \eqno(2.13) $$
To deal with correlation functions for {\em composite}\/ operators it 
is often convenient to introduce suitable additional source terms. For 
example, correlation functions involving density fluctuation operators
can be generated by adding a term
$$ [\phi\rho] := \int_q \phi_q \rho_q \quad {\rm where} \quad
   \rho_q = \int_k \psi_{k-q/2}^* \psi_{k+q/2}            \eqno(2.14) $$
to the action, and taking functional derivatives with respect to the
source field $\phi_q$.
\par
\pp Concerning the momentum integrals we note that in condensed matter
physics there is always a natural cutoff for large momenta:
For lattice systems, such as electrons in a 
metal, momentum space is compact anyway (Brillouin zone). Continuum 
systems such as $^3$He have a physical cutoff given roughly by the 
inverse atomic length scale. 
\par

\bigskip

{\bf 2.2. MODE ELIMINATION AND EFFECTIVE ACTIONS} \par
\medskip
\pp The non-interacting single-particle propagator 
$$ G_0(k) = {1 \over ik_0 - \eps_{\bk} + \mu}             \eqno(2.15) $$
is singular for $k_0 \to 0$ and $\eps_{\bk} \to \mu$. 
The {\em Fermi surface}\/ of the non-interacting system
$$ \FS = \{\bk: \eps_{\bk} = \mu \}                       \eqno(2.16) $$
separates the Fermi sea $\cF = \{\bk: \eps_{\bk} < \mu\}$ from its
complement $\bar\cF$ in momentum space.
\par
\pp We now integrate out "{\em fast modes}\/" $\psi_k$, $\psi_k^*$ 
with momenta $\bk$ far from the Fermi surface $\FS$, such that only 
"{\em slow modes}\/", with momenta whose distance $d(\bk,\FS)$ from 
the Fermi surface is smaller than a certain cutoff $\Lam$, remain to 
be integrated (see Fig.\ 2.2). 
This yields 
$$ \cZ = \int \cD^{<\Lam}[\psi,\psi^*] \>
   e^{S^{\Lam}[\psi,\psi^*]}                              \eqno(2.17) $$
where the {\em effective action} $S^{\Lam}$ depends on slow modes only,
and is given by
$$ e^{S^{\Lam}[\psi,\psi^*]} := \int \cD^{>\Lam}[\psi,\psi^*] \>
   e^{S[\psi,\psi^*]}                                     \eqno(2.18) $$
Here $\cD^{>\Lam}[\psi,\psi^*]$ and $\cD^{<\Lam}[\psi,\psi^*]$
denote integration of fast and slow modes, respectively. 
Restricting the source variables $\chi_k$, $\chi^*_k$ to momenta
$\bk$ with $d(\bk,\FS) < \Lam$, the generating functional $\cG$ can
be expressed in terms of the effective action as
$$ \cG[\chi,\chi^*] = 
   \log \left\{ \> \cZ^{-1} \int \cD^{<\Lam}[\psi,\psi^*] \>
   e^{S^{\Lam}[\psi,\psi^*] + 
   \int_k^{<\Lam}(\psi^*_k\chi_k + \chi^*_k\psi_k)}     
   \right\}                                               \eqno(2.19) $$
Green functions with momenta inside the $\Lam$-shell defined by
$d(\bk,\FS) < \Lam$ can be generated from this restricted functional
by functional derivatives as in (2.8).
\par
\pp We have not yet specified our "distance" $d(\bk,\FS)$.
In fact, there are several possibilities. For isotropic systems, one
may simply choose the Euclidean distance in momentum space,
$d(\bk,\FS) = ||\bk|-k_F|$, where $k_F$ is the Fermi momentum, i.e.\
the radius of the Fermi sphere. For anisotropic systems,
especially in cases where the Fermi surface is close to the Brillouin
zone boundary, it is better to define a distance as $d(\bk,\FS) =
|\eps_{\bk}-\mu|$; in this case $\Lam$ has the dimension of energy
instead of momentum. In the mathematical literature distances
$d(k,\FS) = [k_0^2 + (\eps_{\bk}-\mu)^2]^{1/2}$ in $(k_0,\bk)$-space
are used \cite{FMRT}.
A choice of a distance that depends only on $\bk$ (not on $k_0$) allows
for an interpretation of the effective action as an action for a system 
with a restricted set of single-particle states.
\par
 \pp The effective action contains quadratic, quartic and higher order
monomials in the fields which remain to be integrated:
$$ S^{\Lam}[\psi,\psi^*] = 
   \int_k^{<\Lam} [G^{\Lam}(k)]^{-1} \psi_k^* \psi_k -  
   {\textstyle{1 \over 4}} \int_{k,k',q}^{<\Lam} \Gam^{\Lam}(k,k';q) 
   \> \psi^*_{k-q/2} \psi^*_{k'\!+q/2} \psi_{k'\!-q/2} \psi_{k+q/2} $$
\vskip -6mm
$$ + \cO[(\psi^*\psi)^3]  \hskip 8cm                     \eqno(2.20) $$
where all {\em fermionic} momenta (not the momentum transfer $\bq$) 
are now restricted by the cutoff $\Lam$.
All the terms in $S^{\Lam}$ can be expanded in powers of the bare 
interactions, and may be represented by Feynman diagrams for 
connected n-particle Green functions with external lines amputated.
Internal lines are integrated only over momenta outside the 
$\Lam$-shell.
In particular, the quadratic term in $S^{\Lam}$ is given by
$[G^{\Lam}(k)]^{-1} = G_0^{-1}(k) - \Sg^{\Lam}(k)$, where 
$G_0^{-1}(k) = ik_0 - \eps_{\bk} + \mu$ and $\Sg^{\Lam}(k)$ is
the self-energy with internal lines restricted to momenta far from
the Fermi surface. Note that in the quadratic part no one-particle
reducible diagrams contribute, because $\bk$ in (2.20) must be inside 
the $\Lam$-shell, while momenta on internal lines must be outside.
The quartic term is given by the two-particle vertex function
$\Gam^{\Lam}(k,k';q)$,
i.e.\ the two-particle Green function with external propagators 
amputated (no one-particle reducible terms exist in this case).
Higher order terms are given by amputated n-particle Green functions.
Note that for odd $n > 1$ one-particle reducible diagrams contribute.
For more details on the perturbation expansion and its diagrammatic 
representation, see Refs. \cite{FT,BG,CFS,SHA94}.
\par
\pp In general the Fermi surface is shifted by interactions (with a
change of shape, if the system is anisotropic). To obtain
a well behaved perturbation series, one must not expand around $S_0$,
but rather around another suitably chosen quadratic action with a
Fermi surface already in its interacting position. This is achieved
by adding a counterterm of the form $\int_k \delta\mu_{\bk} \psi_k^*
\psi_k$ to $S_0$, and subtracting it from $S_I$, i.e.\ one defines
$$ S'_0 = S_0 + \int_k \delta\mu_{\bk} \> \psi_k^* \psi_k  
   \> , \quad
   S'_I = S_I - \int_k \delta\mu_{\bk} \> \psi_k^* \psi_k  \eqno(2.21) $$
The perturbation expansion is then carried out around $S'_0$.
The actual value of $\delta\mu_{\bk}$ is of course not known a 
priori, but can be determined (order by order) by a self-consistency
condition.\footnote{To show rigorously that this procedure really 
works, to any order in perturbation theory, is a subtle problem that 
has been recently solved by Feldman et al.\ \cite{FST}.}
A $\bk$-dependent quadratic counterterm can also be used to adjust
the Fermi velocity (s.\ below).
\par
\pp In this work we focus on normal state properties, and therefore 
consider energy scales well above scales dominated by possible symmetry
breaking.
We note, however, that symmetry breaking can be treated by adding
suitable quadratic counterterms. To deal with superconductivity, for
example, anomalous quadratic terms $\psi\psi$ and $\psi^*\psi^*$ must 
be included in $S'_0$ (and accordingly subtracted in $S'_I$) \cite{FMRT}.
\par
\pp The mode elimination can be iterated, by calculating $S^{\Lam'}$ 
with $\Lam' < \Lam$ in terms of $S^{\Lam}$.
At a generic scale $\Lam$, the effective action $S^{\Lam}$ can be 
decomposed in a quadratic part $S_0^{\Lam}$ and a rest 
$S_I^{\Lam}$ (containing interactions and quadratic counterterms)
such that $S_0^{\Lam}$ has the Fermi surface of the interacting 
system.
\par
\pp In the following we will frequently consider {\em isotropic}\/ 
systems for pedagogical reasons.
In this case the Fermi surface is spherically symmetric irrespective of
interactions and the counterterm in (2.21) reduces to a constant shift
$\delta\mu$ of the chemical potential.
It is not hard to extend the formalism to anisotropic situations, while
concrete calculations become of course more tedious.
\par
\smallskip
\pp Expanding the kernel of the (purely) quadratic part $S_0^{\Lam}$
of the action around the Fermi surface, one gets 
$$ S_0^{\Lam} = \int_k^{<\Lam} (Z_{\bk_F}^{\Lam})^{-1} \psi_k^* 
   [ik_0 - \bar v_{\bk_F}^{\Lam} k_r - \bar\Sg^{\Lam}(k)] \psi_k     
                                                         \eqno(2.22) $$
Here momenta $\bk$ are represented by the pair $(k_r,\bk_F)$ where 
$\bk_F$ is the projection of $\bk$ on the Fermi surface, while $k_r$
is its oriented distance (defined positive for $\bk$ outside the Fermi
surface, and negative inside), i.e.
$$ k_r := |\bk| - k_F                                    \eqno(2.23) $$
for isotropic systems.
The {\em field renormalization factor}
$$ Z_{\bk_F}^{\Lam} = \left. 
   \left[ 1 - \partial\Sg^{\Lam}(k_0,\bk)/\partial(ik_0) 
   \right]^{-1} \right|_{(0,\bk_F)}                      \eqno(2.24) $$
is a positive constant $<1$.
The effective {\em Fermi velocity}\/ $\bar\bv_{\bk_F}^{\Lam}$ is  
related to the self-energy by
$$ \bar\bv_{\bk_F}^{\Lam} = \bar\bv_{\bk_F}^{c\Lam} + 
   \delta\bar\bv_{\bk_F}^{\Lam}                          \eqno(2.25) $$
where
$$ \bar\bv_{\bk_F}^{c\Lam} = Z_{\bk_F}^{\Lam} 
   \left. \big[ 
   \bv_{\bk} + \partial\Sg^{\Lam}(k_0,\bk)/\partial\bk 
   \big] \right|_{(0,\bk_F)}                             \eqno(2.26) $$
and $\delta\bar\bv_{\bk_F}^{\Lam} = 
Z_{\bk_F}^{\Lam} \> \delta\bv_{\bk_F}^{\Lam}$ is a (renormalized)
counterterm.
Here $\bv_{\bk} = \nabla\eps_{\bk}$ is the bare velocity of particles
with momentum $\bk$.
Note that the limit $\Lam \to 0$ does not commute (in general) with 
the $\bk$-derivative of $\Sg^{\Lam}$ on the Fermi surface, because
particle-hole excitations with infinitesimal excitation energy can
yield a finite contribution to $\partial\Sg/\partial\bk|_{(0,\bk_F)}$.
The velocity $\bar\bv_{\bk_F}^{c\Lam}$ determines the renormalized
{\em current}\/ operator (s.\ Sec.\ 3.3). 
A counterterm $- \int_k \delta v_{\bk_F}^{\Lam} k_r \> \psi_k^* \psi_k$
has been added to $S_0$ to make $\bar\bv_{\bk_F}^{\Lam}$ converge to
the true Fermi velocity in the limit $\Lam \to 0$.
The residual self-energy $\bar\Sg^{\Lam}(k)$ vanishes faster than 
linearly for $k_0 \to 0$, $\bk \to \FS$. 
For isotropic systems, $Z_{\bk_F}^{\Lam} = Z^{\Lam}$ and 
$\bar v_{\bk_F}^{\Lam} = \bar v_F^{\Lam}$ are constant all over the 
Fermi surface.
\par
\smallskip
\pp The factor $(Z^{\Lam})^{-1}$ in the quadratic part of the action 
can be eliminated by introducing {\em renormalized fields}
$$ (\bar\psi_k,\bar\psi_k^*) = 
   (Z^{\Lam})^{-1/2} (\psi_k,\psi_k^*)                   \eqno(2.27) $$
as the new (functional) integration variables.\footnote
{The corresponding Jacobian is a constant and cancels in
$\cG[\chi,\chi^*]$.}
Of course $Z^{\Lam}$ now shows up in interaction and source terms, 
but can be absorbed in renormalized source fields
$$ (\bar\chi_k,\bar\chi_k^*) = 
   (Z^{\Lam})^{1/2} (\chi_k,\chi_k^*)                    \eqno(2.28) $$
in the {\em renormalized vertex functions}
\vskip -3mm
$$ \bar\Gam^{\Lam}(\{k_i\};\{k_i'\}) = 
   (Z^{\Lam})^n \Gam^{\Lam}(\{k_i\};\{k_i'\})            \eqno(2.29) $$
for n-particle interactions, and in a renormalized chemical potential
shift $\delta\bar\mu^{\Lam} = Z^{\Lam} \delta\mu^{\Lam}$.
Thus the {\em renormalized} effective action becomes
$$ \bar S^{\Lam}[\bar\psi,\bar\psi^*] = 
   \int_k^{<\Lam} \bar\psi_k^* 
   [ik_0 - \bar v_F^{\Lam} k_r - \bar\Sg^{\Lam}(k)] \bar\psi_k 
   - \int_k^{<\Lam} (\delta\bar\mu^{\Lam} - 
     \delta\bar v_F^{\Lam} k_r) 
   \> \bar\psi_k^* \bar\psi_k  \hskip 2cm                            $$
\vskip -4mm 
$$ - {\textstyle {1 \over 4}} 
   \int_{k,k',q}^{<\Lam} \bar\Gam^{\Lam}(k,k';q) 
   \bar\psi^*_{k-q/2} \bar\psi^*_{k'\!+q/2} 
   \bar\psi_{k'\!-q/2} \bar\psi_{k+q/2} + 
   \cO((\bar\psi^*\bar\psi)^3)                           \eqno(2.30) $$
In general the vertex function $\bar\Gam^{\Lam}(k,k';q)$ depends on
three independent $(d+1)$ dimensional energy-momentum variables.
However, many details of these dependences are actually irrelevant
for the low-energy physics, which depends only on the behavior of
$\bar\Gam^{\Lam}(k,k';q)$ in certain limits. Hence the vertex
function in the effective action can be replaced by renormalized
{\em coupling functions}\/ $\bar g^{\Lam}$ with less variables.
For example, in most cases the coupling function
$$ \bar g^{\Lam}_{\bk\bk'}(q) = 
   \lim_{k \to (0,\bk)} \lim_{k' \to (0,\bk')}
   \bar\Gam^{\Lam}(k,k';q)                               \eqno(2.31) $$
contains enough information. Neglecting all irrelevant terms one
thus finds
$$ \bar S^{\Lam}[\bar\psi,\bar\psi^*] = 
   \int_k^{<\Lam} \bar\psi_k^* 
   [ik_0 - \bar v_F^{\Lam} k_r] \bar\psi_k 
   - \int_k^{<\Lam} 
    (\delta\bar\mu^{\Lam} - \delta\bar v_F^{\Lam} k_r) 
   \> \bar\psi_k^* \bar\psi_k  \hskip 4cm                            $$
\vskip -4mm 
$$ - {\textstyle {1 \over 4}} 
   \int_{k,k',q}^{<\Lam} \bar g^{\Lam}_{\bk\bk'}(q) \>
   \bar\psi^*_{k-q/2} \bar\psi^*_{k'\!+q/2} 
   \bar\psi_{k'\!-q/2} \bar\psi_{k+q/2}                  \eqno(2.32) $$
In many specific cases (Fermi liquids etc.) the number of relevant
variables in $\bar g^{\Lam}$ can be further reduced, especially by
replacing (regular) coupling functions by their asymptotic values on 
the Fermi surface.
\par
\pp The renormalized source term reads 
$\int_k^{<\Lam} (\bar\psi_k^*\bar\chi_k + \bar\chi_k^*\bar\psi_k)$.
The exact (unrenormalized) correlation functions can be obtained as
functional derivatives of the renormalized generating functional 
$$ \bar\cG[\bar\chi,\bar\chi^*] := 
   \cG[\bar\chi/(Z^{\Lam})^{1/2},\bar\chi^*/(Z^{\Lam})^{1/2}] 
                                                         \eqno(2.33) $$ 
with respect to the renormalized source fields $\bar\chi$ and 
$\bar\chi^*$, followed by a multiplication with the respective 
Z-factors:
$$ G_n(k_1,...,k_n;k_1',...,k_n') = (Z^{\Lam})^n \>
   \bar G_n^{\Lam}(k_1,...,k_n;k_1',...,k_n')            \eqno(2.34) $$
where $\bar G_n^{\Lam}$ are the {\em renormalized}\/ Green functions 
given by 
$$ (2\pi)^{d+1} \delta(k_1'\!+...+\!k_n'\!-\!k_1\!-...-\!k_n) \>
   \bar G_n^{\Lam}(k_1,...,k_n;k_1',...,k_n') =                      $$
\vskip -5mm
$$ \left.
   {\delta^n \over \delta \bar\chi_{k_1}^* \dots \delta \bar\chi_{k_n}^*} 
   {\delta^n \over \delta \bar\chi_{k_n'} \dots \delta  \bar\chi_{k_1'}}
   \bar\cG[\chi,\chi^*] \right|_{(\bar\chi,\bar\chi^*) = 0} 
                                                         \eqno(2.35) $$ 
Note that renormalized correlation functions depend on the (variable)
cutoff $\Lam$, while unrenormalized correlation functions depend
only on the cutoff determined by the microscopic theory.
\par
\smallskip
\pp What have we gained so far? The renormalized effective action 
$\bar S^{\Lam}$ has usually a much more complicated form than the bare
action $S$. So why analyse low energy physics in terms of effective
actions instead of the bare one? The point is that for small $\Lam$ a new 
small {\em expansion parameter}\/ has emerged: $\Lam/k_F$. In many cases
this allows one to express the exact leading and subleading low-energy
behavior in terms of renormalized couplings even if these are not small.
Expressing thermodynamics and correlation functions in terms of the
effective parameters defining $\bar S^{\Lam}$, and expanding in powers
of small energy scales (temperature, frequency etc.), one realizes that
only few terms in $\bar S^{\Lam}$ contribute in leading order, a finite
number of additional terms to the first subleading order, and so on.
The asymptotic infrared behavior of the system can thus be expressed
in terms of relatively few parameters.
\par
\pp The effective couplings $\bar g^{\Lam}$ can be classified as
{\em relevant}, {\em marginal}, or {\em irrelevant}, depending on 
whether their importance relative to the quadratic part of the action
grows, remains invariant, or decreases in the limit $\Lam \to 0$.\footnote
{Note that this classification depends on the choice of $\bar 
S_0^{\Lam}$! Relevant couplings may sometimes be avoided by a clever
choice of the "non-interacting" part $\bar S_0^{\Lam}$.}
\par
\pp More-than-two body interactions with a finite low-energy limit are 
usually irrelevant. 
More precisely, the contributions from finite (n+1)-particle interactions 
are suppressed by a factor $\Lam$ with respect to those from n-particle 
interactions.
As an illustration, let us estimate the order of magnitude of the
three self-energy diagrams in Fig.\ 2.3 by naive power-counting.
The shaded boxes represent effective two-particle and three-particle
interactions. Internal lines correspond to propagators 
$G_0^{\Lam}(k) = [ik_0 - \bar v_F^{\Lam} k_r - \bar\Sg^{\Lam}(k)]^{-1}$.
Momenta on internal lines must lie inside the thin $\Lambda$-shell 
around the Fermi surface defined by $d(\bk,\FS) < \Lam$, corresponding 
to small excitation energies $\bar v_F^{\Lam} k_r$. 
The propagators become big (of order $(\bar v_F^{\Lam}\Lam)^{-1}$), if the 
energy variables are also small, i.e.\ $k_0 < \bar v_F^{\Lam}\Lam$.
Hence, energy-momentum variables leading to a big propagator fill
a volume proportional to $\Lam^2$ in $(k_0,\bk)$-space.
Notice that this volume is independent of the space dimension $d$
because the propagator is singular on a $(d\!-\!1)$-dimensional surface,
which reduces to a point-singularity (as in standard critical phenomena)
only for $d=1$. Thus, according 
to naive power-counting, the value of a diagram with $l$ internal 
lines and $r$ integrations (equal to the number of loops) is 
proportional to $\Lam^{2r-l}$. In particular, the
first two diagrams (containing only two-particle interactions) in 
Fig.\ 2.3 are of order $\Lam$, while the last one (containing a 
3-particle interaction) is of order $\Lam^2$. It is easy to see that
the replacement of an n-particle interaction in a diagram by an
(n+1)-particle interaction enhances the number of internal lines and
integrations by one, leading to an extra factor $\Lam$ in the 
power-counting estimate.
Actually it turns out that in dimensions $d>1$ the naive 
power-counting often over-estimates the value of a diagram, since 
geometrical phase-space restrictions reduce the actual integration 
volume, leading to additional $\Lam$-powers in most cases.
\par
\pp The above power-counting holds only for interactions that remain 
finite in the low-energy limit. However,
it is easy to see that in a Wilson renormalization scheme effective 
$n$-particle interactions of order $\Lam^{2-n}$ may be generated.
An example for a contribution to an effective three-particle interaction
with a low-energy limit of order $\Lam^{-1}$ is shown in Fig.\ 2.4;
the internal line may have momenta outside but close to the $\Lam$-shell
even if all external lines carry momenta inside the $\Lam$-shell.
Inserting the effective interaction in Fig.\ 2.4 into the third 
self-energy diagram in Fig.\ 2.3, one obtains contribution of order 
$\Lam$ within naive power-counting, i.e.\ that three-particle interaction 
seems to be as important as finite two-particle interactions.
Fortunately, in practice (especially for the purposes of the present 
article) one can usually avoid dealing with these many-particle 
interactions, for different reasons in different cases.
In particular, in one-dimensional systems the Wilson RG can be replaced 
by a field-theoretic RG in the low-energy regime, with a suitable 
effective two-body action that incorporates all higher order 
interactions. The book-keeping of generated terms is different in the 
field-theoretic RG,\footnote
{For a short discussion on this point, see Shankar \cite{SHA94}.}
making effective many-particle interactions ($n>2$) generally irrelevant.
In higher dimensions additional $\Lam$-powers coming from geometrical
phase space contraints suppress the effects of many-particle interactions 
more strongly than the above simple power-counting would predict.
A deeper understanding of the role of effective many-particle 
interactions in Wilson's RG version for Fermi systems has been reached 
in the mathematical literature \cite{FT,BG}. 
In the following we will not consider effective $n$-particle 
interactions with $n>2$ any more, assuming that they are either 
irrelevant or somehow effectively incorporated in one-particle and 
two-particle terms in the low-energy action. 
\par

\bigskip

{\bf 2.3. LOW-ENERGY COUPLING-SPACE} \par
\medskip
\pp Let us now try to become acquainted with the huge coupling space
for two-particle interactions in Fermi systems, and find a suitable 
classification of couplings. 
In general, two-particle interactions scatter energy momenta $k_1$ and 
$k_2$ into $k'_1$ and $k'_2$, where
energy-momentum conservation imposes the restriction $k_1 + k_2 =
k'_1 + k'_2$, which can be manifestly built in by parametrizing
the vertex function in terms of three variables $k$, $k'$ and $q$
such that $k_1 = k+q/2$, $k_2 = k'\!-q/2$, $k'_1 = k-q/2$ and
$k'_2 = k'\!+q/2$.
\par
\pp For small cutoffs $\Lam$, the restriction of all momenta of 
fermions to a thin shell around the Fermi surface combined with total
momentum conservation in scattering processes leads to drastic
geometric constraints on the {\em angles}\/ between the momenta 
of the two ingoing and outgoing particles, in addition to the radial
constraint directly imposed by the cutoff. To see this, let us
consider the limit $\Lam \to 0$, where all the particles must be
situated {\em on}\/ the Fermi surface, which we assume to be spherical
for simplicity.
\par
\pp Let us start with {\em two-dimensional}\/ systems. Here it is easy
to see that there are three distinct types of possible (i.e.\ momentum
conserving) scattering processes, which can be parametrized by a
single angle each (see Fig.\ 2.5):
\par
\medskip
\begin{tabular}{ll}
\pp "{\em forward\/}" (F) scattering: & 
    $\bk'_1 = \bk_1$ and $\bk'_2 = \bk_2$ \\
\pp "{\em exchange\/}" (E) scattering: &
    $\bk'_1 = \bk_2$ and $\bk'_2 = \bk_1$ \\
\pp "{\em Cooper\/}" (C) scattering: &
    $\bk_1 + \bk_2 = 0$.
\end{tabular}
\vskip -14mm
                                                      $$ \eqno(2.36) $$
\vskip 7mm
Actually forward and exchange scattering are equivalent for spinless
fermions or fermions with the same spin projection $\sg_1 = \sg_2$,
since the vertex is antisymmetric with respect to interchange of
incoming (or outgoing) particles. Forward and exchange scattering
are however distinct for $\sg_1 \neq \sg_2$.
Forward and exchange scattering can be uniquely parametrized by the 
angle between the momenta of the incoming particles 
$\theta = \angle(\bk_1,\bk_2)$,
while Cooper scattering is parametrized by the angle defined by the 
momentum transfer $\phi = \angle(\bk_1,\bk_1')$. Whereever the vertex 
function $\bar\Gam^{\Lam}$ is regular near the Fermi surface, 
it can be replaced by the three coupling functions 
$\bar g^{\Lam}_F(\theta)$, $\bar g^{\Lam}_E(\theta)$ and 
$\bar g^{\Lam}_C(\phi)$, which are given by the three Fermi 
surface limits of $\bar\Gam^{\Lam}(k,k';q)$ according to the above 
classification.
\par
\pp The above classification of scattering processes represents a 
canonical generalization of the {\em g-ology}\/ classification in 
{\em one-dimensional}\/ systems \cite{SOL} to two dimensions. 
Specializing to one dimension, where only two angles ($0$ and $\pi$) 
exist, one finds the following correspondence \cite{MD}: \par
\smallskip
\hskip 5cm  $g_F(0) = g_4$, \quad 
            $g_F(\pi) = g_2$  \par
\hskip 5cm  $g_E(0) = g_4$, \quad 
            $g_E(\pi) = g_1$  \par
\hskip 5cm  $g_C(0) = g_2$, \quad 
            $g_C(\pi) = g_1$  \par
\vskip -18mm
                                                      $$ \eqno(2.37) $$
\vskip 8mm
\par
where the numbering $g_1,..,g_4$ follows the usual g-ology convention 
(here we have suppressed bars and $\Lam$'s for easier readability). 
Note that processes which are generically distinct in $d = 2$ may become 
equivalent for the special angles $0$ and $\pi$, which are the only 
angles in $d = 1$. The g-ology coupling $g_3$ does not appear here
since it describes umklapp processes which have not been considered.
Umklapp processes exist only in lattice systems; at low energy they 
are important only in special cases (at specific fillings). 
\par
\pp In {\em three-dimensional}\/ systems, two particles whose momenta 
span an angle $\theta$ may scatter into new states with
momenta spanning the same angle $\theta$ (and their sum pointing in
the same direction of course). For $\theta \neq \pi$ this leaves a
one-dimensional degree of freedom, which can be parametrized by
the angle $\phi$ spanned by the initial and the final momentum of one
of the scattering particles. In two dimensions, $\phi$
could be either $0$ or $\theta$, corresponding to "forward" and "exchange"
scattering, respectively. In three dimensions these two extreme cases
are continuously connected in one class of scattering processes,
which we call "{\em normal}\/" scattering. For $\theta = \pi$ one
has Cooper scattering, with two degrees of freedom for the
tranferred momentum; for the isotropic case considered here the
corresponding coupling strength can however be parametrized by a single 
angle $\phi$, e.g. the one spanned by the initial and the final 
momentum of one of the particles.
Thus, in three dimensions we have two distinct classes of two-particle
interactions on the Fermi surface:
\par
\bigskip
\begin{tabular}{ll}
\pp\pp "{\em normal\/}" (N) scattering: & 
    $\angle(\bk_1,\bk_2) = \angle(\bk'_1,\bk'_2) \neq \pi$ \\
\pp\pp "{\em Cooper\/}" (C) scattering: &
    $\angle(\bk_1,\bk_2) = \angle(\bk'_1,\bk'_2) = \pi$.
\end{tabular}
\vskip -12mm
$$                                                       \eqno(2.38) $$
\vskip 5mm
\par
with coupling functions $\bar g^{\Lam}_N(\theta,\phi)$ and 
$\bar g^{\Lam}_C(\phi)$, respectively. 
The coupling $\bar g^{\Lam}_F(\theta) = \bar g^{\Lam}_N(\theta,0)$, 
describing {\em forward}\/ scattering, plays a special role in Fermi 
liquid theory.
\par
\smallskip
\pp The above analysis has led to a purely {\em kinematic}\/ 
classification of all scattering processes that are geometrically 
possible near the Fermi surface. 
To see how the corresponding effective interactions behave in the 
low-energy limit one must estimate the {\em phase-space}\/ for 
scattering processes, i.e.\ estimate Feynman diagrams. 
The qualitative behavior of effective interactions depends on the
properties of the bare interactions, such as signs and regularity
properties (bounded or singular).
\par
\pp As a specific example, let us discuss an isotropic Fermi system
with short-range interactions in three dimensions.\footnote{
Interactions which are short-ranged in real space are bounded in
momentum space.}
Our aim is to understand the behavior of effective two-particle 
couplings $\bar g^{\Lam}$ in the low-energy regime and to reveal 
possible sources of instabilities. 
As a first step one would try to calculate $\bar g^{\Lam}$
in perturbation theory by evaluating Feynman diagrams for the 
two-particle vertex, with momenta on internal propagators outside
a $\Lam$-shell around the Fermi surface. The diagrams contributing 
to second order in the coupling constant are listed in Fig.\ 2.6.
For almost all $k$, $k'$, and $q$ a direct expansion
of $\bar\Gam^{\Lam}(k,k';q)$ in powers of bare couplings has finite
coefficients even in the limit $\Lam \to 0$. In these cases one can
actually integrate all momenta down to the Fermi surface in one shot. 
However, for $k' = -k$ (the Cooper channel), already the second order 
diagram (a) in Fig.\ 2.6 leads to a logarithmically divergent 
contribution to $\bar\Gam^{\Lam}_{kk';q}$ in the limit $\Lam \to 0$. 
This is a case for the renormalization group transformation. 
\par
\pp To calculate the flow of $\bar\Gam^{\Lam}(k,k';q)$
with $k'=-k$ under infinitesimal reductions of the cutoff, 
let us parametrize interactions that scatter $p$ and $-p$ into $p'$ 
and $-p'$ by a coupling function $\bar g^{\Lam}_{pp'}$. 
To obtain simple (i.e.\ instructive) analytic results, we assume 
that $\bar g^{\Lam}_{pp'}$ depends only on the angle $\phi$ between
initial and final momenta, not on frequencies or moduli of momenta, 
i.e. $\bar g^{\Lam}_{pp'} = \bar g^{\Lam}_C(\phi_{\bp\bp'})$. 
We will see that this assumption is self-consistent in the low-energy 
limit in the sense that no other dependences will be generated by the
flow. 
Note that we are interested in the flow close to the Fermi surface,
i.e.\ $\Lam \ll k_F$, where 3-particle interactions are already
irrelevant. Hence, to second order in the coupling, we can calculate
the flow by computing the Feynman diagrams in Fig.\ 2.6 for ingoing
external energy-momenta $p$ and $-p$, with internal momenta 
restricted to the infinitesimal region between a $\Lam$-shell and a 
$\Lam'$-shell with $\Lam' = \Lam - d\Lam$. 
\par
\pp We first consider the contribution due to the diagram (a), i.e.\
the diagram that produces the logarithmic divergence in perturbation
theory. The Feynman rules yield a contribution
$$ d\bar g^{\Lam}_{pp'} = 
   \int_{\Lam'}^{\Lam} {d^{d+1}k \over (2\pi)^{d+1}} \>
   \bar g^{\Lam}_{pk} \> \bar g^{\Lam}_{kp'} \>
   {1 \over ik_0 - \xi_{\bk}} \> {1 \over -ik_0 - \xi_{-\bk}}  
                                                        \eqno(2.39) $$
where $\xi_{\bk} = \eps_{\bk} - \mu$ and the $\bk$-integration is 
restricted to the above-mentioned infinitesimal region.
The k-integration can be decomposed in integrals over $k_0$, $k_r$
and the solid angle $\Omega_{\bk}$, where $k_r$ is restricted by the
condition $\Lam' < |k_r| < \Lam$.
Since $\bar g^{\Lam}_{pk} = \bar g^{\Lam}(\phi_{\bp\bk})$ is independent
of modula and frequencies, and $\xi_{\bk} = \bar v_F^{\Lam} k_r$ in the 
low-energy limit, the integrations over $k_0$ and $k_r$ can be carried
out explicitly, yielding (at $d=3$)
$$ d\bar g^{\Lam}_C(\phi_{\bp\bp'}) = 
   {d\Lam \over \Lam} {k_F^2 \over (2\pi)^3 \bar v_F^{\Lam}}
   \int d\Omega_{\bk} \>
   \bar g^{\Lam}_C(\phi_{\bp\bk}) \> \bar g^{\Lam}_C(\phi_{\bk\bp'})
                                                          \eqno(2.40) $$
Inserting a partial wave decomposition $ \bar g^{\Lam}_C(\phi) = 
\sum_{l=0}^{\infty} \> \bar g^{\Lam}_l P_l(\cos\phi)$, and using the
addition theorem for the Legendre polynomials $P_l$, one obtains
decoupled flow equations for each angular momentum sector:
$$ {d\bar g^{\Lam}_l \over d\log\Lam} \equiv
   \beta_l(\bar v_F^{\Lam},\bar g^{\Lam}_l) =
   {1 \over 2\pi^2 (2l+1)} \>
   {k_F^2 \over \bar v_F^{\Lam}} \> (\bar g^{\Lam}_l)^2   \eqno(2.41) $$
We note that for spinless fermions the coupling function must be 
antisymmetric in the initial (and final) momenta, i.e.\ only odd
angular momenta $l$ contribute here. Reinserting spin one obtains 
separate equations for the singlet channel, with even $l$, and the 
triplet channel, with odd $l$.
\par
\pp It is easy to see that the diagrams (b) and (c) yield a  
contribution of order $(d\Lam)^2$ to the flow of the Cooper couplings,
which is negligible.
The flow equation (2.41), or its two-dimensional analogue, has been 
derived by Feldman and Trubowitz \cite{FT}, and by Shankar 
\cite{SHA91}. Of course the behavior of the two-particle vertex for 
small total momenta of the particles has been investigated long
before. A scaling equation which is equivalent to the above flow 
equation can be found already in the book by Abrikosov et al.\ 
\cite{AGD}.
\par
\pp The above flow for $\bar g^{\Lam}_l$ makes negative couplings 
grow in the low-energy limit, signalling an instability towards a
qualitatively different phase. This is the renormalization group 
version of the {\em Cooper instability}\/.
Positive couplings scale logarithmically to zero, i.e.\ they disappear 
in the low-energy limit, albeit very slowly. 
\par
\pp We emphasize that the flow equation (2.41) is valid only to 
leading order in $\Lam/k_F$ (since irrelevant terms have been neglected) 
and to second order in the interaction. 
Including subleading orders in $\Lam/k_F$ one finds
that different angular momenta couple and that non-Cooper couplings 
influence the flow. It has been shown that these corrections always
generate some negative couplings $\bar g^{\Lam}_l$ in the low-energy
limit, even if one has started with a purely repulsive interaction
\cite{FT,SHA94}.
Ultimately, in the limit $\Lam/k_F \to 0$, these couplings will be 
governed by the flow (2.41) and thus drive a Cooper instability. 
This renormalization group result substantiates an old, more intuitive 
argument by Kohn and Luttinger \cite{KL65}. 
A quantitative numerical analysis of the flow equation shows that 
in isotropic systems with purely repulsive interaction the instability
is strongest in the p-wave ($l=1$) channel \cite{Tpriv}.
\par
\smallskip
\pp Perturbation theory for $\bar\Gam^{\Lam}(k,k';q)$ also reveals 
a singularity in the limit of small momentum transfer, $q \to 0$. 
In contrast to the Cooper instability, this singularity is not a
divergence (for imaginary frequencies), but only a non-uniqueness 
of limits: the limits $q_0 \to 0$, $\bq \to 0$ and $\Lam \to 0$ do
not commute. This singularity does not signal any instability of
the system, and a standard perturbation expansion is consistent in
the sense that higher orders in the coupling are small if the 
interaction is weak.
Note, however, that this latter statement holds only for the Euclidean
theory, i.e.\ for Green functions with imaginary times or frequencies:
perturbative contributions blow up, if one tries to continue to 
{\em real}\/ frequencies. 
These divergencies are associated with collective modes in the 
interacting system (zero sound) and can be treated by a suitable
resummation of Feynman diagrams (introducing particle-hole irreducible
vertices). 
More on this will follow in the section on Fermi liquid theory.
\par

\vfill\eject

\def\bA{{\bf A}}
\def\bc{{\bf c}}
\def\bk{{\bf k}}
\def\bQ{{\bf Q}}
\def\bq{{\bf q}}
\def\bP{{\bf P}}
\def\bp{{\bf p}}
\def\b0{{\bf 0}}
\def\bi{{\bf i}}
\def\bj{{\bf j}}
\def\br{{\bf r}}
\def\bs{{\bf s}}
\def\bS{{\bf S}}
\def\bv{{\bf v}}
\def\eps{\epsilon}
\def\up{\uparrow}
\def\down{\downarrow}
\def\bra{\langle}
\def\ket{\rangle}
\def\FS{\partial{\cal F}}
\def\Re{{\rm Re}}
\def\Im{{\rm Im}}
\def\xik{\xi_{\bk}}
\def\cO{{\cal O}}
\def\cD{{\cal D}}
\def\cF{{\cal F}}
\def\cG{{\cal G}}
\def\cZ{{\cal Z}}
\def\Lam{\Lambda}
\def\dbm{\delta\bar\mu}
\def\scr{\scriptstyle}
\def\sg{\sigma}
\def\Sg{\Sigma}

\vspace*{1cm}
\centerline{\large 3. EXACT CONSERVATION LAWS}
\vskip 1cm
\pp In this section we will discuss important exact conservation
laws, namely charge and spin conservation, and derive the associated 
Ward identities. 
We emphasize that most of these identities are well-known, at least 
for the case of continuum systems (see, for example, Nozi\`eres
\cite{NOZ}).
Nevertheless we find it useful to provide a concise collection of
those identities which play a role in our article, and present
derivations from general principles that do not depend on the
assumptions of Fermi liquid theory.
An important consequence of Ward identities is the cancellation of
the field renormalization $Z^{\Lam}$ in the response of a Fermi system
to low-energy long-wavelength fields that couple to the density or
current (see 3.3).
More stringent conservation laws obeyed by forward scattering 
processes and the associated Ward identities will be treated later, 
in Secs.\ 5 and 6.
\par

\bigskip

{\bf 3.1. CHARGE CONSERVATION} \par
\medskip
\pp We start by deriving relations following from charge (or particle
number) conservation. The charge-density fluctuation operator is 
defined as
$$ \rho(\bq) = \sum_{\bk\sg} \> 
   a^{\dag}_{\bk-\bq/2,\sg} \> a_{\bk+\bq/2,\sg}          \eqno(3.1) $$
For $\bq = \b0$, $\rho(\b0) = N$ is the particle number operator. 
The (imaginary) time evolution of charge-density fluctuations is given 
by $\rho(\tau,\bq) = e^{K\tau} \rho(\bq) e^{-K\tau}$ where 
$K = H - \mu N$ is the grand-canonical Hamiltonian.
The equation of motion for $\rho(\tau,\bq)$ can be written as
$$ \partial_{\tau} \rho(\tau,\bq) =
   [H,\rho(\tau,\bq)]                                     \eqno(3.2) $$
since $[N,\rho(\bq)] = 0$.
As a consequence of charge conservation, i.e.\ $[N,H] = 0$, the right 
hand side of (3.2) vanishes for $\bq \to 0$. In all cases of interest
there is a current operator $\bj(\bq)$ such that the equation of 
motion for $\rho(\bq)$ assumes the form of a {\em continuity equation}
$$ \partial_{\tau} \rho(\tau,\bq) = 
   - \> \bq \cdot \bj(\tau,\bq)                              \eqno(3.3) $$
at least for small $\bq$. In general, the current operator depends
explicitly on interactions. 
Inserting the decomposition $H = H_0 + H_I$ of the Hamiltonian
into the commutator in (3.2), one obtains a corresponding 
decomposition of the current operator, $\bj = \bj_0 + \bj_I$.
For continuum systems with a free-particle dispersion relation 
$\eps_{\bk} = \bk^2/2m \>$ and a pure density-density interaction
$H_I = {1 \over 2V} \sum_{\bq} \> g(\bq) \> \rho(\bq) \rho(-\bq)$,
one has $\bj_I = 0$ and
$$ \bj(\bq) = \bj_0(\bq) = \sum_{\bk\sg} \> (\bk/m) \>
   a^{\dag}_{\bk-\bq/2,\sg} \> a_{\bk+\bq/2,\sg}        \eqno(3.4) $$
For a lattice system a continuity equation of the form (3.3) holds 
only for small $\bq$ (much smaller than the inverse lattice spacing), 
which is however the most interesting case. 
In the absence of hopping terms in the interaction $H_I$, one still 
has $\bj_I = 0$. For small $\bq$, the commutator of $\rho(\bq)$ with
$H_0$ gives rise to a continuity equation with a current operator of 
the form
$$ \bj(\bq) = \bj_0(\bq) = 
   \sum_{\bk\sg} \> \nabla_{\bk} \eps_{\bk} \>
   a^{\dag}_{\bk-\bq/2,\sg} \> a_{\bk+\bq/2,\sg}        \eqno(3.5) $$
which generalizes (3.4) to systems with a non-quadratic dispersion
relation $\eps_{\bk}$.
\par
\pp It is often useful to collect density and current operators in
a (d+1)-dimensional vector $j^{\mu} = (\rho,\bj)$, where 
$\mu = 0,1,\dots,d$, which will also be referred to as a "current" 
operator.
\par
\smallskip
\pp The continuity equation (3.2) implies {\em Ward identities}\/ 
for correlation (or Green) functions involving charge-currents.
Two such correlation functions are particularly important: The
(charge) current-current correlation function
$$ J^{\mu\nu}(q) = - \> {\textstyle {1 \over V}} \>
   \bra j^{\mu}(q) \> j^{\nu}(-q) \ket                   \eqno(3.6) $$
and the (charge) current vertex part
$$ \Gamma^{\mu}_{\sg}(p;q) = \bra j^{\mu}(q) \> 
   a_{p-q/2,\sg} \> a^{\dag}_{p+q/2,\sg} \ket_{tr}       \eqno(3.7) $$
The above expressions are short-hand notations for thermal expectation 
values of (imaginary) time ordered operator products, with a 
subsequent time-to-frequency Fourier-transformation. The index
"tr" means truncation of external legs (in a diagrammatic language),
i.e.\ division by $G_{\sg}(p+q/2)G_{\sg}(p-q/2)$.
\par
\pp The Ward identity for $J^{\mu\nu}$ can be obtained by deriving
the ($\mu=0$)-component of the current-current correlator in 
time-representation
$$ J^{\mu\nu}(\tau\!-\!\tau',\bq) = 
   - \> {\textstyle {1 \over V}} \> 
   \bra {\cal T} \> j^{\mu}(\tau,\bq) \> j^{\nu}(\tau',-\bq) \ket    
                                                         \eqno(3.8) $$
with respect to $\tau$. One gets
$$ \partial_{\tau} J^{0\nu}(\tau\!-\!\tau',\bq) = 
   {\textstyle {1 \over V}} \> \bra {\cal T} \> 
   \bq \cdot \bj(\tau,\bq) \> j^{\nu}(\tau',-\bq) \ket \>
   - \> {\textstyle {1 \over V}} \> \delta(\tau\!-\!\tau') 
   \bra [j^0(\tau,\bq), j^{\nu}(\tau,-\bq)] \ket         \eqno(3.9) $$
where the first term on the right hand side is due to the 
time-derivative of $j^0(\tau,\bq)$, as given by the 
continuity equation, while the second one has been generated by the 
time-derivative of the time-ordering operator $\cal T$. Fourier
transforming back to frequencies yields the Ward identity
$$ (iq_0,\bq)_{\mu} J^{\mu\nu}(q) =
   {\textstyle {1 \over V}} \> \bra [j^0(\bq), j^{\nu}(-\bq)] \ket =:
   c^{\nu}(\bq)                                         \eqno(3.10) $$
where $(iq_0,\bq)_0 = iq_0$ while $(iq_0,\bq)_j = - q_j$ for 
$j = 1,\dots,d$, and summation over repeated Greek indices is assumed.
Reflection invariance implies obviously
$$ c^0(\bq) = 0                                         \eqno(3.11) $$
Inserting the expression (3.5) for the current operator into the 
commutator in (3.10), one obtains
$$ \bc(\bq) := {\textstyle {1 \over V}} \> 
   \bra [j^0(\bq), \bj(-\bq)] \ket =
   V^{-1} \sum_{\bk\sg} \> \bv_{\bk} \> 
   \bra n_{\bk-\bq/2,\sg} - n_{\bk+\bq/2,\sg} \ket      \eqno(3.12) $$
where $\bv_{\bk} = \nabla_{\bk} \eps_{\bk}$ is the velocity of 
non-interacting particles and $n_{\bk\sg} = a^{\dag}_{\bk\sg} 
a_{\bk\sg}$. 
For models without a cutoff for single-particle momenta\footnote{
Effective low-energy theories such a the g-ology model for 
one-dimensional systems may require a cutoff for single-particle
momenta in order to be well defined. In this case one must take
boundary terms in momentum space into account (see the section on
one-dimensional systems).}, 
the summation variable in (3.12) can be shifted to get
$$ \bc(\bq) = V^{-1} \sum_{\bk\sg} \>
   [(\bq \cdot \nabla_{\bk}) \bv_{\bk}] \> \bra n_{\bk\sg} \ket
                                                        \eqno(3.13) $$
for small $\bq$. For continuum systems with $\eps_{\bk} = \bk^2/2m$
one obtains the simpler expression
$$ \bc(\bq) = {n \over m} \> \bq                        \eqno(3.14) $$
with the particle density $n=N/V$, which is valid for any $\bq$.
\par
\smallskip
\pp For the charge vertex part, the continuity equation (3.3) implies
the Ward identity
$$ (iq_0,\bq)_{\mu} \Gamma^{\mu}_{\sg}(p;q) =
   G^{-1}_{\sg}(p+q/2) - G^{-1}_{\sg}(p-q/2)            \eqno(3.15) $$
which can be obtained by applying a time-derivative to $\bra j^0 a 
a^{\dag}\ket$ in time-represen\-ta\-tion (with respect to the 
time-variable associated with $j^0$), and Fourier-transforming the 
resulting equation of motion. The right-hand-side in (3.15) is due
to the derivative of the time-ordering $\cal T$.
\par
\pp The limit $q \to 0$ is related to the low-energy long-wavelength
response of the system, and is thus particularly important. Hence, we
define
$$ \Gamma^{\mu,\br}_{\sg}(p) = 
   \lim_{{q \to 0 \atop \bq/q_0 = \br}}
   \Gamma^{\mu}_{\sg}(p;q)                              \eqno(3.16) $$
In general the limit $q = (q_0,\bq) \to 0$ is not unique, but depends 
on the ratio $\br = \bq/q_0$.
The Ward identity for $\Gamma^{\mu}_{\sg}(p;q)$ implies
$$ \Gamma^{0,0}_{\sg}(p) = 
   \lim_{q_0 \to 0} \lim_{\bq \to 0} \Gamma^0_{\sg}(p;q) =
    {\partial G^{-1}_{\sg}(p) \over \partial (ip_0)} =
   1 - {\partial \Sg_{\sg}(p) \over \partial (ip_0)}    \eqno(3.17) $$
\vskip -3mm and \vskip -3mm
$$ {\bf\Gamma}^{\infty}_{\sg}(p) =
   \lim_{\bq \to 0} \lim_{q_0 \to 0} {\bf\Gamma}_{\sg}(p;q) =
   - {\partial G^{-1}_{\sg}(p) \over \partial\bp} =
   \bv_{\bp} + {\partial\Sg_{\sg}(p) \over \partial\bp} \eqno(3.18) $$
\par
\smallskip
\pp The Feynman diagrams associated with a perturbation expansion of the 
vertex part $\Gamma^{\mu}_{\sg}(p;q)$ may be one-interaction-reducible,
i.e.\ they can be split in two pieces by cutting a single interaction
line. The sum over all one-interaction-irreducible Feynman diagrams
defines the {\em irreducible}\/ vertex part
$$ \Lambda^{\mu}_{\sg}(p;q) = \bra j^{\mu}(q) \> 
   a_{p-q/2,\sg} \> a^{\dag}_{p+q/2,\sg} \ket_{tr}^{irr} \eqno(3.19) $$
which will play an important role below. 
For systems with pure density-density interactions
$H_I = {1 \over 2V} \sum_{\bq} \> g(\bq) \> \rho(\bq) \rho(-\bq)$,
i.e.\ for most physically relevant systems, the vertex parts 
$\Gamma^{\mu}$ and $\Lambda^{\mu}$ are obviously related by the
Dyson equation
$$ \Gamma^{\mu}_{\sg}(p;q) = \Lambda^{\mu}_{\sg}(p;q) +
   J^{\mu 0}(q) g(\bq) \Lambda^{0}_{\sg}(p;q)            \eqno(3.20) $$
The Ward identities for $\Gamma^{\mu}$ and $J^{\mu\nu}$ imply
$$ (iq_0,\bq)_{\mu} \Lambda^{\mu}_{\sg}(p;q) =
   G^{-1}_{\sg}(p\!+\!q/2) - G^{-1}_{\sg}(p\!-\!q/2)     \eqno(3.21) $$
The inclusion of spin density-density interactions does not modify the
above relations. A Ward identity of the form (3.21) holds also for 
fermions coupled to an abelian gauge-field, where $\Lambda^{\mu}$ is
the irreducible fermion-gauge-field vertex (see Sec.\ 9)).
\par

\bigskip

{\bf 3.2. SPIN CONSERVATION} \par
\medskip
\pp We now derive relations following from spin conservation. 
The logic leading to a continuity equation and Ward identities is
the same as that for charge conservation.
The spin-density fluctuation operator can be defined as
$$ \bs(\bq) = 
   \sum_{\bk} \sum_{\sg,\sg'} \> \bs_{\sg\sg'} \>
   a^{\dag}_{\bk-\bq/2,\sg} \> a_{\bk+\bq/2,\sg'}       \eqno(3.22) $$
where $\bs = (s^x,s^y,s^z) = {1 \over 2} (\sg^x,\sg^y,\sg^z)$ 
with the Pauli matrices $\sg^a$, $a=x,y,z$.
The total spin of the system is given by $\bS = \bs(\b0)$.
If the Hamiltonian conserves spin, i.e.\ $[\bS,H] = 0$, the equation
of motion for $\bs(\tau,\bq)$ assumes the form of a continuity equation
$$ \partial_{\tau} s^a(\tau,\bq) =
   - \> \bq \cdot \bj^a(\tau,\bq)                       \eqno(3.23) $$
at least for small $\bq$. 
For each given system the spin-current operator $\bj^a(\bq)$ can be 
calculated explicitly from the commutator $[H,s^a(\bq)]$.
The decomposition $H = H_0 + H_I$ of the Hamiltonian yields a 
corresponding decomposition $\bj^a = \bj^a_0 + \bj^a_I$ of the current 
operator.
The part $\bj^a_I$ vanishes in most cases of interest (especially for
pure charge density-density interactions), while $\bj^a_0$ can be 
written as
$$ \bj^a_0(\bq) = \sum_{\bk} \sum_{\sg,\sg'} \>
   \nabla_{\bk} \eps_{\bk} \> s^a_{\sg\sg'} \>
   a^{\dag}_{\bk-\bq/2,\sg} \> a_{\bk+\bq/2,\sg'}       \eqno(3.24) $$
For a continuum system with $\eps_{\bk} = \bk^2/2m$, one has
$\nabla_{\bk} \eps_{\bk} = \bk/m$, and the above equations hold for 
any $\bq$.
\par
\pp Spin-density and spin-current operators are conveniently collected
in a (d+1)-dimens\-ional vector $j^{a\mu} = (s^a,\bj^a)$, 
$\mu = 0,1,\dots,d$, which will also be referred to as (spin) "current".
\par
\pp The continuity equation (3.23) implies Ward identities for 
correlation functions involving spin-currents. In close analogy to
the section on charge conservation, we consider the (spin) 
current-current correlation function
$$ J^{a\mu,b\nu}(q) = - \> {\textstyle {1 \over V}} \>
   \bra j^{a\mu}(q) \> j^{b\nu}(-q) \ket                \eqno(3.25) $$
and the (spin) vertex part
$$ \Gamma^{a\mu}_{\sg\sg'}(p;q) = \bra j^{a\mu}(q) \> 
   a_{p-q/2,\sg} \> a^{\dag}_{p+q/2,\sg'} \ket_{tr}     \eqno(3.26) $$
where truncation of legs ("tr") in (3.7) amounts to dividing the
expectation value by $G_{\sg}(p-q/2) G_{\sg'}(p+q/2)$.
\par
\pp The Ward identity for $J^{a\mu,b\nu}(q)$ reads
$$ (iq_0,\bq)_{\mu} J^{a\mu,b\nu}(q) = 
   \> {\textstyle {1 \over V}} \>
   \bra [s^a(\bq), j^{b\nu}(-\bq)] \ket =: 
   c^{ab\nu}(\bq)                                       \eqno(3.27) $$
Using the commutation relations 
$[s^a,s^b] = i\sum_c \eps_{abc} \> s^c$,
where $\eps_{abc}$ is the (totally antisymmetric) Levi-Civita tensor,
one can reduce the commutator in (3.27) to
$$ c^{ab\nu}(\bq) = {\delta_{ab} \over 4} \> c^{\nu}(\bq) + 
   i \sum_c \eps_{abc} \> {\textstyle {1 \over V}}
   \bra j^{c\nu}(\b0) \ket                              \eqno(3.28) $$
where $c^{\nu}(\bq)$, given by (3.11) and (3.13), has already appeared
in connection with charge conservation. 
Note that ${1 \over V} \bra s^a(\b0) \ket$ vanishes in the absence of
a total magnetization, and ${1 \over V} \bra \bj^a(\b0) \ket$ vanishes
generally in equilibrium.
\par
\pp For the spin vertex part, the continuity equation implies the
Ward identity
$$ (iq_0,\bq)_{\mu} \Gamma^{a\mu}_{\sg\sg'}(p;q) = \>
   s^a_{\sg\sg'} \> [G^{-1}_{\sg'}(p+q/2) - G^{-1}_{\sg}(p-q/2)]
                                                        \eqno(3.29) $$
if spin-rotation invariance remains unbroken. In complete analogy to
(3.17) and (3.18), this allows one to relate the spin vertex part for
$q \to 0$ to derivatives of the self-energy with respect to frequency
or momentum.
In analogy to (3.19) one can define an {\em irreducible}\/ spin vertex 
part $\Lambda^{a\mu}_{\sg\sg'}(p;q)$, which, for systems with pure 
(charge or spin) density-density interactions also obeys a Ward 
identity of the form (3.29).
\par

\bigskip

{\bf 3.3. RENORMALIZATION OF RESPONSE FUNCTIONS} \par
\medskip
\pp The response to small external fields, coupled to charge or spin 
densities and currents, belongs to the most important experimentally
accessible properties of a system. 
The combination of renormalization group ideas and conservation laws 
puts powerful constraints on the structure of the effective low-energy 
theory for response functions.
\par
\pp Let us consider the response to a field $A = (\phi,\bA)$ coupled 
to charge density and current as an example.
We consider spinless fermions to keep the number of indices small.
\par
\pp We couple the system to the external field $A$ by adding a term
$$ S_A[\psi,\psi^*;A] = - \int_q A_{\mu}(q) j^{\mu}(q) =
   - \int_q \> [\phi(q)\rho(q) - \bA(q)\bj(q)]          \eqno(3.30) $$
to the action, where $j^{\mu}(q)$ is the current operator constructed
from the Grassmann variables $\psi,\psi^*$ (instead of annihilation 
and creation operators), e.g. 
$\rho(q) = \int_k \psi^*_{k-q/2} \psi_{k+q/2}$.
Adding $S_A$ to the action in the functional integral (2.7), the 
generating functional $\cG[\psi,\psi^*]$ is generalized to a 
functional $\cG[\psi,\psi^*;A]$, whose functional derivatives with
respect to $A_{\mu}$ generate expectation values involving current 
operators $j^{\mu}$. 
\par
\pp In most cases of interest, $S_A$ is quadratic in the fermion 
fields, because $\rho(q)$ is quadratic and $\bj(q)$ is quadratic at 
least for small $\bq$. In all these cases, $j^{\mu}(q)$ can be 
written as
$$ j^{\mu}(q) = \int_p \Gamma_0^{\mu}(p;q) \>
   \psi^*_{p-q/2} \psi_{p+q/2}                          \eqno(3.31) $$
where $\Gamma_0^0(p;q) = 1$ for all $q$, and ${\bf\Gamma}_0(p;q) = 
\bv_{\bp}$ at least for small $\bq$. 
Being quadratic in $\psi$ and $\psi^*$, $S_A$ can be combined with 
$S_0[\psi,\psi^*]$ to give
$$ S_0[\psi,\psi^*;A] = \int_{k,k'}
   \psi^*_k \left[ \hat{\delta}(k'\!-\!k) G_0^{-1}(k) - 
   A_{\mu}(k'\!-\!k) \Gamma_0^{\mu}[(k\!+\!k')/2;k'\!-\!k] 
   \right] \psi_{k'}                                    \eqno(3.32) $$
where $\hat{\delta}(k'\!-\!k) = (2\pi)^{d+1} \delta(k'\!-\!k)$.
\par
\smallskip
\pp Integrating out fermionic fields $\psi_k,\psi^*_k$ outside a
$\Lam$-shell in momentum space, one obtains an effective action
$S^{\Lam}[\psi,\psi^*;A]$ which depends on fields $\psi_k,\psi^*_k$
with $d(\bk,\FS) < \Lam$, and on $A$. The effective action can be 
expanded in powers of $\psi$, $\psi^*$ and $A$. Being interested
in the linear response to small fields, one can drop all terms
beyond quadratic order in $A$. Furthermore, for small $\Lam$, 
monomials $(\psi^*\psi)^n A$ are irrelevant for $n \geq 2$, and
monomials $(\psi^*\psi)^n A^2$ are irrelevant for all $n \geq 1$.
It is easy to see that these terms would yield contributions of
order $\Lam/k_F$ or smaller to the response functions. 
Thus, in the low-energy limit, we need only keep terms of the form
$A^2$ and $\psi^*\psi A$ in addition to terms present already 
without $A$.
In particular, the quartic term in $\psi,\psi^*$ is not affected by
$A$, and it is sufficient to reconsider the quadratic part of 
$S^{\Lam}$, which can be written as
$$ S_0^{\Lam}[\psi,\psi^*;A] \> = \>
   \int_k^{<\Lam} \psi^*_k \> [G_0^{-1}(k) - \Sg^{\Lam}(k)] \> \psi_k
   \> - \> \int_q A_{\mu}(q) J^{\mu\nu\Lam}(q) A_{\nu}(q)           $$
\vskip -5mm
$$ - \> \int_{k,k'}^{<\Lam} \psi_k^* \> A_{\mu}(k'\!-\!k) 
   \Gamma^{\mu\Lam}[(k\!+\!k')/2;k'\!-\!k] \> \psi_{k'} \>
   + \> \cO(\psi^*\psi\>A^2)                            \eqno(3.33) $$
where we have included the contribution of order $A^2$, generated
by the mode elimination. The vertex $\Gamma^{\mu\Lam}$ is given by
the usual Feynman diagrams for the charge vertex part, defined in 
(3.7), where momenta on internal lines are restricted to values 
outside the $\Lam$-shell. The kernel $J^{\mu\nu\Lam}(q)$ is given
by diagrams for the current-current correlator, defined in (3.6),
again with internal momenta outside the $\Lam$-shell. 
Expanding the self-energy $\Sg^{\Lam}$ and introducing renormalized 
fields $\bar\psi_k = \psi_k/(Z^{\Lam}_{\bk_F})^{1/2}$, one obtains 
the renormalized action
$$ \bar S_0^{\Lam}[\bar\psi,\bar\psi^*;A] \> = \>
   \int_k^{<\Lam} \bar\psi^*_k \> 
   [ik_0 - \bar v_{\bk_F}^{\Lam}k_r - \bar\Sg^{\Lam}(k)] \> \bar\psi_k 
   \> - \> \int_q A_{\mu}(q) J^{\mu\nu\Lam}(q) A_{\nu}(q)           $$
\vskip -5mm
$$ - \> \int_{k,k'}^{<\Lam} \bar\psi_k^* \> A_{\mu}(k'\!-\!k) 
   \> (Z^{\Lam}_{\bk_F} Z^{\Lam}_{\bk'_F})^{1/2} \>
   \Gamma^{\mu\Lam}[(k\!+\!k')/2;k'\!-\!k] \> \bar\psi_{k'} \>
                                                        \eqno(3.34) $$
We have not assumed isotropy here, i.e.\ the field renormalization
factor $Z^{\Lam}_{\bk_F}$ and the Fermi velocity 
$\bar v_{\bk_F}^{\Lam}$ may depend on $\bk_F$.
\par
\pp So far, all steps are valid for arbitrary $q$, provided that
$\Lam \ll k_F$. We now consider the limit $q \to 0$, i.e.\ the
response to fields with a low frequency and a small wave number.
There are two simplifications in that limit, reducing considerably
the number of unknown terms in the effective low-energy action.
\par
\pp Firstly, $J^{\mu\nu\Lam}(q)$ has a {\em unique}\/ limit 
$J^{\mu\nu\Lam}(0) \equiv J^{\mu\nu\Lam}_{inc}$ for $q \to 0$ 
(independent of the ratio $\bq/q_0$) at $\Lam > 0$, determined by
"incoherent" contributions. 
Without cutoff $J^{\mu\nu}(q)$ generally has no unique limit for $q \to 0$ 
due to singular contributions of "coherent" particle-hole excitations with 
arbitrarily small momentum transfer $\bq$ (as is well-known from
Fermi liquid theory, see Sec.\ 4). However, particle-hole excitations
across the $\Lam$-shell are not possible, if $|\bq|$ is smaller that
the width of the shell $2\Lam$.
The Ward identity (3.10) then implies that $J^{00\Lam}_{inc} = 0$
and $J^{0j\Lam}_{inc} = J^{j0\Lam}_{inc} = 0$.
Note that the Ward identities derived in 3.1 and 3.2 hold also for
$J^{\mu\nu\Lam}(q)$, $\Gamma^{\mu\Lam}$, $G^{\Lam}$ and other 
correlation functions with a cutoff $\Lam >0$.\footnote
{The expression (3.12) for the function $\bc(\bq)$ has to be modified
to \\ $ \bc^{\Lam}(\bq) := V^{-1} \sum_{\bk\sg} \> \bv_{\bk} \> 
   \big[ \Xi^{>\Lam}(\bk\!+\!\bq/2) \bra n_{\bk-\bq/2,\sg} \ket -
     \Xi^{>\Lam}(\bk\!-\!\bq/2) \bra n_{\bk+\bq/2,\sg} \ket \big] $
if states outside the $\Lam$-shell are excluded, where $\Xi^{>\Lam}$
is the characteristic function of the allowed momenta.} 
In fact these quantities
can be viewed as exact correlation functions in a theory where 
the interaction does not scatter particles inside the $\Lam$-shell
(intermediate states corresponding to internal lines in Feynman
diagrams inside the $\Lam$-shell are not accessible).
\par
\pp The second simplification is due to the Ward identity (3.15) for 
the vertex part $\Gamma^{\mu}$.
Again, the reason for the non-uniqueness of the limit $q \to 0$ 
of $\Gamma^{\mu}(p;q)$, i.e.\ particle-hole excitations with 
arbitrarily small momentum transfer $\bq$, is eliminated for any 
$\Lam > 0$. Hence, the Ward identity (3.15) implies
$$ \lim_{q \to 0} \Gamma^{0\Lam}(p;q) = 
   1 - {\partial \Sg^{\Lam}(p) \over \partial (ip_0)} \>
   \longrightarrow \> 1/Z^{\Lam}_{\bp_F}                \eqno(3.35) $$
\vskip -3mm and \vskip -3mm
$$ \lim_{q \to 0} {\bf\Gamma}^{\Lam}(p;q) = 
   \bv_{\bp} + {\partial\Sg^{\Lam}(p) \over \partial\bp} \> 
   \longrightarrow \>
   \bar\bv^{c\Lam}_{\bp_F}/Z^{\Lam}_{\bp_F}             \eqno(3.36) $$
where the latter limit $p \to (0,\bp_F)$ is obtained from the 
expressions for $Z^{\Lam}_{\bp_F}$ and $\bar\bv^{c\Lam}_{\bp_F}$ in 
terms of $\Sg^{\Lam}$ (see (2.24) and (2.26)). 
Inserting these relations in (3.34), we obtain the following 
simple result for the quadratic part of the effective low-energy
action
$$ \bar S_0^{\Lam}[\bar\psi,\bar\psi^*;A] \> = \>
   \int_k^{<\Lam} \bar\psi^*_k \> 
   [ik_0 - \bar v_{\bk}^{\Lam}k_r - \bar\Sg^{\Lam}(k)] \> \bar\psi_k \>
   - \> \int_q A_{\mu}(q) \> \bar j^{\mu\Lam}(q) \>
   - \> \int_q A_j(q) \> J^{jj'\Lam}_{inc} \> A_{j'}(q) 
                                                          \eqno(3.37) $$
with a renormalized current operator
$$ \bar j^{\mu\Lam}(q) = 
   \int_k^{<\Lam} (1,\bar\bv_{\bk}^{c\Lam})^{\mu} 
   \> \bar\psi^*_{k-q/2} \bar\psi_{k+q/2}                 \eqno(3.38) $$
Note that the {\em field renormalizations have cancelled out}. 
We emphasize that (3.37) is valid only for fields $A_{\mu}(q)$ with 
small $q$, and $\Lam$ chosen such that $2|\bq| < \Lam \ll k_F$! 
\par
\pp In a Fermi liquid, which will be discussed in some detail in Sec.\ 4, 
$\bar j^{\mu\Lam}(q)$ is the quasi-particle current (including backflow),
and $\bar S_0^{\Lam}[\bar\psi,\bar\psi^*;A]$, supplemented by residual
quasi-particle interactions, is nothing but the Landau model for the
response of an interacting Fermi system to weak and slowly varying
external fields \cite{NOZ}. 
To describe the response of charged fermions to electromagnetic fields
$A^{\mu}$ one must of course add the diamagnetic term ($\propto \int 
\rho \bA^2$) to the microscopic action. 
\par

\vfill\eject

\def\bA{{\bf A}}
\def\bc{{\bf c}}
\def\bk{{\bf k}}
\def\bQ{{\bf Q}}
\def\bq{{\bf q}}
\def\bP{{\bf P}}
\def\bp{{\bf p}}
\def\b0{{\bf 0}}
\def\bi{{\bf i}}
\def\bj{{\bf j}}
\def\bn{{\bf n}}
\def\br{{\bf r}}
\def\bs{{\bf s}}
\def\bS{{\bf S}}
\def\bv{{\bf v}}
\def\alf{\alpha}
\def\eps{\epsilon}
\def\up{\uparrow}
\def\down{\downarrow}
\def\bra{\langle}
\def\ket{\rangle}
\def\sDelta{{\scriptstyle \Delta}}
\def\FS{\partial{\cal F}}
\def\Re{{\rm Re}}
\def\Im{{\rm Im}}
\def\xik{\xi_{\bk}}
\def\cO{{\cal O}}
\def\cD{{\cal D}}
\def\cF{{\cal F}}
\def\cG{{\cal G}}
\def\cZ{{\cal Z}}
\def\gam{\gamma}
\def\Gam{\Gamma}
\def\Lam{\Lambda}
\def\lam{\lambda}
\def\dbm{\delta\bar\mu}
\def\scr{\scriptstyle}
\def\sg{\sigma}
\def\Sg{\Sigma}

\vspace*{1cm}
\centerline{\large 4. FERMI LIQUIDS}
\vskip 1cm
\pp In this section we describe a renormalization group picture of 
microscopic Fermi liquid theory.
Renormalization (group) ideas and language have been used to a certain
extent already in many early works on Fermi liquid theory.
The idea to derive a systematic low-energy expansion by integrating out
high-energy degrees of freedom had been worked
out for normal and superfluid Fermi systems in the 1980's \cite{SR}. 
As a subject in its own right the rephrasing of Fermi liquid theory 
in modern renormalization group language has attracted interest
only quite recently \cite{SHA94,POL,DC}.
Besides providing a faster way of understanding many aspects of 
Fermi liquid theory (without adding new results) the renormalization 
group is also a powerful tool for obtaining rigorous results on the 
Fermi liquid behavior of specific systems.\footnote
{Significant rigorous results on the Fermi liquid behavior of
interacting Fermi systems have been obtained very recently by
Feldman et al.\ \cite{FMRT,FKL}. In particular, it has been shown that
two-dimensional Fermi systems without Cooper instability exhibit a
finite discontinuity in the momentum distribution function within a 
finite radius of convergence in coupling space.}
\par
\pp The following presentation of Fermi liquid theory from a 
renormalization group point of view absorbs many ideas from the
above-mentioned earlier works on this subject. 
We have clarified several points that we found so far either unclear
or incomplete, especially concerning the flow of coupling functions
and the derivation of response functions in Fermi liquids.
We consider all orders in perturbation theory, but do not address
the issue of convergence and non-perturbative effects \cite{FMRT,FKL}.
For pedagogical reasons we consider only spinless fermions with 
short-range interactions. 
Including spin or (sufficiently screened) long-range interactions
is not hard. 
\par

\bigskip

{\bf 4.1. MICROSCOPIC FERMI LIQUID THEORY:} \par
\medskip
\pp A Fermi liquid can be characterized by the following asymptotic
behavior of one- and two-particle Green (or vertex) functions
in the low-energy limit $k_0 \to 0$ at $T=0$ \cite{NOZ,AGD}. \par
\pp The single-particle propagator behaves as
$$ G(k) = {Z_{\bk_F} \over 
   ik_0 - v^*_{\bk_F} k_r - \bar\Sg(k)}                   \eqno(4.1) $$
where $\bar\Sg(k)$ vanishes faster than linearly for $k \to (0,\bk_F)$.
Here $k_r$ measures the distance of $\bk$ from the Fermi surface of
the interacting system, which is defined by the position of the pole
in $G(k)$. This behavior implies the existence of {\em Landau
quasi-particles}, i.e.\ gapless fermionic single-particle excitations 
with a linear dispersion relation. The {\em spectral weight}\/ 
$Z_{\bk_F} \in \> ]0,1]$ of these excitations can be obtained from the 
self-energy as
$$ Z_{\bk_F} = \left. 
   \left[ 1 - \partial\Sg(k_0,\bk)/\partial(ik_0) \right]^{-1}
   \right|_{(0,\bk_F)}                                    \eqno(4.2) $$
and the quasi-particle or Fermi {\em velocity}\/ is given by
$$ \bv^*_{\bk_F} = Z_{\bk_F} \left.
   \big[ \bv_{\bk} + \partial\Sg(k_0,\bk)/\partial\bk \big]
   \right|_{(0,\bk_F)}                                    \eqno(4.3) $$
where $\bv_{\bk} = \nabla\eps_{\bk}$ is the bare velocity in the
absence of interactions. The momentum distribution function in the
ground state has a finite discontinuity $Z_{\bk_F}$ across the Fermi 
surface.
For isotropic systems $Z_{\bk_F} = Z$ and $v^*_{\bk_F} = v^*_F$ are 
constant all over the Fermi surface.
\par
\pp The two-particle vertex $\Gam(k_1,k_2;k'_1,k'_2)$ is finite 
in the limit $k_i \to (0,\bk_{Fi})$, $k'_i \to (0,\bk'_{Fi})$.\footnote
{Note our notation 
$\tilde\Gam(k_1,k_2;k'_1,k'_2) = 
 (2\pi)^{d+1} \delta(k'_1\!+\!k'_2\!-\!k_1\!-\!k_2) \>
 \Gam(k_1,k_2;k'_1,k'_2)$ and $\Gam(k,k';q) = 
 \Gam(k\!-\!q/2,k'\!+\!q/2;k\!+\!q/2,k'\!-\!q/2)$.}
For $\bk'_{F1} \neq \bk_{F1},\bk_{F2}$ this limit is unique and 
defines the quasi-particle {\em scattering amplitudes}
$$ a(\bk_{F1},\bk_{F2};\bk'_{F1},\bk'_{F2}) \> = \> 
   [Z_{\bk_{F1}} Z_{\bk_{F2}} Z_{\bk'_{F1}} Z_{\bk'_{F2}}]^{1/2}
   \lim_{k_i \to (0,\bk_{Fi})} \> \lim_{k'_i \to (0,\bk'_{Fi})}
   \Gam(k_1,k_2;k'_1,k'_2)                          \eqno(4.4) $$
For $\bk'_{F1} = \bk_{F1}$ or $\bk'_{F1} = \bk_{F2}$ the above limit
is not unique. Due to the antisymmetry of the vertex function under
exchange of the outgoing (or incoming) particles it is sufficient to
analyze one of these cases. Consider, for example, the vertex 
function for $k'_1 \sim k_1 \sim (0,\bk_F)$ and $k'_2 \sim k_2 
\sim (0,\bk'_F)$. 
The non-uniqueness of the limit $q = k'_1 - k_1 \to 0$ is due to 
particle-hole pairs with momentum $q$ as in the second diagram on the
right hand side of Fig.\ 4.1, where the 
irreducible vertices $\Gam_{irr}$ are defined by excluding 
subdiagrams with such particle-hole pairs.
The Feynman diagrams contributing to the irreducible 
vertex $\Gam_{irr}$ are regular in the limit $q \to 0$.
The product of propagators $G(p\!-\!q/2) G(p\!+\!q/2)$ in Fig.\ 4.1 
can be effectively replaced by \cite{NOZ,AGD}
$$ 2\pi\delta(p_0) \> Z_{\bp_F}^2 \> 
   {\Theta(\xi^*_{\bp+\bq/2}) - \Theta(\xi^*_{\bp-\bq/2}) \over
   iq_0 - \bv^*_{\bp_F}\!\cdot\!\bq} \> + \> \phi(p)      \eqno(4.5) $$
where $\xi^*_{\bk} = v^*_F k_r$ and $\phi(p)$ is a regular function 
of $p$. The first term in (4.5) vanishes for $\bq \to \b0$ at finite
$q_0$ while it yields finite contributions for $q \to 0$ with 
$|\bq|/q_0 \neq 0$. Two special ways of taking the limit $q \to 0$, 
i.e.
$$ \Gamma^0(\bk_F,\bk'_F) = 
   \lim_{q_0 \to 0} \lim_{\bq \to \b0} 
   \Gam(\bk_F,\bk'_F;q)                                  \eqno(4.6a) $$
\vskip -4mm
$$ \Gamma^{\infty}(\bk_F,\bk'_F) = 
   \lim_{\bq \to \b0} \lim_{q_0 \to 0}
   \Gam(\bk_F,\bk'_F;q)                                  \eqno(4.6b) $$
play a prominent role in Fermi liquid theory (the notation refers to
a more general limit $\Gam^{\br}$ where $\bq/q_0 \to \br$ as 
$q \to 0$). The first one defines Landau's quasi-particle 
{\em interaction} $f_{\bk_F\bk'_F}$ as
$$ Z_{\bk_F} Z_{\bk'_F} \Gamma^0(\bk_F,\bk'_F) = 
   f_{\bk_F\bk'_F}                                        \eqno(4.7) $$
and the second the {\em forward scattering amplitude} 
$a(\bk_F,\bk'_F)$ via
$$ Z_{\bk_F} Z_{\bk'_F} \Gamma^{\infty}(\bk_F,\bk'_F) = 
   a(\bk_F,\bk'_F) =
   \lim_{\bk_{F1},\bk'_{F1} \to \bk_F} \>
   \lim_{\bk_{F2},\bk'_{F2} \to \bk'_F}
   a(\bk_{F1},\bk_{F2};\bk'_{F1},\bk'_{F2})               \eqno(4.8) $$
For small $q$ and $k,k'$ close to the Fermi surface, the vertex 
function $\Gamma(k,k';q)$ can be reconstucted from the quasi-particle
interaction by solving the linear integral equation
$$ \bar\Gamma(\bk_F,\bk'_F;q) = f_{\bk_F\bk'_F} +
   \int_{\bk''_F} f_{\bk_F\bk''_F} \> 
   {\bn_{\bk''_F}\!\cdot\!\bq \over 
   iq_0 - \bv^*_{\bk''_F}\!\cdot\!\bq} \> 
   \bar\Gamma(\bk''_F,\bk'_F;q)                           \eqno(4.9) $$
for the renormalized vertex $\bar\Gamma(\bk_F,\bk'_F;q) = 
Z_{\bk_F} Z_{\bk'_F} \> \Gamma(\bk_F,\bk'_F;q)$,
where the integral extends over the Fermi surface.\footnote
{The notation $\int_{\bk_F} \dots \>$ is an abbreviation for 
 $(2\pi)^{-d} \int_{\FS} dS(\bk_F) \dots \>$, where $dS(\bk_F)$ 
 is the infinitesimal surface element at $\bk_F$.}
\par
\pp For isotropic systems, $f_{\bk_F\bk'_F}$ and $a(\bk_F,\bk'_F)$ 
depend only on the angle $\theta$ between $\bk_F$ and $\bk'_F$.
Multiplying by the density of quasi-particle excitations at the
Fermi level, $N^*_F$, one obtains dimensionless functions 
$F(\theta) = N^*_F \, f(\theta)$ and $A(\theta) = N^*_F \, a(\theta)$.
Expanding in spherical harmonics, $F$ and $A$ can be expressed in 
terms of a discrete set of Landau parameters $F_l$ and $A_l$, 
respectively, where $l$ labels the various harmonics.
\par

\bigskip

{\bf 4.2. EFFECTIVE ACTION AND FIXED POINT} \par
\medskip
\pp The low-energy effective action of a (possibly anisotropic) Fermi 
system has the form
$$ \bar S^{\Lam}[\bar\psi,\bar\psi^*] = 
   \int_k^{<\Lam} \bar\psi_k^* 
   [ik_0 - \bar v_{\bk_F}^{\Lam} k_r - \bar\Sg^{\Lam}(k)] \bar\psi_k 
   - \int_k^{<\Lam} 
   (\delta\bar\mu^{\Lam}_{\bk_F} - \delta\bar v_{\bk_F}^{\Lam} k_r) 
   \> \bar\psi_k^* \bar\psi_k \hskip 2cm                             $$
\vskip -4mm 
$$ - {\textstyle {1 \over 4}} 
   \int_{k,k',q}^{<\Lam} \bar\Gam^{\Lam}(k,k';q) \> 
   \bar\psi^*_{k-q/2} \bar\psi^*_{k'\!+q/2} 
   \bar\psi_{k'\!-q/2} \bar\psi_{k+q/2}                  \eqno(4.10) $$
for $\Lam \ll k_F$, provided there are no instabilities with an
energy scale outside the $\Lam$-shell (cf.\ Sec.\ 2). 
More-than-two-particle interactions are irrelevant for the low-energy
behavior and have already been omitted. 
The {\em residual}\/ self-energy $\bar\Sg^{\Lam}(k)$ vanishes faster 
than linearly for $k_0,k_r \to 0$ (linear terms have been absorbed in
$Z_{\bk_F}^{\Lam}$ and $\bar v_{\bk_F}^{\Lam}$).
\par
\pp In a Fermi liquid, {\em all renormalizations are finite}\/ in the
limit $\Lam \to 0$. The field renormalization tends to a finite
constant
$$ Z^{\Lam}_{\bk_F} \to Z_{\bk_F} \in \> ]0,1]           \eqno(4.11) $$
which is easily identified as the quasi-particle weight. 
The renormalized Fermi velocity converges to the quasi-particle 
velocity
$$ \bar v_{\bk_F}^{\Lam} \to v^*_{\bk_F} > 0             \eqno(4.12) $$
\pp The counterterm $-\delta\bar\mu^{\Lam}_{\bk_F}$ had been introduced
to compensate shifts of the Fermi surface induced by the interaction.
For $\Lam \ll k_F$ these shifts are of order $\Lam$, generated by 
tadpole diagrams (other diagrams lead to shifts of order $\Lam^2$).
Hence, this counterterm can be omitted from the effective action, 
if the latter is normal ordered (with respect to the Fermi sea).
Normal ordering also compensates the velocity counterterm 
$\delta\bar v_{\bk_F}^{\Lam} k_r$.
\par
\pp The behavior of the vertex function in the limit $\Lam \to 0$ is 
quite subtle. According to (4.4), one has
$$ \bar\Gam^{\Lam}(\bk_{F1},\bk_{F2};\bk'_{F1},\bk'_{F2}) \> \to
   \> a(\bk_{F1},\bk_{F2};\bk'_{F1},\bk'_{F2})  \quad
   {\rm for} \quad \bk'_{F1} \neq \bk_{F1},\bk_{F2}      \eqno(4.13) $$
On the other hand, if $\bk'_{F1}$ coincides with $\bk_{F1}$ or 
$\bk_{F2}$, $\bar\Gam^{\Lam}$ does not tend to the scattering
amplitude $a$, but to the Landau function $f$:
$$ \bar\Gam^{\Lam}(\bk_F,\bk'_F;\bk_F,\bk'_F) = 
  -\bar\Gam^{\Lam}(\bk_F,\bk'_F;\bk'_F,\bk_F) \> \to \>
   f_{\bk_F\bk'_F}                                       \eqno(4.14) $$
Note that the limits $\Lam \to 0$ and $\bk'_{F1} \to \bk_{F1}$ (or 
$\bk_{F2}$) do not commute!\footnote
{This problem has led to some confusion in previous works on RG and
Fermi liquid theory; for example, in Ref.\ \cite{SHA94} it seems that
the "fixed point coupling $u(\bk_{F1},\bk_{F2};\bk'_{F1},\bk'_{F2})$",
obtained in the limit $\Lam \to 0$, converges to the Landau-function 
in the forward scattering limit, in contrast to (4.8).} 
This singularity has the same origin as the non-uniqueness of the 
limit $q \to 0$ discussed already in 4.1. To see that,
consider the vertex function $\bar\Gam^{\Lam}(k,k';q)$ for
$k \sim (0,\bk_F)$, $k' \sim (0,\bk'_F)$ and $q \sim 0$. For $|\bq| 
< 2\Lam$, particle-hole pairs of propagators as in Fig.\ 4.1 do not 
yield any singular contributions to $\bar\Gam^{\Lam}(k,k';q)$, 
because there are no particle-hole excitations across the $\Lam$-shell
with $|\bq| < 2\Lam$. 
Hence, for $\Lam \ll k_F$ and $k,k'$ close to the Fermi surface (and 
energy) one has $\bar\Gam^{\Lam}(k,k';q) \sim f_{\bk_F\bk'_F}$ if 
$|\bq| < 2\Lam$. 
\par
\pp The flow of the vertex (or coupling) function is obviously
singular for $\bk'_{F1} \!\sim\! \bk_{F1}$ (or $\bk_{F2}$). For 
$|\bk'_{F1}\!-\!\bk_{F1}| < 2\Lam$, particle-hole pairs do not yield
any singular contribution to $\bar\Gam^{\Lam}$, which therefore seems
to flow towards Landau's $f$-function. But then, reducing $2\Lam$
below $|\bk'_{F1}\!-\!\bk_{F1}|$, the coupling function undergoes a 
rapid flow towards $a(\bk_{F1},\bk_{F2};\bk'_{F1},\bk'_{F2}) \sim 
a(\bk_{F1},\bk_{F2})$, i.e.\ the $\beta$-functional is strongly 
$\Lam$-dependent and singular for $|\bk'_{F1}\!-\!\bk_{F1}| \sim \Lam 
\ll k_F$!
Fixed points are reached only pointwise on the Fermi surface, but not
uniformly. 
\par
\pp {\em Assuming}\/ the above convergence properties, one can express
the low-energy behavior of the system in terms of $Z_{\bk_F}$,
$v^*_{\bk_F}$ and $a$ (or $f$) by expressing physical
quantities in terms of $\bar S^{\Lam}$ and expanding in powers of
small energy scales ($T$, $k_0$ etc.) and $\Lam/k_F$. 
The leading and in some cases also the first subleading low-energy
behavior can be obtained from a small (i.e.\ tractable) set of 
Feynman diagrams \cite{SHA94,SR}.
\par
\pp As observed already by Landau, residual scattering processes with 
finite momentum transfers (i.e.\ non-forward scattering processes) do 
not contribute to the {\em leading}\/ 
low-energy behavior of thermodynamics, single-particle excitations
and long-wavelength response functions (due to the restriction of 
ingoing and outgoing momenta to the thin $\Lam$-shell). 
To this leading order one may therefore effectively drop the 
interactions with momentum transfers $|\bq| > 2\Lam$ from the action 
$\bar S^{\Lam}$.
In the limit $\Lam \ll k_F$ one then obtains a scale-invariant
or {\em fixed point}\/ action
$$ \bar S^*[\bar\psi,\bar\psi^*] = 
   \int_k^{<\Lam} \bar\psi_k^* (ik_0 - v_{\bk_F}^* k_r) \bar\psi_k 
   - {\textstyle {1 \over 2}} 
   \int_{k,k',q}^{<\Lam} f_{\bk_F\bk'_F} \> 
   \bar\psi^*_{k-q/2} \bar\psi^*_{k'\!+q/2} 
   \bar\psi_{k'\!-q/2} \bar\psi_{k+q/2}                  \eqno(4.15) $$
\par
where, in addition to the restriction of fermionic momenta to the
$\Lam$-shell, the momentum transfers $\bq$ are now restricted by the 
{\em condition}\/ $|\bq| < 2\Lam$! 
The prefactor ${1 \over 2} = {2 \over 4}$ in the interaction part is
the result of a summation of interaction terms with $k'_1 \sim k_1$ 
and their conjugate under exchange, $k'_1 \sim k_2$.
Solving the action $\bar S^*[\bar\psi,\bar\psi^*]$ by Hartree 
mean-field theory (neglecting fluctuations) 
leads to Landau's energy functional
$$ \delta E[\delta\bar n] = 
   \int_{\bk} v^*_{\bk_F} k_r \> \delta\bar n_{\bk} +
   {\textstyle {1 \over 2}} \int_{\bk,\bk'} f_{\bk_F\bk'_F} \> 
   \delta\bar n_{\bk} \delta\bar n_{\bk'}                \eqno(4.16) $$
where $\delta\bar n_{\bk}$ is the quasi-particle distribution function.
The renormalized single-particle propagator is then 
determined by the quadratic part (only) 
of the fixed point action $\bar S^*[\bar\psi,\bar\psi^*]$ as
$$ \bar G(k) \sim \bar G^*_0(k) = 
   {1 \over ik_0 - v_{\bk_F}^* k_r}                      \eqno(4.17) $$
This is the quasi-particle propagator, which describes stable 
quasi-particle excitations with a velocity $v_{\bk_F}^*$.
\par
\pp For three-dimensional systems the irrelevance of fluctuations
and the ensuing irrelevance of residual interactions for the single
particle propagator have been already established by Landau's estimate
of the quasi-particle decay-rate. A detailed analysis of the action
(4.15) in Sec.\ 7 will show that these results hold actually in any
dimension $d > 1$. 
\par

\bigskip

\vfill\eject

{\bf 4.3. RESPONSE FUNCTIONS} \par
\medskip
\pp We now derive the low-energy and long-wavelength density and 
current response of a Fermi liquid. 
Adding the renormalized source term from subsection 3.3 to the fixed 
point action $\bar S^*[\bar\psi,\bar\psi^*]$, we immediately find 
that the linear response functions for perturbations coupling to the 
density or current in a Fermi liquid are described by the effective 
action
$$ \bar S^*[\bar\psi,\bar\psi^*,A] = 
   \bar S^*[\bar\psi,\bar\psi^*] -
   \int_q A_{\mu}(q) \> \bar j^{\mu}(q) -
   \int_q A_j(q) \> J^{*jj'}_{inc} \> A_{j'}(q)         \eqno(4.18) $$
with a {\em quasi-particle current}
$$ \bar j^{\mu}(q) = \int_k^{<\Lam} 
   (1,\bv^{c*}_{\bk_F})^{\mu} \> 
   \bar\psi^*_{k-q/2} \bar\psi_{k+q/2}                  \eqno(4.19) $$
for $|\bq| \ll k_F$ and small $q_0$, where
$\bv^{c*}_{\bk_F} = \lim_{\Lam \to 0} \bar\bv^{c\Lam}_{\bk_F}$ and
$J^{*jj'}_{inc} = \lim_{\Lam \to 0} J^{jj'\Lam}_{inc}$. 
We have thus obtained a relatively short microscopic justification 
of the {\em Landau model}\/ for the response of a Fermi system to
low-energy long-wavelength perturbations.\footnote{It is instructive
to compare with the derivation in Nozi\`eres' book \cite{NOZ}.}
Note that the source field $A_{\mu}(q)$ remains unrenormalized, as
a consequence of charge conservation. Consequently {\em no 
renormalization factors appear in the (long wavelength) response 
functions}. For Galilean invariant systems there is another
simplification: $J^{*jj'}_{inc}$ vanishes.
\par
\pp Powercounting \cite{SHA94,SR} shows that a solution of 
$\bar S^*[\psi,\psi^*,A]$ within the random phase approximation (RPA) 
yields the exact leading low-energy behavior of response functions 
(expressed in terms of fixed point parameters).
The RPA effective interaction $D$ is equal to the renormalized vertex 
function for small $q$, i.e.\
$$ D_{\bk\bk'}(q) = \bar\Gam(\bk_F,\bk'_F;q)            \eqno(4.20) $$
and can be obtained from $f_{\bk_F\bk'_F}$ by solving the integral
equation (4.9). The current-current response function $J^{\mu\nu}(q)$
(including the density-density response as $J^{00}$) is then given by
$$ J^{\mu\nu}(q) = J^{*\mu\nu}_{inc} +
   \bar J^{\mu\nu}_0(q) \> + \>
   \int_{k,k'} \lam^{\mu}(\bk) \>
   \bar G^*_0(k\!-\!q/2) \bar G^*_0(k\!+\!q/2)           \hskip 3cm $$
\vskip -6mm
$$ \hskip 2cm \times \> D_{\bk\bk'}(q) \>
   \bar G^*_0(k'\!-\!q/2) \bar G^*_0(k'\!+\!q/2) \> 
   \lam^{\nu}(\bk')                                     \eqno(4.21) $$
where $\lam^{\mu}(\bk) = (1,\bv^{c*}_{\bk_F})^{\mu}$ and 
$$ \bar J^{\mu\nu}_0(q) = \int_k \lam^{\mu}(\bk) \>
   \bar G^*_0(k\!-\!q/2) \bar G^*_0(k\!+\!q/2) \>
   \lam^{\nu}(\bk)                                      \eqno(4.22) $$
We recall from Sec.\ 3.3 that $J^{*\mu\nu}_{inc}$ vanishes, if either 
$\mu$ or $\nu$ are zero.
\par
\pp There is no explicit solution of the RPA integral equation for
general $f_{\bk_F\bk'_F}$, but various important generic properties 
of $J^{\mu\nu}(q)$ have been established \cite{NOZ,BP}. 
Among those is the existence of propagating collective modes, known
as {\em zero sound}\/ \cite{LAN}. The dynamical charge conductivity
$\sg(\omega)$ (related to the current-current response by the
Kubo formula), has a $\delta$-peak at zero frequency (at $T=0$),
implying a vanishing charge DC-resistivity. 
A simple expression can be obtained for the compressibility 
$\kappa = dn/d\mu$, which is related to $J^{00}(q)$ via the limit
$$ \kappa = - \lim_{\bq \to \b0} \lim_{q_0 \to 0} 
   J^{00}(q)                                            \eqno(4.23) $$
In an isotropic system, $\kappa$ depends on residual interactions 
only via the Landau parameter $F_0$:
$$ \kappa = {N^*_F \over 1 + F_0}                       \eqno(4.24) $$
For further implications of (4.21) the reader is referred to any 
textbook on Fermi liquid theory \cite{NOZ,BP}.
\par

\bigskip

{\bf 4.4. SUBLEADING CORRECTIONS} \par
\medskip
\pp So far we have discussed the {\em leading}\/ low-energy behavior,
which was completely determined by the fixed point action $\bar S^*$
in (4.15) and (4.18). To this leading order, very simple 
{\em scaling}\/ relations hold for all quantities, such as
$$ G_{\bk_F}(sk_0,sk_r) = s^{-1} \> G_{\bk_F}(k_0,k_r) 
   \quad {\rm and} \quad 
   J^{\mu\nu}(sq_0,s\bq) = J^{\mu\nu}(q_0,\bq)          \eqno(4.25) $$
Here $G_{\bk_F}(k_0,k_r) = G(k_0,\bk)$ where $\bk_F$ is a projection
of $\bk$ on the Fermi surface.
\par
\pp At small but finite energy scales, there are {\em subleading}\/
corrections to this simple behavior. 
For example, the single-particle propagator has the form (see (4.1))
$$ G(k) = {Z_{\bk_F} \over 
   ik_0 - v^*_{\bk_F} k_r - \bar\Sg(k)}                 \eqno(4.26) $$
with a residual self-energy $\bar\Sg(k)$ of order $\cO(k_0^2,k_r^2)$ 
in three dimensions, describing quasi-particle decay and a small
incoherent background (with a spectral weight that vanishes 
quadratically in the low-energy limit). For real frequencies $\xi$ 
close to the quasi-particle pole, one thus obtains
$$ G(k) = {Z_{\bk_F} \over \xi - \xi^*_{\bk} \pm i\gam_{\bk}}
   \quad {\rm for} \quad \xi^*_{\bk} {_> \atop ^<} 0    \eqno(4.27) $$
with a quasi-particle dispersion relation 
$\xi^*_{\bk} = v^*_{\bk_F} k_r + \Re\bar\Sg(\xi^*_{\bk},\bk)$  
and a decay rate 
$\gam_{\bk} = |\Im\bar\Sg(\xi^*_{\bk},\bk)|$ of order $k_r^2$.
\par
\pp These subleading corrections are partially due to scattering 
processes with arbitrary momentum transfers, generated by the 
quasi-particle scattering amplitude $a$, and also due to irrelevant
terms contained in $\bar\Sg^{\Lam}$, which had been discarded
in $\bar S^*$. 
Sometimes subleading corrections can be expressed exactly in terms
of fixed point parameters. In particular, in three-dimensional
systems the decay rate $\gam_{\bk}$ is given by Fermi's Golden Rule
with the scattering amplitude $a$ as matrix elements \cite{BP}.
This corresponds to a single Feynman diagram for the self-energy
\cite{SR}, i.e.\ diagram (b) in Fig.\ 2.3.\footnote
{In two dimensions, more diagrams are needed \cite{EHR}.}
\par
\pp Subleading corrections due to residual scattering processes are 
also responsible for the damping of collective modes and, in particular,
for a finite (i.e.\ non-zero) resistivity $\rho(T)$ at finite 
temperature (if not already made finite by impurities). Umklapp
processes in metals are usually an "irrelevant" perturbation
(except for special Fermi surface geometries) in the sense that
they die out in the zero energy limit. Nevertheless they are very
important at any finite temperature, where, in the absence of 
impurities and phonons, they are the only way to make the DC 
resistivity finite. In conventional three-dimensional metals these
terms lead to a contribution
$$ \rho_u(T) \propto T^2                                \eqno(4.28) $$
to the electrical DC resistivity \cite{ZIM}.
\par

\bigskip

{\bf 4.5. INSTABILITIES} \par
\medskip
\pp There are several consistency conditions that a Fermi liquid must
satisfy for stability. Two types of instabilities are
generic in the sense that they do not require low dimensionality or
special Fermi surface geometries: The Pomerantchuk instability and
the Cooper instability.
\par
\pp The {\em Pomerantchuk instability}\/ \cite{POM} occurs when the 
energy of the (putative) Fermi liquid can be lowered by deforming the 
Fermi surface. 
This can happen if the Landau function $f_{\bk_F\bk'_F}$ exceeds 
certain limits.
For three-dimensional isotropic systems, a Pomerantschuk instability
occurs if $F_l < - (2l+1)$ for any of the harmonics $l$ \cite{NOZ}. 
In particular, $F_0 < -1$ would imply a negative compressibility, 
i.e.\ an instability with respect to phase separation. 
A stable isotropic Fermi liquid must therefore satisfy the condition 
$$ F_l > - (2l+1) \quad {\rm for \> all} \>\> l          \eqno(4.29) $$
\par
\pp The {\em Cooper instability}\/ \cite{COP} occurs when the system
can lower its energy by forming bound pairs in the presence of 
effective attractive interactions. A perturbative signal for this is a 
run-away flow of the vertex function in the Cooper channel 
$\bar g^{\Lam}_C(\bk_F,\bk'_F) =$ 
$\bar\Gam^{\Lam}(\bk_F,-\bk_F;\bk'_F,-\bk'_F)$.
As discussed already in Sec.\ 2, in isotropic systems this happens
if at least one of the partial wave amplitudes $\bar g^{\Lam}_l$ 
becomes negative. 
Since the $\beta$-function for $\bar g^{\Lam}_C(\bk_F,\bk'_F)$ 
vanishes only if $\bar g^{\Lam}_C(\bk_F,\bk'_F)$ itself vanishes, 
the scattering amplitude in a stable Fermi liquid must satisfy the 
condition
$$ a(\bk_F,-\bk_F;\bk'_F,-\bk'_F) = 0                   \eqno(4.30) $$
According to an argument by Kohn and Luttinger \cite{KL65}, all 
interacting Fermi systems will undergo a Cooper instability (if no 
other) at sufficiently low energy scales, even if the bare 
interactions are purely repulsive. Their intuitive argument has
been substantiated by systematic low-density expansions \cite{FL,KW,KC},
as well as by recent explicit calculations of the Cooper flow with 
the inclusion of irrelevant and higher (than second) orders in 
perturbation theory \cite{SHA94,Tpriv}.
In real systems the energy scale for the Kohn-Luttinger instability 
seems to be extremely small and thus far below the scale where Fermi 
liquid low-energy behavior or other (than Kohn-Luttinger) symmetry
breaking mechanisms set in.
\par

\vfill\eject

\def\pp{\hskip 5mm}
\def\bA{{\bf A}}
\def\bc{{\bf c}}
\def\bk{{\bf k}}
\def\bQ{{\bf Q}}
\def\bq{{\bf q}}
\def\bP{{\bf P}}
\def\bp{{\bf p}}
\def\b0{{\bf 0}}
\def\bi{{\bf i}}
\def\bj{{\bf j}}
\def\bn{{\bf n}}
\def\br{{\bf r}}
\def\bs{{\bf s}}
\def\bS{{\bf S}}
\def\bv{{\bf v}}
\def\alf{\alpha}
\def\eps{\epsilon}
\def\up{\uparrow}
\def\down{\downarrow}
\def\bra{\langle}
\def\ket{\rangle}
\def\sDelta{{\scriptstyle \Delta}}
\def\FS{\partial{\cal F}}
\def\Re{{\rm Re}}
\def\Im{{\rm Im}}
\def\xik{\xi_{\bk}}
\def\cO{{\cal O}}
\def\cD{{\cal D}}
\def\cF{{\cal F}}
\def\cG{{\cal G}}
\def\cZ{{\cal Z}}
\def\Lam{\Lambda}
\def\lam{\lambda}
\def\om{\omega}
\def\dbm{\delta\bar\mu}
\def\scr{\scriptstyle}
\def\sg{\sigma}
\def\Sg{\Sigma}

\vspace*{1cm}
\centerline{\large 5. FORWARD SCATTERING}
\vskip 1cm
\pp In many systems the low-energy physics is dominated by forward
scattering, i.e.\ scattering processes with small momentum transfers
$\bq$. Forward scattering obviously plays a prominant role in systems
with long-range interactions, where the scattering amplitudes diverge
for small momentum transfers. However, forward scattering also 
determines several low-energy properties of systems with short-range
interactions, and may lead to quite subtle effects especially in
low dimensions. For example, forward scattering governs the leading 
low-energy long-wavelength response of a Fermi liquid as well as the 
breakdown of Fermi liquid theory in one-dimensional systems.
\par
\pp In this section we will derive several important 
relations for low-energy correlation functions in forward scattering
dominated systems. These relations follow from the conservation of 
the velocities of particles in scattering processes with small momentum
transfers, combined with the global conservation of charge and spin.
They can also be viewed as {\em asymptotic Ward identities}\/ associated 
with the separate conservation of charge and spin near each given point 
of the Fermi surface in a scattering process with $\bq \sim \b0$ 
\cite{CDM94}. 
\par
\pp An alternative way of treating forward scattering non-perturbatively,
by {\em bosonization}, has been pioneered by Haldane \cite{HAL92}.
This method, which leads to the same results as the Ward identity
approach, will be reviewed in 5.3.
\par
\pp In the following we consider only scattering processes with 
small momentum transfers $|\bq| \ll k_F$.
We assume that the low-energy physics can be described by an effective 
{\em forward scattering action}\/ $S^F$ that involves only 
interactions with small momentum transfers $|\bq| \ll k_F$.
In cases where the low-energy physics is indeed governed by (residual)
forward scattering only, $S^F$ can be constructed from the bare 
action $S$ as follows. 
Elimination of high-energy states maps $S$ to an effective
action $\bar S^{\Lam}$, where $\Lam \ll k_F$ is a band cutoff defining 
a thin shell around the Fermi surface (see Sec.\ 2). 
In forward scattering dominated systems, renormalized interactions
with large momentum transfers are negligible compared to forward 
scattering in the low-energy limit $\Lam \ll k_F$.
We emphasize that in many cases of interest (e.g.\ most Luttinger 
liquids) large-$\bq$ interactions contribute significantly to the map
$S \mapsto \bar S^{\Lam}$, but become irrelevant in the calculation 
of correlation functions in terms $\bar S^{\Lam}$. 
Interactions with large momentum transfers now being negligible, we can
impose an explicit {\em momentum transfer cutoff}\/ of order $\Lam$
and shift the bandwidth cutoff to some larger value $\Lam_0 \gg \Lam$,
without changing the qualitative physics. This step is technically
convenient because the smallness of momentum transfers can be best
exploited if scattering processes are not contrained by a band cutoff.
A quantitatively correct way of pushing the band cutoff back to
larger values is to choose an effective action $S^F$ with small momentum 
transfers $|\bq| < \Lam \ll k_F$ (here $\Lam$ is generally
a {\em smooth}\/ cutoff) and a band cutoff $\Lam_0 \gg \Lam$ 
such that mode elimination from $\Lam_0$ to $\Lam$ would map
$S^F$ back to $\bar S^{\Lam}$.\footnote{In practice one rather analyses
classes of effective actions (with general coupling functions) and 
relates the unknown parameters to suitable physical quantities, which
are accessible via analytic or numerical solutions (for microscopic 
models) or by experiment (for real systems).}
Schematically, $S$, $\bar S^{\Lam}$ and $S^F$ are thus related by
$$ S \> \stackrel{\rm RG \>}{\longrightarrow} 
   \underbrace{\>\> 
    \bar S^{\Lam} \>}_{|\bq| < \Lam \ll k_F} 
   \stackrel{\> \rm RG}{\longleftarrow} 
   \underbrace{\>\> 
    S^F \>}_{|\bq| < \Lam \ll \Lam_0,k_F}    
                                                           \eqno(5.1) $$
All these actions describe the same low-energy physics, but only $S^F$ 
is a suitable starting point for analyzing low-energy scattering 
processes to all orders in the (renormalized) couplings. 
An instructive one-dimensional example for (5.1) is:
$S =$ repulsive non-half-filled Hubbard model, $\bar S^{\Lam} =$ 
g-ology model with band cutoff and back-scattering scaled to zero, 
$S^F =$ Luttinger model (with momentum transfer cutoff).
Note that the band cutoff $\Lam_0$ can be ignored completely if the
single-particle dispersion relation of the effective model is
bounded from below. If it is unbounded (as in the g-ology model),
$\Lam_0$ must be kept to make the model well defined. 
\par\smallskip
\pp Here we consider effective forward scattering actions of the form
$$ S^F = \sum_{\sg} \int_k 
   \psi^*_{k\sg} (ik_0 - \xi_{\bk}) \psi_{k\sg} \> - \>
   {\textstyle {1 \over 2}} \sum_{\sg\sg'} \int_{k,k',q} 
   g_{\bk\bk'}^{\sg\sg'}(q) \>
   \psi^*_{k-q/2,\sg} \psi_{k+q/2,\sg} 
   \psi^*_{k'\!+q/2,\sg'} \psi_{k'\!-q/2,\sg'}            \eqno(5.2) $$
where $\xi_{\bk}$ is a (possibly renormalized) dispersion relation,
with the chemical potential subtracted, i.e.\ $\xi_{\bk}$ vanishes 
on the Fermi surface. The momentum transfers $\bq$ are restricted by 
the condition $|\bq| \ll k_F$. We have suppressed bars (indicating
renormalization) and explicit cutoffs in $S^F$ to keep the notation
readable, but keep of course in mind that $\xi_{\bk}$ and 
$g_{\bk\bk'}^{\sg\sg'}(q)$ in (5.2) differ from the bare dispersion 
and coupling function.
The effective coupling $g_{\bk\bk'}^{\sg\sg'}(q)$ may depend on momentum 
and energy transfers (including retarded interactions), but we restrict 
our analysis to systems where no relevant dependences on the energies 
$k_0$ and $k'_0$ occur. 
Clearly, $S^F$ can describe the low-energy behavior only if 
interactions with large momentum transfers do not generate spontaneous 
symmetry breaking or other dramatic modifications of the effective 
low-energy action. 
Most of the subsequent analysis applies also to Fermi systems coupled
to a gauge field, to be treated in more detail in Sec.\ 9.
\par

\bigskip

\vfill\eject

{\bf 5.1. GLOBAL CHARGE AND SPIN CONSERVATION} \par
\medskip
\pp The effective action $S^F$, (5.2), conserves charge and the z-component
of spin; for $g_{\bk\bk'}^{\sg\sg}(q) = g_{\bk\bk'}^{\sg,-\sg}(q)$ it is
fully spin-rotation invariant. $S_F$ thus respects the corresponding 
conservation laws (assumed to be not broken) of the 
underlying microscopic system. The correlation functions derived from
$S^F$ must therefore satisfy the Ward identities associated with these
conservation laws. In Sec.\ 3 we have already derived Ward identities for 
systems with pure density-density interactions, which apply to most 
familiar microscopic models. 
In the following it will be useful, however, to derive Ward identities 
for (renormalized) correlation functions within the effective low-energy 
theory $S^F$. 
To this end we must now deal with other than density-density interactions,
since the effective coupling function $g_{\bk\bk'}^{\sg\sg'}(q)$ depends
generically on $\bk$ and $\bk'$. In general, such interactions would 
lead to complicated currents with quartic (in the Fermi operators) 
interaction-dependent terms. Fortunately, the current operators
associated with the effective forward scattering action $S^F$ involve 
only a {\em quadratic}\/ interaction dependent term (known as 
{\em "back-flow"}\/ in Fermi liquids).
\par
\smallskip
\pp Let us address {\em charge}\/ conservation first. Since we derive the 
Ward identities in a Hamiltonian language (as in Sec.\ 3), we assume for
a moment that $g_{\bk\bk'}^{\sg\sg'}(q)$ depends only on $\bq$, not on
$q_0$. The resulting identities are clearly not restricted by this 
assumption, and may be derived more generally via path integrals.
\par
\pp To construct a conserved current, we must evaluate the commutator
$[\rho(\bq),H]$, where $\rho(\bq)$ is the density fluctuation operator 
(cf.\ Sec.\ 3.1). The non-interacting part of the Hamiltonian yields
a contribution $[\rho(\bq),H_0] = \bq\cdot\bj_0(\bq)$ with a current
operator
$\bj_0(\bq) = \sum_{\bk,\sg} \bv_{\bk} \> a^{\dag}_{\bk-\bq/2,\sg} \>
a_{\bk+\bq/2,\sg}$ for small $\bq$, while the interaction part reads
$$ [\rho(\bq),H_I] = {\textstyle {1 \over 2 V}} 
   \sum_{\bk\bk'} \sum_{\sg\sg'} \sum_{\bq'}   
   g_{\bk\bk'}^{\sg\sg'}(\bq') \>  \hskip 7cm                          $$
\vskip -4mm
$$ \times
   \Big[ ( a^{\dag}_{\bk-\bq-\bq'\!/2,\sg} \> a_{\bk+\bq'\!/2,\sg} -
           a^{\dag}_{\bk-\bq'\!/2,\sg} \> a_{\bk+\bq+\bq'\!/2,\sg} )
        \> a^{\dag}_{\bk'\!+\bq'\!/2,\sg'} \> a_{\bk'\!-\bq'\!/2,\sg'} $$
\vskip -4mm
$$    + \> a^{\dag}_{\bk-\bq'\!/2,\sg} \> a_{\bk+\bq'\!/2,\sg} \>
         ( a^{\dag}_{\bk'\!-\bq+\bq'\!/2,\sg'} \> 
           a_{\bk'\!-\bq'\!/2,\sg'} -
           a^{\dag}_{\bk'\!+\bq'\!/2,\sg'} \> 
           a_{\bk'\!+\bq-\bq'\!/2,\sg'} )  \Big]            \eqno(5.3) $$
For pure density-density interactions, this interaction term vanishes.
Arbitrary interactions lead obviously to a rather cumbersome expression,
which is quartic in the Fermi operators.
For our forward scattering model (5.2), however, the expression can be
reduced to a much simpler quadratic form. Firstly, non-diagonal terms
in the brackets $(\dots)$ can be neglected because they give rise to
non-diagonal (with four different momenta) quartic contributions to 
the current operator, which are irrelevant in the low-energy limit.
With the remaining diagonal contributions one obtains
 $$ [\rho(\bq),H_I] = 
   {\textstyle {1 \over 2 V}} \sum_{\bk\sg} \sum_{\bk'\sg'}
   g_{\bk\bk'}^{\sg\sg'}(-\bq) \>
   (n_{\bk-\bq/2,\sg} - n_{\bk+\bq/2,\sg}) \>
   a^{\dag}_{\bk'\!-\bq/2,\sg'} \> a_{\bk'\!+\bq/2,\sg'}               $$
\vskip -4mm
$$ + \> {\textstyle {1 \over 2 V}} \sum_{\bk\sg} \sum_{\bk'\sg'}
   g_{\bk\bk'}^{\sg\sg'}(\bq) \>
   a^{\dag}_{\bk-\bq/2,\sg} \> a_{\bk+\bq/2,\sg} \>
   (n_{\bk'\!-\bq/2,\sg'} - n_{\bk'\!+\bq/2,\sg'})          \eqno(5.4) $$
To linear order in $\bq$ this can be written as  
$[\rho(\bq),H_I] = \bq \cdot \bj_I(\bq)$ with a current operator
$$ \bj_I(\bq) = \sum_{\bk,\sg} \bv^I_{\bk\sg}(\bq) \>
    a^{\dag}_{\bk-\bq/2,\sg} \> a_{\bk+\bq/2,\sg}           \eqno(5.5) $$
where
$$ \bv^I_{\bk\sg}(\bq) = 
    - V^{-1} \sum_{\bk'\sg'} g_{\bk\bk'}^{\sg\sg'}(\bq) \> 
    {\partial \bra n_{\bk'\sg'} \ket \over \partial\bk'}    \eqno(5.6) $$
In the last equation, we have replaced the operator $n_{\bk\sg}$ by its
expectation value. To justify this step, let us decompose the 
$\bk'$-integral in a radial integral (over $k'_r$) and a Fermi surface
integral (over $\bk'_F$). Since $S^F$ involves only small momentum
transfers, $\partial n_{\bk'\sg'} / \partial\bk'$ is non-zero only close
to the Fermi surface (for any low-energy state of the interacting system).
A possible dependence of the coupling function 
$g_{\bk\bk'}^{\sg\sg'}(\bq)$ on $k'_r$ can therefore be neglected (it
is clear that such a dependence is irrelevant in the low-energy limit
anyway). Furthermore, since $n_{\bk'\sg'} = 1$ deep inside the Fermi
sea and $n_{\bk'\sg'} = 0$ far outside, one has
$\int dk'_r \> g_{\bk\bk'}^{\sg\sg'}(\bq) \> \partial n_{\bk'\sg'} / 
\partial\bk' = - g_{\bk\bk'_F}^{\sg\sg'}(\bq) \> \hat\bn_{\bk'_F}$,
where $\hat\bn_{\bk'_F}$ is a normal (with respect to the Fermi 
surface) unit vector at $\bk'_F$.
The $k'_r$ integral over the operator expression thus yields a c-number,
and the same c-number is obtained when replacing $n_{\bk'\sg'}$ by its
expectation value.
\par
\pp We have thus derived a continuity equation of the form (3.3) with
an interaction-dependent current operator 
$$ \bj(\bq) = \bj_0(\bq) + \bj_I(\bq)                       \eqno(5.7) $$
\par
\pp The continuity equation implies Ward identities for the 
current-current correlator 
$ J^{\mu\nu}(q) = - {1 \over V} \> \bra j^{\mu}(q) \> j^{\nu}(-q) \ket $
and the vertex part 
$\Gamma^{\mu}_{\sg}(p;q) = \bra j^{\mu}(q) \> a_{p-q/2,\sg} \> 
a^{\dag}_{p+q/2,\sg} \ket_{tr}$ 
precisely as in Sec.\ 3, with only one modification: the interaction
part $\bv^I_{\bk\sg}(\bq)$ has to be added to the velocity $\bv_{\bk}$ 
entering the function $\bc(\bq)$, (3.12).
\par
\pp We now derive a very useful Ward identity for the {\em irreducible}\/
vertex part
$$ \Lam^{\mu}_{\sg}(p;q) = \bra j^{\mu}_0(q) \> 
   a_{p-q/2,\sg} \> a^{\dag}_{p+q/2,\sg} \ket_{tr}^{irr}    \eqno(5.8) $$
which is constructed with the {\em non-interacting}\/ (or 
"quasi-particle") part of the current operator, in contrast to the 
function
$$ {\Lam}'^{\mu}_{\sg}(p;q) = \bra j^{\mu}(q) \> 
   a_{p-q/2,\sg} \> a^{\dag}_{p+q/2,\sg} \ket_{tr}^{irr}    \eqno(5.9) $$  
defined with the full current (in Sec.\ 3 no such distinction was
needed, since there $\bj_I$ vanished).  
The vertex part $\Gamma^{\mu}$ and its irreducible counterpart 
${\Lam'}^{\mu}$ are obviously related by the Dyson equation (see also
Fig. 5.1)
$$ \Gamma^{\mu}_{\sg}(p;q) =  {\Lam'}^{\mu}_{\sg}(p;q) +
   {\textstyle {1 \over V}} \sum_{\bk'\sg'} \sum_{\bk''\sg''}
   J^{\mu,\bk'\sg'}(q) \> g_{\bk'\bk''}^{\sg'\sg''}(\bq) \>
   \Lam^{\bk''\sg''}_{\sg}(p;q)                            \eqno(5.10) $$
where 
$$ J^{\mu,\bk\sg}(q) = 
   - \int {\textstyle{dk_0 \over 2\pi}} \> \bra j^{\mu}(q) \> 
   a^{\dag}_{k+q/2,\sg} \> a_{k-q/2,\sg} \ket  =
   \int {\textstyle{dk_0 \over 2\pi}} \>
   \Gamma^{\mu}_{\sg}(k;q) \> 
   G_{\sg}(k\!+\!q/2) \> G_{\sg}(k\!-\!q/2)                \eqno(5.11) $$ 
and
$$ \Lam^{\bk\sg'}_{\sg}(p;q) = 
   \int {\textstyle{dk_0 \over 2\pi}} \> \bra 
   a^{\dag}_{k-q/2,\sg'} \> a_{k+q/2,\sg'} \>
   a_{p-q/2,\sg} \> a^{\dag}_{p+q/2,\sg} \ket_{tr}^{irr}   \eqno(5.12) $$
The Ward identity (3.15) for $\Gamma^{\mu}$ yields
$$ (iq_0,\bq)_{\mu} J^{\mu,\bk\sg}(q) = \sum_{\sg} 
   (\bra n_{\bk-\bq/2,\sg} \ket - \bra n_{\bk+\bq/2,\sg} \ket)  
   \to - \bq \cdot {\partial \bra n_{\bk\sg} \ket \over 
   \partial\bk}                                            \eqno(5.13) $$
Hence, applying $(iq_0,\bq)_{\mu}$ to the Dyson equation (5.10) one
obtains
$$ (iq_0,\bq)_{\mu} \Gamma^{\mu}_{\sg}(p;q) =
   (iq_0,\bq)_{\mu} {\Lam'}^{\mu}_{\sg}(p;q) +
   \bq \cdot \sum_{\bk'\sg'} \bv^I_{\bk'\sg'}(\bq) 
   \Lam^{\bk'\sg'}_{\sg}(p;q)  =
   (iq_0,\bq)_{\mu} {\Lam}^{\mu}_{\sg}(p;q)                \eqno(5.14) $$
i.e. the "back-flow" terms cancel. The Ward identity for $\Gamma^{\mu}$ 
thus implies an identity of the same form for $\Lam^{\mu}$:
$$ (iq_0,\bq)_{\mu} \Lam^{\mu}_{\sg}(p;q) =
   G^{-1}_{\sg}(p\!+\!q/2) - G^{-1}_{\sg}(p\!-\!q/2)       \eqno(5.15) $$
This result generalizes (3.21) to more general coupling functions
than just $g(\bq)$.
\par
\smallskip
\pp From {\em spin}\/ conservation one can derive analogous relations.
In the case of full spin-rotation invariance one obtains
$$ (iq_0,\bq)_{\mu} \Lam^{a\mu}_{\sg\sg'}(p;q) = \>
   s^a_{\sg\sg'} \> 
   [G^{-1}_{\sg'}(p\!+\!q/2) - G^{-1}_{\sg}(p\!-\!q/2)]    \eqno(5.16) $$
for the irreducible spin vertex part, defined with the non-interacting
spin current operator $\bj^a_0(\bq) = \sum_{\bk} \sum_{\sg,\sg'} \>
\bv_{\bk} \> s^a_{\sg\sg'} \> a^{\dag}_{\bk-\bq/2,\sg} \> 
a_{\bk+\bq/2,\sg'}$ (cf.\ Sec.\ 3.2.).
If only $S^z$ is conserved, (5.16) holds only for $a=z$.
\par   
        
\bigskip

{\bf 5.2. VELOCITY CONSERVATION} \par
\medskip
\pp For small momentum transfers $\bq$, the velocity of a particle
varies little in a scattering process: 
$\sDelta \bv_{\bk} = \bv_{\bk+\bq} - \bv_{\bk} \sim 
(\bq \cdot \nabla_{\bk}) \bv_{\bk}$. In particular, for a continuum
system with $\eps_{\bk} = \bk^2/2m$ one has 
$\sDelta v_{\bk}/v_{\bk} \sim |\bq|/k_F$ near the Fermi surface.
To leading order in the small parameter $|\bq|/k_F$, the velocity is
conserved in the scattering process. This asymptotic conservation law
leads to systematic cancellations and other asymptotic Ward identities, 
which hold (at least) to leading order in $|\bq|/k_F$ and to any order
in the coupling constant.
\par

\bigskip

{\bf 5.2.1. Loop cancellation and response functions} \par
\medskip
\pp An important consequence of velocity conservation is
{\em loop-cancellation}: 
For small momentum transfers, {\em Feynman diagrams involving fermionic 
loops with more than two insertions cancel each other}, i.e. the
sum over permutations of various orderings of the insertions attached
to the loop is smaller than what one would expect on the basis of
naive power-counting.
This cancellation has been noticed long ago in the Luttinger model,
where loops cancel completely for any $\bq$ \cite{DL}, while in higher 
dimensions loop-cancellation has been assumed and exploited at least 
implicitly in many works on singular forward scattering. 
An explicit proof in higher dimensions has been constructed recently by
Kopietz et al. \cite{KHS} and by one of us \cite{MET95}. 
Here we will provide a slightly different argument, avoiding the 
decomposition of momentum space in small sectors. 
\par
\pp Consider a fermionic loop with $N$ insertions as shown in Fig.\ 5.2. 
The insertions may be external density or current vertices, or may
be due to interactions that connect the loop to the rest of a bigger
Feynman diagram. The value of the diagram can be written as an
integral over the function
$$ f(q_1,\dots,q_N) = 
   \int_{k} h(\bk_1,\dots,\bk_N;q_1,\dots,q_N) \>
   G_0(k_1) \dots G_0(k_N)                                 \eqno(5.17) $$
where $k_1 := k$, and $k_2,\dots,k_N$ are given in terms of 
$k$ and the momentum transfers $q_{\nu}$ by momentum conservation. 
The fermion lines are associated with the propagators
$G_0(k) = [ik_0 - \xi_{\bk}]^{-1}$ derived from the quadratic part 
of the effective action $S^F$.
The function $h(\bk_1,\dots,\bk_N;q_1,\dots,q_N)$ depends in general on 
the rest of the Feynman diagram; $h$ is independent of $\bk_{\nu}$ if all 
insertions on the loop are pure density-vertices (external or due to 
density-density interactions). 
Eventually the momenta $q_{\nu}$ associated with interaction lines are
to be integrated, while those associated with external density or current 
vertices remain fixed.
In addition to the diagram in Fig.\ 5.2, one has contributions from all 
possible permutations of the insertions attached to the loop. 
\par
\pp For $|\bq_{\nu}| \ll k_F$, all the momenta $\bk_{\nu}$ on the 
fermion lines are almost equal, i.e. $\bk_{\nu} \approx \bk$ for all 
$\nu$. Hence, all the velocities on the lines in the loop are
approximately equal to $\bv_{\bk} = \nabla\xi_{\bk}$ and the function 
$h(\bk_1,\dots,\bk_N;q_1,\dots,q_N)$ can be replaced by 
$h(\bk,\dots,\bk;q_1,\dots,q_N)$, to leading order in $|\bq_{\nu}|/k_F$.
Corrections are proportional to $\bq_{\nu}$ multiplied by the gradients of
$\bv_{\bk}$ and $h(\bk_1,\dots,\bk_N;q_1,\dots,q_N)$ in $\bk_{\nu}$-space. 
\par
\pp From now on, one can simply follow the proof of loop-cancellation
in the one-dimens\-ional Luttinger model, where the Fermi velocity is a 
constant and the function $h$ is independent of $\bk_{\nu}$ \cite{BOH}. 
The basic idea is to consider a loop with $N-1$ insertions with momenta
$q_1,\dots,q_{N-1}$ arranged in an arbitrary way, and sum over all 
possible positions of an additional insertion with momentum $q_N$
(summing thus only a subset of permutations of all the insertions).
Using the basic relation
$$ G_0(k+q/2) \> G_0(k-q/2) = 
   {G_0(k-q/2)- G_0(k+q/2) \over
   iq_0 - \bv_{\bk}\cdot\bq + \cO(|\bq|^3)}                \eqno(5.18) $$
for the product of propagators associated with the two lines connected
to the extra-insertion, one finds that there is a complete cancellation
of terms in the sum over all positions, to leading order in $\bq_{\nu}$.
Note that (5.18) is exact without corrections of order $|\bq|^3$ in the
Luttinger model and also in (unrenormalized) continuum systems with
quadratic dispersion $\eps_{\bk} = \bk^2/2m$.
In Appendix A we show explicitly how these cancellations work
for the case $N=3$. 
\par
\pp Note that the above argument does not always imply that the sum over
all permutations vanishes in the limit $q_{\nu} \to 0$, because
the corrections of order $|\bq_{\nu}|$ in the expansion of the velocities 
and the function $h$ are multiplied by factors which are generally
singular for small $q_{\nu}$. 
What we have shown, however, is that cancellations reduce the 
contribution of loops with respect to what one would get from naive
power-counting estimates. The naive degree of divergence of
the function $f(q_1,\dots,q_N)$ in the limit $q_{\nu} \to 0$ is 
given by $D_f = D_h + N - 2$. Here $D_h$ is the behavior of the
function $h$, each of the $N$ propagators increases $D_f$ by one, while
the $k$-integral reduces the degree of divergence by two (the 
codimension of the Fermi surface in energy-momentum space).
However, by virtue of the above cancellation, at least 
for systems with short-range interactions (where $g_{\bk\bk'}(q)$
is finite for $q \to 0$) loops with $N > 2$ and small momentum transfers
are irrelevant in any dimension, while according to naive power-counting 
they would yield finite contributions to the correlation functions.
As in one-dimensional systems, the cancellation guarantees that their
contribution is suppressed by a positive power of $\Lam$.
The situation is not so clear for systems where the coupling function or
scattering amplitude is singular for small momentum transfers, 
especially if strong singularities persist even after screening by
polarization effects has been taken into account; this latter problem
occurs in Fermi systems coupled to a gauge field (see Sec.\ 9). In such
cases the above argument demonstrates only a cancellation of leading
singularities appearing in single Feynman diagrams. A recent explicit
low-order calculation \cite{KFWL} confirms in fact this cancellation
and suggests the irrelevance of loops with $N>2$ even in gauge systems.
A general proof of irrelevance, however, would require a better control
of corrections due to velocity fluctuations.
\par
\pp More detailed results are available for the simple loop
$$ l(q_1,\dots,q_N) = 
   \int_k  G_0(k_1) \dots G_0(k_N)                        \eqno(5.19) $$
which is relevant for cases where the function $h$ depends only on
$q_1,\dots,q_N$, not on $\bk$. It is easy to show that \cite{HK}
$$ l(0,\dots,0) \equiv \int_k G_0(k) \dots G_0(k) = 
   - {1 \over (N\!-\!1)!}
   \left.{\partial^{N-2} N(\xi) \over \partial\xi^{N-2}}
   \right|_{\xi=0}                                        \eqno(5.20) $$
where $N(\xi)$ is the density of states per spin (of the non-interacting
system).
Note that the limit $q_{\nu} \to 0$ is not unique and $l(0,\dots,0)$
as defined above reflects the so-called static limit where the
frequency components $q_{\nu 0}$ go to zero first. Note also that
in this particular limit each single loop with $N>2$ is already much 
smaller than the power-counting estimate, and no further cancellations 
occur in the sum over permutations. In one dimension with a linearized 
$\eps_{\bk}$ and in two dimensions with $\eps_{\bk} = \bk^2/2m$ the
density of states is a constant and $l(0,\dots,0)$ thus vanishes for
$N>2$.
For two-dimensional systems with $\eps_{\bk} = \bk^2/2m$ the problem
of evaluating $l(q_1,\dots,q_N)$ has been recently reduced to elementary
integrals for any $N$ \cite{FKST}. 
An explicit (possibly lengthy) evaluation of these integrals with an 
analysis of the detailed behavior for small $q_{\nu}$ has not yet been 
done.
\par
\smallskip
\pp As a consequence of loop-cancellation, the effective interaction
(including polarization) and response functions for 
small $\bq$ are given by the {\em random phase approximation}\/ (RPA),
even in cases (such as Luttinger liquids) where contributions beyond
RPA are not simply suppressed by a small phase space and yield singular
corrections to other quantities such as the single-particle Green 
function.
The {\em effective interaction}\/ $D$ is obtained by summing the 
diagrams illustrated in Fig.\ 5.3, which is equivalent to the integral
equation
$$ D_{\bk\bk'}^{\sg\sg'}(q) = g_{\bk\bk'}^{\sg\sg'}(q) +
   \sum_{\sg''} \int_{k''} g_{\bk\bk''}^{\sg\sg''}(q)
   G_0(k''\!-\!q/2) G_0(k''\!+\!q/2) 
   D_{\bk''\bk'}^{\sg''\sg'}(q)                           \eqno(5.21) $$
In general, the bubbles in Fig.\ 5.3 must be dressed by self-energy
and vertex corrections. However, these corrections involve fermionic
loops with more than two insertions and must therefore cancel each 
other for small $\bq$ and $\Lam$, at least to leading order in these 
small parameters.
\par
\pp The charge current-current response function $J^{\mu\nu}(q)$ 
for small $q$ can be constructed from the effective interaction in the 
way illustrated in Fig.\ 5.4, i.e.\
$$ J^{\mu\nu}(q) = J^{\mu\nu}_0(q) \> + \> 
   \sum_{\sg\sg'} \int_{k,k'} \lam^{\mu}(\bk) \> 
   G_0(k\!-\!q/2) G_0(k\!+\!q/2)                          \hskip 3cm $$
\vskip -6mm
$$ \hskip 2cm \times \> D_{\bk\bk'}^{\sg\sg'}(q) \> 
   G_0(k'\!-\!q/2) G_0(k'\!+\!q/2) \> \lam^{\nu}(\bk')    \eqno(5.22) $$
where $\lam^{\mu}(\bk) = (1,\bv_{\bk})$ and 
$$ J^{\mu\nu}_0(q) = 2 \int_k \lam^{\mu}(\bk) \>
   G_0(k\!-\!q/2) G_0(k\!+\!q/2) \> \lam^{\nu}(\bk)       \eqno(5.23) $$
The factor $2$ is due to the spin sum. Again, vertex and self-energy 
corrections cancel each other in the loops. 
\par
\pp Similarly, the spin current-current response function 
$J^{a\mu,b\nu}(q)$ can be written as
$$ J^{z\mu,z\nu}(q) = J^{z\mu,z\nu}_0(q) \> + \> 
   \sum_{\sg\sg'} \int_{k,k'} \lam^{z\mu}_{\sg\sg}(\bk) \>
   G_0(k\!-\!q/2) G_0(k\!+\!q/2)                           \hskip 3cm $$
\vskip -6mm
$$ \hskip 2cm \times \> D_{\bk\bk'}^{\sg\sg'}(q) \>
   G_0(k'\!-\!q/2) G_0(k'\!+\!q/2) \>
   \lam^{z\nu}_{\sg'\sg'}(\bk')                           \eqno(5.24) $$
for small $q$, where $\lam^{z\mu}_{\sg\sg'}(\bk) = 
s^z_{\sg\sg'}(1,\bv_{\bk})^{\mu}$ and 
$$ J^{z\mu,b\nu}_0(q) = \sum_{\sg} 
   \int_k \lam^{z\mu}_{\sg\sg}(\bk) \> G_0(k\!-\!q/2) 
   G_0(k\!+\!q/2) \> \lam^{z\nu}_{\sg\sg}(\bk)            \eqno(5.25) $$
\par
\bigskip

{\bf 5.2.2. Density and current vertex} \par
\medskip
\pp Asymptotic velocity conservation implies a simple asymptotic Ward
identity for the irreducible charge vertex $\Lam^{\mu}_{\sg}(p;q)$, 
defined in eq.\ (5.8).
\par
\pp The perturbation expansion for $\Lam^{\mu}$ can be represented by
Feynman diagrams as in Fig.\ 5.5, where the zeroth-order vertex is
$\lam^{\mu}(\bk) = (1,\bv_{\bk})^{\mu}$. We recall that one-interaction
reducibe diagrams do not contribute.
If the external momentum transfer $\bq$ as well as the momentum 
transfers $\bq_{\nu}$ due to interactions are all small, one has the 
following simplifications (to leading order in $\bq$ and $\bq_{\nu}$).
Diagrams containing fermion loops with more than two insertions cancel 
each other as a consequence of loop cancellation. An example for two
such cancelling diagrams is given in Fig.\ 5.6. 
In the remaining diagrams the vertex $\lam^{\mu}(\bk)$ must lie on
the fermion line connecting the two fermionic external points of
$\Lam^{\mu}_{\sg}(p;q)$ (carrying momenta $p \pm q/2$). 
The momentum $\bk$ passing through
the vertex $\lam^{\mu}(\bk)$ differs from $\bp$ only by certain
linear combinations of the small momentum transfers $\bq_{\nu}$
caused by the interactions. Thus, to leading order in $\bq_{\nu}$
we can replace $\lam^{\mu}(\bk)$ by $\lam^{\mu}(\bp)$ in each diagram
contributing to $\Lambda^{\mu}(p;q)$. This implies that the current
vertex ${\bf\Lambda}$ is related to the density vertex $\Lambda^0$
by the simple asymptotic identity \cite{CDM94,ILA}
$$ {\bf\Lambda}_{\sg}(p;q) = 
   \bv_{\bp} \Lambda^0_{\sg}(p;q)                         \eqno(5.26) $$
Combining this with the Ward identity (5.15) associated with charge
conservation, one obtains \cite{CDM94,ILA}
$$ \Lambda^0_{\sg}(p;q) = 
   {G^{-1}_{\sg}(p\!+\!q/2) - G^{-1}_{\sg}(p\!-\!q/2) \over
   iq_0 - \bv_{\bp}\cdot\bq}                              \eqno(5.27) $$
These asymptotic Ward identities express the charge vertex 
$\Lambda^{\mu}$ uniquely in terms of the propagator $G$. They are
exact in the one-dimensional Luttinger model \cite{DL}, while in
general they hold only asymptotically for small momentum transfers, 
with the additional proviso that singular interactions do not 
overpower the loop-cancellation.
\par
\pp For the z-component of the irreducible spin vertex, one obtains
essentially the same results, i.e.\
$$ {\bf\Lambda}^z_{\sg\sg'}(p;q) = 
   \bv_{\bp} \Lambda^{z0}_{\sg\sg'}(p;q)                 \eqno(5.28) $$
\vskip -3mm and \vskip -3mm
$$ \Lambda^{z0}_{\sg\sg'}(p;q) = s^z_{\sg\sg'} \>
   {G^{-1}_{\sg'}(p\!+\!q/2) - G^{-1}_{\sg}(p\!-\!q/2) \over
   iq_0 - \bv_{\bp}\cdot\bq}                             \eqno(5.29) $$
\par
\pp Corrections to the asymptotic identities (5.27) and (5.29) can be
parametrized by adding a (generally unknown) complex function $Y(p;q)$
of order $\cO(\bq^2,\bq_{\nu}^2)$ to the denominator $iq_0 - 
\bv_{\bp}\cdot\bq$ on the right hand side \cite{CDM94,ILA}.
\par
\pp It is easy to see that the above Ward identities for $\Lam$ imply
that polarization bubbles are not dressed by interactions (as follows
already from loop cancellation). For example, consider the charge 
density polarization, which can always be written as
$$ \Pi(q) = \sum_{\sg} \int_p \Lam^0_{\sg}(p;q) \> 
   G_{\sg}(p\!+\!q/2) \> G_{\sg}(p\!-\!q/2)              \eqno(5.30) $$
Inserting (5.27) and performing the $p_0$-integral, one obtains
$$ \Pi(q) = \sum_{\sg} \int_{\bp} 
   {\bra n_{\bp-\bq/2,\sg} \ket - \bra n_{\bp+\bq/2,\sg} \ket
   \over iq_0 - \bv_{\bp}\cdot\bq} =
   \Pi_0(q)                                              \eqno(5.31) $$
where $\Pi_0(q)$ is the bubble for the non-interacting system.
The last equation holds since 
$\bra n_{\bp-\bq/2,\sg} \ket - \bra n_{\bp+\bq/2,\sg} \ket =
 - \bq \cdot \partial \bra n_{\bp\sg} \ket / \partial\bp$ for small
$\bq$ and $\int dp_r \> \partial \bra n_{\bp\sg} \ket / \partial\bp =
 - \hat\bn_{\bp_F}$ for any momentum distribution in the forward 
scattering model.
\par

\bigskip

{\bf 5.2.3. Single-particle propagator} \par
\medskip
\pp The self-energy correction due to small-$\bq$ scattering obeys
the Dyson equation represented diagrammatically in Fig.\ 5.7. 
Algebraically, this equation reads
$$ \Sg_{\sg}(p) = - \sum_{\sg'} \int_{p'} 
   D_{\bp\bp}^{\sg\sg'}(p-p') \> G_{\sg}(p') \> 
   \Lam^0_{\sg'\sg}[(p\!+\!p')/2;p'\!-\!p]               \eqno(5.32) $$
\vskip -3mm where \vskip -3mm
$$ \Lam^0_{\sg'\sg}(p;q) = \bra \rho_{\sg'}(q) 
   a_{p-q/2,\sg} a^{\dag}_{p+q/2,\sg} \ket_{tr}^{irr}    \eqno(5.33) $$
is the irreducible density vertex for fermions with spin orientation 
$\sg$, with $\rho_{\sg}(\bq) = \sum_{\bk} a^{\dag}_{\bk-\bq/2,\sg} 
a_{\bk+\bq/2,\sg}$.
The effective interaction $D$ is given in terms of the coupling
function $g_{\bk\bk'}^{\sg\sg'}(q)$ by Eq.\ (5.21). 
Note that (5.32) holds only if the contributing momentum transfers
$\bq = \bp - \bp'$ are small compared to the scale on which 
$g_{\bk\bk'}^{\sg\sg'}(q)$ varies as a function of $\bk$ and $\bk'$.
\par
\pp The asymptotic Ward identies (5.27) and (5.29) for the charge and
spin vertex imply the identity
$$ \Lam^0_{\sg'\sg}(p;q) = \delta_{\sg'\sg} \>
   {G^{-1}_{\sg}(p\!+\!q/2) - G^{-1}_{\sg}(p\!-\!q/2) \over
   iq_0 - \bv_{\bp}\cdot\bq}                             \eqno(5.34) $$
Note that $\Lam^0_{\sg,-\sg}$ must vanish as a direct consequence of
loop cancellation.
Now $\Lam^0_{\sg'\sg}(p;q)$ can be eliminated from (5.32) in favor
of $G_{\sg}$. Combining the result of this substitution with the exact
Dyson equation $G = G_0 + G_0 \Sg G$, one obtains a linear integral 
equation that determines $G$ in terms of $D$ \cite{CDM94}:
$$ (ip_0 - \xi_{\bp}) G_{\sg}(p) \> = \> 1 \> - \>
   \int_{p'} {D_{\bp\bp}^{\sg\sg}(p\!-\!p') \over 
   ip_0 \!-\! ip'_0 - \bv_{(\bp+\bp')/2}\cdot(\bp\!-\!\bp')} \>
   G_{\sg}(p') \> + \> X_{\sg}(p)G_{\sg}(p)              \eqno(5.35) $$
where
\vskip -3mm
$$ X_{\sg}(p) =  \int_{p'} {D_{\bp\bp}^{\sg\sg}(p\!-\!p') \over 
   ip_0 \!-\! ip'_0 - \bv_{(\bp+\bp')/2}\cdot(\bp\!-\!\bp')}   
                                                         \eqno(5.36) $$
\par
\pp Let us now specialize to rotationally invariant systems, to arrive at
a more explicit solution. In this case the propagator depends only on
two variables, the energy $p_0$ and the radial momentum component
$p_r = |\bp|-k_F$, i.e.\ we can write $G(p) = G(p_0,p_r)$. Furthermore
$|\bv_{\bk}| = v_F$ is now constant, and for small $\bq = \bp - \bp'$ we
can use the relation
$$ \bv_{(\bp+\bp')/2}\cdot(\bp-\bp') \> = \> 
   v_F (p_r-p'_r) + \cO(\bq^2)                           \eqno(5.37) $$
For each momentum $\bp$ near the Fermi surface, the momentum transfer 
$\bq$ can be decomposed in a radial (or normal) component $\bq_r$ and 
a tangential component $\bq_t$, via
$$ \bq_r = q_r {\bf \hat p} \> , \quad 
   q_r = \bq \cdot {\bf \hat p} \> , \quad
   \bq_t = \bq - \bq_r                                   \eqno(5.38) $$
where ${\bf \hat p}$ is a unit vector parallel to $\bp$ (i.e.\ normal
to the Fermi surface near $\bp$).
For small $\bq_t$, we have $p_r - p'_r = q_r + \cO(\bq_t^2)$, and
$|\bq_t|/k_F$ is the angle between $\bp$ and $\bp'$.
For isotropic systems and small $\bq$, the effective interaction
$D_{\bp\bp}^{\sg\sg}(q_0,\bq)$ can be parametrized by $q_0$, $q_r$ and 
$q_t = |\bq_t|$, i.e.
$$ D_{\bp\bp}^{\sg\sg}(q_0,\bq) = D(q_0,q_r,q_t)  \quad
   {\rm for} \quad \bq \sim \b0                          \eqno(5.39) $$
Note that $q_t \geq 0$ by definition, while $q_r$ can also be negative.
The crucial point is that small tangential momentum transfers enter the
integrand in Eqs.\ (5.35) and (5.36) only via the effective interaction 
$D$. We therefore define a $q_t$-averaged effective interaction
$$ \bar D^{\Lam_t}(q_0,q_r) = {S_{d-1} \over (2\pi)^{d-1}}
   \int_0^{\Lam_t} dq_t \> q_t^{d-2} \> 
    D(q_0,q_r,q_t)                                       \eqno(5.40) $$
where $S_{d-1} = 2\pi^{d/2}/\Gamma(d/2)$ is the surface of the
$d$-dimensional unit-sphere, and $\Lam_t$ is a cutoff for tangential
momentum transfers. The factor $q_t^{d-2}$ is due to the phase space
in the $(d\!-\!1)$-dimensional tangential space spanned by $\bq_t$. 
We will see that leading low-energy terms are often independent of 
the cutoff $\Lam_t$, confirming the dominance of forward scattering. 
In the following we will usually suppress the $\Lam_t$-dependence of 
$\bar D$ in our notation.
Note that in one dimension, where no tangential degrees of freedom
exist, eq.\ (5.40) reduces to $\bar D = D$. 
\par
\pp In terms of $\bar D$, the function $X_{\sg}(p)$ in (5.36) can be 
written as
$$ X(p) = \int_{q_0,q_r} {\bar D(q_0,q_r) \over 
   iq_0 - v_F q_r}                                       \eqno(5.41) $$
where $q_0 = p_0-p'_0$ and $q_r = p_r-p'_r$. This is obviously a 
constant ($p$-independent), which can been absorbed by shifting the 
chemical potential in $\xi_{\bk} = \eps_{\bk} - \mu$ (keeping thus 
the density and the Fermi surface fixed). 
The integral equation for the propagator now simplifies to \cite{CDM94}
$$ (ip_0 - v_F p_r) \> G(p_0,p_r) \> = \> 1 \> - \> 
   \int_{p_0',p_r'} {\bar D(p_0\!-\!p_0',p_r\!-\!p_r') \over 
   ip_0 - ip_0' - v_F (p_r\!-\!p_r')} \> G(p'_0,p'_r)    \eqno(5.42) $$
where we have omitted the (now unimportant) spin variables.
This equation has precisely the same form as the equation obtained by 
Dzyaloshinkii and Larkin \cite{DL} for the propagator of the 
one-dimensional Luttinger model, with $D$ substituted by $\bar D$. 
Thus, to leading order in small momentum transfers, the single-particle 
propagator is given by a {\em one-dimensional}\/ equation of motion
corresponding to a ficticious Luttinger model with an effective 
interaction given by $\bar D(q_0,q_r)$.
\par
\smallskip
\pp The solution of the integral equation (5.42) proceeds as in one 
dimension \cite{DL}. 
While Eq.\ (5.42) holds also at small finite temperatures (with
$p'_0$ to be summed over the Matsubara frequencies), we present the 
solution only at zero temperature. 
Continuing (5.42) to real frequencies yields
$$ (\xi - v_F p_r) \> G(\xi,p_r) \> = \> 1 \> + \> 
   \int_{\xi',p_r'} {i \bar D(\xi\!-\!\xi',p_r\!-\!p_r') \over 
   \xi - \xi' - v_F (p_r \!-\! p_r')} \> G(\xi',p'_r)    \eqno(5.43) $$
We transform this equation to real space-time by defining
$$ G(t,r) = \int {d\xi \over 2\pi} \int {dp_r \over 2\pi}
   \> G(\xi,p_r) \> e^{ip_r r - i\xi t}                  \eqno(5.44) $$
Note that $G(t,r)$ is {\em not}\/ the $(d+1)$-dimensional Fourier
transform of $G(\xi,\bp)$. The integral equation (5.43) transfroms to 
a partial differential equation for $G(t,r)$, namely
$$ (\partial_t + v_F \partial_r) G(t,r) = 
   \delta(t)\delta(r) + K(t,r) G(t,r)                    \eqno(5.45) $$
where $K(t,r)$ is the Fouriertransform of 
$\bar D(\om,q_r)/(\om \!-\! v_Fq_r)$. 
This differential equation is solved by the ansatz
$$ G(t,r) = e^{L(t,r)-L_0(r-v_Ft)} G_0(t,r)              \eqno(5.46) $$ 
where $L(t,r)$ is a solution of the inhomogeneous linear differential 
equation
$$ (\partial_t + v_F \partial_r) L(t,r) = K(t,r)         \eqno(5.47) $$
i.e.\ the Fouriertransform of $L(\om,q_r) = 
i\bar D(\om,q_r)/(\om-v_Fq_r)^2$. More explicitly, one obtains
$$ L(t,r) = \int {d\om \over 2\pi} \int {dq_r \over 2\pi} \>
   {i\bar D(\om,q_r) \over [\om-v_Fq_r+i0^+s(\om)]^2} \>
   e^{iq_r r - i\om t}                                   \eqno(5.48) $$
where $s(.)$ is the sign-function. The function $L_0(r\!-\!v_Ft)$ is 
a solution of the corresponding homogeneous equation, which is chosen 
such that boundary conditions in the complex $t$-plane are satisfied: 
Analyticity of $G(t,r)$ in the second and forth quadrant of the
$t$-plane and
$$ G(0^+,r) - G(0^-,r) = -i\delta(r)                     \eqno(5.49) $$
The non-interacting propagator in $(t,r)$-representation reads
$$ G_0(t,r) = {1 \over 2\pi} \> 
   {1 \over r - v_Ft + i0^+s(t)}                         \eqno(5.50) $$
and satisfies of course the boundary conditions. Analyticity of $G(t,r)$
is automatically guaranteed (even for $L_0(r \!-\! v_Ft) = 0$), since 
$\bar D$ is a time-ordered function, while the condition (5.49) can be 
satisfied by choosing 
$$ L_0(r \!-\! v_Ft) = L(0,0) =: L_0                     \eqno(5.51) $$
i.e.\ a constant.
The extension of the above solution to anisotropic systems is 
straightforward: one simply has to replace $v_F$ by $v_{\bk_F}$ and 
add a label $\bk_F$ to $\bar D$, L and G. 
\par
\pp The {\em momentum distribution}\/ $n_{\bp}$ is obtained by Fourier 
transforming $G(0^-,r)$, i.e.
$$ n_{\bp} = -iG(0^-,\bp) =
   -i \int_{-\infty}^{\infty} dr \> 
   G(0^-,r) \> e^{-i p_r r}                              \eqno(5.52) $$ 
The {\em density of states}\/ (per spin) can be calculated by Fourier
transforming $G(t,0)$, i.e.
$$ N(\xi) = \pi^{-1} \> \left| \> \Im \int_{-\infty}^{\infty} dt 
   \> G(t,0) \> e^{i\xi t} \> \right|                     \eqno(5.53) $$
In Appendix B we derive an expression for $L(t,r)$ in terms of the 
spectral function $\bar\Delta$ associated with $\bar D$. With that
representation one can check general analytic properties such as the
reality of $n_{\bp}$. 
\par
\smallskip
\pp In $d>1$ the above result for the propagator $G$ is plagued by 
unphysical exponential singularities in the special limit $r,t \to 
\infty$ with $r/t \to v_F$, corresponding to the quasi-particle regime
$p_r,\xi \to 0$ with $\xi/p_r \to v_F$. These singularities are 
associated with the double pole in $\om = v_F q_r$ in 
the expression (5.48) for the exponent $L(t,r)$. In the one-dimensional
Luttinger model this double pole is reduced to a simple pole by the
effective interaction in the numerator, which vanishes linearly for
$\om \to v_F q_r$ in that case. 
In general, however, $\bar D(\om,q_r)$ is finite for $\om = v_F q_r$, 
but the double pole should then be cut off by neglected terms of order 
$\bq^2$ in the denominator, especially in the Ward identities (5.27)
and (5.29). Although these quadratic corrections are usually subleading, 
they become important when the leading term $\om - v_F q_r$ in the 
denominator vanishes. 
Indeed, for $r,t \to \infty$ with $r/v_Ft = 1$, the function $L(t,r)$ 
is dominated by Fourier components $L(\om,q_r)$ with $\om \sim v_F q_r$,
which is precisely the situation where corrections should be important, 
and would cut off the pole in the Ward identities. 
In all other cases one integrates over many values for the ratio 
$\om/v_F q_r$, and quadratic corrections are thus negligible. 
In particular, the results for the momentum distribution and the 
density of states, obtained from $G(0,r)$ and $G(t,0)$, respectively,
are stable.
\par
\smallskip
\pp Asymptotic Ward identities and a solution for the fermion 
propagator in Fermi systems with gauge-fields will be discussed in 
Sec.\ 9.
\par 

\bigskip

{\bf 5.3. BOSONIZATION} \par
\medskip
\pp An alternative way of treating systems whose low-energy physics
is dominated by forward scattering is {\em bosonization}\/, i.e.\
the representation of the Hamiltonian and of fermionic creation and 
annihilation operators in terms of (bosonic) density fluctuation 
operators. 
This technique has been originally invented for the analysis of
one-dimensional Fermi systems \cite{ML,LP,MAT}, where it proved to be an
extremely valuable non-perturbative tool \cite{SOL,HAL81,VOI95}.
An early generalization of bosonization to higher dimensions by
Luther \cite{LUT79} could deal only with very special scattering 
processes, and therefore remained essentially unnoticed.
Recently, however, Haldane \cite{HAL92} invented a different
bosonization approach in dimensions $d \geq 1$, which turned out to
be more successful. Haldane's idea of d-dimensional bosonization 
has been elaborated in detail by Houghton and Marston \cite{HM}, and 
subsequently by many others \cite{HKM,CF,KHVbos}.
A formulation of Haldane's bosonization in terms of functional 
integrals instead of operators and Hamiltonians has been proposed
by Kopietz and Sch\"onhammer \cite{KS}, and further elaborated by
Kopietz et al.\ \cite{KHS}. 
Mathematically oriented readers are also referred to a paper by
Fr\"ohlich et al.\ \cite{FGMbos} on a bosonization approach in terms
of gauge forms.
\par
\pp We will now briefly review the main ideas of the bosonization 
approach to d-dimensional Fermi systems. As already pointed out by
several authors \cite{CD94,HKM,KHS}, the results obtained by bosonization 
are equivalent to those obtained from asymptotic Ward identities.
Since spin plays no important role in the following, we consider
spinless fermions to make the notation as simple as possible.
\par
\smallskip
\pp Common to all versions of d-dimensional bosonization is a 
decomposition of momentum space in disjoint sectors $K_{\alf}$,
$\alf = 1,\dots,M$, which yields a corresponding partition of the
Fermi surface $\FS$ in "patches" $\FS_{\alf} = K_{\alf} \cap \FS$.
Let $\Lam_{\alf}$ be a length in momentum space such that 
$\Lam_{\alf}^{d-1}$ is the ($(d\!-\!1)$-dimensional) area of the 
patch $\FS_{\alf}$. The size of the patches must be choosen small 
enough to make sure that the Fermi surface is almost flat and the 
Fermi velocity almost constant within each patch; for most systems
one thus has to require $\Lam_{\alf} \ll k_F$.
The condition of constant velocities is equivalent to the velocity
conservation underlying the Ward identity approach.
The central objects in the bosonization method are the {\em density 
fluctuation operators on sectors}
$$ \rho_{\alf}(\bq) = \sum_{\bk} 
   \chi_{\alf}(\bk\!-\!\bq/2) \> \chi_{\alf}(\bk\!+\!\bq/2) \>
   a^{\dag}_{\bk-\bq/2} \> a_{\bk+\bq/2}                \eqno(5.54) $$
where $\chi_{\alf}(\bk)$ is the characteristic function of the sector 
$K_{\alf}$ (i.e.\ one if $\bk \in K_{\alf}$ and zero else).
For $|\bq| \ll \Lam_{\alf}$, the operators $\rho_{\alf}(\bq)$ obey
the commutation relation \cite{HAL92,HM}
$$ [\rho_{\alf}(\bq), \rho_{\alf'}(\bq')] \> \sim \>
   \delta_{\alf\alf'} \delta_{\bq,-\bq'} \>
   {V \over (2\pi)^d} \> \Lam_{\alf}^{d-1} \> 
   \bn_{\alf}\!\cdot\!\bq                               \eqno(5.55) $$
on a (restricted) Hilbert space of states that differ from the Fermi
sea only in the vicinity of the Fermi surface within a distance 
$\lam_{\alf} \ll \Lam_{\alf}$.
Here $\bn_{\alf}$ is a unit vector normal to the patch $\FS_{\alf}$
and $V$ is the volume of the system.
Possibly existing band cutoffs must be much larger than $|\bq|$.
In one-dimensional systems one needs only two sectors corresponding
to left-moving and right-moving particles, respectively; in that 
case the two "patches" are simply the two Fermi points, and the "area"
$\Lam_{\alf}^{d-1}$ in (5.55) has to be substituted by a factor one.
Notice that in $d=1$ eq.\ (5.55) is an exact operator identity. 
In $d>1$, instead, it holds only asymptotically for states with a 
support $\lam_{\alf} \ll \Lam_{\alf} \ll k_F$.
\par
\pp The density fluctuation operators $\rho_{\alf}(\bq)$ can be 
related to canonically normalized boson annihililation and creation
operators $b_{\alf}(\bq)$ and $b^{\dag}_{\alf}(\bq)$ with
$\bn_{\alf}\!\cdot\!\bq > 0$ via
$$ \rho_{\alf}(\bq) = 
   \sqrt{\Omega_{\alf} |\bn_{\alf}\!\cdot\!\bq|}
   \left[ \Theta(\bn_{\alf}\!\cdot\!\bq) \> b_{\alf}(\bq) +
   \Theta(-\bn_{\alf}\!\cdot\!\bq) \> b^{\dag}_{\alf}(-\bq)
   \right]                                              \eqno(5.56) $$
where $\Omega_{\alf} \equiv V \tilde\Omega_{\alf} = (V/(2\pi)^d) \>
\Lam_{\alf}^{d-1}$. The bosonic commutation relation
$$ [b_{\alf}(\bq), b^{\dag}_{\alf'}(\bq')] =
   \delta_{\alf\alf'} \delta_{\bq\bq'}                  \eqno(5.57) $$
is equivalent to (5.55).
\par
\pp Correlation functions involving fermionic creation and 
annihilation operators (especially the single-particle propagator)
can be expressed as expectation values of the operators \cite{HM}
$$ \psi_{\alf}(t,\br) = 
   \sqrt{\lam \tilde\Omega_{\alf}} \> 
   e^{i\bk_{F\alf}\cdot\br} \>
   e^{i\tilde\Omega_{\alf}^{-1} \sqrt{4\pi} \phi_{\alf}(t,\br)}
   \> O_{\alf}                                          \eqno(5.58) $$
and the corresponding conjugate $\psi^{\dag}_{\alf}(t,\br)$,
where $\phi_{\alf}(\br)$ is the Fourier transform of
$$ \phi_{\alf}(\bq) = \rho_{\alf}(\bq)/[i\sqrt{4\pi} \> 
   \bn_{\alf}\!\cdot\!\bq]                              \eqno(5.59) $$
and $\lam$ is an ultraviolet cutoff; $\bk_{F\alf}$ is a suitably
chosen Fermi momentum belonging to the patch $\FS_{\alf}$.
The ordering operator $O_{\alf}$ is chosen such that $\psi_{\alf}(\br)$ 
and $\psi_{\alf'}(\br)$ obey fermionic anticommutation relations also
for $\alf \neq \alf'$.
\par
\smallskip
\pp The main point of the bosonization idea is the observation that the 
Hamiltonian (or action) of systems with scattering processes restricted
to small momentum transfers can be expressed as a {\em quadratic
form}\/ in $\rho_{\alf}(\bq)$, to leading order in $\bq$. 
Combining this with the bosonization (5.58) of fermionic operators, 
any correlation function can be calculated explicitly. 
\par
\pp Consider the spinless forward scattering Hamiltonian
$$ H^F = H_0 + H_I = 
   \sum_{\bk} \xi_{\bk} a^{\dag}_{\bk} a_{\bk} +
   {\textstyle {1 \over 2V}} \sum_{\bk,\bk',\bq} g_{\bk\bk'}(\bq)
   \> a^{\dag}_{\bk-\bq/2} \> a_{\bk+\bq/2} \>
   a^{\dag}_{\bk'+\bq/2} \> a_{\bk'-\bq/2}               \eqno(5.60) $$
which is the Hamiltonian corresponding to a spinless version of the 
forward scattering action $S^F$ in Eq.\ (5.2), with a non-retarded
coupling function $g_{\bk\bk'}(\bq)$. Momentum transfers are assumed
to satisfy the condition $|\bq| \ll k_F$.
\par
\pp For $\Lam_{\alf} \ll k_F$ and excitations restricted to the
vicinity of the Fermi surface, the kinetic part of the Hamiltonian
can be written as \cite{HAL92,HM}
$$ H_0 = {\textstyle {1 \over 2}} \sum_{\alf} \sum_{\bq}
   {\bv_{\alf}\!\cdot\!\bq \over \Omega_{\alf} |\bn_{\alf}\!\cdot\!\bq|} 
   \> :\!\rho_{\alf}(\bq)\rho_{\alf}(-\bq)\!: \> = \>
   \sum_{\alf} \sum_{\bq} 
   \Theta(\bn_{\alf}\!\cdot\!\bq) \> \bv_{\alf}\!\cdot\!\bq 
   \> b^{\dag}_{\alf}(\bq) b_{\alf}(\bq)                 \eqno(5.61) $$
where $\bv_{\alf}$ is the velocity $\bv_{\bk} = \nabla\xi_{\bk}$ for 
states near the patch $\FS_{\alf}$, and the double dots indicate 
normal ordering with respect to the Fermi sea.
The condition $\Lam_{\alf} \ll k_F$ makes sure that the velocity 
$\bv_{\bk}$ can be replaced by a constant $\bv_{\alf}$ within each
sector $K_{\alf}$.
\par
\pp For interactions with momentum transfers restricted by $|\bq| \ll
\Lam_{\alf}$, the interaction part of the Hamiltonian can be written
as
$$ H_I = {\textstyle {1 \over 2V}} 
   \sum_{\alf,\alf'} \sum_{\bq} g_{\alf\alf'}(\bq) 
   :\!\rho_{\alf}(\bq) \> \rho_{\alf'}(-\bq)\!:          \eqno(5.62) $$
where $g_{\alf\alf'}(\bq)$ is obtained from $g_{\bk\bk'}(\bq)$ by
averaging over $\bk \in \FS_{\alf}$ and $\bk' \in \FS_{\alf'}$.
Via (5.56) this can be also be expressed as a quadratic form of the
canonical Bose creation and annihilation operators 
$b^{\dag}_{\alf}(\bq)$ and $b_{\alf}(\bq)$.
The condition $|\bq| \ll \Lam_{\alf}$ is necessary to make scattering
of particles from one sector to another (not taken into account in 
(5.42)) rare. 
\par
\pp For interactions satisfying $|\bq| \ll k_F$ one can choose the 
size of the sectors $K_{\alf}$ such $|\bq| \ll \Lam_{\alf} \ll k_F$. 
Using (5.61) and (5.62), the whole Hamiltonian $H_0 + H_I$ can then be 
written as a quadratic form in $\rho_{\alf}(\bq)$, i.e.\
$$ H^F = \sum_{\alf,\alf'} \sum_{\bq}
   \left[{\textstyle 
   {\bv_{\alf}\!\cdot\bq \over 
    2\Omega_{\alf} |\bn_{\alf}\!\cdot\bq|} \delta_{\alf\alf'} 
   + {1 \over 2V} g_{\alf\alf'}(\bq) } \right]
   :\!\rho_{\alf}(\bq) \rho_{\alf'}(-\bq)\!:            \eqno(5.63) $$
or, equivalently, in $b^{\dag}_{\alf}(\bq)$ and $b_{\alf}(\bq)$. 
Clearly these results hold only to leading order in $\bq$, since
corrections due to scattering into different sectors have been 
neglected.
\par
\smallskip
\pp Due to the quadratic structure of the bosonized representation
of the Hamiltonian $H^F$, the dynamics of the operators 
$\rho_{\alf}(\bq)$ can be calculated exactly. Since the (total)
density fluctuation operator $\rho(\bq)$ is given by the sum
$\sum_{\alf} \rho_{\alf}(\bq)$ for $|\bq| \ll \Lam_{\alf}$, the
dynamical density-density response for small $\bq$ is thus easily 
obtained \cite{HAL92,HM}. The result is the same as that obtained
in subsection 5.2.1 as a corollary of loop cancellation: RPA. 
\par
\pp Using the bosonic representation of fermionic operators, the
single-particle propagator can be expressed in terms of the 
expectation value
$$ \bra b_{\alf}(q) b^{\dag}_{\alf}(q) \ket =
   {i \over \om - \bv_{\alf}\!\cdot\!\bq + i0^+s(\om)} +
   {\Lam_{\alf}^{d-1} \over (2\pi)^d} \> \bn_{\alf}\!\cdot\!\bq \>
   {i D_{\alf\alf}(q) \over 
    [\om - \bv_{\alf}\!\cdot\!\bq + i0^+s(\om)}]^2      \eqno(5.64) $$
where $D_{\alf\alf'}(q)$ is obtained from the RPA effective 
interaction $D_{\bk\bk'}(q)$, eq.\ (5.21), by averaging over 
$\bk \in \FS_{\alf}$ and $\bk' \in \FS_{\alf'}$.
The propagator $G(p)$ is then obtained \cite{HM,HKM} as the Fourier 
transform of 
$$ G_{\alf}(t,\br) = 
   {\delta_{\Lam_{\alf}^{-1}}(\br_t) \> e^{i\bk_{F\alf}\cdot\br} 
   \over \bn_{\alf}\!\cdot\!\br - v_{\alf}t + i0^+s(t)} \>
   \exp \left[ \int_q \> [e^{i(\bq\cdot\br - \om t)} - 1] \>
   {i D_{\alf\alf}(q) \over [\om - 
   \bv_{\alf}\!\cdot\!\bq + i0^+s(\om)]^2} \right]      \eqno(5.65) $$
for $\bp \in K_{\alf}$, where $\br_t$ is the component of $\br$ 
orthogonal to $\bn_{\alf}$ and $\delta_{\Lam^{-1}}(.)$ is a 
broadened $\delta$-function with width $\Lam^{-1}$. 
The behavior of $G(p)$ close to the Fermi surface is determined by
$G_{\alf}(t,\br)$ for large $\br$; the tangential component $\br_t$, 
being limited by $\Lam_{\alf}^{-1}$, can thus be set to zero 
\cite{HKM}.
Decomposing $\bq$ in normal and tangential components as in 5.2.3,
one thus recovers the result (5.46)-(5.51) for $G$ obtained from the 
asymptotic Ward identities.
An advantage of the Ward identity approach is that it avoids the 
patch construction with its somewhat artificial complications due to
overcompleteness of states and inter-patch scattering processes.
\par
\smallskip
\pp An alternative formulation of d-dimensional bosonization via
functional integrals has been proposed by Kopietz and Sch\"onhammer
\cite{KS}, who generalized earlier work on functional bosonization
of the Luttinger model \cite{FOG,LC} to higher dimensions.
Unlike the Hamiltonian formulation the path integral language is
applicable also to retarded interactions. In addition, the path
integral approach seems to be also more suitable for a discussion of
corrections to the leading (free boson) results.
\par
\pp Starting point is an action for interacting fermions written as
$$ S[\psi,\psi^*] = \int_k \psi^*_k (ik_0 - \xi_{\bk}) \psi_k -
   {\textstyle {1 \over 2}} \sum_{\alf\alf'} \int_q 
   g_{\alf\alf'}(q) \> \rho_{\alf}(q) \rho_{\alf'}(-q)  \eqno(5.66) $$
\vskip -3mm where \vskip -3mm
$$ \rho_{\alf}(q) = \int_k \chi_{\alf}(\bk) \> 
   \psi^*_{k-q/2} \psi_{k+q/2}                          \eqno(5.67) $$
and $g_{\alf\alf'}(q)$ is an average of $g_{\bk\bk'}(q)$ over 
$\bk \in \FS_{\alf}$ and $\bk' \in \FS_{\alf'}$. 
As long as the variation of $g_{\bk\bk'}(q)$ as a function of $\bk$ 
and $\bk'$ is small on the scale set by $\Lam_{\alf}$, this averaging
does not introduce significant errors.  
Note that in (5.67) only $\bk$ is restricted to the sector $K_{\alf}$,
while in (5.54) both fermionic momenta $\bk\pm\bq/2$ had to be in
$K_{\alf}$. 
\par
\pp Introducing an auxiliary field $\phi_{\alf}(q)$ that mediates the 
interaction, the partition function can be written as a functional 
integral over
$\psi_k$, $\psi^*_k$ and $\phi_{\alf}(q)$, with an action \cite{KHS,KS}
$$ S[\psi,\psi^*,\phi] = 
   \int_k \psi^*_k (ik_0 - \xi_{\bk}) \psi_k -
   i\sum_{\alf} \int_q \phi_{\alf}(-q) \rho_{\alf}(q) -
   {\textstyle {1 \over 2}} \sum_{\alf\alf'} \int_q
   [g^{-1}(q)]_{\alf\alf'} \phi_{\alf}(q) \phi_{\alf'}(-q)  
                                                         \eqno(5.68) $$
where $g^{-1}(q)$ is the inverse of the matrix $g(q)$ defined by the
matrix elements $g_{\alf\alf'}(q)$.
\par
\pp The integration over the fermionic fields $\psi$ and $\psi^*$ is
now Gaussian and can thus be carried out exactly. This yields an
effective action \cite{KHS,KS}
$$ S[\phi] = S_{kin}[\phi] - 
   {\textstyle {1 \over 2}} \sum_{\alf\alf'} \int_q
   [g^{-1}(q)]_{\alf\alf'} \phi_{\alf}(q) \phi_{\alf'}(-q)  
                                                         \eqno(5.69) $$
\vskip -3mm where \vskip -3mm
$$ S_{kin}[\phi] = 
   {\rm tr} \log [\hat 1 \!-\! \hat G_0 \hat\Phi]        \eqno(5.70) $$
Here the trace is over all momenta and frequencies, and $\hat G_0$
and $\hat\Phi$ are infinite matrices in momentum-frequency space,
with matrix elements given by
$[\hat G_0]_{kk'} = \delta_{kk'} G_0(k)$ and $[\hat\Phi]_{kk'} = 
i \sum_{\alf} \chi_{\alf}(\bk) \> \phi_{\alf}(k\!-\!k')$.
Expanding the logarithm, the kinetic part of the action can also be
written as an infinite sum
$S_{kin}[\phi] = \sum_{n=1}^{\infty} S_{kin,n}[\phi]$ where
$$ S_{kin,n}[\phi] = 
   - {1 \over n} \> {\rm tr} [\hat G_0 \hat \Phi]^n =:
   {1 \over n} \sum_{\alf_1,\dots,\alf_n} \int_{q_1,\dots,q_n}
   U_n^{\alf_1,\dots,\alf_n}(q_1,\dots,q_n) \>
   \phi_{\alf_1}(q_1) \dots \phi_{\alf_n}(q_n)           \eqno(5.71) $$
Diagrammatically, the n-th order term corresponds to a fermionic
loop with n insertions due to fields $\phi_{\alf_1}(q_1),\dots,
\phi_{\alf_n}(q_n)$ summed over all possible permutations.
\par
\pp In general, $S_{kin}[\phi]$ is a complicated functional, 
containing arbitrarily high powers in $\phi$. 
Significant simplifications occur however in systems dominated by
forward scattering, where only interactions with small momentum 
transfers are important. In that case the contribution from 
fields $\phi_{\alf}(q)$ with large $\bq$ is negligible, since 
$\phi_{\alf}(q)$ mediates interactions with momentum transfer $\bq$. 
As a consequence of the {\em loop cancellation}\/ discussed already 
in 5.2.1, loops with more than two insertions are irrelevant for small 
$\bq$ to leading order in $\bq$, and the series expansion of 
$S_{kin}[\phi]$ can be truncated at second order. 
The first order term vanishes for all $q \neq 0$, while the quadratic 
part yields a kinetic term \cite{KHS,KS}
$$ S_{kin}[\phi] = 
   {\textstyle {1 \over 2}} \sum_{\alf\alf'} \int_q
   \Pi_0^{\alf}(q) \> \phi_{\alf}(q) \phi_{\alf}(-q)     \eqno(5.72) $$
where $\Pi_0^{\alf}(q)$ is the bare polarisation bubble with momenta
on the fermion lines restricted to the sector $K_{\alf}$, i.e.
$$ \Pi_0^{\alf}(q) = {\Lam_{\alf}^{d-1} \over (2\pi)^d} \>
   {\bn_{\alf}\!\cdot\!\bq 
   \over iq_0 - \bv_{\alf}\!\cdot\!\bq}                  \eqno(5.73) $$
The $q$-integral in (5.72) and in subsequent results is restricted
by a cutoff function $\chi_{\alf}(\bk_{F\alf}+\bq)$, which is chosen
such that the degrees of freedom are correctly counted \cite{KOPhs}.
For the one-dimensional Luttinger model the truncation leading to
(5.72) is exact to all orders in $\bq$ \cite{LC}.
\par
\pp Due to the loop cancellation, the action $S[\phi]$ is 
now quadratic in the field $\phi$, i.e.\ its partition function can
be calculated exactly. 
\par
\pp Instead of $\phi$, one may also choose its conjugate field, the 
density fluctuation field $\tilde\rho$ as dynamical variable,
especially to make contact with the Hamiltonian version of 
bosonization. One then obtains an effective action \cite{KHS,KS} 
$$ \tilde S[\tilde\rho] = {\textstyle {1 \over 2}}
   \sum_{\alf\alf'} \int_q \left\{
   [\Pi_0^{\alf}(q)]^{-1} \delta_{\alf\alf'} - g_{\alf\alf'}(q)
   \right\} \tilde\rho_{\alf}(q) \tilde\rho_{\alf'}(-q)  \eqno(5.74) $$
which is a quadratic form in the bosonic collective field 
$\tilde\rho_{\alf}(q)$. For frequency-independent interactions the
action $\tilde S[\tilde\rho]$ is equivalent to the bosonized Hamiltonian
in Eq.\ (5.63). 
For the small-$\bq$ density-density response, which can be expressed 
as an expectation value 
$\sum_{\alf\alf'} \bra\tilde\rho_{\alf}(q)\tilde\rho_{\alf'}(-q)\ket$
with the action $\tilde S[\tilde\rho]$, one obviously recovers the 
RPA-result.
\par
\pp The single-particle propagator $G$ has been calculated by Kopietz
et al.\ \cite{KS,KOPhs} directly from $S[\phi]$, without introducing 
the $\tilde\rho$-field. 
The expectation value $\bra\psi\psi^*\ket$ can be written as a 
functional integral over $\psi$, $\psi^*$ and $\phi$ with the
action $S[\psi,\psi^*,\phi]$ in (5.68). Performing the Gaussian
integral over $\psi$ and $\psi^*$, the propagator $G(k)$ is obtained
as an average
$$ G(k) = \bra [\hat G]_{kk} \ket_{S[\phi]}              \eqno(5.75) $$
Here $\hat G$ is an infinite matrix in momentum-frequency space,
defined by the Dyson equation 
$[\hat G^{-1}]_{kk'} = [\hat G_0^{-1}]_{kk'} - [\hat\Phi]_{kk'}$,
where $[\hat G_0^{-1}]_{kk'} = (2\pi)^{d+1} \delta(k\!-\!k') G_0(k)$,
and $S[\phi]$ is given by (5.69).
The matrix $\hat G$ is the propagator of a non-interacting system in
a dynamical random field $\phi$. If $\phi_{\alf}(q)$ is finite only 
for small momenta $\bq$, the diagonal elements $[\hat G]_{kk}$ can be 
expressed explicitly as \cite{KOPhs}
$$ [\hat G]_{kk} = \sum_{\alf} \chi_{\alf}(\bk) 
   \int\!d\br \int_0^{\beta} d\tau \>
   e^{-i(\bk-\bk_{F\alf})\br + ik_0\tau} 
   \exp \left[ i\int_q {e^{i(\bq\cdot\br - q_0\tau)} - 1 \over 
   iq_0 - \bv_{\alf}\!\cdot\!\bq} \phi_{\alf}(q) \right]
   G_0^{\alf}(\tau,\br)                                 \eqno(5.76) $$
where $G_0^{\alf}(\tau,\br)$ is the Fourier transform of 
$(ik_0 - \bv_{\alf}\!\cdot\!\bk)^{-1}$.
Approximating $S_{kin}[\phi]$ by the quadratic form in (5.72), 
the calculation of $G(k)$ from (5.75) has thus been reduced to a
Gaussian integral over $\phi$, that can be carried out explicitly
\cite{KS,KOPhs}. The result is equivalent to the result (5.65) from 
the Hamiltonian version of bosonization and also to the one obtained
from asymptotic Ward identities. 
\par
\smallskip
\pp We emphasize that d-dimensional bosonization, like the asymptotic 
Ward identity method, can treat only scattering processes with small 
momentum transfers $\bq$. 
The results are meaningful only if scattering
processes with large momentum transfers are negligible with respect
to forward scattering. The calculation of corrections to the leading
order result is difficult. Kopietz et al.\ \cite{KHS} have analyzed 
certain corrections to the RPA result for the density-density response 
function by including non-quadratic terms in the expansion of the
effective kinetic action $S_{kin}[\phi]$. However, scattering processes
into other sectors $K_{\alf}$ seem to be hard to treat, and thus
only a partial contribution to the subleading corrections in a 
small-$\bq$ expansion could be calculated explicitly. Expressed as
a correction to the polarization insertion $\Pi$, this contribution 
turned out to be linear in the RPA effective interaction $D$.
Corrections to loop cancellation (and other asymptotic Ward identities)
to finite order in $D$ can of course also be calculated from the 
corresponding Feynman diagrams. 
\par

\vfill\eject

\def\pp{\hskip 5mm}
\def\bA{{\bf A}}
\def\bc{{\bf c}}
\def\bk{{\bf k}}
\def\bQ{{\bf Q}}
\def\bq{{\bf q}}
\def\bP{{\bf P}}
\def\bp{{\bf p}}
\def\b0{{\bf 0}}
\def\bi{{\bf i}}
\def\bj{{\bf j}}
\def\bn{{\bf n}}
\def\br{{\bf r}}
\def\bs{{\bf s}}
\def\bS{{\bf S}}
\def\bv{{\bf v}}
\def\alf{\alpha}
\def\eps{\epsilon}
\def\up{\uparrow}
\def\down{\downarrow}
\def\bra{\langle}
\def\ket{\rangle}
\def\para{\parallel}
\def\FS{\partial{\cal F}}
\def\Re{{\rm Re}}
\def\Im{{\rm Im}}
\def\xik{\xi_{\bk}}
\def\cO{{\cal O}}
\def\cD{{\cal D}}
\def\cF{{\cal F}}
\def\cG{{\cal G}}
\def\cZ{{\cal Z}}
\def\gam{\gamma}
\def\Gam{\Gamma}
\def\Lam{\Lambda}
\def\lam{\lambda}
\def\dbm{\delta\bar\mu}
\def\scr{\scriptstyle}
\def\sg{\sigma}
\def\Sg{\Sigma}

\vspace*{1cm}
\centerline{\large 6. ONE-DIMENSIONAL LUTTINGER LIQUID}
\vskip 1cm
\pp In one-dimensional interacting Fermi systems Fermi liquid theory 
is not valid. Its breakdown is signalled already in second order 
perturbation theory. 
The perturbatively calculated reduction of the quasi-particle weight
by interactions diverges logarithmically at the Fermi surface. 
The problem of treating these divergencies has been first solved by a 
weak coupling renormalization group method applied to an effective
low-energy theory known as {\em "g-ology"}\/ model \cite{SOL}. 
Assuming a scaling ansatz for the vertex functions, one approaches
the low-energy regime by rescaling the fields and the coupling constants 
(a small number of "g's"). 
The consistency of the ansatz is verified order by order in perturbation
theory.
Depending on the values of the bare couplings the renormalized couplings 
flow either to strong coupling, and hence out of the perturbatively 
controlled regime, or to a fixed point Hamiltonian, the exactly soluble 
Luttinger model \cite{ML,TL}.
In the latter case the system is a {\em "Luttinger liquid"}, i.e.\
a normal (not symmetry-broken) metallic phase characterized by 
i) a continuous momentum distribution with a power-law singularity at
the Fermi surface, the exponent $\eta$ being non-universal;
ii) a single-particle density of states which vanishes as $\omega^{\eta}$
near the Fermi energy, i.e.\ absence of fermionic quasi-particles;
iii) finite charge and spin density response for long wavelengths and
the existence of collective bosonic charge and spin density modes;
iv) power-law singularities in various superconducting and short wave
length (e.g.\ $|\bq| = 2k_F$) density correlation functions;
v) separation of spin and charge degrees of freedom \cite{SOL,VOI95,
SCH95}.
\par
\pp Luttinger liquid behavior is not confined to special weak-coupling 
models but may also govern strongly interacting systems. Although in
the latter case a perturbative renormalization group calculation of the
correlation functions is not adequate, the low-energy properties are still
uniquely characterized by a small number of parameters, which are directly 
related to simple physical quantities.
This leap beyond weak coupling, which follows the spirit of Fermi liquid 
theory, has been pioneered by Haldane \cite{HAL81}, who also introduced 
the suggestive term "Luttinger liquid". 
The one-dimensional Hubbard model, which is exactly soluble by the Bethe
ansatz method \cite{LW}, is most probably a Luttinger liquid for {\em any} 
coupling strength, except for half-filling \cite{FK,HT,KY,SCH90}. 
This conjecture is well established for weak coupling, while for strong 
coupling it is supported by the structure of the low lying excitations 
\cite{FK,HT}, the absence of non-analyticities in the exact ground state
energy at finite coupling \cite{SCH90}, by studies of the infinite 
coupling limit \cite{OS}, as well as by numerical evaluations of finite
systems \cite{PS90}. The Luttinger liquid parameters (and hence all
critical exponents) for the one-dimensional Hubbard model have been 
obtained exactly for arbitrary coupling strength 
\cite{FK,HT,KY,SCH90}.\footnote{Note, however, that the low-energy
scale controlled by Luttinger liquid behavior may shrink for increasing
coupling strength; this has been demonstrated very clearly for the
Hubbard model by Penc, Mila and Shiba \cite{PMS}.}
\par
\smallskip
\pp Fermi liquid theory and one-dimensional Luttinger liquid theory 
are commonly formulated in two distinct languages. 
Fermi liquid theory (i.e.\ its microscopic basis) is derived in
terms of Feynman diagrams and Ward identities \cite{NOZ,AGD}, while 
Luttinger liquid theory is usually worked out using the bosonization 
technique \cite{ML,HAL81,LP,MAT}. Nevertheless, it has often been 
emphasized that both liquids are in many respects very similar 
\cite{CH}.
In fact Luttinger liquid theory can also be constructed by standard
many-body techniques, without bosonization, by exploiting the Ward
identities associated with the peculiar conservation laws in
one-dimensional Fermi systems \cite{DL,ES,MD}. 
The choice of a common language for Fermi and Luttinger liquid theory
allows for a direct comparison of both theories, making common features
and differences more clear. 
\par
\pp The use of Ward identities makes evident the crucial role played 
by conservation laws in one-dimensional Fermi systems (especially in
Luttinger liquids).
In addition to the usual charge and spin conservation, the discrete 
structure of the Fermi surface in one dimension allows for a more 
stringent conservation law: separate charge (and possibly spin) 
conservation in low-energy-scattering processes for particles near the
left and right Fermi points, respectively. 
Only the {\em sepa\-rate} conservation of charge (spin) on each Fermi 
point guarantees the requirement of finite charge (spin) density 
response in "normal" one-dimensional metals.
The velocities associated with the corresponding conserved currents 
provide a complete parametrization of the low energy physics \cite{MD}. 
The conservation of charge and spin on each Fermi point in a 
one-dimensional Luttinger liquid implies velocity conservation as
defined and discussed for d-dimensional systems in Sec.\ 5.
\par
\smallskip
\pp There are several good reviews on one-dimensional Fermi systems: 
The understanding reached by the end of the seventies has been 
reviewed by Solyom \cite{SOL} and Emery \cite{EME}. Recent up-to-date
reviews have been presented by Voit \cite{VOI95} and Schulz 
\cite{SCH95}.
The following is not a comprehensive review on one-dimensional 
Luttinger liquids. Our main goal here is to show how Luttinger liquid
theory fits in with the more general framework constructed for 
d-dimensional systems in Secs.\ 2,3 and 5.
\par

\bigskip

{\bf 6.1. THE G-OLOGY MODEL} \par
\medskip
\pp The generic low-energy physics of a one-dimensional Fermi system 
with short-range interactions
is incorporated in the {\em g-ology model}\/ \cite{SOL}, a continuum
model with a linear dispersion relation and two-particle interactions.
The kinetic term
$$ H_0 = {\sum_{\bk,\sg}}^{<\Lam} \>
    v_F k_r \> a^{\dag}_{\bk\sg} a_{\bk\sg}               \eqno(6.1) $$
describes a linear band as illustrated in Fig.\ 6.1. 
Here $k_r = |\bk|-k_F$ measures the distance from the Fermi "surface"
(consisting of  two Fermi points in one dimension) and the momentum 
sum is restricted by $|k_r| < \Lam$.
We keep vector notation (bold face) for momenta even in one dimension 
to distinguish them from bi-vectors $k = (k_0,\bk)$ which include the 
energy variable $k_0$. 
The only parameter characterizing $H_0$ is the Fermi velocity $v_F$, 
i.e.\ the slope of the band.
\par
\pp The interaction $H_I$ is given by
$$ H_I = H_1 + H_2 + H_3 + H_4                            \eqno(6.2) $$
where
$$ H_1 = {\textstyle {1 \over V}} 
   \sum_{\bf q} \sum_{\sg \sg'} g^{\sg \sg'}_1
   \rho^+_{\sg}({\bf q}) \rho^-_{\sg'}(-{\bf q}) \quad \>        
   \hskip 44truemm                                       \eqno(6.3a) $$
\vskip -3mm
$$ H_2 = {\textstyle {1 \over V}}
   \sum_{\bf q} \sum_{\sg \sg'} g^{\sg \sg'}_2
   \rho_{+,{\sg}}({\bf q}) \rho_{-,{\sg'}}(-{\bf q}) \quad \> 
   \hskip 38truemm                                       \eqno(6.3b) $$
\vskip -3mm
$$ H_3 = {\textstyle {1 \over 2V}}
   \sum_{\bf q} \sum_{\sg \sg'} g^{\sg \sg'}_3
   [\rho^+_{\sg}({\bf q}) \rho^+_{\sg'}(-{\bf q}) +
   \rho^-_{\sg}({\bf q}) \rho^-_{\sg'}(-{\bf q})] \quad \>      
   \hskip 7truemm                                        \eqno(6.3c) $$
\vskip -3mm
$$ H_4 = {\textstyle {1 \over 2V}}
   \sum_{\bf q} \sum_{\sg \sg'} g^{\sg \sg'}_4
   [\rho_{+,{\sg}}({\bf q}) \rho_{+,{\sg'}}(-{\bf q}) +
   \rho_{-,{\sg}}({\bf q}) \rho_{-,{\sg'}}(-{\bf q})]      
                                                         \eqno(6.3d) $$
where V is the volume of the system and
$$ \rho_{\alf\sg}(\bq) = {\sum_{\bk}}^{<\Lam} \>
   \chi_{\alf}(\bk\!-\!\bq/2) \> \chi_{\alf}(\bk\!+\!\bq/2) \>
   a^{\dag}_{\bk-\bq/2,\sg} a_{\bk+\bq/2,\sg}            \eqno(6.4a) $$
\vskip -3mm
$$ \rho^{\pm}_{\sg}(\bq) = {\sum_{\bk}}^{<\Lam} \>
   \chi_{\pm}(\bk\!-\!\bq/2) \> \chi_{\mp}(\bk\!+\!\bq/2) \>
   a^{\dag}_{\bk-\bq/2,\sg} a_{\bk+\bq/2,\sg}            \eqno(6.4b) $$
Here $\chi_{\alf}(\bk)$ is the characteristic function of the two
{\em sectors}\/ $K_{\alf}$ in momentum space corresponding to 
{\em right-}moving ($\alf = +$) and {\em left-}moving ($\alf = -$) 
particles (cf.\ Sec.\ 5.4).
Below we will often use the spin symmetric and antisymmetric linear
combinations
$$ \rho_{\alf}(\bq) = 
   \rho_{\alf\up}(\bq) + \rho_{\alf\down}(\bq) 
   \quad {\rm and} \quad
   s^z_{\alf}(\bq) = 
   [\rho_{\alf\up}(\bq) - \rho_{\alf\down}(\bq)]/2        \eqno(6.5) $$
\par
\pp All parts of the Hamiltonian are understood to be normal ordered 
with respect to the ground state of $H_0$, i.e.\ the Fermi sea.
The interaction $H_I$ generates the various types of scattering 
processes listed in Fig.\ 6.2: small momentum transfer processes 
($H_2$ and $H_4$), back-scattering ($H_1$) and umklapp-scattering 
($H_3$).\footnote
{Note that in $H_1$ and $H_3$ the momentum transfer is $|\bq| \sim
 2k_F$.}
The coupling constants may be spin-dependent:
$$ g^{\sg\sg'}_i = 
   g_{i\parallel} \delta_{\sg \sg'} +
   g_{i\perp} \delta_{\sg,-\sg'} \>, 
   \quad i=1,\dots,4                                      \eqno(6.6) $$
i.e.\ $g_{i\parallel}$ refers to parallel and $g_{i\perp}$ to opposite 
spins, respectively. For later reference we define symmetric and
antisymmetric linear combinations
$$ g^c_i   = (g_{i\para} + g_{i\perp})/2 \quad {\rm and} \quad
   g^s_i = (g_{i\para} - g_{i\perp})/2                    \eqno(6.7) $$
where the indices "c" and "s" are abbreviations for "charge" and
"spin", respectively.
\par
\pp The terms $H_{1\para}$ and $H_{2\para}$ describe obviously the 
same process (since fermions with equal spin projection are 
indistinguishable). 
Hence, we can (and will) set $g_{1\para} = 0$ without loss of 
generality. Due to momentum conservation, umklapp-processes ($H_3$) 
can be relevant only if $4k_F$ is equal to a reciprocal lattice
vector (so that all scattering particles can be near the Fermi points),
as is the case, for example, for a Hubbard model at half-filling. 
\par

\bigskip

{\bf 6.2.\ RENORMALIZATION GROUP AND CUTOFFS} \par
\medskip
\pp The g-ology model is the generic effective low-energy theory for 
one-dimensional interacting Fermi systems.
A g-ology Hamiltonian (or action) is the general outcome of a Wilson 
renormalization procedure (integrate out far-from-Fermi-surface 
momentum states) applied to a microscopic one-dimensional model, 
such as the one-dimensional Hubbard model. 
Non-linear dependences in the band structure and momentum-dependences
in the couplings $g_1 \dots g_4$ are irrelevant for the leading 
low-energy behavior. 
The model parameters are related to the effective low-energy action
$\bar S^{\Lam}$ from Sec.\ 2 as follows:
\begin{eqnarray*}
  v_F & = & v_F^{\Lam} \\
  g_1 & = & \bar\Gam^{\Lam}(\bk_F,-\bk_F;-\bk_F,\bk_F) \\
  g_2 & = & \bar\Gam^{\Lam}(\bk_F,-\bk_F;\bk_F,-\bk_F) \\
  g_3 & = & \bar\Gam^{\Lam}(-\bk_F,-\bk_F;\bk_F,\bk_F) \\
  g_4 & = & \bar\Gam^{\Lam}(\bk_F,\bk_F;\bk_F,\bk_F)
\end{eqnarray*}
\vskip -1.5cm $$                                          \eqno(6.8) $$
Here we have suppressed spin variables and energy variables (which
are all set equal to zero) in the vertex functions.
\par
\pp Perturbation theory for one-dimensional Fermi systems is plagued
by several logarithmic infrared divergencies in one- and two-particle
correlation functions, which cannot be treated by a simple resummation 
of certain subsets of Feynman diagrams. A first understanding and
successful treatment of these divergencies has been achieved in the
1970's by Solyom and coworkers \cite{SOL} by applying perturbative
renormalization group methods to the g-ology model. 
Making a scaling ansatz for the vertex functions, he approached the 
low-energy limit by rescaling fields and coupling constants. 
The consistency of the ansatz has been verified a posteriori order 
by order in perturbation theory. 
One may view this approach as a slightly modified version of the
field-theoretic renormalization group \cite{DJ,AMI}; consistency of
the scaling ansatz is guaranteed by the renormalizability of 
the g-ology model \cite{MD,DM}.
Depending on the values (and signs) of the bare couplings, and on
$4k_F$ being equal or different from a reciprocal lattice vector, the 
renormalized couplings may either follow a run-away trajectory or
reach a fixed point (at weak coupling for weak bare interactions).
The former behavior signals a dramatic change of the effective 
low-energy Hamiltonian (or action) compared to the bare one, usually
associated with dynamical generation of gaps \cite{LE,SOL}. 
In the latter one has reached a fixed point Hamiltonian given by the 
Luttinger model \cite{TL}, where only interactions with small momentum
transfers survive, i.e. $g_1^* = g_3^* = 0$.
The Luttinger model is exactly soluble \cite{ML},
and the low-energy physics associated with this fixed point can be 
calculated without further use of perturbative renormalization
group methods (see Section 6.4). 
\par
\pp It is however important to realize that all the peculiarities 
of Luttinger liquid behavior have clear perturbative signals. 
An instructive example is the anomalous scaling behavior of the 
single-particle propagator
$$ G(sk_0,sk_r) = s^{\eta-1} G(k_0,k_r)                   \eqno(6.9) $$
where $\eta > 0$ is a (non-universal) constant, which implies a
power-law singularity of the momentum distribution function near
$\bk_F$ and a density of states vanishing as $\omega^{\eta}$. 
This anomalous scaling is signalled by a logarithmic singularity
in the perturbatively calculated quasi-particle weight
$Z = [1-\partial\Sg/\partial(ik_0)]^{-1}$ at the Fermi level (cf.\
eq.\ (2.24)) in second order perturbation theory. 
Within a perturbative renormalization group scheme one obtains a
wave function renormalization obeying
$$ {d \log Z^{\Lam} \over d \log\Lam} \> = \>
   {(\bar g^*_{2\para})^2 + (\bar g^*_{2\perp})^2  
   \over 8\pi^2 v_F^2} + \cO(\bar g^{*3})                \eqno(6.10) $$
for the Luttinger model (where $g_1 = g_3 = 0$), where the quadratic
terms are obtained from 2-loop contributions to the self-energy.
In the low-energy limit, one has thus
$$ Z^{\Lam} \to 0 \quad {\rm for} \quad \Lam \to 0       \eqno(6.11) $$
Integrating the flow for $Z^{\Lam}$ one obtains anomalous scaling
behavior as in (6.9) with an anomalous dimension
$$ \eta \> = \> 
   {(\bar g^*_{2\para})^2 + (\bar g^*_{2\perp})^2  
   \over 8\pi^2 v_F^2} + \cO(\bar g^{*3})                \eqno(6.12) $$
By constrast, in a Fermi liquid $Z^{\Lam} \to Z > 0$, and there is
no anomalous scaling dimension (i.e.\ $\eta = 0$).
In both (Fermi and Luttinger) liquids the long wavelength 
response-functions do not acquire anomalous scaling dimensions and
both describe a normal metallic phase.
\par
\pp Note that concrete perturbative renormalization group calculations
for the Luttinger or g-ology model are easiest within the 
field-theoretic version of the RG or, what is almost the same, by 
using Solyom's scaling ansatz. 
Wilson's RG version becomes relatively cumbersome in practical 
calculations beyond one-loop order,
while characteristic differences between Luttinger and Fermi liquid 
behavior show up only at two-loop order.
\par
\smallskip
\pp The effective g-ology model obtained by integrating out high-energy
states has a {\em band-width cutoff}\/ $\Lam$, and no explicit 
momentum transfer cutoff. Momentum transfers are restricted only 
indirectly by the finite band-width cutoff.
For a non-perturbative analysis it is however more convenient to 
restrict the momentum transfers $\bq$ in the interaction part (6.3) to
values 
$|\bq| < \Lam_q$ in $H_2$ and $H_4$ and $||\bq|-2k_F| < \Lam_q$ in
$H_1$ and $H_3$, with a {\em "momentum transfer cutoff"} 
$\Lam_q \ll \Lam$ \cite{SOL}.
This can be done either ad hoc, or via the mapping described in the 
introductory part of Sec.\ 5.
The qualitative structure of the infrared asymptotics is of course 
not altered by different choices of the cutoff procedure.
In practice not even quantitative results depend on the cutoff 
procedure, if the low-energy physics is expressed in terms of 
parameters which are directly related to physical observables.
\par
\pp Note that the scale $k_F$ does not appear in the g-ology model.
Important is only the {\em existence} of the two Fermi points, not 
their {\em distance} in momentum space. Thus the inverse cutoff 
$\Lambda_q^{-1}$ is the only length scale in the model; $\Lambda_q$ 
replaces the natural cutoff given by the Brillouin zone boundary or 
by non-linear terms in $\eps_{\bk}$ in the underlying microscopic 
system. 
\par

\bigskip

{\bf 6.3. CHARGE/SPIN CONSERVATION AND WARD IDENTITIES} \par
\medskip
{\bf 6.3.1. Global charge and spin conservation} \par
\smallskip
\pp The g-ology Hamiltonian conserves charge and the spin component
in z-direction (the quantization axis). Spin components in any 
direction are conserved only if the coupling constants satisfy the 
relation $g_{2\perp} \!-\! g_{1\perp} = g_{2\para} \!-\! g_{1\para}$, 
which can be obtained from the condition $[H,\bS] = 0$.
As discussed in detail in Secs.\ 3 and 5, such conservation laws give
rise to important Ward identities.
\par
\pp In general, the g-ology interactions are not of density-density 
type. The current operators depend therefore explicitly on the
coupling constants. 
We begin with charge (or particle) conservation and its consequences,
and address spin conservation briefly afterwards. 
Calculating the commutator of the g-ology
Hamiltonian with the charge density fluctuation operator at small $\bq$,
i.e. $\rho(\bq) = \rho_+(\bq) + \rho_-(\bq)$,
one obtains a continuity equation with a current operator given
by \cite{MD}
$$ \bj(\bq) = 
   v_c \> [\rho_+(\bq) - \rho_-(\bq)]                     \eqno(6.13) $$
with a coupling-dependent velocity 
$$ v_c = v_F + (g^c_4 - g^c_2)/\pi                        \eqno(6.14) $$
The current-operator depends only on {\em forward}\/ scattering 
couplings, because $\rho(\bq)$ commutes with $H_1$ (with 
$g_{1\para} \!=\! 0$) and $H_3$. 
Hence, we can now simply apply the Ward identities derived from global
charge conservation for the d-dimensional forward scattering action in
Sec.\ 5.1 to the one-dimensional case. 
In particular, the correlation function 
$J^{\mu\nu} = - {1 \over V} \bra j^{\mu} j^{\nu} \ket$ 
and the current vertex
$\Gam^{\mu} = \bra j^{\mu} a a^{\dag} \ket_{tr}$, 
obey the Ward identities (3.10) and (3.15), respectively \cite{MD}, 
with a function $\bc(\bq)$ given by
$$ \bc(\bq) = {2 \over \pi} v_c \> \bq                    \eqno(6.15) $$
for the g-ology model. Note that the formula (3.13) for $\bc(\bq)$
does not apply here, as a consequence of the band-width cutoff;
the more general expression (3.12) yields a finite result due to this 
cutoff, not as a consequence of a finite gradient of $\bv_{\bk}$ in 
momentum space.
The irreducible current vertex $\Lam^{\mu}_{\sg}(p;q) = 
 \bra j^{\mu}_0(q) \> a_{p-q/2,\sg} \> a^{\dag}_{p+q/2,\sg} 
 \ket^{irr}_{tr}$ 
with the non-interacting current operator
$$ \bj_0(\bq) = v_F \> [\rho_+(\bq) - \rho_-(\bq)]        \eqno(6.16) $$
obeys the Ward identity (5.15).
Note that $\bj_0$ has been constructed with $v_F$, i.e. the bare 
velocity in the g-ology model.
\par
\smallskip
\pp Conservation of the spin component in z-direction has analogous 
consequences. A continuity equation relates the z-component of the
spin density $s^z(\bq) = s^z_+(\bq) + s^z_-(\bq)$ to a spin current 
given by
$$ \bj^z(\bq) = v_s \> [s^z_+(\bq) - s^z_-(\bq)]          \eqno(6.17) $$
with a velocity
$$ v_s = v_F + (g^s_4 - g^s_2)/\pi                        \eqno(6.18) $$
Again only forward scattering couplings are involved.
The spin correlation function 
$J^{z\mu,z\nu} = - {1 \over V} \bra j^{z\mu} j^{z\nu} \ket$ 
and the spin current vertex
$\Gam^{z\mu} = \bra j^{z\mu} a a^{\dag} \ket_{tr}$, 
obey the Ward identities (3.27) and (3.29), 
respectively \cite{MD}, with $\bc(\bq)$ given by (6.15).
The irreducible spin current vertex
$\Lam^{z\mu}_{\sg\sg'} = \bra j^{z\mu}_0 a_{\sg} a^{\dag}_{\sg'} 
\ket_{tr}^{irr}$ with the non-interacting spin current operator
$$ \bj^z_0(\bq) = v_F \> [s^z_+(\bq) - s^z_-(\bq)]        \eqno(6.19) $$
obeys the Ward identity (5.16) for $a=z$.
\par
\medskip

{\bf 6.3.2. Separate left/right conservation laws} \par
\smallskip
\pp In the absence of umklapp processes, there is an additional 
conservation law: {\em charge near the left and right Fermi point is 
conserved separately}. 
This case is by no means academic since the presence of umklapp 
processes, being subject to the matching condition $4\bk_F = \bQ$ 
($\bQ$ a reciprocal lattice vector), is the exception rather than the 
rule. For example, the g-ology describing the low-energy physics of 
the Hubbard model involves umklapp terms only at half-filling. 
The separate left/right charge conservation yields additional (besides
those implied by usual charge conservation) significant constraints 
on the structure of correlation functions and the renormalization 
group \cite{MD,DM}.
\par 
\pp Separate {\em left/right spin conservation}\/ is spoiled by the 
back-scattering process $H_{1\perp}$, which is generically present in a 
model of spin-${1 \over 2}$ fermions. However, in many cases of interest 
(in particular in the one-dimensional Hubbard model with repulsive 
interaction \cite{SCH95}) the back-scattering amplitude scales to 
zero at low energies, {\em restoring}\/ thus separate left/right spin 
conservation asymptotically, and yielding further constraints on the 
asymptotic low-energy theory.
This is indeed the case of the Luttinger liquid discussed in the next
Section 6.4.
\par
\smallskip
\pp Let us start by exploring the consequences of separate 
conservation of charge on each Fermi point.
In addition to conservation of the global charge, i.e.\ left
and right summed, separate left/right conservation implies also the
conservation of the charge {\em difference}, i.e.\ right minus
left. Correspondingly, the operator
$$ \tilde\rho(\bq) = \rho_+(\bq) - \rho_-(\bq)            \eqno(6.20) $$
obeys a continuity equation
$\partial_{\tau}\tilde\rho(\tau,\bq) = [H,\tilde\rho(\tau,\bq)] =
 - \bq\!\cdot\!\tilde\bj(\tau,\bq)$ with a current operator given
by \cite{MD}
$$ \tilde\bj(\bq) = 
   \tilde v_c \> [\rho_+(\bq) + \rho_-(\bq)]              \eqno(6.21) $$
and a velocity
$$ \tilde v_c = v_F + (g^c_4 + g^c_2)/\pi                 \eqno(6.22) $$
In analogy to the Thirring model one may refer to $\tilde j^{\mu} =
(\tilde\rho,\tilde\bj)$ as the {\em axial charge}\/ current 
\cite{JON}.
\par
\pp The continuity equation for $\tilde j^{\mu}$ implies the Ward 
identity
$$ (iq_0,\bq)_{\mu} \tilde J^{\mu\nu}(q) = 
   \tilde c^{\nu}(\bq) = 
   \delta_{\nu 1} {2 \over \pi} \tilde v_c \> \bq         \eqno(6.23) $$
for $\tilde J^{\mu\nu}(q) = - {1 \over V} \bra \tilde j^{\mu}
\tilde j^{\nu} \ket$ and
$$ (iq_0,\bq)_{\mu} \tilde\Gam^{\mu}_{\sg}(p;q) = s(\bp)
   \left[ G_{\sg}^{-1}(p\!+\!q/2) - G_{\sg}^{-1}(p\!-\!q/2)
   \right]                                                \eqno(6.24) $$
for $\tilde\Gam^{\mu} = \bra \tilde j^{\mu} a a^{\dag} \ket_{tr}$,
where $s(\bp)$ is the sign of $\bp$ (i.e. $s(\bp) = 1$ for right-moving
particles and $s(\bp) = -1$ for left-moving particles).
\par
\pp Combining the Ward identities from global and axial conservation
laws determines $J^{\mu\nu}$ (and $\tilde J^{\mu\nu}$) completely.
Results will be discussed in 6.4.
Note that the velocities $v_c$ and $\tilde v_c$ appearing in the two
conserved currents correspond to Haldane's \cite{HAL81} velocities 
$v_J$ and $v_N$, which control the current ($J$) and density ($N$)
excitations (cf.\ eqs.\ (6.32) and (6.34)).
\par
\pp In the absence of backscattering, the z-component of spin is 
conserved separately on each Fermi point. In that case one has an
additional continuity equation for an {\em axial spin}\/ current
$\tilde j^{z\mu} = (\tilde s^z,\tilde\bj^z)$ where
$$ \tilde s^z(\bq) = s^z_+(\bq) - s^z_-(\bq)              \eqno(6.25) $$
\vskip -3mm and \vskip -3mm
$$ \tilde\bj^z(\bq) = 
   \tilde v_s \> [s^z_+(\bq) + s^z_-(\bq)]                \eqno(6.26) $$
with a velocity
$$ \tilde v_s = v_F + (g^s_4 + g^s_2)/\pi                 \eqno(6.27) $$
This yields Ward identities for correlation and vertex functions in
complete analogy to those from axial charge conservation. 
The identities from global and axial spin conservation determine
$J^{z\mu,z\nu}$. Results will be discussed in 6.4.
\par

\bigskip

{\bf 6.4. LUTTINGER LIQUID FIXED POINT} \par
\medskip
\pp In many one-dimensional Fermi systems both umklapp- and 
backscattering are irrelevant in the low-energy limit.\footnote
{Backscattering is always marginal at tree (0-loop) level, but usually
(marginally) irrelevant due to fluctuations (1-loop or higher) for 
systems with repulsive interactions.}
In these cases only {\em forward scattering}\/, i.e.\ scattering
with small momentum transfers survives. The corresponding couplings
$g_2$ and $g_4$ reach finite non-universal fixed points $g^*_2$ and
$g^*_4$ in the low-energy limit \cite{SOL}. 
The $\beta$-function at the fixed points vanishes due to 
cancellations imposed by the peculiar left/right conservation laws 
\cite{MD,DM}, not simply as a consequence of vanishing phase space as 
in a Fermi liquid!
\par
\pp The fixed point Hamiltonian can be written as
$$ H^* = H_0 + {\textstyle{1 \over 2V}} 
   \sum_{\sg,\sg'} \sum_{\alf,\alf' = \pm 1} \sum_{\bq}
   f_{\alf\alf'}^{\sg\sg'} \> 
   \rho_{\alf\sg}(\bq) \rho_{\alf'\sg'}(-\bq)           \eqno(6.28) $$
with $H_0$ as in (6.1) and a {\em Landau function}
$$ f_{\alf\alf}^{\sg\sg'} = g_4^{*\sg\sg'} 
   \quad {\rm and} \quad
   f_{\alf,-\alf}^{\sg\sg'} = g_2^{*\sg\sg'}            \eqno(6.29) $$ 
This is the Hamiltonian of the {\em Luttinger model} (actually a 
simple generalization of the original Tomonaga-Luttinger model 
\cite{TL} obtained by including spin). 
As proposed by Haldane \cite{HAL81}, one-dimensional systems whose 
leading low-energy physics is governed by a Luttinger model fixed 
point are called {\em Luttinger liquids}. 
Note that the fixed point action $\bar S^*$ corresponding to $H^*$ 
is identical to the fixed point action (4.15) of a Fermi liquid, 
adapted to the one-dimensional case with only two Fermi points 
$\bk_F = \pm k_F$!
Differences between Fermi and Luttinger liquid behavior arise only
due to the enhanced phase space for forward scattering in one
dimension.
\par
\pp With a momentum transfer cutoff $\Lam_q \ll \Lam$, the Luttinger
model can be solved exactly and completely in the sense that all
correlation functions can be obtained \cite{SOL}. 
This can be achieved either by exploiting the Ward identities 
associated with the peculiar conservation laws of the system 
\cite{DL,MD,ES}, or by bosonization \cite{ML,HAL81,LP}.
The Luttinger model Hamiltonian $H^*$ conserves charge and the
z-component of spin separately on each Fermi point. The Ward
identities obtained from the continuity equations associated with
these conservation laws yield a complete system of equations
for any correlation function.
\par
\pp The solution of the Luttinger model can also be obtained as
a special case of the (more general) results for d-dimensional
systems in Sec.\ 5. The properties derived there to leading order
in a small-momentum-transfer expansion for systems dominated by
forward scattering are exact to all orders in $\bq$ in the
Luttinger model. 
The reason for this is that here the velocity of the fermions 
is conserved {\em exactly}\/ in each scattering process.\footnote
{Velocity conservation is equivalent to left/right conservation
laws for charge and spin together with the linearity of the 
dispersion relation in the Luttinger model.} 
In particular, the loop cancellation derived in 5.1 and the 
density-current relations ${\bf\Lam}(p;q) = \bv_{\bp} \Lam^0(p;q)$
obtained in 5.2 hold exactly in the Luttinger model.
\par
\smallskip
\pp As a consequence of loop cancellation, polarization insertions 
remain undressed and RPA results for the long-wavelength charge and 
spin response are exact \cite{DL,SOL}. 
Let us discuss the results for the density-density response and the
dynamical conductivity.
\par
\pp Continuing analytically from Matsubara frequencies $q_0$ to real 
frequencies $\omega$, RPA (or the Ward identities) yield the well-known 
result for the charge density-density response \cite{SOL}, which
can be expressed in terms of the velocities $v_c$ and $\tilde v_c$
associated with global and axial charge conservation as \cite{MD}

$$ N(\omega,\bq) := J^{00}(\omega,\bq) = {2 \over \pi} \> 
   {v_c \bq^2 \over \omega^2 - (u_c\bq)^2}              \eqno(6.30) $$
\vskip -3mm where \vskip -3mm
$$ u_c = \sqrt{v_c \tilde v_c}                          \eqno(6.31) $$
This response function has poles in $\omega = \pm u_c\bq$, 
implying the existence of charge density modes (zero sound) with 
velocity $u_c$ in the system. The existence of these modes is a direct 
consequence of left/right charge conservation:
Noting that $\bj = v_c \tilde\rho$ and $\tilde\bj = \tilde v_c \rho$, 
the continuity equations $i\partial_t \rho = \bq\cdot\bj$ and 
$i\partial_t \tilde\rho = \bq\cdot\tilde\bj$ can be combined to 
a harmonic oscillator equation for $\rho(t,\bq)$, namely 
$\partial_t^2 \rho(t,\bq) + v_c \tilde v_c \> \bq^2 \rho(t,\bq) = 0$, 
describing undamped harmonic oscillations with frequency $u_c|\bq|$ 
where $u_c = \sqrt{v_c \tilde v_c}$. 
\par
\pp The {\em compressibility}\/ $\kappa = \partial n/\partial \mu$ is
obtained from (6.30) as
$$ \kappa = {2 \over \pi \tilde v_c}                    \eqno(6.32) $$
\pp Using the relation $\sg(q) = (i\omega/\bq^2) \> N(q)$ 
for the dynamical {\em conductivity}, one finds
$$ \sg(\omega,\bq) = 
   {2 \over \pi} v_c {i\omega \over \omega^2 - (u_c\bq)^2}
                                                        \eqno(6.33) $$
The absorptive part in a homogeneous field is therefore 
$$ Re \> \sg(\omega) = 2v_c \delta(\omega)              \eqno(6.34) $$
i.e.\ a delta-peak with weight $2v_c$. Note that the velocities $v_c$ 
and $\tilde v_c$ are in a one-to-one correspondence to directly 
observable physical quantities: $v_c$ determines the weight of the 
Drude peak in the conductivity and $\tilde v_c$ the compressibility 
of the system. 
\par
\pp The compressibility and the weight of the Drude peak are both 
finite, indicating that the g-ology model without (or with irrelevant)
Umklapp terms describes a {\em stable metal}. 
By contrast, if Umklapp terms are relevant the compressibility is
singular and the system becomes insulating.
This latter behavior occurs in the one-dimensional Hubbard model
with repulsive interactions at half-filling \cite{SCH95}. 
\par
\pp The spin density-density response is obtained as \cite{SOL,MD}
$$ S(\omega,\bq) := J^{z0,z0}(\omega,\bq) = {1 \over 2\pi} \> 
   {v_s \bq^2 \over \omega^2 - (u_s\bq)^2}              \eqno(6.35) $$
\vskip -3mm with \vskip -3mm
$$ u_s = \sqrt{v_s \tilde v_s}                          \eqno(6.36) $$
where $v_s$ and $\tilde v_s$ are the velocities associated with the 
conserved global and axial spin currents. 
The poles in $S(\omega,\bq)$ imply the existence of spin density
modes with a velocity $u_s$. 
The spin {\em susceptibility}\/ obtained from (6.35) is simply
$$ \chi = {2\mu_B^2 \over \pi \tilde v_s}               \eqno(6.37) $$
where $\mu_B$ is the Bohr magneton.
\par
\pp Note that the above results for the charge response remain valid
even in the presence of backscattering, while the results for the 
spin response are not affected by umklapp scattering.
\par
\smallskip
\pp The single-particle propagator can be obtained from the solution 
for $G$ derived for d-dimensional forward scattering systems in 
Sec.\ 5. This solution is exact for the Luttinger model.
Notice that the Ward identities (5.27) and (5.29) are indeed exact in
the Luttinger model, and could be derived, without invoking loop
cancellation, by directly using the results of Section 6.3.2 
obtained from separate left and right conservation laws (which imply
exact loop cancellation) \cite{MD}.
Hence, $G$ is determined by the integral equation (cf.\ (5.42))
$$ (p_0 - v_F p_r) \> G(p_0,p_r) \> = \> 1 \> - \> 
   \int_{p_0',p_r'} {D(p_0\!-\!p_0',p_r\!-\!p_r') \over 
   i(p_0\!-\!p_0') - v_F (p_r\!-\!p_r')} \> 
   G(p'_0,p'_r)                                         \eqno(6.38) $$
where $D(q_0,q_r) = D_{\alf\alf}^{\sg\sg}(q_0,\bq)$ is
the RPA effective interaction between particles with parallel 
spin near the same Fermi point, and $q_r = \bn_{\alf}\cdot\bq$. 
In one dimension, Eq.\ (5.21) for $D$ can be written as
$$ D_{\alf\alf'}^{\sg\sg'}(q) = f_{\alf\alf'}^{\sg\sg'} + 
   \sum_{\sg''} \sum_{\alf''} 
   f_{\alf\alf''}^{\sg\sg''} \> \Pi_0^{\alf''}\!(q) \>
   D_{\alf''\alf'}^{\sg''\sg'}(q)                       \eqno(6.39) $$
where
$$ \Pi_0^{\alf}(q) = 
   {1 \over 2\pi} \> {q_r \over iq_0 - v_F q_r}          \eqno(6.40) $$
The above equation for $G$ has been first derived by Dzyaloshinskii
and Larkin \cite{DL}.
Equation (6.39) for the matrix $D_{\alf\alf'}^{\sg\sg'}(q)$ can be
easily solved; the "diagonal" element can be written explicitly as
$$ D(q_0,q_r) = (iq_0 - v_F q_r) \pi \sum_{\nu = c,s} \left[ 
   {(2-\eta_{\nu}) (u_{\nu}-v_F) \over iq_0 - u_{\nu} q_r} +
   {   \eta_{\nu}  (u_{\nu}+v_F) \over iq_0 + u_{\nu} q_r}
   \right]                                               \eqno(6.41) $$
where $u_{\nu} = (v_{\nu} \tilde v_{\nu})^{1/2}$ is the velocity of 
the collective modes ($\nu = c,s$), and
$$ \eta_{\nu} = (K_{\nu} + K_{\nu}^{-1} - 2)/4 
   \quad {\rm where} \quad 
   K_{\nu} = (v_{\nu}/\tilde v_{\nu})^{1/2}              \eqno(6.42) $$
Inserting the analytic continuation of $D$ to real frequencies 
into the expression (5.48) for $L$, one obtains
$$ L(t,r) \> = \> \log(r - v_Ft + is(t)/\Lam_q) \hskip 8cm $$
\vskip -8mm
$$ - \sum_{\nu=c,s} \Big[(1/2 + \eta_{\nu}/2) 
   \log(r - u_{\nu}t + is(t)/\Lam_q) +
   (\eta_{\nu}/2) \log(r + u_{\nu}t + is(t)/\Lam_q)\Big] \eqno(6.43) $$
and $L_0 = L(0,0) = \eta\log\Lam_q$, where $\eta = \eta_c + \eta_s$. 
For large $r$ and/or large $t$ one thus finds \cite{SOL}
$$ G(t,r) = {1 \over 2\pi \Lam_q^{\eta}}
   \prod_{\nu=c,s}
   {1 \over (r - u_{\nu}t + is(t)/\Lam_q)^{1/2+\eta_{\nu}/2}}
   {1 \over (r + u_{\nu}t - is(t)/\Lam_q)^{\eta_{\nu}/2}}        
                                                         \eqno(6.44) $$
Note that the exponents are uniquely determined by $K_{\nu}$, i.e.\ 
by the ratios $v_{\nu}/\tilde v_{\nu}$.
For non-interacting particles $K_c = K_s = 1$. The value $K_s = 1$
is maintained for spin-rotation invariant interactions, while
$K_c <1$ ($>1$) for repulsive (attractive) forces.
The Fourier-transform $G(\omega,k_r)$ of $G(t,r)$ is not an 
elementary function.
The result (6.44) implies that the density of single-particle 
excitations vanishes as $\omega^{\eta}$ at low energy, and the
momentum distribution function near $k_F$ obeys a power law 
$$ n_{\bk} - n_{\bk_F} \> \propto \> 
   - s(k_r)|k_r|^{\eta}                                  \eqno(6.45) $$
Spin-rotationally invariant microscopic models such as the Hubbard
model lead generically to $\eta_s = 0$, $\eta_c > 0$ and $u_c \neq u_s$. 
In this case, for $\bk$ outside the Fermi-surface and $u_c > u_s$, say, 
the $\bk$-resolved spectral function $\rho(\omega,\bk)$ for single 
particle excitations, has power-law divergences for $\omega \to 
u_c k_r$ and for $\omega \to u_s k_r$ with $\omega > u_s k_r$,
vanishes for $-u_c k_r < \omega < u_s k_r$, and has finite values also
for $\omega < -u_c k_r$ \cite{MS,VOI93}.
The two peaks in the spectral function at $\omega = u_c k_r$ and
$\omega = u_s k_r$ indicate that the extra charge and spin associated
with an additional electron inserted into the system propagate with
different velocities, a phenomenon called "spin-charge-separation".
This is in striking contrast to the behavior of a Fermi liquid, where 
the spectral function has the form $\rho(\omega,\bk) \propto 
\delta(\omega-\xik)$ in the low-energy limit. 
Landau quasi-particle excitations are absent in the Luttinger liquid.
\par
\pp The velocity ratios $v_{\nu}/\tilde v_{\nu}$ (i.e.\ $K_{\nu}$) 
determine also the anomalous scaling dimensions of $2k_F$-density 
correlations and Cooper pair correlation functions in a Luttinger 
liquid \cite{VOI95,SCH95}.
\par
\smallskip
\pp The (bosonic) charge and spin density modes corresponding to the
poles in the charge and spin response functions are the only 
low-energy excitations in a Luttinger liquid. They determine the
leading low-temperature contribution to the specific heat
$$ c_V = {\pi \over 6} (u_c^{-1} + u_s^{-1}) \> T        \eqno(6.46) $$
A computation of the free energy associated with $\bar S^*$ (given
by RPA due to loop cancellation) shows that there are no other
contributions. 
\par
\pp In the Luttinger model, the charge and spin density modes are
exact eigenstates (i.e.\ undamped), and any excited state of the 
model is a superposition of these elementary excitations.
This important fact becomes particularly explicit in the bosonized
form of the Luttinger model obtained first by Mattis and Lieb 
\cite{ML}. The bosonization procedure described in Sec.\ 5 is exact 
for the Luttinger model. The expression (5.63) for the Hamiltionian
as a quadratic form in density fluctuation operators turns into
$$ H^* = {\textstyle{1 \over 2V}} 
   \sum_{\sg,\sg'} \sum_{\alf,\alf'} \sum_{\bq} 
   \Big[ 2\pi \bv_{\alf}\!\cdot\!\bq \> 
   \delta_{\sg\sg'} \delta_{\alf\alf'} + 
   f_{\alf\alf'}^{\sg\sg'} \Big] \>
   \rho_{\alf\sg}(\bq) \rho_{\alf'\sg'}(-\bq)            \eqno(6.47) $$
for fermions with spin in one dimension. 
Transforming to canonical boson annihilation and creation operators
$b_{\alf\sg}(\bq)$ and $b^{\dag}_{\alf\sg}(\bq)$ with 
$\bn_{\alf}\!\cdot\!\bq > 0$ via (cf.\ (5.56))
$$ \rho_{\alf\sg}(\bq) = \sqrt{V/2\pi}
   \left[\Theta( \bn_{\alf}\!\cdot\!\bq) b_{\alf\sg}(\bq) +
   \Theta(-\bn_{\alf}\!\cdot\!\bq) b^{\dag}_{\alf\sg}(-\bq)
   \right]                                              \eqno(6.48) $$
one obtains a quadratic form in $b$ and $b^{\dag}$. Off-diagonal
terms are easily eliminated by a linear (Bogoliubov) transformation
from $b$ and $b^{\dag}$ to new boson operators $\beta$ and 
$\beta^{\dag}$.
In each subspace with fixed particle number and total momentum,
the Luttinger model Hamiltonian can thus be transformed into 
\cite{ML,HAL81,SOL}
$$ H^* = \sum_{\nu=c,s} \sum_{\bq} u_{\nu} |\bq|
   \beta^{\dag}_{\nu}(\bq) \beta_{\nu}(\bq)             \eqno(6.49) $$
Any excited state is thus obviously a superposition of non-interacting
elementary bosonic excitations corresponding to charge or spin 
density modes in the underlying interacting Fermi system.
In particular, one has "spin-charge separation" in the sense that 
charge and spin excitations are independent from each other.
\par
\smallskip
\pp Let us briefly compare the above results to those of Fermi
liquid theory.
Concerning the leading low-energy long-wavelength response 
functions there is no difference between Fermi and Luttinger liquids.
In both liquids the response is governed by the same fixed point
action $\bar S^*$ and the residual forward scattering can be treated 
in RPA. However, in a Luttinger liquid RPA works due to more subtle
reasons than in a (two- or three-dimensional) Fermi liquid. In the
latter corrections to RPA are suppressed simply by the reduced phase
space for residual scattering processes, while in the former RPA
emerges due to cancellations (of self-energy and vertex corrections)
imposed by the (asymptotic) conservation laws!
\par
\pp Marked differences between Fermi and Luttinger liquids appear
in the single-particle propagator $G$, which determines the momentum 
distribution function and the spectral density for single-particle
excitations. In a Fermi liquid residual interactions modify the
propagator only on a subleading level (leading to a small 
quasi-particle decay etc.), while in a Luttinger liquid forward
scattering affects the leading low energy behavior.
Another distinctive feature of Luttinger liquids is the singular
behavior of density correlations with momenta near $2k_F$ 
\cite{SOL,VOI95}.
\par
\pp The similarity of the fixed point Hamiltonians in Fermi and 
Luttinger liquids (cf.\ (4.15) and (6.28)) as well as the common
RPA structure of response functions suggests the identification of
$f_{\alf\alf}^{\sg\sg'} = g_4^{\sg\sg'}$ and 
$f_{\alf,-\alf}^{\sg\sg'} = g_2^{\sg\sg'}$ as a Landau function
in one-dimensional systems. One should, however, be aware of the
following ambiguity: In a Luttinger liquid, the Landau function
is not uniquely determined from the low-energy behavior of the
correlation functions, in constrast to the situation in a Fermi
liquid, where (4.7) yields a unique identification. In fact,
all correlation functions in the one-dimensional Luttinger model 
are uniquely para\-metrized by the four velocities 
$v_{\nu}$, $\tilde v_{\nu}$ ($\nu = c,s$) in the low-energy limit, 
where, in terms of $v_F$ and $g$'s (see Sec.\ 6.3)
$$        v_{\nu} = v_F + (g_4^{\nu} - g_2^{\nu})/\pi \quad , \quad 
   \tilde v_{\nu} = v_F + (g_4^{\nu} + g_2^{\nu})/\pi        
                                                          \eqno(6.50) $$
An equivalent parametrization can be given in terms of $K_{\nu}$ and
the sound velocities $u_{\nu}$.
Obviously all the velocities $v_{\nu}$, and hence all correlation 
functions are invariant under the shift 
$$ v_F \mapsto v_F + \delta v_F \quad , \quad
   g_4^{\sg\sg} \mapsto g_4^{\sg\sg} - \pi \delta v_F     \eqno(6.51) $$
Auxiliary functions such as the effective interaction $D(\omega,q_r)$ 
depend on the choice of $v_F$. 
The ambiguity in the identification of a Landau function in $d=1$ can
be removed by imposing the Landau sum-rule for the scattering 
amplitudes \cite{NOZ,BP}, which is equivalent to the condition 
$D(0,q_r) = 0$ in eq.\ (6.41).  
In general, one finds
$$ D(0,q_r) = 2 \pi v_F - {\pi v_F^2 \over 2} 
   \sum_{\nu} \left( v_{\nu}^{-1} + \tilde v_{\nu}^{-1}
                                                \right)   \eqno(6.52) $$ 
Hence, only for the special and unique choice
$$ v_F = \Big[ {\textstyle{1 \over 4}} \sum_{\nu} 
   \left( v_{\nu}^{-1} + \tilde v_{\nu}^{-1} \right) 
   \Big]^{-1}                                             \eqno(6.53) $$
$D$ satisfies the condition $D(0,q_r) = 0$ associated with the Landau
sum-rule. These remarks will play a role in the analysis of the
crossover from Luttinger to Fermi liquid behavior as a function of
dimensionality in Sec.\ 7.
\par

\bigskip

{\bf 6.5. INSTABILITIES} \par
\medskip
\pp Instabilities of the Luttinger liquid closely parallel those of 
a Fermi liquid. 
Stability with respect to deformations of the Fermi "surface" 
requires 
$$ (g_4^{\nu} \pm g_2^{\nu})/\pi v_F \> > \> -1         \eqno(6.54) $$
or (equivalently) $\tilde v_{\nu} > 0$ and $v_{\nu} > 0$. Violation
of one of these conditions leads to a one-dimensional analogue of
the {\em Pomerantchuk instabilities}\/ known for Fermi liquids 
(cf.\ (4.29)).
For example, for $(g_4^c + g_2^c)/\pi v_F < -1$ the compressibility
becomes negative and the liquid will undergo phase-separation.
\par
\pp Depending on the signs of the coupling constants, backscattering
processes may or may not generate an instability of the Luttinger 
liquid towards a state with a spin-gap and enhanced charge-density wave
or Cooper-pair correlations (both may be enhanced; the sign of $K_c-1$ 
determines which correlation is stronger) \cite{SOL}. 
This behavior may signal a {\em Spin-Peierls instability}\/ or a 
one-dimensional analogue of the {\em Cooper instability},
although genuine off-diagonal long-range order is prevented here by 
the strong order-parameter fluctuations in one dimension.
Similarly, umklapp scattering may lead to the formation of a charge
gap. This latter instability is an analogue of {\em density-wave 
instabilities}\/ (or Mott transition) in higher dimensions.
\par

\vfill\eject

\def\vm{\vskip -4mm}
\def\bk{{\bf k}}
\def\bQ{{\bf Q}}
\def\bq{{\bf q}}
\def\bP{{\bf P}}
\def\bp{{\bf p}}
\def\b0{{\bf 0}}
\def\bi{{\bf i}}
\def\bj{{\bf j}}
\def\bJ{{\bf J}}
\def\bn{{\bf n}}
\def\br{{\bf r}}
\def\bR{{\bf R}}
\def\bv{{\bf v}}
\def\binf{{\bf\infty}}
\def\eps{\epsilon}
\def\up{\uparrow}
\def\down{\downarrow}
\def\bra{\langle}
\def\ket{\rangle}
\def\sDelta{{\scriptstyle \Delta}}
\def\FS{\partial{\cal F}}
\def\Re{{\rm Re}}
\def\Im{{\rm Im}}
\def\xik{\xi_{\bk}}
\def\cO{{\cal O}}
\def\cD{{\cal D}}
\def\cF{{\cal F}}
\def\cG{{\cal G}}
\def\cZ{{\cal Z}}
\def\Lam{\Lambda}
\def\lam{\lambda}
\def\dbm{\delta\bar\mu}

\def\xip{\xi_{\bp}}
\def\xik{\xi_{\bk}}
\def\xikq{\xi_{\bk+\bq}}
\def\tilk{\tilde k}
\def\tilth{\tilde\theta}
\def\tilxi{\tilde\xi}
\def\dph{\Delta^{ph}}
\def\dpp{\Delta^{pp}}
\def\nsim{\sim \hskip -4truemm / \>}
\def\Dt{\tilde D}
\def\tht{\tilde\theta}
\def\omt{\tilde\omega}
\def\q0t{\tilde q_0}
\def\qrt{\tilde q_r}
\def\qtt{\tilde q_t}
\def\xit{\tilde \xi}
\def\tt{\tilde t}
\def\Lt{\tilde L}
\def\cdotr{\!\cdot\!}
\def\om{\omega}
\def\sg{\sigma}
\def\Sg{\Sigma}

\vspace*{1cm}
\centerline{\large 7. SHORT-RANGE INTERACTIONS IN D DIMENSIONS}
\vskip 1cm
\pp This section is devoted to a quantitative analysis of residual
low-energy scattering processes in normal d-dimensional Fermi systems 
with {\em short-range}\/ interactions. Our starting point is an 
effective low-energy action of the form
$$ \bar S^{\Lam} = \sum_{\sg} \int_k 
   \psi^*_{k\sg} (ik_0 - \xi_{\bk}) \psi_{k\sg} \> - \>
   {\textstyle {1 \over 2}} \sum_{\sg\sg'} \int_{k,k',q} 
   g_{\bk\bk'}^{\sg\sg'}(\bq) \>
   \psi^*_{k-q/2,\sg} \psi_{k+q/2,\sg} 
   \psi^*_{k'\!+q/2,\sg'} \psi_{k'\!-q/2,\sg'}              \eqno(7.1) $$
where $\xi_{\bk} = v_{\bk_F} k_r$ is a linearized dispersion relation 
and $g_{\bk\bk'}^{\sg\sg'}(\bq)$ a renormalized coupling function. 
In a system with short-range interactions, $g$ is assumed to be a 
{\em regular} function of $\bk$, $\bk'$ and $\bq$.
To keep the notation readable, we have suppressed bars indicating 
renormalization and also the cutoff-dependence on the right hand side
of (7.1). 
Of course we have to keep in mind that a field renormalization 
$Z^{\Lam}$ has been performed in passing from some microscopic model 
to the effective low-energy action. 
We assume that $\Lam$ is much smaller than $k_F$, and will analyze the 
effect of the residual two-particle scattering processes. Explicit
results will be derived mainly for rotationally invariant continuum 
systems (without umklapp processes) with a spherical Fermi surface and 
a constant Fermi velocity $v_{\bk_F} = v_F$.
\par
\pp Let us assume that the residual interactions in the Cooper channel
are purely repulsive such that we need not bother about the Cooper 
instability (or backscattering instability in $d=1$). 
We will refer to $\bar S^{\Lam}$ in (7.1) with finite (non-singular) 
coupling functions and vanishing or repulsive Cooper couplings as the 
{\em regular normal model}.
This model describes the low-energy physics of all those Fermi systems 
with short-range interactions where high-energy degrees of freedom do 
not generate an instability (such as symmetry breaking, pair-formation, 
etc.).
\par
\pp In two or three dimensions the regular normal model leads to 
Fermi liquid behavior with fermionic quasi-particle excitations 
(at least to each finite order in perturbation theory), while in one 
dimension Luttinger liquid behavior is found.
In any dimension the leading low-energy long-wavelength response 
is given exactly by the random phase approximation in the limit $\Lam 
\to 0$. 
The momentum distribution function and the structure of single-particle 
excitations are however dramatically different in $d=1$, compared to 
$d=2$ or $d=3$.
\par
\pp In this situation it is interesting to analyze the crossover
from one-dimensional Luttinger liquid behavior to Fermi liquid 
behavior in higher dimensions as a function of {\em continuous 
dimensionality}.\footnote
{Analytic continuation of dimensionality is a common technique in
statistical mechanics and field-theory, especially as a means to 
treat strong coupling problems by performing an $\eps$-expansion
around a critical dimension of the system \cite{DJ,AMI}.}
In particular, one would like to know which {\em critical dimension}\/ 
$d_c$ separates Fermi from Luttinger liquid behavior, and whether
new non-Fermi liquid fixed points can be found within a suitable
$\eps$-expansion.
It turns out that there are in fact {\em two}\/ characteristic 
dimensions: 
In any dimension $d$ above the critical dimension $d_c = 1$ the 
{\em leading}\/ low-energy 
behavior is of Fermi liquid type, but for $d \leq d_c' = 2 $ the 
{\em subleading}\/ corrections differ from the simple behavior known 
for three-dimensional Fermi systems. 
In systems with short-range interactions there are no non-Fermi liquid 
fixed points in $d = 1+\eps$ dimensions.
\par
\pp We will reach a detailed understanding of the influence of 
residual interactions on the low-energy physics of d-dimensional
systems in three steps: i) perturbation theory, ii) RPA, and iii)
a resummation of forward scattering along the lines of Sec.\ 5.
Forward scattering is favored in dimensions $1 \leq d < 2$ by an
angle-dependent phase-space factor (see the angle-integration in
eq.\ (7.2)). 
As a byproduct we obtain some technical material such as analytic
results for the d-dimensional particle-hole bubble, that can be useful 
in future investigations of Fermi systems in continuous dimensionality
(especially $\eps$-expansions).
\par
\pp A short account of the main results of this section has been
published in Ref.\ \cite{CDM94}. 
In the following we work exclusively with real frequencies at zero 
temperature.
\par

\bigskip
 
{\bf 7.1. CONTINUATION TO NON-INTEGER DIMENSIONS} \par
\medskip
\pp The continuation of the theory to non-integer dimensions is obtained 
as usual by analytic continuation of Feynman diagrams, defined for general 
integer $d$, in the complex $d$-plane. 
For our purposes it will be sufficient to continue momentum integrals of 
functions $f(\bk)$ which depend on $\bk$ only via $|\bk|$ and an angle 
$\theta$ between $\bk$ and another momentum which is fixed. In these cases 
one can use
$$ \int d^d\bk \dots \> = \> 
   S_{d-1} \int_0^{\infty} d|\bk| \> |\bk|^{d-1} 
   \int_0^{\pi} d\theta \> (\sin\theta)^{d-2} \dots         \eqno(7.2) $$
where 
$$ S_d = 2\pi^{d/2}/\Gamma(d/2)                             \eqno(7.3) $$ 
is the surface of the d-dimensional unit sphere. 
In the limit $d \to 1$ one has $S_{d-1} \sim d-1$, and thus
$$ S_{d-1} (\sin\theta)^{d-2} \> \to \> 
   \delta(\theta) + \delta(\theta-\pi) 
   \quad {\rm for} \quad d \to 1                            \eqno(7.4) $$ 
as expected. For fermionic momenta (not momentum transfers) near the 
Fermi surface it is convenient to decompose
$$ |\bk| = k_F + k_r                                        \eqno(7.5) $$
where $k_r$ (the oriented distance from the Fermi surface) is much 
smaller than $k_F$. Approximating $|\bk|^{d-1} \sim k_F^{d-1}$, (7.2) 
may then be simplified to
$$ \int d^d\bk \dots \> = \> S_{d-1} k_F^{d-1} \int dk_r
   \int_0^{\pi} d\theta \> (\sin\theta)^{d-2} \dots         \eqno(7.6) $$
where the $k_r$-integration is restricted either by a cutoff $\Lam$ or
by small external momentum or energy variables.
Corrections are suppressed by a factor of order $(d\!-\!1)k_r/k_F$ in
the integral (7.6).
\par

\bigskip

{\bf 7.2. PERTURBATIVE RESULTS} \par
\medskip
\pp We will first present explicit results for the particle-hole and
particle-particle bubble in $d$ dimensions, and then for the second
order self-energy.
\par

\bigskip

{\bf 7.2.1. Particle-hole bubble} \par
\medskip
\pp The {\em particle-hole bubble}\/ (or bare polarization insertion) 
$$ \Pi_0(q) = - i \int_k G_0(k) G_0(k+q)                    \eqno(7.7) $$
is essentially the dynamical density-density correlation function of
the non-inter\-acting system, and will be important for subsequent explicit 
calculations of the self-energy $\Sg$ and the effective interaction 
$D$. At zero temperature, the particle-hole bubble can be written as
$$ \Pi_0(\om,\bq) = \int {d^d\bk \over (2\pi)^d} \>
   {\Theta(\xikq) - \Theta(\xik) \over 
   \om - (\xikq-\xik) + i0^+s(\om)}                         \eqno(7.8) $$
where $\xik = \eps_{\bk} - \mu$. The spectral density of particle-hole
excitations is given by
$$ \dph_0(\om,\bq) = 
   - {1 \over \pi} s(\om) \> \Im\Pi_0(\om,\bq)              \eqno(7.9) $$ 
This is a positive quantity for $\om > 0$ and negative for 
$\om < 0$ (as usual for spectral functions of bosons). 
For positive energies, $\dph_0$ can be written as
$$ \dph_0(\om,\bq) = \int {d^d\bk \over (2\pi)^d} \>
   \Xi(\xik < 0 < \xikq) \>
   \delta[\om - (\xikq-\xik)]                              \eqno(7.10) $$
while $\dph_0(-\om,\bq) = - \dph_0(\om,\bq)$. 
Here $\Xi(condition) = 1$ if the condition is satisfied, and 
$\Xi(condition) = 0$ otherwise.
\par
\smallskip
\pp For a linear dispersion, $\xik = v_F k_r$, several analytic results
can be obtained for the low energy behavior (small $\om$) of 
$\Pi_0(\om,\bq)$, in any dimension $d$. Using 
$$ \xikq = v_F(|\bk+\bq|-k_F) = 
   v_F \> \big[ (|\bk|^2 + |\bq|^2 + 
   2|\bk||\bq| \cos\theta)^{1/2} - k_F \big]               \eqno(7.11) $$
where $\theta$ is the angle spanned by $\bk$ and $\bq$, and introducing
spherical coodinates, $\Pi_0(\om,\bq)$ can be written as a two-fold 
integral over the variables $|\bk|$ and $\theta$. 
\par
\pp For the {\em imaginary} part $\dph_0$ the integration is simplified 
by the presence of the $\delta$-function. For small $\om$ (compared
to $v_Fk_F$) but general $\bq$ one obtains
$$ \dph_0(\om,\bq) \> \sim \> 
   {S_{d-1} \over (2\pi)^d} {2 \over d\!-\!1} 
   {k_F^d \over v_F} {1 \over |\bq|}
   (1+|\bq|/2k_F)^{d-3 \over 2} 
   \big[ 1-(\om/v_F|\bq|)^2 \big]^{d-3 \over 2} 
   \hskip 3cm  $$ 
\vskip -7mm
$$ \left[ (1-|\bq|/2k_F+\om/2v_Fk_F)^{d-1 \over 2} - 
   (1-|\bq|/2k_F-\om/2v_Fk_F)^{d-1 \over 2} 
   \Xi(\om/v_F < 2k_F-|\bq|) \right]                    \eqno(7.12) $$
for $\max(0,|\bq|-2k_F) < \om/v_F < |\bq|$, while $\dph_0(\om,\bq)
= 0$ for other $\om > 0$, and $\dph_0(-\om,\bq) = 
- \dph_0(\om,\bq)$. For $|\bq| < 2k_F$ and not close to $2k_F$, this
simplifies to 
$$ \dph_0(\om,\bq) \sim {S_{d-1} \over (2\pi)^d} 
   {k_F^{d-1} \over v_F^2} \> {\om \over |\bq|} \>
   [1-\om^2/(v_F|\bq|)^2]^{d-3 \over 2} 
   [1-(|\bq|/2k_F)^2]^{d-3 \over 2} 
   \Xi(|\om| < v_F|\bq|)                                \eqno(7.13) $$
\pp In general (non-integer) dimensions, the {\em real} part of 
$\Pi_0(\om,\bq)$ cannot be expressed in terms of elementary functions,
even if $\om$ is small.
Hence we will only present some useful results obtained for various
limits. 
The behavior of $\Pi_0(\om,\bq)$ for small $\om$ and $\bq$ is
particularly important. Since for $|\bq| \sim 0$ the excitation 
energies $\xikq - \xik \sim |\bq| \cos\theta$
depend only on $|\bq|$ and $\theta$, the integration over $|\bk|$
is easily carried out. For $\cos\theta > 0$ one has contributions from
$k_r \in [-|\bq|\cos\theta,0]$, while for $\cos\theta < 0$ values
$k_r \in [0,-|\bq|\cos\theta]$ contribute, yielding
$$ \Pi_0(\om,\bq) \sim {S_{d-1} \over (2\pi)^d} \> k_F^{d-1}
   \int_0^{\pi} d\theta {|\bq| \cos\theta \over 
   \om - v_F|\bq| \cos\theta + i0^+s(\om)} (\sin\theta)^{d-2} 
                                                           \eqno(7.14) $$
This depends on $\om$ and $\bq$ only via the ratio $\om/|\bq|$.
Results from a numerical evaluation of (7.14) in various dimensions 
are shown in Fig.\ 7.1.
For $|\bq|/\om \to 0$, $\Pi_0(\om,\bq)$ vanishes, while in the
opposite limit $\om/|\bq| \to 0$ it assumes the negative real value 
$$ \Pi_0(0,\bq) \sim - {S_d \over (2\pi)^d} 
   {k_F^{d-1} \over v_F} = - N_F                           \eqno(7.15) $$    
where $N_F$ is the density of states per spin at the Fermi level.
For a linear dispersion, $\Pi_0(\om,\bq)$ diverges in the limit 
$\om/v_F|\bq| \to 1$ in any dimension $d \leq 3$. In $d < 3$, the
leading singularity is given by
$$ \Pi_0(\om,\bq) \sim {S_{d-1} \over (2\pi)^d} \> 2^{d-3 \over 2}
   (B_d^{\pm} + i\pi) k_F^{d-1} v_F^{-1} \>
   |\om/v_F|\bq| - 1|^{d-3 \over 2} \quad
   {\rm for} \quad {\om \over v_F|\bq|} \to 1+0^{\pm}      \eqno(7.16) $$
where
$$ B_d^+ = B[(d\!-\!1)/2),(3\!-\!d)/2] \> , \quad
   B_d^- = B_d^+ \cos[(3\!-\!d)/2]                         \eqno(7.17) $$
where $B(x,y) = \Gamma(x)\Gamma(y)/\Gamma(x+y)$ is Euler's Beta-function 
(a special function, not to be confused with the $\beta$-function from 
the renormalization group).
Note that $B_d^-$ is positive in $d>2$ and negative in $d<2$.
An exception is the real part of $\Pi_0(\om,\bq)$ for $\om < 
v_F|\bq|$ in $d = 2$, since $B_2^- = 0$; in this case one finds
$$ \Re\Pi_0(\om,\bq) \> \sim \> - k_F/2\pi v_F \quad
   {\rm for \>\> all} \quad \om < v_F|\bq| \quad
   {\rm in} \>\> d=2                                       \eqno(7.18) $$ 
\par
\pp Note that the cutoff $\Lam$ does not appear in the above results,
since the $\bk$-integral is already limited by other small variables
such as $\om/v_F$ or $|\bq|$.
\par
\pp In Appendix C we list results for the particle-hole bubble in d
dimensions obtained with a quadratic (not linearized) dispersion 
relation $\eps_{\bk} = \bk^2/2m$.
\par

\bigskip

{\bf 7.2.2. Particle-particle bubble} \par
\medskip
\pp The particle-particle bubble is defined by
$$ K_0(p) = i \int_k G_0(k) G_0(p-k)                       \eqno(7.19) $$
This quantity plays a role in the T-matrix approximation for systems 
with local interactions \cite{FHN,ER92}. It also enters into a 1-loop 
calculation of the $\beta$-function for Cooper couplings.
\par
\pp At zero temperature, the particle-particle bubble can be written as
$$ K_0(\om,\bp) = \int {d^d\bk \over (2\pi)^d}
   {\Theta(\xi_{\bp-\bk}) - \Theta(-\xik) \over 
   \om - (\xi_{\bp-\bk}+\xik) + i0^+s(\om)}                \eqno(7.20) $$
The spectral density for two-particle (two-hole) excitations is given
by
$$ \dpp_0(\om,\bp) = 
   - {1 \over \pi} s(\om) \Im K_0(\om,\bp)                 \eqno(7.21) $$
This quantity is positive for $\om > 0$ (two particles) and negative 
for $\om < 0$ (two holes). Another way of writing $\dpp_0$ is
$$ \dpp_0(\om,\bp) = \int {d^d\bk \over (2\pi)^d}
   [\Theta(\xik)\Theta(\xi_{\bp-\bk}) - 
    \Theta(-\xik)\Theta(-\xi_{\bp-\bk})] 
   \delta(\om-\xik-\xi_{\bp-\bk})                          \eqno(7.22) $$
\pp We now present analytic results for $\dpp_0(\om,\bp)$ for small 
$\om$ and a linear dispersion relation $\xik = v_F k_r$. For
$\om < v_F(|\bp|-2k_F)$, $\dpp_0(\om,\bp)$ vanishes. For 
$\om > v_F(|\bp|-2k_F)$, one finds
$$ \dpp_0(\om,\bp) \sim 
   s(\om) {S_d \over 2(2\pi)^d} {k_F^{d-1} \over v_F}     \eqno(7.23a) $$
\vskip -4mm
if $|\om| > |\bp|$, while
$$ \dpp_0(\om,\bp) \sim {S_{d-1} \over (2\pi)^d} {k_F^{d-1} \over v_F}
   \left[(1+\om/2v_Fk_F)^2 - (|\bp|/2k_F)^2 \right]^{d-3 \over 2} 
   \int_0^{\om/v_F|\bp|} dx \, (1-x^2)^{(d-3)/2}          \eqno(7.23b) $$ 
if $|\om| < |\bp|$. The latter result simplifies in particular cases,
namely to
$$ \dpp_0(\om,\bp) \sim {S_{d-1} \over (2\pi)^d} {k_F^{d-1} \over v_F}
   \int_0^{\om/v_F|\bp|} (1-x^2)^{(d-3)/2}                \eqno(7.24a) $$ 
for small $|\bp|$,
$$ \dpp_0(\om,\bp) \sim {S_{d-1} \over (2\pi)^d} {k_F^{d-1} \over v_F}
   \left[(1+\om/2v_Fk_F)^2 - (|\bp|/2k_F)^2 \right]^{d-3 \over 2} 
   {\om \over |\bp|}                                      \eqno(7.24b) $$
for $|\bp| \sim 2k_F$, and to
$$ \dpp_0(\om,\bp) \sim {S_{d-1} \over (2\pi)^d} {k_F^{d-1} \over v_F}
   [(1 - (|\bp|/2k_F)^2]^{d-3 \over 2} \> 
   {\om \over |\bp|}                                      \eqno(7.24c) $$
if $|\bp|$ is neither close to $0$ nor close to $2k_F$. We emphasize
that these results hold only for $\om$ small compared to $v_Fk_F$. 
\par
\pp Note that (7.23a) implies that $|\dpp_0(\om,\b0)|$ is a constant
at small $\om$ in any dimension.
Hence (by Kramers-Kronig relations), the real
part of the particle-particle bubble $\Re K_0(\om,\b0)$ is
logarithmically divergent for $\om \to 0$:
$$ \Re K_0(\om,\b0) \sim 
   - {S_d \over (2\pi)^d} {k_F^{d-1} \over v_F} \log(\om/v_F\Lam)
                                                           \eqno(7.25) $$
where $\Lam$ is a cutoff. This leads to a finite contribution to the
$\beta$-function for Cooper couplings, as long as (renormalized) Cooper
couplings are non-zero (see Sec.\ 2).
\par
\pp In Appendix C we list results for the particle-particle bubble in 
d dimensions obtained with a quadratic (not linearized) dispersion 
relation $\eps_{\bk} = \bk^2/2m$.
\par

\bigskip

{\bf 7.2.3. Second order self-energy} \par
\medskip
\pp The {\em first order}\/ self-energy does not yield any many-body 
effects, but merely shifts the chemical potential $\mu$, and possibly 
the Fermi velocity $v_F$. Hence we concentrate immediately on the 
{\em second order}\/ self-energy, which will be calculated for a 
{\em constant coupling}\/ $g$ acting between particles with {\em opposite 
spins}. 
The restriction to a momentum-independent coupling corresponds to a local 
interaction in real space, and is made to provide us with explicit results
for one concrete case.\footnote
{Obviously, a constant coupling between parallel spins would yield no 
contribution, as a consequence of the Pauli principle.} 
This case is however a generic
representative for short-range interactions, yielding all the typical
dynamical many-body effects described by a self-energy in a normal
Fermi system: quasi-particle decay, reduction of quasi-particle weight,
and complete destruction of the quasi-particle in one dimension.
\par
\pp In a system with interactions between opposite spins (only), there
is only one contribution to the second order self-energy, represented
diagrammatically in Fig.\ 7.2. Algebraically, this corresponds to
$$ \Sg(p) = g^2 \int_{k,q} G_0(p-q) G_0(k) G_0(k+q)        \eqno(7.26) $$
We concentrate on the imaginary part, $\Im\Sg(p)$, from which the
real part can be constructed via the Kramers-Kronig (or spectral) 
representation
$$ \Sg(\xi,\bp) = -\pi^{-1} \int_{-\infty}^{\infty} d\xi' \>
   {s(\xi')\Im\Sg(\xi',\bp) \over \xi-\xi' + i0^+s(\xi)}   \eqno(7.27) $$
In general the imaginary part determines the real part only up to an
energy-independent constant, which however in our case must vanish, since
$\Sg(p)$ in (7.26) obviously vanishes for $\xi \to \infty$.
For constant couplings, $\Im\Sg$ can be conveniently expressed in 
terms of the spectral density of particle-hole excitations as
$$ \Im \Sg (\xi,\bp) = 
   \left\{ \begin{array}{r@{\quad {\rm for} \quad}l}
   -\pi g^2 \int {d^d\bk \over (2\pi)^d} \> \dph_0(\xi-\xik,\bp-\bk) \>
   \Xi(0< \xik < \xi) &
   \xi > 0 \\
   -\pi g^2 \int {d^d\bk \over (2\pi)^d} \> \dph_0(\xi-\xik,\bp-\bk) \> 
   \Xi(\xi < \xik < 0) &
   \xi < 0
   \end{array} \right.                                     \eqno(7.28) $$

or, alternatively, in terms of the spectral density of two-particle (or
two-hole) excitations
$$ \Im \Sg (\xi,\bp) = 
   \left\{ \begin{array}{r@{\quad {\rm for} \quad}l}
   -\pi g^2 \int {d^d\bk \over (2\pi)^d} \> \dpp_0(\xi+\xik,\bp+\bk) \>
   \Xi(\xi+\xik > 0 > \xik) &
   \xi > 0 \\
   -\pi g^2 \int {d^d\bk \over (2\pi)^d} \> \dpp_0(\xi+\xik,\bp+\bk) \> 
   \Xi(\xi+\xik < 0 < \xik) &
   \xi < 0
   \end{array} \right.                                     \eqno(7.29) $$

\pp A general analytic evaluation of these integrals is difficult. Hence,
we have calculated $\Im\Sg$ numerically, and confirmed only the most
striking properties by analytic derivations.
Typical results for $\Im\Sg(\xi,\bp)$ are shown in Fig.\ 7.3.
In three dimensions we observe the well-known quadratic energy dependence, 
$\Im\Sg(\xi,\bp) \propto \xi^2$, without any special feature at 
$\xi = \xip = v_F p_r$. 
For $1 < d \leq 2$ contributions of order $\xi^2$ are superposed by larger 
$\xip$-dependent terms. 
In $d<2$, for small $p_r = |\bp|-k_F$ and $\xi$, the self energy 
scales as 
$$ \Im\Sg(\xi,\bp) = 
   |p_r|^d \> \Im\tilde\Sg(\xi/p_r)                        \eqno(7.30) $$
and diverges in $\xi = \xip$ as \cite{CDM94}
$$ \Im\Sg(\xi,\bp) \> \sim \> - C_d \> g^2 k_F^{d-1} v_F^{-(d+1)} 
    s(\xip) \> \xi^2 |\xi - \xip|^{d-2}                    \eqno(7.31) $$
where $C_d = 2^{-d-4}\pi^{1-2d} S_{d-1}^2 B[2\!-\!d,(d\!-\!1)/2]$ is 
a constant depending only on dimensionality. In $d=2$ there is a weak
logarithmic divergence for $\xi \to \xip$. This singularity
is exclusively due to {\em forward scattering}\/ (i.e.\ small momentum
transfers $\bq \!\sim\! 0$) of particles with {\em opposite spin}\/ and 
almost {\em parallel momenta}. Hence, for more general spin- 
and momentum-dependent couplings $g_{\bk\bk'}^{\sg\sg'}(\bq)$ one 
would obtain the same singularity, with a coupling 
$g_{\bk\bk}^{\sg,-\sg}(\b0) = g_F^{\sg,-\sg}(\theta=0)$ replacing $g$.
Forward scattering of particles with {\em antiparallel momenta}\/ yields 
a contribution proportional to $g^2|\xi-\xip|^d \> \theta(|\xi|-|\xip|)$ 
in $d<2$, with $g$ to be replaced by $g_{\bk,-\bk}^{\sg\sg'}(\b0) =  
g_F^{\sg\sg'}(\theta=\pi)$ for more general couplings.\footnote
{Analytic results for the second order self-energy within a
one-dimensional g-ology model are given in Appendix D.}
\par
\pp In $d < 2$, the single-particle propagator $G$ is drastically affected 
by the singular contributions of forward scattering to the second order 
self-energy. 
\par
\pp Let us consider the one-dimensional case first. In $d = 1$, forward 
scattering between particles with opposite momenta ($H_2$, in g-ology
notation) yields a contribution 
proportional to $g^2(\xi-\xip)\theta(|\xi|-|\xip|)$ to $\Im\Sg$, which, 
via Kramers-Kronig, yields a real part proportional to 
$g^2 \xi \log|\xi|$ for $\bp$ on the Fermi surface. 
This leads to a wave function renormalization
$Z = (1 - \partial\Sg/\partial\xi)^{-1} \propto g^2/\log|\xi| \to 0$ as 
$\xi \to \xip = 0$, which is a well-known perturbative signal for the
breakdown of Fermi liquid theory in one dimension \cite{SOL}, i.e. a
power-law behavior of the wave function renormalization with a 
non-universal exponent $\eta$.
Forward scattering between particles with parallel momenta ($H_4$, in
g-ology notation) does not 
contribute to the wave function renormalization, but nevertheless 
destroys the quasiparticle pole in the propagator, leading to {\em 
spin-charge separation}\/ \cite{VOI95,SCH95}. 
In contrast to common wisdom this latter effect also has a clear 
perturbative signal \cite{CDM94}: In $d \to 1$, (7.31) reduces to 
$$ \Im\Sg(\xi,\bp) \> \sim \> - {g^2 \over 8\pi v_F^2} 
   s(\xip) \> \xi^2 \delta(\xi - \xip)                     \eqno(7.32) $$
yielding, by Kramers-Kronig, a real part 
$$ \Re\Sg(\xi,\bp) \> \sim \> {g^2 \over 8\pi^2 v_F^2} 
   {\xip^2 \over \xi - \xip}                               \eqno(7.33) $$
Inserting this into $G = (\xi-\xip-\Sg)^{-1}$ one obtains a propagator
which has two poles instead of one, i.e.\ the spectral function becomes
a sum of two $\delta$-functions with weight $1/2$ each:
$$ \rho(\xi,\bp) = 
   {1 \over 2} \delta(\xi-\xip^+) + {1 \over 2} \delta(\xi-\xip^-)
                                                           \eqno(7.34) $$
where $\xip^- < \xip < \xip^+$.
\par
\pp In $1<d<2$ the perturbatively calculated wave function renormalization 
is finite, but the forward scattering processes of particles with almost 
parallel momenta still have dramatic consequences: 
$\Sg(\xi,\bp)$ has an algebraic divergence proportional to 
$(\xi-\xip)^{d-2}$ for $\xi \to \xip$, leading to two well-separated peaks 
of comparable weight in the spectral function, as shown for $d = 1.5$ in 
Fig.\ 7.4. 
The width of the peaks is finite in $d>1$, but smaller than the distance
between them.\footnote
{A similar behavior has been found recently near zero-curvature points 
on special anisotropic Fermi surfaces in two dimensions \cite{FO}.}
Hence, second order perturbation theory seems to indicate destruction 
of the quasiparticle pole due to forward scattering of particles with 
almost parallel momenta in any dimension below two! 
However, the divergence found in $\Sg(\xi,\bp)$ clearly forces us to go 
beyond perturbation theory even for weak coupling constants.
\par

\bigskip

{\bf 7.3. RANDOM PHASE APPROXIMATION} \par
\medskip
\pp Our next step is the calculation of the small-$q$ density-density 
response and the self-energy in random phase approximation (RPA). 
To leading order in $q$, the response function obtained by this 
"approximation" is actually the exact result for the regular normal 
model in any dimension, if the cutoff $\Lam$ is so small that the 
forward scattering interactions 
$f_{\bk\bk'}^{\sg\sg'} = g_{\bk\bk'}^{\sg\sg'}(\b0)$ have already 
reached their fixed point values.
\par

\bigskip

{\bf 7.3.1. Effective interaction} \par
\medskip
\pp The RPA {\em effective interaction} $D$ is an important auxiliary
quantity, which will play a central role not only in RPA calculations.
It can be defined by the sum of diagrams illustrated already in 
Fig.\ 5.2, which is equivalent to the linear integral equation (5.21).
For $\bk$ and $\bk'$ close to the Fermi surface and small $q$ this can 
be simplified to 
$$ D_{\bk\bk'}^{\sg\sg'}(q) = g_{\bk\bk'}^{\sg\sg'}(\b0) +
   \sum_{\sg''} \int_{\bk'' \in \FS} g_{\bk\bk''}^{\sg\sg''}(\b0) \>
   {\bn_{\bk''} \cdotr \bq \over 
   \om - \bv_{\bk''} \cdotr \bq + i0^+s(\om)}\>
   D_{\bk''\bk'}^{\sg''\sg'}(q)                           \eqno(7.35) $$
with a surface integral extending over the Fermi surface $\FS$, where
$\bn_{\bk}$ is a normal unit vector on the Fermi surface in $\bk$. 
Obviously, for small $q$ the effective action depends only via the ratio 
$\bq/\om$ on $q = (\om,\bq)$.
Equation (7.35) cannot be solved in general. There are however
instructive special cases with drastic simplifications. In particular,
$$ D_{\bk\bk'}^{\sg\sg'}(q) = D(q) = 
   {g \over 1 - 2g\Pi_0(q)} \quad {\rm for} \quad 
   g_{\bk\bk'}^{\sg\sg'}(\bq) = g                         \eqno(7.36) $$
where the factor $2$ is due to the spin degeneracy, while
$$ D_{\bk\bk'}^{\sg\sg'}(q) = D^{\sg\sg'}(q) = 
   \left\{ \begin{array}{r@{\quad {\rm for} \quad}l}
   g^2 \Pi_0(q)/[1 - (g\Pi_0(q))^2] &
   \sg = \sg' \\
   g/[1 - (g\Pi_0(q))^2] &
   \sg = - \sg'
   \end{array} \right. \hskip 2cm $$
\vskip -6mm
$$ \hskip 7cm {\rm for} \quad g_{\bk\bk'}^{\sg\sg'}(\bq) = 
   g \delta_{\sg,-\sg'}                                   \eqno(7.37) $$
The latter case, constant coupling between opposite spins, has been
chosen for the perturbative evaluation of the self-energy in the preceding
section. In both special cases the effective interaction depends only
via the modulus $|\bq|$ on $\bq$. If in addition $q$ is small, the
effective interactions depend only on a single variable, e.g. the
ratio $\omt = \om/|\bq|$. 
\par
\pp Numerical results for $D^{\sg\sg}(q)$ for small $q$ as a function 
of $\omt$ in various dimensions are shown in Fig.\ 7.5, for the 
case (7.37) with $g = 2$ and $v_F = 1 = k_F$.
In any dimension there is a singularity of the form 
(for $\omt > 0$)
$$ D_c(q) = {Z_c \over \omt - u_c + i0^+}                  \eqno(7.38) $$
with $u_c > v_F$ associated with a propagating density mode.
Apart from a $\delta$-function in $\om = u_c |\bq|$, one has 
$\Im D^{\sg\sg}(q) = 0$ for $\om > v_F |\bq|$, 
as follows directly from the absence of particle-hole
excitations with energies larger than $v_F |\bq|$. 
For $\omt \to 0$, $\Im D^{\sg\sg}(q)$ vanishes linearly, and in
$1< d < 3$ as $(v_F-\omt)^{(3-d)/2}$ for $\omt \to v_F-0^+$. 
Close to one dimension, the effective interaction has a damped 
singularity at $\om = u_s |\bq|$ with a velocity $u_s < v_F$. 
For $d \to 1$ the damping vanishes, and one is left with an additional 
pole as in (7.38), but now in $\omt = u_s$, while all the spectral 
weight at other energies ($\om \neq u_c |\bq|, u_s |\bq|$) has 
disappeared.
Qualitatively the same behavior is found for other positive coupling 
strengths, too. 
The only difference in the case (7.36) is the absence of the second
singularity at $\om = u_s |\bq|$.
\par

\bigskip

{\bf 7.3.2. Density-density response} \par
\medskip
\pp The RPA charge density-density response (or correlation function) 
$N(q)$ can be constructed from the effective interaction in the way 
illustrated in Fig.\ 5.3, i.e.
$$ N(q) = 2\Pi_0(q) + \sum_{\sg\sg'} \int_{k,k'}
   D_{\bk\bk'}^{\sg\sg'}(q) \> G_0(k-q/2) G_0(k+q/2) 
   G_0(k'\!-q/2) G_0(k'\!+q/2)                             \eqno(7.39) $$
For small $q$ the energy-momentum integrals can be reduced to Fermi
surface averages as before. For the two simple special cases introduced
in 7.3.1 one obtains
$$ N(q) = {2 \Pi_0(q) \over 1 - 2g\Pi_0(q)}  \quad {\rm for} \quad
   g_{\bk\bk'}^{\sg\sg'}(\bq) = g                          \eqno(7.40) $$
and
$$ N(q) = {2 \Pi_0(q) \over 1 - g\Pi_0(q)}   \quad {\rm for} \quad 
   g_{\bk\bk'}^{\sg\sg'}(\bq) = g\delta_{\sg,-\sg'}
                                                           \eqno(7.41) $$ 
For $g > 0$ these functions have a pole in $\om = u_c |\bq|$, 
describing a propagating charge density mode, which is Landau's 
\cite{LAN} zero-sound mode in our special system. 
\par
\pp Similar results are easily obtained for the spin density-density
response. 
At least one propagating (charge or spin) density mode exists in any 
Fermi liquid with short-range interactions \cite{MER}.
\par

\bigskip

{\bf 7.3.3. RPA self-energy} \par
\medskip
\pp We now calculate the self-energy in random phase approximation, 
concentrating in particular on the leading contributions from scattering
with {\em small}\/ momentum transfers, which led to striking singularities
in perturbation theory in dimensions $d<2$. Diagrammatically, the RPA
self-energy is described by Fig.\ 7.6, representing a propagator dressed 
once by the effective interation $D$. Algebraically, this reads
$$ \Sg(p) = 
   i \int_{p'} \> D_{\bk\bk}^{\sg\sg}(p\!-\!p') \> G_0(p') 
   \quad {\rm where} \quad \bk = {\bp+\bp' \over 2}        \eqno(7.42) $$
In particular, the imaginary part can be expressed as
$$ \Im\Sg(\xi,\bp) = 
   \left\{ \begin{array}{r@{\quad {\rm for} \quad}l}
   \int {d^d\bp' \over (2\pi)^d} \> 
   \Im D_{\bk\bk}^{\sg\sg}(\xi-\xi_{\bp'},\bp-\bp') \>
   \Xi(0< \xi_{\bp'} < \xi) &
   \xi > 0 \\
   -\int {d^d\bp' \over (2\pi)^d} \> 
   \Im D_{\bk\bk}^{\sg\sg}(\xi-\xi_{\bp'},\bp-\bp') \> 
   \Xi(\xi < \xi_{\bp'} < 0) &
   \xi < 0
   \end{array} \right.                                     \eqno(7.43) $$
\pp We are interested in the contributions from small momentum transfers
$\bq = \bp\!-\!\bp'$ to $\Im\Sg(p)$. A small $\bq$ can be decomposed in
normal and tangential components with respect to the Fermi surface in 
$\bp \sim \bp' \sim \bk$, 
and the effective interaction can be parametrized as (see Eq.\ (5.21))
$$ D_{\bk\bk}^{\sg\sg}(\om,\bq) \> \sim \>
   D(\om,q_r,q_t) \quad {\rm for} \quad 
   \bq = \bp\!-\!\bp' \sim \b0                              \eqno(7.44) $$
We now consider contributions from scattering processes with tangential
momentum transfers restricted by $q_t < \Lam_t$ to the self-energy, 
where $\Lam_t \ll k_F$. 
Decomposing the $\bp'$-integral in a radial and an angular integral as 
in (7.6) and using $\sin\theta \sim \theta \sim q_t/k_F$, one obtains
$$ \Im\Sg^{\Lam_t}(\xi,\bp) =   
   \int_0^{\xi/v_F} {dp'_r \over 2\pi} \> 
   \Im \bar D^{\Lam_t}(\xi\!-\!v_F p'_r,p_r\!-\!p'_r)        \eqno(7.45) $$
where $\bar D^{\Lam_t}$ is the $q_t$-averaged effective interaction, 
defined by (cf.\ (5.22))
$$ \bar D^{\Lam_t}(\om,q_r) = {S_{d-1} \over (2\pi)^{d-1}}
   \int_0^{\Lam_t} dq_t \> q_t^{d-2} \> 
    D(\om,q_r,q_t)                                         \eqno(7.46) $$
\par
\pp We will now show that for $p \to (0,\bk_F)$ the leading contribution 
to $\Im\Sg^{\Lam_t}(p)$ obeys a scaling law with a scaling function
$\Im\tilde\Sg(\xi/p_r)$ that is independent of $\Lam_t$ in $d < 2$.
To see this, we note that for short-range interactions the effective 
interaction depends on $(\om,\bq)$ only via the ratio $\bq/\om$, if
$q$ is small, i.e.\ (7.44) reduces to
$$ D_{\bk\bk}^{\sg\sg}(\om,\bq) \sim D(q_r/\om,q_t/|\om|)
   \quad {\rm for} \quad \bq \sim \b0                      \eqno(7.47) $$
Introducing a rescaled variable $\qtt = q_t/|\om|$, one thus obtains
$$ \Im \bar D^{\Lam_t}(\om,q_r) \> = 
   {S_{d-1} \over (2\pi)^{d-1}} |\om|^{d-1} 
   \int_0^{\Lam_t/|\om|} d\qtt \> \qtt^{d-2} \> \Im D(q_r/\om,\qtt)    $$
\vskip -4mm
$$ \stackrel{\om,q_r \to 0}{\longrightarrow} \>
   |\om|^{d-1} \Im\Dt(q_r/\om)                             \eqno(7.48) $$
in $d<2$, where
$$ \Im\Dt(q_r/\om) = {S_{d-1} \over (2\pi)^{d-1}} \int_0^{\infty} 
    d\qtt \> \qtt^{d-2} \> \Im D(q_r/\om,\qtt)             \eqno(7.49) $$
Note that this integral is convergent in $d<2$ (but not in $d \geq 2$),
since $\Im D(q_r/\om,q_t/|\om|)$ vanishes linearly for 
$\om \to 0$ at fixed $\bq$, and thus the integrand is proportional to
$\qtt^{d-3}$ for large $\qtt$. Inserting (7.48) in (7.45), one obtains
$$ \Im\Sg^{\Lam_t}(\xi,\bp) \> \sim \> 
   s(p_r)|p_r|^d \> \Im\tilde\Sg(\xi/p_r)
   \quad {\rm for} \quad p \to (0,\bk_F)                   \eqno(7.50) $$
\vskip -4mm
where 
$$ \Im\tilde\Sg(\tilxi) =  
   \int_0^{\tilxi/v_F} d\tilde p'_r \> 
   |\tilxi-v_F\tilde p'_r|^{d-1} \>
   \Im\Dt[(1\!-\!\tilde p'_r)/(\tilxi\!-\!v_F\tilde p'_r)] \eqno(7.51) $$
\par
\pp The fact that the tangential cutoff $\Lam_t$ does not appear in 
the leading small-$p$ scaling function implies in particular that
contributions from momentum transfers with $q_t > \Lam_t$ scale to 
zero more rapidly than $|p_r|^d$ in $d<2$, however small $\Lam_t$ may be.
The closer $(\xi,\bp)$ is to the Fermi surface, the smaller is the typical
size of $q_t$ contributing to the self-energy. Note that the radial
momentum transfer $q_r$ is also confined to a small interval, e.g. to
$[p_r\!-\!\xi,p_r]$ for $\xi>0$.
In this sense the low-energy behavior of the RPA self-energy is dominated
by small momentum transfers $|\bq|$ in $d<2$. 
We note that special Cooper processes with momentum transfers
$|\bq| \sim 2k_F$ would give rise to a contribution of the same order.
However, we ignore this contribution, since their scattering amplitudes
must vanish in the low-energy limit in a normal phase. 
\par
\pp In Fig.\ 7.7 we show $\Im\Sg(\xi,\bp)$ at $p_r = 0.1k_F$
for a constant coupling $g=2$ between opposite spins (as in Fig.\ 7.3),
in $d = 1.5$ dimensions. The divergence in $\xi = \xip$ found in second
order perturbation theory has disappeared. Instead one finds two finite
peaks above and below $\xip$, which are obviously due to low energy charge 
and spin density fluctuations (signalled by the corresponding peaks in
the effective interaction $D$). In contrast to the perturbative result,
$\tilde\Sg(\tilxi)$ is now a bounded function. 
Hence, in $1<d<2$ dimensions and for $\bp$ sufficiently close to the 
Fermi surface, the RPA self-energy does not destroy the quasi-particle 
pole in the propagator, but leads only to a damping proportional to 
$|p_r|^d$, which is larger than the quadratic behavior known for 
three-dimensional Fermi liquids, but smaller than the quasi-particle 
energy $\xip = v_F p_r$.
\par

\bigskip

{\bf 7.4. RESUMMATION OF FORWARD SCATTERING} \par
\medskip
\pp We have learned from the perturbative and RPA analysis that in the
absence of Cooper scattering (or "back-scattering", in $d=1$) the 
self-energy of the regular normal model in $d < 2$ dimensions is 
dominated by scattering processes with small momentum transfers 
$\bq \sim \b0$, i.e.\ forward scattering. 
We now sum these dominant processes to all orders in the (residual)
interaction by applying the results derived in Sec.\ 5.
In one dimension this resummation is necessary to obtain the correct
Luttinger liquid behavior of the single-particle propagator.
Hence, such a resummation is also required to get a quantitative control
of the crossover from one-dimensional Luttinger liquid behavior to 
Fermi liquid behavior in dimensions $d>1$.
\par
\pp Focussing on the role of forward scattering, we can introduce a
momentum transfer cutoff $\Lam_q \ll \Lam, k_F$, and restrict momentum
transfers by $|\bq| < \Lam_q$ (as done already in Secs.\ 5 and 6). 
Actually we will frequently work with
two distinct cutoffs $\Lam_r$ and $\Lam_t$ for radial and tangential
momentum transfers, respectively, and discuss their role separately.
In explicit calculations we will usually cut off smoothly with an
exponential factor, e.g. $e^{-|\bq|/\Lam_q}$, which is technically more
convenient than a sharp cutoff.
\par
\pp In Sec.\ 5 we have shown that a resummation of forward scattering
to all orders in the coupling leads to the effective one-dimensional 
equation of motion 
$$ (\xi - v_F p_r) \> G(\xi,p_r) \> = \> 1 \> + \> 
   \int_{\xi',p_r'} {i \bar D(\xi\!-\!\xi',p_r\!-\!p_r') \over 
   \xi - \xi' - v_F (p_r \!-\! p_r')} \> G(\xi',p'_r)     \eqno(7.52) $$
for the propagator $G$, which is valid to leading order in $\Lam_q$
(see (5.25)). Here the $q_t$-averaged effective interaction $\bar D$ 
is constructed from the RPA effective interaction $D$ as in (7.46).
As shown in (5.28)-(5.33), the solution for $G(\xi,p_r)$ is given
by the Fourier transform of 
$$ G(t,r) = e^{L(t,r) - L_0} \> G_0(t,r)                  \eqno(7.53) $$
where $L(t,r)$ is the Fourier transform of 
$L(\om,q_r) = i \bar D(\om,q_r)/(\om-v_Fq_r)^2$ and $G_0(t,r)$ the
Fourier transform of $G_0(\xi,p_r)$; the constant $L_0$ is given by
$L(t\!=\!0,r\!=\!0)$.
\par
\pp In the RPA calculation in 7.3 we have shown that the 
{\em imaginary}\/ part 
$\Im \bar D$ is asymptotically independent of the tangential cutoff
$\Lam_t$ for $q_r,\om \ll \Lam_t$ in $d<2$ dimensions, since
$\Im D(\om,q_r,q_t)$ vanishes for $v_F|\bq| \gg \om$. By contrast, 
for the {\em real} part an arbitrary choice of effective couplings 
$g_{\bk\bk'}^{\sg\sg'}(\bq)$ will generally lead to a finite limit for 
$v_F|\bq| \gg \om$, and the dependence on $\Lam_t$ does not disappear 
(except, of course, in one dimension).
However, the effective low-energy couplings are not really independent.
For a Fermi liquid it has been shown that the antisymmetry of the 
two-particle vertex poses a constraint on the Landau function 
$f_{\bk\bk'}^{\sg\sg'} = g_{\bk\bk'}^{\sg\sg'}(\b0)$, which is 
equivalent to the Landau sum-rule for Landau parameters \cite{BP}.
As a consequence, effective interactions constructed from Landau
functions as couplings satisfy \cite{BP}
$$ D^{\sg\sg}_{\bk\bk}(\om,\bq) \to 0 \quad {\rm for} 
   \quad (\om,\bq) \to (0,\b0) \quad {\rm with} \quad 
   |\bq/\om| \to \infty                                   \eqno(7.54) $$
For the behavior of $D$ in a one-dimensional Luttinger liquid in that
limit see the final paragraph of Sec.\ 6.4.
Equation (7.54) implies that the real part of the effective interaction 
vanishes, too, for $v_F|\bq| \gg \om$ (recall that for short-range
interactions $D$ depends only on the ratio $\bq/\om$). 
Hence, the asymptotic behavior of $\bar D$ is given by the 
$\Lam_t$-independent expression
$$ \bar D(\om,q_r) \sim |\om|^{d-1} \Dt(q_r/\om) =
   |q_r|^{d-1} |\omt|^{d-1} \Dt(\omt^{-1})  \quad 
   {\rm for} \quad \om,q_r \to 0                          \eqno(7.55) $$
where $\omt = \om/q_r$ and
$$ \Dt(q_r/\om) = {S_{d-1} \over (2\pi)^{d-1}} 
   \int_0^{\infty} d\qtt \> \qtt^{d-2} \> D(q_r/\om,\qtt) \eqno(7.56) $$
The latter integral is convergent in $d<2$ since $D(q_r/\om,\qtt)$ 
vanishes linearly for $\om \to 0$ at fixed small $\bq$, for forward 
scattering interactions that satisfy the Landau sum-rule.
\par
\smallskip
\pp We introduce an exponential cutoff $\Lam_r =: \Lam$ for {\em radial}\/ 
momentum transfers in the expression (5.48) for $L(\om,q_r)$. 
Using the asymptotic scaling behavior of $\bar D$, one has
$$ L(\om,q_r) = |q_r|^{d-1} e^{-|q_r|/\Lam} 
   {i |\omt|^{d-1} \Dt(\omt^{-1})
   \over [\om - v_F q_r + i0^+ s(\om)]^2 }                 \eqno(7.57) $$
A cutoff for $q_r$ is necessary for short-range interactions in any 
dimension (including $d=1$) to make the Fourier transform $L(t,r)$ 
well-defined. 
Changing integration variables
in the Fourier transformation from $\om$ and $q_r$ to $\omt$ and $q_r$, 
the $q_r$ integration can be carried out analytically, leading to
$$ L(t,r) = {1 \over (2\pi)^2}  
   \int_{-\infty}^{\infty} d\omt \> 
   {i |\omt|^{d-1} \Dt(\omt^{-1}) \over 
   [\omt - v_F + i0^+ s(\omt)]^2} \> 
   I_{d,\Lam}(r - \omt t)                                  \eqno(7.58) $$
where the function $I_{d,\Lam}$ is given by
$$ I_{d,\Lam}(y) = 
   \int_{-\infty}^{\infty} dq_r |q_r|^{d-2}
   e^{iq_r y - |q_r|/\Lam} = 
   \Gamma(d\!-\!1) {2\cos[(d\!-\!1){\rm arctg}(y\Lam)]  
   \over (y^2 + \Lam^{-2})^{(d-1)/2}}                      \eqno(7.59) $$
\par
\smallskip
\pp For $r,t \to 0$, one obtains from (7.58)
$$ L_0 = L(0,0) = \Lam^{d-1} \tilde L_0                    \eqno(7.60) $$ 
\vskip -2mm
where
$$ \tilde L_0 = {\Gamma(d\!-\!1) \over 2\pi^2} 
   \int_{-\infty}^{\infty} d\omt \>
   {i |\omt|^{d-1} \Dt(\omt^{-1}) \over 
   [\omt - v_F + i0^+ s(\omt)]^2}                          \eqno(7.61) $$
which is a cutoff-independent number. Note that the limit $r,t \to 0$ is 
unique, i.e.\ it is not important whether $t$ or $r$ tends to zero first. 
\par
\pp Appendix B contains expressions for $L(t,r)$ and $L_0$ in 
terms of the spectral function $\tilde\Delta$ associated with $\Dt$.
\par
\smallskip
\pp For $r,t \to \infty$, $\tt = t/r$ fixed and arbitrary, the function 
$L(t,r)$ scales as 
$$ L(t,r) \sim |r|^{1-d} \Lt(\tt)                          \eqno(7.62) $$ 
\vskip -3mm
where
$$ \Lt(\tt) = {1 \over (2\pi)^2}  
   \int_{-\infty}^{\infty} d\omt \>
   {i |\omt|^{d-1} \Dt(\omt^{-1}) \over [\omt-v_F+i0^+s(\omt)]^2}
   \> \tilde I_d(1-\omt\tt)                                \eqno(7.63) $$
\vskip -2mm
and
$$ \tilde I_d(y) = 
   {2\Gamma(d\!-\!1) \over |y|^{d-1}} \cos[(d\!-\!1)\pi/2] \eqno(7.64) $$
which is independent of the cutoff $\Lam$. 
\par
\smallskip
\pp We now consider the functions $L(0,r)$ and $L(t,0)$,
which determine the momentum distribution function and the density
of states for single-particle excitations, respectively.
\par
\pp The function $L(t,r)$ has a unique limit for $t \to 0$, which can be
written as
$$ L(0,r) = \tilde L_0 \> (r^2 + \Lam^{-2})^{(1-d)/2} 
   \> \cos[(d\!-\!1){\rm arctg}(\Lam r)]                   \eqno(7.65) $$
where $\tilde L_0$ is the constant defined in (7.61).
For $|r| \gg \Lam^{-1}$, one finds
$$ L(0,r) \sim 
   \tilde L_0 \> \cos[(d\!-\!1)\pi/2] \> |r|^{1-d}         \eqno(7.66) $$
The asymptotic behavior of $G(0^{\pm},r)$ for large $|r|$ is thus given
by 
$$ G(0^{\pm},r) \> \sim \> \exp \left[ \tilde L_0 
   \cos[(d\!-\!1)\pi/2] |r|^{1-d} - \tilde L_0 \Lam^{d-1} \right]  
   G_0(0^{\pm},r)                                          \eqno(7.67) $$
and the leading large-$|r|$ behavior is
$$ G(0^{\pm},r) \to e^{-L_0} G_0(0^{\pm},r)  \quad {\rm for} \quad
   |r| \to \infty                                          \eqno(7.68) $$
The {\em momentum distribution}\/ $n_{\bp}$ is obtained by Fourier 
transforming $G(0^-,r)$ as in (5.34), i.e.
$n_{\bp} = -i \int_{-\infty}^{\infty} dr \>  G(0^-,r) \> e^{-i p_r r}$.
From (7.68) we see that the discontinuity in $n_{\bp}$ is reduced by a
factor 
$$ \bar Z = e^{-L_0}                                       \eqno(7.69) $$ 
Here we have to remember that the above propagator and momentum 
distribution function are actually renormalized quantities, calculated
for an effective low-energy action $\bar S^{\Lam}$. 
A wave function renormalization $Z^{\Lam}$ had been performed already 
in passing from a microscopic description to the effective action in 
(7.1), while the above $\bar Z$ is the additional renormalization due 
to residual forward scattering. Thus, the total renormalization $Z$, 
which is equal to the discontinuity of the unrenormalized (i.e.\ 
physical) momentum distribution function, can be written as
$$ Z = Z^{\Lam} \bar Z                                     \eqno(7.70) $$
In Appendix B it is shown that the constant $L_0$ is a positive real 
number, which guarantees that $\bar Z \leq 1$, as it should be.
\par
\smallskip
\pp The limit $r \to 0$ in $L(t,r)$ is also unique. In particular,
for $|t| \to \infty$ (i.e. $v_F|t| \gg \Lam^{-1}$) one obtains the
scaling behavior
$$ L(t,0) \sim \hat L |t|^{1-d}                            \eqno(7.71) $$
where 
$$ \hat L = {\Gamma(d\!-\!1) \over 2\pi^2}
   \int_{-\infty}^{\infty} d\omt \>
   {i\Dt(\omt^{-1}) \over [\omt - v_F + i0^+s(\omt)]^2}
   \> \cos[(d\!-\!1)\pi/2]                                 \eqno(7.72) $$
which is finite and independent of $\Lam$. 
The asymptotic behavior of $G(t,0)$ is thus
$$ G(t,0) \> \sim \> e^{\hat L |t|^{1-d} - L_0} \>
   G_0(t,0)                                                \eqno(7.73) $$
In $d>1$, $L(t,0)$ vanishes for $|t| \to \infty$, i.e.\ eventually
$$ G(t,0) \> \to \> e^{-L_0} G_0(t,0) \quad {\rm for} \quad
   |t| \to \infty                                          \eqno(7.74) $$
Fourier transforming $G(t,0)$ as in (5.35) yields the {\em density of 
states}\/ (per spin), $ N(\xi) = - \pi^{-1} s(\xi) \> 
\Im \int_{-\infty}^{\infty} dt \> G(t,0) e^{i\xi t}$.
Hence, the factor $\bar Z = e^{-L_0}$ determines the reduction of the 
density of states at the Fermi level $\om = 0$. 
Since the same factor determined the reduction of the discontinuity, 
the Fermi velocity $v_F$ is obviously not renormalized by the residual 
interactions.
\par

\bigskip

{\bf 7.5. CRITICAL DIMENSIONS AND DIMENSIONAL CROSSOVER} \par
\medskip
{\bf 7.5.1. Leading low-energy behavior} \par
\medskip
\pp The above results imply that to {\em leading}\/ order in a low-energy 
expansion, Fermi liquid behavior is not destroyed by residual scattering
in any dimension $ d > d_c = 1$. This follows from the behavior of the
function $L(t,r)$ for large distances, which according to (7.62) scales
as 
$$ L(s t,s r) = s^{1-d} L(t,r)                            \eqno(7.75) $$
and hence vanishes for $r,t \to \infty$ in dimensions $d>1$. Strictly
speaking, this conclusion is justified only if $r/v_Ft \neq 1$ when 
taking the large-distance limit, because the scaling function $\Lt(\tt)$
in (7.63) diverges for $v_F\tt = 1$. Thus, in the special limit 
$r,t \to \infty$ with $r/v_Ft = 1$, the function
$L(t,r)$ does not vanish and may even diverge. This behavior can be
traced back to the neglect of higher order terms in the small-$\bq$
expansion, especially in deriving the asymptotic Ward identities 
(5.11) and (5.13) for the charge and spin density vertices.
As discussed already in the final paragraph of 5.3, in the limit
$r,t \to \infty$ with $r/v_Ft = 1$ the function $L(t,r)$ is dominated 
by Fourier components $L(\om,q_r)$ with $\om \sim v_F q_r$, where 
subleading corrections to the asymptotic Ward 
identities should be important, because they cut off the poles in
(5.11) and (5.13). In all other cases one integrates over many 
values for the ratio $\om/v_F q_r$, and corrections are generally 
negligible. 
In short, the singular behavior of our result for $G(t,r)$ in the 
special large-distance limit with $r/t \to v_F$ is an artefact produced 
by truncating the small-$\bq$ expansion at leading order. 
Since explicit expressions for the corrections are not available, one
cannot calculate the behavior in that special limit quantitatively. 
\par
\pp Equation (7.75) and the above arguments imply that
$$ G(t,r) \> \to \> e^{-L_0} G_0(t,r) \quad {\rm for} \quad
   r,t \to \infty \quad {\rm in} \quad d > d_c = 1        \eqno(7.76) $$
at least for $r \neq v_F t$, and most plausibly also for $r = v_F t$.
Fourier transforming, and "unrenormalizing" the propagator by 
multiplying with $Z^{\Lam}$, we obtain
$$ G(p) \> \to \> {Z \over \xi - v_F p_r + i0^+s(\xi)} \quad 
   {\rm for} \quad p \to (0,\bk_F) \quad
   {\rm in} \quad d > d_c = 1                             \eqno(7.77) $$ 
where $Z = Z^{\Lam} e^{-L_0}$. By contrast, in one dimension $G(p)$ is 
given by the very different Luttinger liquid form (see Sec.\ 6). 
Hence, the random phase approximation to the self-energy in 7.3 gave 
the correct criterion on the stability of the quasi-particle pole in 
$d$ dimensions, while second order perturbation theory led to 
misleading results. 
\par
\pp An earlier analysis by Ueda and Rice \cite{UR} also indicated the
possible stability of the quasi-particle pole with respect to residual
interactions in dimensions above but close to one. Their argument
was based on an $\eps$-expansion around a one-dimensional system, with
$\eps = d-1$. The coupling space was restricted to $g_1$- and $g_2$-type
couplings (in g-ology notation), while $g_4$ and the generation of
other couplings in higher dimensions was ignored. With these simplifying
assumptions, Ueda and Rice calculated the flow of $g_1$ and $g_2$ to
one-loop order in $d = 1 + \eps$, finding that for $\eps>0$ these
couplings scale either to zero, or to strong coupling. The former result
was interpreted as a Fermi liquid fixed point, and the latter as a signal
for an instability towards ordered phases. Embedding this calculation 
in our more general context we would say that Ueda and Rice have correctly
calculated the flow of Cooper couplings, which indeed flow either to 
zero or to large values. However the role of forward scattering couplings
in $d>1$, which do not scale to zero but nevertheless do not destroy
the quasi-particle pole, has not been addressed in this earlier work.
\par
\pp We now analyze how the wave function renormalization $Z$ vanishes in
the limit $d \to 1$. We recall from (7.70) that $Z$ is a product of two
factors, namely $Z^{\Lam}$ for arriving from a microscopic model at the
effective action $\bar S^{\Lam}$, and the factor $\bar Z = e^{-L_0}$ 
from residual scattering within $\bar S^{\Lam}$. Of course only the 
latter is controlled non-perturbatively by our results.
Remember also that $Z$ measures the jump in the momentum distribution 
function and the weight of the quasi-particle peak in the spectral 
function. 
Above we have shown that
$$ \bar Z =  Z/Z^{\Lam} = e^{-\Lt_0 \Lam^{d-1}}           \eqno(7.78) $$
where $\Lt_0$ is a cutoff independent number given by (7.61).
The right hand side of (7.78) is the renormalization due to residual 
scattering in the effective action, and the expression is valid to all
orders in the coupling constants, provided that $\Lam \ll k_F$.
From Eq.\ (C.12) we see that $\bar Z$ is a finite number between zero 
and one in $d>1$. For $d \to 1$, however, $\Lt_0$ diverges because the
prefactor $\Gamma(d\!-\!1) \sim (d-1)^{-1}$ diverges, and consequently
$\bar Z$ (and hence $Z$) vanishes, as expected. Quantitatively,
one finds
$$ \bar Z \to e^{-(\eta/\eps)\Lam^{\eps}} \quad
   {\rm for} \quad d \to 1                                \eqno(7.79) $$
where $\eps = d-1$ is the deviation from one-dimension, and
$$ \eta := \lim_{d \to 1} \> (d\!-\!1) \Lt_0 = 
   {1 \over 2\pi^2} \int_{-\infty}^{\infty} d\omt \>
   {i |\omt|^{d-1} \Dt(\omt^{-1}) \over 
   [\omt - v_F + i0^+ s(\omt)]^2}                         \eqno(7.80) $$
is nothing but the anomalous scaling dimension determined by the 
Luttinger liquid one approaches in the limit $d \to 1$! 
This latter identification can be obtained directly from the solution of 
the one-dimensional model (see Appendix E). 
Alternatively, the identification can be made by $\eps$-expanding our
result (7.67) for $G(0,r)$:
Expanding, for $r>0$, 
$\> \Lt_0 = \eta/\eps + \cO(1)$, 
$\> r^{1-d} = 1 - \eps \log r + \cO(\eps^2)$,
$\> \cos[(d\!-\!1)\pi/2] = 1 + \cO(\eps^2)$, and
$\> \Lam^{d-1} = 1 + \eps\log\Lam + \cO(\eps^2)$, one finds
$\> L(0,r) - L_0 = - \eta\log(r\Lam) + \cO(\eps)$, and thus 
$$ G(0,r) \propto r^{-1-\eta}  \quad {\rm in} \quad d = 1 \eqno(7.81) $$
We finally note that the constant $\hat L$ in (7.72) becomes equal to 
$\Lt_0$ for $d \to 1$, and is thus also given by $\eta/\eps$.
\par

\bigskip

{\bf 7.5.2. Subleading corrections} \par
\medskip
\pp For $|r| \gg \Lam^{-1}$ and $|t| \gg (v_F\Lam)^{-1}$, the 
(renormalized) propagator can be written as
$$ G(t,r) = e^{|r|^{1-d}\Lt(t/r) - L_0} G_0(t,r)          \eqno(7.82) $$
where the scaling function $\Lt(\tt)$ is given by (7.63), and the constant
$L_0$ by (7.60) and (7.61). 
For $|r| \gg \Lam^{-1}$ at $t=0^{\pm}$, the asymptotic behavior of $G$ is 
given by (7.67), and for $|t| \gg (v_F\Lam)^{-1}$ at $r=0$ by (7.73). 
In 7.5.1 we have discussed only the {\em leading}\/ low-energy (or 
long-distance) behavior of $G$, obtained in the limit where only the 
constant $L_0$ survives in the exponent. At any finite energy (or,
distance), the $(t,r)$-dependent term yields {\em subleading}\/ 
corrections. 
For dimensions close to $d = 1$, these vanish very slowly as a function
of energy (or distance), and {\em in the limit $d \to 1$ they merge with 
the leading terms}, giving rise to a different (Luttinger-type) leading 
behavior.
\par
\pp To discuss these subleading corrections, let us write the exact (and 
unrenormalized) propagator in the vicinity of the quasi-particle pole 
in a form customary in Fermi liquid theory, as  
$$ G(p) \> = \> 
   {Z \over \xi - \xip^* \pm i\gamma_{\bp}}               \eqno(7.83) $$
where $\xip^*$ is the quasi-particle energy and $\gamma_{\bp}$ the 
quasi-particle decay rate (see eq.\ (4.27)).
In three dimensions, subleading corrections to the Fermi
liquid fixed point scale quadratically to zero in the low-energy limit,
i.e.
$\> \xip^* - v_F p_r = \cO(p_r^2)$ and
$\> \gamma_{\bp} = \cO(p_r^2)$ \cite{NOZ}.
Perturbation theory and RPA as described in 7.2 and 7.3 yield the same
behavior in any dimension $d > 2$. In exactly two dimensions, logarithmic
corrections to the quadratic behavior are known to occur \cite{HSW,BL}.
Below two dimensions, perturbative and RPA results have indicated a 
scaling of corrections with a power $d$ instead of two. 
This is confirmed by our resummation of forward scattering: 
We have found large-distance corrections of order $|t|^{1-d}$ or 
$|r|^{1-d}$, which, by Fourier transform, correspond to low-energy 
corrections with a scaling power $d$, i.e.
$$ \left. \begin{array}{rcl}
    \xip^* - v_F p_r & = & \cO(p_r^d)       \\ 
    \gamma_{\bp}   & = & \cO(p_r^d)
   \end{array} \right. 
   \quad {\rm in} \quad 1 < d < 2                         \eqno(7.84) $$ 
Thus, a normal Fermi system with short-range interactions has 
{\em two characteristic dimensions}, where the low-energy behavior 
undergoes qualitative changes. 
In any dimension above the {\em critical dimension}\/ 
$d_c = 1$ one has asymptotically stable quasi-particles, but below 
$d = 2$ their decay rate and other subleading corrections are bigger 
than the quadratic behavior known from three-dimensional Fermi systems.
\par
\smallskip
\pp A phase-diagram in the dimensionality-coupling plane summarizing
the behavior of the regular normal model can be found in the 
conclusions (Sec.\ 10).
\par

\vfill\eject

\def\pp{\hskip 5mm}
\def\vm{\vskip -4mm}
\def\ba{{\bf a}}
\def\bk{{\bf k}}
\def\bQ{{\bf Q}}
\def\bq{{\bf q}}
\def\bP{{\bf P}}
\def\bp{{\bf p}}
\def\b0{{\bf 0}}
\def\bi{{\bf i}}
\def\bj{{\bf j}}
\def\bJ{{\bf J}}
\def\bn{{\bf n}}
\def\br{{\bf r}}
\def\bR{{\bf R}}
\def\bv{{\bf v}}
\def\bx{{\bf x}}
\def\binf{{\bf\infty}}
\def\eps{\epsilon}
\def\up{\uparrow}
\def\down{\downarrow}
\def\bra{\langle}
\def\ket{\rangle}
\def\sDelta{{\scriptstyle \Delta}}
\def\FS{\partial{\cal F}}
\def\Re{{\rm Re}}
\def\Im{{\rm Im}}
\def\xik{\xi_{\bk}}
\def\cO{{\cal O}}
\def\cD{{\cal D}}
\def\cF{{\cal F}}
\def\cG{{\cal G}}
\def\cZ{{\cal Z}}
\def\Gam{\Gamma}
\def\gam{\gamma}
\def\Lam{\Lambda}
\def\lam{\lambda}
\def\dbm{\delta\bar\mu}

\def\xip{\xi_{\bp}}
\def\xik{\xi_{\bk}}
\def\xikq{\xi_{\bk+\bq}}
\def\tilk{\tilde k}
\def\tilth{\tilde\theta}
\def\tilxi{\tilde\xi}
\def\dph{\Delta^{ph}}
\def\dpp{\Delta^{pp}}
\def\nsim{\sim \hskip -4truemm / \>}
\def\Dt{\tilde D}
\def\tht{\tilde\theta}
\def\omt{\tilde\omega}
\def\q0t{\tilde q_0}
\def\qrt{\tilde q_r}
\def\qtt{\tilde q_t}
\def\xit{\tilde \xi}
\def\tt{\tilde t}
\def\Lt{\tilde L}
\def\cdotr{\!\cdot\!}
\def\om{\omega}
\def\sg{\sigma}
\def\Sg{\Sigma}

\vspace*{1cm}
\centerline{\large 8. LONG-RANGE DENSITY-DENSITY INTERACTIONS}
\vskip 1cm

\pp So far we have restricted our analysis to Fermi systems with 
pure short-range interactions, corresponding to bounded coupling 
functions $g_{kk';q}$ in momentum space. 
In this and the following section we analyze the low-energy physics 
of systems with {\em long-range}\/ interactions. 
Sec.\ 9 deals with fermions coupled to an abelian gauge field, 
which gives rise to an effective long-range current-current 
interaction between fermions.
\par
\pp Here we consider a class of long-range {\em density-density}\/ 
interactions, with coupling functions that diverge in the forward
scattering channel.
To be specific, we analyze interactions with an arbitrary power-law 
decay in real space. 
The corresponding coupling functions in momentum space depend only on 
the momentum transfer $\bq$, not on $\bk$ and $\bk'$ (nor on 
spin), where the $\bq$-dependence is given by a power-law
$$ g(\bq) = g_0/|\bq|^{\gam} \> , \quad g_0 > 0 \>, \quad 
   \gam > 0                                                \eqno(8.1) $$
In real space, this corresponds to a pair-potential $V(\br) \propto
1/|\br|^{d-\gam}$.
This generalization of the Coulomb interaction has been introduced by 
Bares and Wen \cite{BW}, who also first derived the most important
physical consequences of such an interaction. 
In the Coulomb case one has $\gam = d-1$ and $g_0 = \Gam(d\!-\!1)
S_d \> e^2 \>$ in $d$ dimensions, where $e$ is the electron charge.
Multiplying $g(\bq)$ by the bare density of states at the Fermi level,
$N_F$, one may define a dimensionless coupling function
$$ \tilde g(\bq) = N_F \> g(\bq) = (\kappa/|\bq|)^{\gam}   \eqno(8.2) $$
where $\kappa = (N_F g_0)^{1/\gam}$ is a generalized Thomas-Fermi 
screening wave vector \cite{KOPhs}.
\par
\pp Since $g(\bq)$ diverges for $\bq \to \b0$, the importance of 
scattering processes with small momentum transfers is enhanced with
respect to the case of short-range interactions.
The screening vector $\kappa$ sets a scale in momentum space that
separates a weak coupling region for $|\bq| \gg \kappa$ from a strong
coupling region for $|\bq| \ll \kappa$.
For systems with $\kappa \ll k_F$, only interactions with small
momentum transfers $|\bq| \ll k_F$ contribute significantly, such that
the techniques developed for forward scattering dominated systems
derived in Sec.\ 5 are adequate. 
In particular, for continuum systems with a free-particle dispersion
$\eps_{\bk} = \bk^2/2m$ one has $N_F \propto k_F^{d-2}$ and thus 
$\kappa \propto k_F^{(d-2)/\gam}$. 
Hence, one has $\kappa \ll k_F$ in the high-density limit 
$k_F \to \infty \>$ if $d < 2 + \gam$, as for example in the
high-density Coulomb gas in any dimension \cite{KS}. 
\par
\pp In the following we discuss the low-energy small-$\bq$ density-density 
response and the low-energy behavior of the single-particle propagator 
in systems specified by (8.1), concentrating on contributions from
scattering processes with $|\bq| \ll k_F$.
We will rely on the general techniques and results derived in Sec.\ 5.
These techniques have been applied to the present problem in their
Ward identity version \cite{CD94} as well as in bosonization language
\cite{KS,KOPhs,HKMS}.
The results obtained by Bares and Wen \cite{BW} are thereby confirmed 
and extended.
\par
\smallskip
\pp As a consequence of loop cancellation (see Sec.\ 5) in forward
scattering dominated systems, the leading low-energy small-$\bq$ charge 
density-density response function is given by the RPA form
$$ N(q) = {2\Pi_0(\om,\bq) \over 
   1 - 2 g(\bq) \Pi_0(\om,\bq)}                            \eqno(8.3) $$
This function has a real pole in $|\om| = \om_c(\bq)$, corresponding 
to a propagating collective excitation. For $\gam > 0$ and small $\bq$
it is clear that $\Pi_0(\om,\bq)$ must be small and positive to make
$|\om| = \om_c(\bq)$ satisfy the equation $|\bq|^{\gam} = 
2g_0\Pi_0(\om,\bq)$ for the position of the pole. This implies that
$\om_c(\bq)$ is much larger than $v_F|\bq|$ for small $\bq$. For 
$|\om| \gg v_F|\bq|$, the particle-hole bubble has the form
$$ \Pi_0(q) \sim {N_F \over 2d} \> 
   {v_F^2 |\bq|^2 \over \om^2} \>, \quad 
   |\om| \gg v_F|\bq|                                      \eqno(8.4) $$ 
Inserting this in the equation for the pole, one obtains
$$ \om_c(\bq) \> \sim \> \lam |\bq|^{1-\gam/2} \quad 
   {\rm where} \quad  \lam = v_F (N_F g_0/d)^{1/2}         \eqno(8.5) $$
We stress that this simple form holds only for small $\bq$. Obviously
$\om_c(\bq)$ is the dispersion of a {\em gapless}\/ mode as long as 
$\gam < 2$.
For the Coulomb interaction in $d=3$ one has $\gam = 2$, and 
$\om_c(\bq)$ is the well-known plasmon mode with a gap given by the 
plasma frequency $\lam = \omega_p$. 
For $\gam > 2$, the dispersion relation $\om_c(\bq)$ has a gap, too.
\par
\smallskip
\pp We now analyze the low-energy behavior of the single-particle
propagator $G$ and related quantities. 
We focus on the regime $0 < \gam < 2$, where $\om_c(\bq) \to 0$
for $|\bq| \to 0$. 
We would like to see in particular how strongly the single-particle 
propagator is affected by the presence of the gapless collective mode,
and under which conditions Fermi liquid behavior is destroyed.
\par
\pp The RPA effective interaction associated with $g(\bq)$ is
$$ D_{\bk\bk'}^{\sg\sg'}(q) = D(q) = 
   {g(\bq) \over 1 - 2g(\bq)\Pi_0(\om,\bq)} =
   {g_0 \over |\bq|^{\gam} - 2g_0\Pi_0(\om,\bq)}           \eqno(8.6) $$     
This function has the same denominator as $N(q)$, and thus the same
pole in $|\om| = \om_c(\bq)$.
It is convenient to decompose the effective interaction as
$ D(q) = D_c(q) + D_{reg}(q)$, where
$$ D_c(q) \> = \> \lam g_0 |\bq|^{1-3\gam/2} 
   {\om_c(\bq) \over \om^2 - \om_c^2(\bq)} \> = \>
   \lam^2 g_0 {|\bq|^{2-2\gam} \over
   \om^2 - \lam^2 |\bq|^{2-\gam}}                          \eqno(8.7) $$
contains the pole, while $D_{reg}(q)$ is a regular function which is 
finite in the limit $q \to 0$. The effect of the regular part 
$D_{reg}(q)$ on the propagator is qualitatively the same as for 
short-range interactions. 
In particular, it cannot destroy the quasi-particle pole in dimensions
$d>1$. Hence, we now concentrate on the singular part $D_c(q)$ only.
In contrast to the effective interaction for short-range interactions,
$D_c(q)$ does not depend only on the ratio $\bq/\om$. Instead, one has
the scaling behavior
$$ D_c(q) \mapsto b^{-\gam} D_c(q) \quad {\rm for} \quad
   \bq \mapsto b \, \bq \>, \quad \om \mapsto b^{1-\gam/2} \om
                                                          \eqno(8.8) $$
Note that this scaling transformation is anisotropic in $(\om,\bq)$-space
(the dynamical scaling index $z = 1-\gam/2$ differs from one).
\par
\pp Since the interaction $g(\bq)$ is singular only for $\bq \to \b0$, it
is clear that only scattering processes with a small momentum transfer
can have a drastic effect on the propagator. For these processes we can
apply the results derived in Sec.\ 5. 
We define an angular averaged effective interaction
$$ \bar D_c^{\Lam_t}(\om,q_r) = {S_{d-1} \over (2\pi)^{d-1}} \> 
   \lam^2 g_0 \int_0^{\Lam_t} dq_t \> q_t^{d-2}
   {(q_r^2 + q_t^2)^{1-\gam} \over 
   \om^2 - \lam^2 (q_r^2 + q_t^2)^{1-\gam/2}}             \eqno(8.9) $$
where $\Lam_t \ll k_F$ is a cutoff for tangential momentum transfers.
This quantity obeys the scaling relation
$$ \bar D_c^{b\Lam_t}(b^{1-\gam/2}\om,bq_r) = 
   b^{d-1-\gam} \bar D_c^{\Lam_t}(\om,q_r)               \eqno(8.10) $$
Applying the results from Sec.\ 5, the modification of the propagator 
by scattering pro\-cesses with $|\bq| < \Lam \ll k_F$ is given by
Eqs.\ (5.28)-(5.33) with the above $\bar D_c^{b\Lam_t}$ and $\Lam_t =
(\Lam^2 - q_r^2)^{1/2}$. From (5.30) and (8.10) one obtains the scaling
behavior at large distance and long time
$$ L(s^{1-\gam/2}t,sr) = s^{1-d+\gam/2} L(t,r)           \eqno(8.11) $$
Thus, for dimensions above the {\em critical dimension}
$$ d_c^{\gam} = 1 + \gam/2                               \eqno(8.12) $$
scattering processes with small momentum transfers do not modify the 
leading long-dis\-tance asymp\-totics of $G(t,r)$ with respect to
the Fermi liquid behavior. By contrast, for $d \leq d_c^{\gam}$ Fermi
liquid behavior is completely destroyed \cite{CD94,KS,HKMS}. 
This result from a resummation of forward scattering processes confirms 
the earlier result obtained by Bares and Wen \cite{BW}, which was 
based on an RPA calculation of the self-energy and other approximate 
methods.
Note that $d_c^{\gam}$ is smaller than $1+\gam$, which is what one 
might expect from a (very) naive power-counting argument. Counting the
power of the bare coupling $g(\bq)$ misses screening, which weakens the 
singularity, and does not take into account the modified dynamical
scaling, eq.\ (8.8), imposed by the "plasmon" mode.  
\par
\pp For $d < 1+\gam$ the angular averaged effective interaction becomes
independent of $\Lam_t$ for $\om,q_r \to 0$. 
Defining $\qtt = q_t/|q_r|$ and $\omt = \om/|q_r|^{1-\gam/2}$, one can 
write
$$ \bar D_c^{\Lam_t}(\om,q_r) \> \to \> |q_r|^{d-1-\gam} \Dt_c(\omt)
   \quad {\rm for} \quad \om,q_r \to 0                    \eqno(8.13) $$
where the scaling function $\Dt_c(\omt)$ is given by
$$ \Dt_c(\omt) = {S_{d-1} \over (2\pi)^{d-1}} \> 
   \lam^2 g_0 \int_0^{\infty} d\qtt \> \qtt^{d-2}
   {(1 + \qtt^2)^{1-\gam} \over 
   \omt^2 - \lam^2 (1 + \qtt^2)^{1-\gam/2}}               \eqno(8.14) $$
which is independent of tangential cutoffs.
\par
\smallskip
\pp Explicit results have so far been obtained for the propagator
$G(0,r)$ at $t=0$ and especially for the quasi-particle weight $Z$.
\par
\pp The quasi-particle weight is related to the function $L(t,r)$
via $Z = e^{-L_0}$ where $L_0 = L(0,0)$ (cf.\ Sec.\ 7). Using the
spectral representation for the effective interaction $D$ 
(see App.\ C), one can express $L_0$ as \cite{KOPhs}
$$ L_0 = {\lam g_0 \over 2} \int {d^d\bq \over (2\pi)^d} \>
   {|\bq|^{1-3\gam/2} \over 
   [\om_c(\bq) + |\bv_{\bk_F} \cdot \bq|]^2}
   \> + \> {\rm continuum \> contributions}               \eqno(8.15) $$
The first term is exclusively due to the collective mode in the
effective interactions. For $d \leq d_c^{\gam} = 1 + \gam/2$ this term
is infrared divergent, i.e.\ $Z$ vanishes, signalling again the 
breakdown of Fermi liquid behavior in this case \cite{KS}.
For $d < 1 + 2\gam$ the $\bq$-integrals in (8.15) are convergent at
large $\bq$ even without ultraviolet cutoff; the screening vector
$\kappa$ acts as a natural cutoff in this situation \cite{KS}.
Hence, for $d_c^{\gam} < d < 1 + 2\gam$ one obtains a finite 
{\em cutoff-independent}\/ expression for $Z$ (with collective mode
and continuum contributions). The Coulomb interaction satisfies 
these conditions in any dimension, and the result obtained for $Z$
from the above resummation of forward scattering is expected to be
asymptotically exact in the high-density limit. A simple analytic
expression has been derived for the spinless Coulomb gas close to
one dimension \cite{KS}:
$$ Z \sim \sqrt{k_F/\kappa} \> e^{-1/2(d-1)} \quad 
   \mbox{for} \quad d \to 1 \quad
   \mbox{at fixed} \quad \kappa \ll k_F                   \eqno(8.16) $$
\par
\pp In the {\em marginal}\/ case $d = d_c^{\gam}$ the propagator 
$G(0,r)$ obeys a power law decay with an anomalous dimension 
\cite{KOPhs}
$$ \eta = {\sqrt{d} \over 2} (\kappa/k_F)^{d-1}          \eqno(8.17) $$
implying a corresponding power-law behavior of the momentum
distribution function near the Fermi surface as in a one-dimensional
Luttinger liquid. 
This power-law has been first obtained by Bares and Wen \cite{BW} 
from a hydrodynamical calculation of
the orthogonality catastrophe as well as from an exponentiation of a 
logarithmic divergence found in the RPA self-energy. 
\par

\vfill\eject

\def\pp{\hskip 5mm}
\def\vm{\vskip -4mm}
\def\ba{{\bf a}}
\def\bk{{\bf k}}
\def\bQ{{\bf Q}}
\def\bq{{\bf q}}
\def\bP{{\bf P}}
\def\bp{{\bf p}}
\def\b0{{\bf 0}}
\def\bi{{\bf i}}
\def\bj{{\bf j}}
\def\bJ{{\bf J}}
\def\bn{{\bf n}}
\def\br{{\bf r}}
\def\bR{{\bf R}}
\def\bv{{\bf v}}
\def\bx{{\bf x}}
\def\binf{{\bf\infty}}
\def\eps{\epsilon}
\def\up{\uparrow}
\def\down{\downarrow}
\def\bra{\langle}
\def\ket{\rangle}
\def\sDelta{{\scriptstyle \Delta}}
\def\FS{\partial{\cal F}}
\def\Re{{\rm Re}}
\def\Im{{\rm Im}}
\def\xik{\xi_{\bk}}
\def\cO{{\cal O}}
\def\cD{{\cal D}}
\def\cF{{\cal F}}
\def\cG{{\cal G}}
\def\cZ{{\cal Z}}
\def\Gam{\Gamma}
\def\gam{\gamma}
\def\Lam{\Lambda}
\def\lam{\lambda}
\def\dbm{\delta\bar\mu}

\def\xip{\xi_{\bp}}
\def\xik{\xi_{\bk}}
\def\xikq{\xi_{\bk+\bq}}
\def\tilk{\tilde k}
\def\tilth{\tilde\theta}
\def\tilxi{\tilde\xi}
\def\dph{\Delta^{ph}}
\def\dpp{\Delta^{pp}}
\def\nsim{\sim \hskip -4truemm / \>}
\def\Dt{\tilde D}
\def\tht{\tilde\theta}
\def\omt{\tilde\omega}
\def\q0t{\tilde q_0}
\def\qrt{\tilde q_r}
\def\qtt{\tilde q_t}
\def\xit{\tilde \xi}
\def\tt{\tilde t}
\def\Lt{\tilde L}
\def\cdotr{\!\cdot\!}
\def\alf{\alpha}
\def\om{\omega}
\def\sg{\sigma}
\def\Sg{\Sigma}

\vspace*{1cm}
\centerline{\large 9. FERMIONS COUPLED TO A GAUGE FIELD}
\vskip 1cm

\pp The physics of two-dimensional Fermi systems coupled to a $U(1)$
gauge-field has recently become important in two different contexts.
One is the theory of the half-filled Landau level proposed by 
Halperin, Lee and Read \cite{HLR} and by Kalmeyer and Zhang \cite{KZ}, 
where the gauge-field is associated with a ficticious magnetic flux 
attached to the electrons \cite{JAI,LF,WILC}. 
The other is the theory of high-temperature superconductors, where 
gauge-fields have been introduced by Baskaran and Anderson \cite{BA},
and subsequently by Ioffe and Larkin \cite{IL} and Lee and Nagaosa 
\cite{LN}, to describe "spin liquids", i.e. correlated 
electron systems with local constraints imposed by strong repulsive 
forces. 
The singular behavior of the gauge-field propagator at small $q$
generates strong forward scattering of the fermions in these 
systems.\footnote
{A similar strong forward scattering is obtained for Fermi systems
near a Pomerantchuk instability towards phase separation \cite{CDG}.}
The main issue is to what extent Fermi liquid behavior is thereby
destroyed.
\par
\pp Many years ago Holstein, Norton and Pincus \cite{HNP} have already 
pointed out that the transverse component of the electromagnetic field
is not screened in a metal and leads to non-Fermi liquid properties
even in three-dimensional systems (see also Ref.\ \cite{REI}).
However, due to the smallness of the ratio $v_F/c$ in usual metals 
these effects have a tiny energy scale and are thus hard to observe.
By contrast, in the above-mentioned more recent physical contexts,
the dimensionless coupling constant is not small. In addition, the 
reduced dimensionality $d=2$ enhances the phase space for forward 
scattering.
\par
\pp The problem of treating the infrared singularities in a Fermi
system coupled to a gauge-field has been addressed by numerous methods:
Self-consistent approximations \cite{POL94}, 
bosonization \cite{KHM,KOP95}, 
asymptotic Ward identities \cite{ILA,CD94},
eikonal expansion \cite{KSgau},
$1/N$-expansion \cite{POL94,ILA}
and renormalization group \cite{GW,CNS,NW}.
Polchinski \cite{POL94} and Ioffe, Lidsky and Altshuler \cite{ILA} have
introduced an arbitrary number of fermion species (or "flavor number")
$N$ as a control and expansion parameter.
Expansions around both limits $N \to 0$ and $N \to \infty$, as carried
out in much detail by Altshuler, Ioffe and Millis \cite{ILA}, have 
helped to assess the regime of validity of several earlier approaches.
Our review is restricted to pure systems; for the interesting interplay
between disorder and fluctuating gauge-fields the reader should consult
the papers by Aronov and W\"olfle \cite{AW}.
\par

\bigskip

\vfill\eject

{\bf 9.1. ACTION} \par
\medskip
\pp The theory of the {\em half-filled Landau level}\/ leads to an 
action \cite{HLR,KZ}
$$ S = S_0 + S_{CS} + S_I                                  \eqno(9.1) $$
which involves a fermionic field $\psi$ and a $U(1)$ gauge-field 
$a_{\mu}$.
For fully polarized systems in a strong external magnetic field, 
$\psi$ is a one-component (or "spinless") fermionic field.
In euclidean (imaginary time) representation the various terms in 
(9.1) are defined as follows.
The first term
$$ S_0 = \int dx \> \Big\{ 
   \psi^*(x) \> [-\partial_0 + \mu - ia_0] \> \psi(x) 
   \> - \> {1 \over 2m} \> 
   \psi^*(x) \> [-i\nabla - \ba(x)]^2 \> \psi(x) \Big\}   \eqno(9.2) $$
describes non-relativistic free fermions coupled minimally to the
gauge-field $a = (a_0,\ba)$. The integration extends over the 
(2+1)-dimensional euclidean space-time $x = (x_0,\bx)$.
The scalar field $a_0$ and the Chern-Simons term
$$ S_{CS} = (2\pi \tilde\phi)^{-1} \int dx \>
   ia_0(x) \> \nabla\!\times\!\ba(x)                      \eqno(9.3) $$
impose the {\em constraint}\/ 
$\nabla\!\times\!\ba(x) = 2\pi \tilde\phi \> \psi^*(x) \psi(x)$
that links the flux tubes to the electrons. The number of 
flux quanta $\tilde\phi$ attached to each electron is chosen equal
to two in the theory of the half-filled Landau level. 
With this choice the external magnetic field is cancelled by the
average (mean-field) ficticious gauge-field and the Aharonov-Bohm
phase of the flux tubes does not transmute the fermionic statistics,
i.e.\ the composite object given by an electron and its flux tube
is a fermion. 
Electron-electron interactions are described by
$$ S_I = - {\textstyle{1 \over 2}} \int dx dx' \>
   \psi^*(x) \psi(x) \> v(\bx\!-\!\bx') \> 
   \psi^*(x') \psi(x')                                    \eqno(9.4) $$
where $v(\bx)$ is a Coulomb potential proportional to $e^2/|\bx|$,
or an effective short-range interaction in systems where the 
long-range part is screened by charges outside the two-dimensional
electron gas, e.g.\ by a metallic gate.
Using the above-mentioned constraint, $S_I$ can also be expressed
as a quadratic form in the gauge-field $\ba$ instead of a quartic 
monomial in fermionic fields.
\par
\smallskip
\pp The theory of {\em spin liquids} constructed in the context of
high-$T_c$ superconductors has led to the action \cite{BA,IL,LN}
$$ S = \int dx \> \Big\{ 
   \psi^*(x) \> [-\partial_0 + \mu] \> \psi(x) 
   \> - \> {1 \over 2m} \> 
   \psi^*(x) \> [-i\nabla - \ba(x)]^2 \> \psi(x) \Big\}   \eqno(9.5) $$
where $\psi = \left( {\psi_{\up} \atop \psi_{\down}} \right)$ is now
a two-component spinor. This action is sometimes supplemented by a
pure gauge term
$$ S_a = - {1 \over 4g^2} \int dx \> f_{\mu\nu}^2 
   \quad \mbox{where} \quad f_{\mu\nu} = 
   \partial_\mu a_{\nu} - \partial_{\nu} a_{\mu}          \eqno(9.6) $$
which is generated by integrating out high-energy processes
(g is an effective coupling constant).
We emphasize that the action $S$ ($+S_a$) is only a part of what is 
expected to describe the strongly correlated metallic phase in
high-$T_c$ materials, since the holon part \cite{BA} is missing.
Nevertheless, the above action has attracted considerable theoretical 
interest in its own right.
\par 
\pp Note that we use the Coulomb gauge $\nabla\!\cdot\!\ba = 0$
throughout the section.
\par

\bigskip

{\bf 9.2. GAUGE-FIELD PROPAGATOR} \par
\medskip
\pp We will first discuss the structure of the gauge-field propagator
within the random phase approximation (i.e.\ 1-loop order) and will 
then point out that higher order contributions do not lead to 
qualitative modifications.
\par
\pp Within RPA, the gauge-field propagator is given by the Feynman
diagrams shown in Fig.\ 9.1, i.e.\ the polarization insertion $\Pi$ 
(or gauge-field self-energy) is calculated only to 1-loop order.
In Coulomb gauge the longitudinal component of the vector-field $\ba$ 
vanishes and the polarization tensor $\Pi^{\mu\nu}(q)$ can be 
expressed in terms of two functions $\Pi^0(q)$ and $\Pi^t(q)$ coupling
to the scalar field $a_0$ and the transverse vector-field $\ba$, 
respectively.
\par 
\pp To 1-loop order, the scalar component of the polarisation tensor 
is given by the density-density correlator of the non-interacting 
Fermi gas
$$ \Pi^0_0(q) \equiv \Pi_0(q) = 
   \int {d^d\bk \over (2\pi)^d} \>
   {\Theta(\xi_{\bk+\bq/2}) - \Theta(\xi_{\bk-\bq/2}) 
    \over \om - (\xi_{\bk+\bq/2} - \xi_{\bk-\bq/2})
    + i0^+s(\om)}                                          \eqno(9.7) $$
while the transverse component is given by the transverse 
current-current correlator
$$ \Pi^t_0(q) \> = \> {n_s \over m} \> + \> 
   \int {d^d\bk \over (2\pi)^d} \>
   {1 \over m^2} [\bk^2 - (\bk\!\cdot\!\hat\bq)^2] \>
   {\Theta(\xi_{\bk+\bq/2}) - \Theta(\xi_{\bk-\bq/2}) 
    \over \om - (\xi_{\bk+\bq/2} - \xi_{\bk-\bq/2})
    + i0^+s(\om)}                                          \eqno(9.8) $$
where $n_s$ is the density per spin species. The term $n_s/m$ is due
to the tadpole diagram in Fig.\ 9.1. Note that these are expressions
for time-ordered real-frequency quantities.
In two dimensions and for small $q=(\om,\bq)$ with $|\om| \ll v_F|\bq|$ 
one obtains \cite{HLR}
$$ \Pi^0_0(q) \sim - {m \over 2\pi} \> 
   \Big( 1 + i{|\om| \over v_F |\bq|} \Big)                \eqno(9.9) $$
$$ \Pi^t_0(q) \sim 
   -i\gam_0 {|\om| \over |\bq|} + \chi_0 \bq^2            \eqno(9.10) $$
where $\chi_0 = 1/24\pi m$ is the (Landau) diamagnetic susceptibility 
of the electron gas (per spin species) and $\gam_0 = k_F/2\pi$.
\par
\pp Written in a general matrix form, the RPA gauge-field propagator
is determined by the equation
$$ D^{-1} = D_0^{-1} - \Pi_0                              \eqno(9.11) $$
which resummes the geometric series in Fig.\ 9.1.
Instead of solving this equation in the $(d\!+\!1)$-dimensional basis
corresponding to $a^{\mu}$ with $\mu = 0,..,d$, it is easier to 
decompose the vector part in subspaces corresponding to transverse 
and longitudinal fluctuations, where the latter vanish due to the
Coulomb gauge.
\par
\pp For the half-filled Landau-level one has a 2x2-matrix structure 
in a space spanned by scalar field $a_0$ and the transverse 
vector-field $a_t$. The RPA equation for $D$ becomes \cite{HLR}
$$ \left( \begin{array}{cc} 
   D^{00} & D^{0t} \\
   D^{t0} & D^{tt} 
   \end{array} \right)^{-1} = 
   \left( \begin{array}{cc}
   0 & {i|\bq| \over 2\pi \tilde\phi} \\
   - {i|\bq| \over 2\pi \tilde\phi} & 
     {\bq^2 v(\bq) \over (2\pi \tilde\phi)^2}
   \end{array} \right) -
   \left( \begin{array}{cc} 
   \Pi^0_0(q) & 0 \\ 0 & \Pi^t_0(q) 
   \end{array} \right)                                    \eqno(9.12) $$
where the coupling function $v(\bq)$ is the Fourier transform of 
$v(\bx)$ in (9.4). The off-diagonal elements are due to the
Chern-Simons term in the action. Note that the interaction part $S_I$
has been attributed to the "non-interacting" (i.e.\ Gaussian) 
gauge-field propagator $D_0$ by virtue of the constraint linking
the electron density operator to $\nabla\times\ba$.
If $v(\bx)$ is a Coulomb interaction, one has 
$v(\bq) = 2\pi e^2/\eps |\bq|$ where $\eps$ is a dielectric constant.
\par
\pp For small $q$ and $|\om| \ll v_F|\bq|$ the propagator for 
transverse gauge-field fluctuations has the form
$$ D^{tt}(q) \equiv D^t(q) \sim 
   {1 \over i\gam_0 |\om|/|\bq| - \chi(\bq) |\bq|^2}
   \quad \mbox{where} \quad
   \chi(\bq) = \chi_0 + {1 \over 2\pi m \tilde\phi^2} +
   {v(\bq) \over (2\pi \tilde\phi)^2}                     \eqno(9.13) $$
Obviously $D^t(q)$ is highly singular in the limit $q \to 0$ with
$|\om|/|\bq| \to 0$.
By contrast, $D^{t0}$ and $D^{00}$ involve extra factors $|\bq|$ and 
$|\om|/|\bq|$, respectively, in the numerator, and are thus 
comparatively small \cite{HLR}.
If $v(\bx)$ is a Coulomb interaction, one has 
$\chi(\bq) \sim e^2/2\pi\eps \tilde\phi^2 |\bq|$, 
while in the case of external screening the function $\chi(\bq)$ tends
to a constant $\chi(\b0) \equiv \chi$ in the limit $\bq \to 0$.
\par
\pp In the spin liquid case only transverse vector fields enter. 
Hence the gauge-field propagator is purely transverse and given by
$$   [D^t(q)]^{-1} = [D^t_0(q)]^{-1} - 2\Pi^t_0(q)        \eqno(9.14) $$
in RPA. 
The factor two in front of $\Pi^t_0$ is due to the spin-degeneracy. 
In the absence of a pure gauge term one has $[D^t_0(q)]^{-1} = 0$
and thus 
$$ D^t(q) = - {1 \over 2\Pi^t_0(q)} \sim 
   {1 \over 2i\gam_0 |\om|/|\bq| - 2\chi_0 |\bq|^2}       \eqno(9.15) $$
where the latter relation holds for small $q$ with $|\om| \ll v_F|\bq|$.
Including a gauge term as in (9.6) merely modifies the prefactor of
the term proportional to $|\bq|^2$ in (9.15) without changing the
qualitative asymptotic structure.
\par
\smallskip
\pp To obtain the exact gauge-field propagator, one has to replace 
the bare bubbles $\Pi_0$ in Fig.\ 9.1 by dressed polarization 
insertions $\Pi$. These insertions can be expanded in powers of
RPA gauge-field propagators. 
Singular corrections to the bare bubbles can be expected
only from scattering processes with small momentum transfers, where
the RPA propagator diverges. Now a crucial point is that all these
corrections involve fermionic loops with more than two insertions.
Hence, the {\em loop cancellation}\/ theorem derived in Sec.\ 5 
predicts the cancellation of the leading contributions from individual
diagrams.
Indeed, an explicit evaluation of polarization insertions
to 2-loop order by Kim et al.\ \cite{KFWL} has shown that corrections
to the bare bubbles do not change the qualitative small-$q$ behavior 
(for any ratio $|\om|/|\bq|$) although individual diagrams seem to
over-power the 1-loop contributions.
For example, two-loop corrections with a purely transverse gauge
propagator $D^t(q) \sim [i\gam|\om/|\bq| - \chi|\bq|^2]^{-1}$
yield a total correction \cite{KFWL}
$$ \delta\Im\Pi^t(q) \sim 
   {|\om| \over v_F|\bq|} \>
   \bigg[ a \> {|\om|^{2/3} \over \gam^{1/3} \chi^{2/3}} +
         b \> { |\om| \over \chi |\bq|} \bigg]           \eqno(9.16) $$   
for $|\om| \ll v_F|\bq|$ and $|\bq| \ll k_F$, where $a$ and $b$ are 
dimensionless constants. This correction is negligible with respect
to the 1-loop result (9.10). The result (9.16) is a sum of self-energy
and vertex corrections to the bare bubble. The separate sum over
self-energy corrections leads to \cite{KFWL}
$$ \delta\Im\Pi^t_s(q) \sim 
   {k_F \over 2\pi} \> {|\om| \over |\bq|} \> 
   {(|\om|/\chi)^{2/3} \over \gam^{1/3} |\bq|}           \eqno(9.17) $$  
which becomes larger than the 1-loop contributions if $|\om|/|\bq|$
vanishes more slowly than $|\om|^{1/3}$. The leading singularity
in (9.17) is obviously cancelled by the vertex corrections.
\par
\pp Several authors have concluded from various arguments that the 
exact gauge-field propagator has RPA form for small $q$.
The importance of loop cancellation in the gauge problem has been
emphasized in particular by Kopietz et al.\ \cite{KHS}. 
Gan and Won \cite{GW} and Fr\"ohlich et al.\ \cite{FGMgau} have derived
the absence of renormalizations of the RPA propagator from the 
irrelevance of non-quadratic terms in the effective gauge-field action. 
Polchinski \cite{POL94} has derived the stability of the RPA result 
from a self-consistency argument. He neglected vertex corrections,
however, and has therefore been critisized by Kim et al.\ \cite{KFWL}.
\par
\smallskip
\pp In summary, in the limit $q \to 0$ with $|\om|/|\bq| \to 0$
the transverse gauge-field propagator $D^t(q)$ has the singular form
$$ D^t(q) = 
   {1 \over i\gam |\om|/|\bq| - \chi |\bq|^{\zeta}}      \eqno(9.18) $$
where $\zeta = 1$ for half-filled Landau level systems with Coulomb 
interactions and $\zeta = 2$ for spin liquids and half-filled Landau 
level systems with short-range interactions.
The singularity describes an overdamped mode with a relaxation rate 
proportional to $|\bq|^{1+\zeta}$.
In general the coefficients $\gam$ and $\chi$ are effective low-energy 
parameters which include contributions from high-energy fluctuations.
\par
\pp The largest gauge-field fluctuations carry energies
$\om \sim (\chi/\gam) \> |\bq|^{1+\zeta}$.
This energy matches with fermionic excitation energies 
$\xi_{\bp+\bq/2} - \xi_{\bp-\bq/2} \sim v_F q_r$ only if 
$q_r \ll q_t \sim |\bq|$, i.e.\ the dominant scattering processes 
involve {\em small momentum transfers which are essentially parallel to 
the Fermi surface}.\footnote
{Recall from Sec.\ 5 that $q_r = \bq\!\cdot\!\hat\bp$ and $q_t = 
\sqrt{\bq^2 - q_r^2}$ are the radial and tangential components of 
$\bq$, respectively, referred to a momentum $\bp$ close to the Fermi
surface.}
Note also that the fermion-gauge-field vertex favors contributions
where the transverse field (perpendicular to $\bq$) is parallel to the
electron current (i.e.\ normal to the Fermi surface).
\par

\bigskip

{\bf 9.3. FERMION PROPAGATOR} \par
\medskip
\pp The fermion (single-particle) propagator $G$ is strongly affected
by the gauge-field fluctuations. 
This is already evident from the behavior of the RPA result for the 
fermionic self-energy, i.e.\ the first-order (in $D$) contribution shown 
in Fig.\ 9.2.
The singular structure of the propagator $D^t$ for transverse 
gauge-field fluctuations at small momenta leads to a large decay rate
for (putative) quasi-particles. In particular, for $\zeta = 2$ in a
two-dimensional system one finds \cite{LEE,REI}
$$ \tau_{\bk}^{-1} = |\Im\Sg(\xi_{\bk},\bk)| \> \sim \>
   \xi_0^{1/3} \> |\xi_{\bk}|^{2/3}                       \eqno(9.19) $$
where $\xi_0$ is an energy scale given by $v_F^3/\gam\chi^2$ (times
numerical factors). For general small $\xi$ and $\xi_{\bk}$, the
self-energy is proportional to \cite{HLR,BM}
$$ \Im\Sg(\xi,\bk) \propto
   \left\{ \begin{array}{lll}
   |\xi|^{2/3} & {\rm for} & 
    |\xi| > (\chi/\gam)|\xi_{\bk}/v_F|^3  \\
   \xi^2/\xi_{\bk}^4 & {\rm for} &
    |\xi| < (\chi/\gam)|\xi_{\bk}/v_F|^3
   \end{array} \right.                                    \eqno(9.20) $$
For $\zeta = 1$ the RPA self-energy is proportional to 
$\xi \> \log(\eps_F/|\xi|)$ \cite{ILA}.
In any case the self-energy correction modifies the non-interacting 
fermion propagator dramatically, i.e.\ the low-energy structure of the 
bare fermionic excitations is completely destroyed. 
Halperin et al.\ \cite{HLR} have proposed to replace the bare fermions
by renormalized quasi-particles with a dispersion relation 
$\xi^*_{\bk} \propto s(k_r)|k_r|^{3/2}$ for $\zeta = 2$ and by 
$\xi^*_{\bk} \propto k_r |\log|k_r||$ in the case $\zeta = 1$.
\par
\pp The strong modification of the fermion propagator by the RPA
self-energy clearly calls for a controlled treatment of higher order
corrections. This problem and especially the choice of an adequate
method has been discussed somewhat controversially in the literature, 
but the results for $G$ found by different authors do not differ very
much after all.
We will first derive a non-perturbative result for $G$ along the 
lines of Sec.\ 5, i.e.\ a resummation exploiting the smallness of 
relevant momentum transfers, and will also discuss possible flaws of 
this approach in the gauge problem. 
We will then address other methods, especially the large-$N$ expansion 
and scaling (i.e.\ renormalization group) theories.
\par
\smallskip
\pp We now extend results from Sec.\ 5, that had been derived for
systems with purely fermionic degrees of freedom, to the gauge theory. 
Asymptotic Ward identities \cite{CDM94,ILA} and bosonization 
\cite{KHM,KOP95} yield the same result for $G$. 
Here we focus on the Ward identity approach since it allows for a more 
direct comparison with the other diagrammatic methods.
Since spin does not affect the logical structure of the derivation,
we consider spinless fermions for simplicity.
\par
\pp The irreducible current vertex in the gauge theory is defined 
by\footnote
{Note that we work with a real time/frequency representation here, 
 while in Sec.\ 5 a euclidean representation prevailed.}
$$ \Lam^{\mu}(p;q) = 
   - \bra j^{\mu}(q) \> \psi_{p-q/2} \> \psi^*_{p+q/2}
   \ket^{irr}_{tr}                                         \eqno(9.21) $$
with a current operator
$$ \begin{array}{rll} 
   j^0(q) & \equiv & \rho(q) = 
   \int_k \psi^*_{k-q/2} \psi_{k+q/2} \\
   \bj(q) & = & 
   \int_k \bv_{\bk} \> \psi^*_{k-q/2} \> \psi_{k+q/2} + 
   {1 \over m} \!\int_{q'} \rho(q') \> \ba(q\!-\!q')           
   \end{array}                                             \eqno(9.22) $$
The labels "tr" and "irr" indicate truncation of external fermion legs
and irreducibility of Feynman diagrams with respect to cutting a single 
gauge-field propagator, respectively.
Charge conservation (or gauge invariance) implies the well-known exact
Ward identity
$$ q_{\mu} \Lam^{\mu}(p;q) = 
   G^{-1}(p\!+\!q/2) - G^{-1}(p\!-\!q/2)                   \eqno(9.23) $$
As discussed in detail in Sec.\ 5.2, the dominance of forward scattering
suggests the asymptotic {\em density-current relation}
$$ {\bf\Lam}(p;q) \sim \bv_{\bp} \Lam^0(p;q)               \eqno(9.24) $$
to leading order in a small momentum transfer (or scattering angle) 
expansion.
Note that the $\ba$-dependent part of the current operator does not enter 
here, because the expectation value $\bra \ba \> \rho \> \psi\psi^* \ket$ 
vanishes.
Combining (9.23) and (9.24) leads to
$$ \Lam^{\mu}(p;q) = 
   {G^{-1}(p\!+\!q/2) - G^{-1}(p\!-\!q/2) \over
    \om - \bv_{\bp}\!\cdot\!\bq } \> \lam^{\mu}(p)         \eqno(9.25) $$
where $\lam^{\mu}(p) = (1,\bv_{\bp})$ is the bare current vertex. The 
dressed current vertex $\Lam^{\mu}$ is thus a simple functional of the 
propagator $G$.
\par
\pp The fermionic self-energy $\Sg$ is related to $G$, $D^{\mu\nu}$ and 
$\Lam^{\mu}$ by the exact Dyson equation 
$$ \Sg(p) = 
   i \int_{p'} G(p') \> \lam_{\mu}[(p\!+\!p')/2] \> 
   D^{\mu\nu}(q) \> \Lam_{\nu}[(p\!+\!p')/2;p'\!-\!p]      \eqno(9.26) $$
which is illlustrated diagramatically in Fig.\ 9.3. Inserting 
$\Lam^{\mu}$ from (9.25) and using the relation $G^{-1} = G_0^{-1} - \Sg$ 
yields a linear integral equation for $G$,
$$ (\xi - \xi_{\bp}) \> G(p) = 1 + \int_{p'} 
   {i D_{\bp}(p\!-\!p') \over 
   \xi - \xi' - \bv_{(\bp+\bp')/2} \cdot\! (\bp\!-\!\bp')} 
   \> G(p')                                                \eqno(9.27) $$
with a contracted gauge-field propagator
$$ D_{\bp}(q) = 
   \lam_{\mu}(p) \> D^{\mu\nu}(q) \> \lam_{\nu}(p)         \eqno(9.28) $$
This integral equation is the analogon of (5.17) for fermions coupled 
to a gauge-field.
\par
\pp For purely {\em transverse}\/ gauge field fluctuations (which yield 
the most singular contributions) the gauge-field propagator can be 
written as 
$$ D^{jj'}(q) = 
   \Big( \delta_{jj'} - {q_j q_{j'} \over |\bq|^2} \Big) 
   D^t(q) \quad \mbox{for} \quad j,j' = 1,...,d            \eqno(9.29) $$
while $D^{0\mu} = D^{\mu 0} = 0$. In this case one has
$$ D_{\bp}(q) = 
   \bv_{\bp}^2 \> {q_t^2 \over |\bq|^2} \> D^t(q)          \eqno(9.30) $$
where $q_t$ is the tangential component of $\bq$ (with respect to the 
Fermi surface near $\bp$, see Sec.\ 5).
Decomposing the momenta in (9.27) in radial and tangential components
as in Sec.\ 5, one obtains an effective one-dimensional equation of 
motion of the form (5.25), with a $q_t$-averaged gauge-field 
propagator \cite{ILA}
$$ \bar D(\om,q_r) = {S_{d-1} \over (2\pi)^{d-1}} \> 
   v_F^2 \int dq_t \> {q_t^d \over |\bq|^2} \> D^t(q)      \eqno(9.31) $$
The general formal solution of this equation of motion is given by 
(5.28)-(5.33). 
Exactly the same formal expression for the fermionic propagator in the 
gauge theory has been derived by Haldane-bosonization \cite{KHM,KOP95},
and the equivalence with the Ward identity approach has been pointed 
out \cite{CD94,KHM}. A similar expression has been obtained within the
eikonal approximation \cite{KSgau}, which also exploits the smallness
of relevant momentum transfers.
For fermions with spin the solution has precisely the same form, 
i.e.\ spin enters only indirectly via $D^t$.
\par
\pp We now discuss explicit results following from the above formal
solution for $G$, focussing mainly on the case with the strongest
singularities, i.e.\ for $\zeta = 2$ in two dimensions (relevant for 
the half-filled Landau level with screened interactions and the spin 
liquid model). 
In this case the $q_t$-averaged gauge-field propagator has the simple 
asymptotic low-energy form \cite{ILA}
$$ \bar D(\om,q_r) \propto |\om|^{-1/3}                    \eqno(9.32) $$
Since the singularity is dominated by momenta with $q_r \ll q_t$, the
$q_r$-dependence in $\bar D$ is irrelevant. 
Note that $\bar D(\om,q_r)$ does not depend on any momentum transfer
cutoff, since for $\zeta=2$ the integrand in (9.31) decays rapidly as 
a function of $q_t$.
\par
\pp An explicit result for $G(t,r) = \int_{\xi,p_r} G(\xi,p_r) \> 
e^{ip_r r - i\xi t}$ has been reported by Ioffe et al.\ \cite{ILA}:
$$ G(t,r) \sim {1 \over 2\pi} \> {1 \over r - v_F t} \>
   \exp\bigg[{-|r| \over \>
   r_0^{1/3} [|r| - v_F s(r)t]^{2/3}} \bigg]               \eqno(9.33) $$
where $s(.)$ is the sign-function and $r_0$ is a length scale of order 
$\gam\chi^2/v_F^2$. 
Obviously the fermionic density of states $N(\xi)$, given by the
imaginary part of the Fourier transform of $G(t,0)$, is not affected
by the gauge-field. However, the locus of a particle is smeared over 
a distance $r \propto t^{2/3}$, the probability to find it along the
classical path $r = v_F t$ decays exponentially, and the velocity of 
a wave packet vanishes at $t \to \infty$ \cite{ILA}.
Hence, a single fermion cannot propagate in the presence of transverse
gauge-fields (while a particle-hole pair with a small relative 
momentum can). 
The equal-time correlator 
$$ G(0,r) \sim {1 \over 2\pi r} \>  e^{-|r/r_0|^{1/3}}    \eqno(9.34) $$
decays exponentially, i.e.\ the momentum distribution function is 
analytic at the Fermi surface. 
This latter result has also been derived via Haldane-bosonization
\cite{KOP95}.
An analytic momentum distribution function has also been found by 
Khveshchenko and Stamp \cite{KSgau} within their eikonal method.
\par
\pp Altshuler et al.\ \cite{ILA} have also presented an explicit 
result for $G(\xi,\bp)$ in the low-energy limit, with a behavior
corresponding to overdamped fermions with a characteristic energy
that scales as $|p_r|^{3/2}$. In particular, they have found the 
asymptotic behavior
$$ G(\xi,\bp) \propto
   \left\{ \begin{array}{lll}
   -1/v_F p_r & \mbox{for} & 
    \xi_0^{1/3} |\xi|^{2/3} \ll v_F |p_r| \\
    1/\xi_0^{1/3}(i\xi)^{2/3} & \mbox{for} &
    \xi_0^{1/3} |\xi|^{2/3} \gg v_F |p_r|
    \end{array} \right.                                    \eqno(9.35) $$
which is equivalent to the RPA result.
By contrast, the eikonal method seems to yield an exponential decay
of $G(\xi,\bp_F)$ for $\xi \to 0$ \cite{KSgau}.
\par
\pp Kopietz \cite{KOP95} has derived explicit results via bosonization 
for the equal-time propagator $G(0,r)$ in the case $\zeta = 2$ for general
dimensionality $d$. He obtained Fermi liquid behavior 
$G(0,r) \propto 1/r$ for all $d > 3$ (with large subleading corrections
for $d < 6$), anomalous power-law decay $G(0,r) \propto 1/|r|^{1+\eta}$
with a non-universal exponent $\eta$ in $d = 3$, and exponential
decay
$$ G(0,r) \sim {1 \over 2\pi r} \> e^{-|r/r_0|^{1-d/3}}    \eqno(9.36) $$
in dimensions $d < 3$. This identification of critical dimensions and 
the anomalous power-law in $d = 3$ is consistent with earlier
renormalization group results by Gan and Wong \cite{GW}.
\par
\pp Kwon et al.\ \cite{KHM} have analyzed the case $\zeta = 1$ in two
dimensions within bosonization. They find "marginal Fermi liquid
behavior" with a linear energy dependence of the quasi-particle decay
rate, and an anomalous power-law for the equal-time propagator 
$G(0,r)$. 
We note that for $\zeta=1$ the $q_t$-average in (9.31) does not 
converge at the upper end, i.e.\ the original assumption that
momentum transfers be small is not respected by the solution, as
long as no explicit cutoff is introduced.
\par
\smallskip
\pp The validity of results obtained by resumming forward scattering
processes in the gauge theories via asymptotic Ward identities or
bosonization has been questioned by several authors. 
A peculiar feature that distinguishes the gauge theories from other
forward scattering dominated systems is that the most important
processes are characterized by (small) momentum transfers which are
essentially parallel to the Fermi surface, i.e. $q_t \gg q_r$. 
This makes the linearization of fermionic energies
such as $\xi_{\bk_F+\bq} \sim v_F q_r$ used in the resummation
techniques problematic, since the neglected correction of order
$\bq^2 = q_r^2 + q_t^2$ may become comparable to (or even larger than) 
the linear term $v_F q_r$. For example, corrections to the RPA
self-energy in the case $\zeta = 2$ in two dimensions pick up major
contributions from energy-momentum transfers with $\om \sim q_t^3 
\sim q_r^{3/2}$, such that the transverse correction $q_t^2/2m$ to
the linearized energy $v_F q_r$ is not negligible \cite{POL94}.
On the other hand one should note that these corrections often
cancel, such as for example in the symmetrized expression for
particle-hole excitation energies 
$\xi_{\bk+\bq/2} - \xi_{\bk-\bq/2} = \bv_{\bk}\!\cdot\!\bq$ .
This may explain why these corrections do not seem to overpower
the loop cancellation property derived in Sec.\ 5, as 
confirmed for the gauge theories by the explicit two-loop calculation 
of Kim et al.\ \cite{KFWL}.
\par
\pp To equip the problem with a suitable expansion parameter,
Polchinski \cite{POL94} and Ioffe et al.\ \cite{ILA} have 
generalized the above gauge theories (including $S_a$, eq.\ (9.6)) 
by allowing for $N$ fermion 
flavors with arbitrary $N$. The latter group rescales terms in the 
bare action such that the RPA gauge-field propagator has the form
(9.18) with $\gam \mapsto N\gam$ and $\chi \mapsto \chi/N^{1/2}$.
The energy scale $\xi_0 \propto v_F^3/\gam\chi^2$ remains thus
independent of $N$.
\par
\pp According to Ioffe et al.\ \cite{ILA} the results obtained by 
resummation techniques which do not treat tangential momentum transfers
correctly are reliable only in the limit $N \to 0$, because in this 
limit momentum transfers which are almost parallel to the Fermi
surface are no longer favorable.
Based on estimates of general vertex corrections, Altshuler et al.\ 
\cite{ILA} conclude that the density-current relation (9.24) has
important corrections of order $N^{1/2}$ as soon as $N$ differs from
zero. They find that the current vertex ${\bf\Lam}$ has a transverse
component ${\bf\Lam}_t$ given by
$$ {\bf\Lam}_t(p;q) \sim \bq_t \> {v_F q_r \over 2 q_t^2} \>
   \Big[ \sqrt{1 - 2\alf |\om| q_t/q_r^2 + i0^+} - 1 \Big] 
   \> \Lam_r(p;q)                                         \eqno(9.37) $$
for small $N$ and $q_t$, where $\alf = N^{1/2}/(2\pi)^2\gam\chi$ and 
$\Lam_r(p;q) = \hat\bp\!\cdot\!{\bf\Lam}(p;q) \sim  v_F \Lam^0(p;q)$ 
is the radial component of ${\bf\Lam}$.
Such a correction smears the pole in $\om = v_F q_r$ in (9.25), which 
leads to an intrinsic cutoff for exponential singularities such as in 
the result (9.33) for $G$.
\par
\pp Another solvable limit is reached for $N \to \infty$ \cite{ILA,
POL94}. For large $N$ the vertex corrections in the Dyson equation 
(9.26) are suppressed by powers of $1/N$, i.e.\ a self-consistent
random phase approximation is exact for $N \to \infty$. Due to its
weak momentum dependence, the plain RPA self-energy (calculated from
$G_0$) solves also the self-consistent equation \cite{ILA,POL94}.
The propagator $G(\xi,\bp)$ obtained by this large-$N$ expansion 
has several features in common with the one obtained for $N \to 0$.
These common features hold plausibly for any $N$. 
In particular, for $\zeta = 2$ in two dimensions one recovers the
low-energy behavior (9.35), with the same characteristic energy
scale $\xi \propto |p_r|^{3/2}$. 
Discrepancies appear for the precise form of the propagator for 
$\xi \sim |v_F p_r|^{3/2}/\xi_0^{1/2}$.
The extrapolation of results obtained in the large-$N$ limit to the
physical values $N = 1$ or $2$ is of course problematic, especially 
since the importance of vertex corrections at finite $N$ has been 
clearly demonstrated in the two-loop calculation by Kim et al.\ 
\cite{KFWL}.
\par
\smallskip
\pp In addition to the resummation techniques discussed so far,
scaling and renormalization group approaches can provide valuable 
insight into the low-energy structure of Fermi systems coupled to 
transverse gauge-fields.
By a renormalization group analysis 
of the case $\zeta = 2$, Gan and Wong \cite{GW} have identified $d=3$ 
as the critical dimension separating Fermi from non-Fermi liquid
behavior \cite{GW}. At the critical dimension they find that the 
quasi-particle weight vanishes as $\om^{\eta}$, where $\eta$ is a
non-universal exponent. 
Chakravarty et al.\ \cite{CNS} have constructed an expansion in 
$\eps = 3\!-\!d$. In constrast to Gan and Wong these authors obtain 
an anomalous exponent of order $\eps$ for $d<3$, and only logarithmic 
corrections to Fermi liquid behavior in $d=3$.
Nayak and Wilczek \cite{NW} have analyzed the action for the 
half-filled Landau level (in two dimensions) with a generalized 
Coulomb interaction defined by a coupling function $v(\bq) \propto 
1/|\bq|^{\gam}$. Introducing deviations $1-\gam$ from the Coulomb 
case $\gam = 1$ as a small control parameter, they were able to treat
the infrared singularities of the system by a systematic renormalization
group procedure. 
A scenario similar to that of equilibrium critical phenomena near
four dimensions was found.
The case of Coulomb interactions is characterized by marginally
irrelevant (i.e.\ logarithmic) corrections to Fermi liquid theory 
\cite{NW}.
This is consistent with the results from the $1/N$-expansion \cite{ILA},
but inconsistent with the anomalous power-law behavior of the
equal-time propagator found by Kwon et al.\ \cite{KHM}.
If the interactions are shorter-ranged (i.e.\ $\gam < 1$), a 
non-trivial fixed point controls anomalous infrared power-laws, 
analogous to the Wilson-Fisher fixed point in $4-\eps$ dimensions
\cite{NW}. In particular, the quasi-particle weight vanishes as 
$Z \propto \om^{\eta}$ and the effective Fermi velocity as 
$v_F^* \propto \om^{|\eta_{v_F}|}$, where the exponents are given
by $\eta = - \eta_{v_F} = (1\!-\!\gam)/2 + \cO[(1\!-\!\gam)^2]$.
Whereever comparable, these results are consistent with those
obtained by most other groups. It would be interesting to see 
whether the full momentum and energy dependence of $G(\xi,\bp)$
can be calculated within this expansion. 
For short-range interactions ($\gam = 0$) the expansion parameter
$1\!-\!\gam$ is of course not small, but one may hope (as in the theory
of critical phenomena) that the qualitative infrared structure does
not change on the way from $\gam \sim 1$ to $\gam = 0$.
\par
\smallskip
\pp We finally note that the single-fermion propagator $G$ is not a
gauge-invariant quantitiy.
Here we will not address the question of how observable physical 
properties may be related to it. 
A short discussion of that problem is given for example by Kim et al.\
\cite{KFWL}.
In any case the fermion propagator at fixed gauge is an important 
theoretical quantity and its behavior has been the main issue in most
theoretical works on the coupled fermion-gauge-field system.
\par

\bigskip

{\bf 9.4. RESPONSE FUNCTIONS} \par
\medskip
\pp The response of the coupled Fermi gauge-field system is described
by the (dressed) polarization tensor $\Pi^{\mu\nu}$. 
For low frequencies and long wavelengths ($\bq$ small) we have 
already seen in 9.2 that $\Pi^{\mu\nu}$ is not drastically affected 
by the gauge-field fluctuations. The leading low-energy small-$\bq$ 
response behaves therefore in a Fermi liquid fashion. 
This is in fact experimentally observed in the case of half-filled 
Landau level systems (for references, see Kim et al.\ \cite{KFWL}).
A comparison with the non-Fermi liquid properties of high-$T_c$
superconductors is not direct since the holons are not included in
the spinon-gauge-field action (9.5).
\par
\smallskip
\pp Pronounced deviations from Fermi liquid behavior appear for
response functions with momenta near $2k_F$.
In particular, in systems with $\zeta = 2$ the density-density 
correlators acquire anomalous power-law behavior at $2k_F$, and 
interaction vertices $\Gam_{2k_F}$ with external fields at momenta 
$2k_F$ are strongly enhanced \cite{ILA,POL94}.
Non-Fermi liquid results for $2k_F$-functions in the case $\zeta = 1$ 
have also been reported \cite{ILA,KHM}.
\par
\pp As an example, let us compare the results obtained for 
$\Gam_{2k_F}$ in systems with $\zeta = 2$ by different methods.
Neither the leading asymptotic Ward identity (9.25) (equivalent to 
bosonization) nor the omission of vertex corrections seems to provide 
the correct result for systems with finite $N$. The former yields an 
exponential divergence 
$\Gamma_{2k_F}(\om) \propto e^{|\om_0/\om|^{1/3}}$,
where $\om$ is supposed to be the biggest of all energy variables of
the vertex function, and $\om_0$ is a system dependent energy scale
\cite{ILA}. Neglecting vertex corrections, one finds a function 
$\Gamma_{2k_F}(\om)$ that tends to a constant in the low-energy limit 
\cite{ILA}. This latter result is exact in the limit $N \to \infty$.
However, including vertex corrections via $(1/N)$-corrections, 
Altshuler et al.\ \cite{ILA} have obtained a power-law
$$ \Gamma_{2k_F}(\om) \propto \om^{-\sg}                \eqno(9.38) $$
with an exponent
$$ \sg = {1 \over 2N} + {\log^3(N) \over 2\pi^2 N^2}
       + \cO(N^{-2}) 
   \quad \mbox{for} \quad N \gg 1                       \eqno(9.39) $$
A power-law is also obtained by modifying the simple density-current 
relation (9.24) to (9.37) by including small-$N$ corrections of order 
$N^{1/2}$. This yields an exponent \cite{ILA}
$$ \sg = {16\sqrt{2} \over 9\pi} \> {1 \over N^{1/2}} 
       + \cO(1)
   \quad \mbox{for} \quad N \ll 1                       \eqno(9.40) $$
Since a power-law is obtained for both small and large $N$, one may
be confident that $\Gamma_{2k_F}(\om)$ obeys a power-law for any
finite $N$.
\par

\bigskip

\vfill\eject

\def\vm{\vskip -4mm}
\def\bk{{\bf k}}
\def\bQ{{\bf Q}}
\def\bq{{\bf q}}
\def\bP{{\bf P}}
\def\bp{{\bf p}}
\def\b0{{\bf 0}}
\def\bi{{\bf i}}
\def\bj{{\bf j}}
\def\bJ{{\bf J}}
\def\br{{\bf r}}
\def\bR{{\bf R}}
\def\bv{{\bf v}}
\def\binf{{\bf\infty}}
\def\eps{\epsilon}
\def\up{\uparrow}
\def\down{\downarrow}
\def\bra{\langle}
\def\ket{\rangle}
\def\sDelta{{\scriptstyle \Delta}}
\def\FS{\partial{\cal F}}
\def\Re{{\rm Re}}
\def\Im{{\rm Im}}
\def\xik{\xi_{\bk}}
\def\cO{{\cal O}}
\def\cD{{\cal D}}
\def\cF{{\cal F}}
\def\cG{{\cal G}}
\def\cZ{{\cal Z}}
\def\gam{\gamma}
\def\Lam{\Lambda}
\def\lam{\lambda}
\def\dbm{\delta\bar\mu}
\def\xip{\xi_{\bp}}
\def\xik{\xi_{\bk}}
\def\xikq{\xi_{\bk+\bq}}
\def\tilk{\tilde k}
\def\tilth{\tilde\theta}
\def\tilxi{\tilde\xi}
\def\dph{\Delta^{ph}}
\def\dpp{\Delta^{pp}}
\def\Dt{\tilde D}
\def\tht{\tilde\theta}
\def\omt{\tilde\omega}
\def\q0t{\tilde q_0}
\def\qrt{\tilde q_r}
\def\qtt{\tilde q_t}
\def\xit{\tilde \xi}
\def\tt{\tilde t}
\def\Lt{\tilde L}
\def\cdotr{\!\cdot\!}
\def\sg{\sigma}
\def\Sg{\Sigma}

\vspace*{1cm}
\centerline{\large 10. CONCLUSIONS}
\vskip 1cm
\pp We have reviewed the low-energy structure of d-dimensional Fermi 
systems with short-range or long-range interactions, and also of
Fermi systems coupled to a gauge-field.
The analysis was restricted to pure systems and normal (i.e.\ not 
symmetry-broken) phases.
As in Fermi liquid theory, the low-energy physics was assumed to be 
governed by excitations close to a (Fermi) surface in momentum space.
Depending on dimensionality and the nature of the interactions, the
systems belong to distinct "universality classes": conventional Fermi
liquids, unconventional Fermi liquids with enhanced subleading 
corrections, and various types of non-Fermi liquids such as the 
one-dimensional Luttinger liquid.
The main-focus of this article were non-trivial interaction effects
due to (residual) interactions in the {\em forward scattering}\/ 
channel, i.e.\ with small momentum transfers. 
Without leading to spontanous symmetry breaking or dynamical gap 
generation, these interactions affect the qualitative low-energy 
behavior of any Fermi system.
\par
\pp The {\em renormalization group}\/ \`a la Wilson provides a 
well-defined link between microscopic systems and effective low-energy
theories (see Sec.\ 2). 
The central concept here is a family of effective low-energy actions 
$\bar S^{\Lam}$, 
defined on a thin momentum shell of width $\Lam$ around the Fermi surface,
and (in principle) uniquely determined by the microscopic theory via 
integration over high energy degrees of freedom. In practice this 
integration can be carried out only for weak coupling, but our experience
with specific real systems and some exactly solvable models shows that
the qualitative structure of effective actions emerging from a
weak coupling analysis often applies to strong coupling systems, too.
\par
\pp {\em Conservation laws}\/ highly constrain the low-energy behavior 
of a Fermi system. The continuity equations for conserved currents give
rise to {\em Ward identities}\/ which relate different correlation,
response and vertex functions.
In particular, charge and spin conservation guarantee that the field
renormalization $Z$ cancels from the low-energy long-wavelength response
functions. This cancellation, which is crucial for Landau's theory
of the response of a Fermi liquid to external perturbations \cite{NOZ},
holds in non-Fermi liquids, too (see Sec.\ 3).
\par
\pp All Fermi systems seem to share the common feature that residual 
interactions with finite momentum transfers (i.e. of order $k_F$)
either drive an instability towards symmetry breaking\footnote
{In one dimension one may rather have dynamical gap-generation
without symmetry breaking, since the latter is prevented by strong 
fluctuations} or do not play any prominant role at all. 
On the other hand, residual interactions in the forward scattering 
channel modify at least the leading low-energy long-wavelength response 
in normal systems (for example, Landau parameters renormalize the
compressibility and magnetic susceptibility).
In this sense the low-energy theory of interacting Fermi systems is 
never asymptotically free, but contains at least marginal interactions
to deal with.
Fortunately (for the theorist), these residual interactions, which are 
usually not weak in real systems, are easier to treat than bare 
interactions with arbitrary momentum transfers in a microscopic theory.
There are the following two sources of simplifications.
\par
\pp First, the reduced phase space for virtual excitations in the 
effective low-energy theory, which is restricted to the thin 
$\Lam$-shell around the Fermi surface, leads to a suppression of most
virtual processes in the perturbation expansion.
Real decay or scattering processes, such as the decay of a quasi
particle via a single particle-hole excitation,
are also restricted by an intrinsic energy cutoff set by the energy of
the excitations (Landau's argument).
If the dimensionality of the system is not too low and the interactions 
are not too singular, only a very restricted set of Feynman diagrams 
determines the leading and sometimes even the subleading low-energy 
behavior in terms of effective low-energy parameters 
(e.g.\ Landau parameters).\footnote
{We emphasize that this has been explained already without 
extensive use of RG machinary some time ago in the literature, 
see for example Serene and Rainer \cite{SR}.}
\par
\pp Second, forward scattering processes obey a special conservation 
law: The velocity of each scattering particle is conserved in the process.
Equivalently one can say that charge and spin are not only conserved
globally, but even locally in arbitrarily small sectors in momentum
space. 
This conservation law holds only asymptotically in the sense that it
is exact only in the forward scattering limit $\bq \to 0$. 
Only in special models such as the one-dimensional Luttinger model
the velocity conservation is exact even for finite $\bq$.
As shown in detail in Sec.\ 5, this asymptotic conservation law
gives rise to several simplifications, which become particularly 
important in systems where the phase-space reduction in the low-energy 
theory is compensated by singular interactions or low dimensionality.
One of these simplifications is {\em loop cancellation}, i.e.\ Feynman 
diagrams containing fermionic loops with more than two insertions cancel
each other or at least leading singularities of single diagrams cancel.
This implies in particular that the polarization insertion $\Pi(q)$ 
is not dressed by forward scattering processes for small $\bq$. 
Consequently the random phase approximation describes the effect of
forward scattering processes on the leading low-energy long-wavelength 
response not only in Fermi liquids, but also in systems where 
quasi-particle excitations are destroyed by forward scattering.
Furthermore, the asymptotic velocity conservation yields an asymptotic
{\em density-current relation}\/ between the irreducible density vertex 
and the irreducible current vertex, i.e.
${\bf\Lam}(p;q) \sim \bv_{\bp} \Lam^0(p;q)$.
Combining this relation with the exact Ward identity reflecting total 
charge (or spin) conservation, leads to an {\em asymptotic Ward 
identity}\/ that expresses $\Lam^0(p;q)$ and ${\bf\Lam}(p;q)$ uniquely 
in terms of the single-particle propagator $G$.
Plugging this into a Dyson equation, we have derived an expression for 
$G$ which sums the dressing by forward scattering processes to all 
orders in the coupling constant.
An alternative way of exploiting the asymptotic conservation laws
obeyed by forward scattering is d-dimensional {\em bosonization}\/
as proposed by Haldane \cite{HAL92}, which leads to the same results.
\par
\smallskip
\pp In a {\em Fermi liquid}\/ (Sec.\ 4) residual interactions (within a
thin $\Lam$-shell in momentum space) do not affect the leading asymptotic
behavior of the single-particle propagator in the low-energy limit.
In particular, the wave function renormalization $Z^{\Lam}$ has a 
finite limit $Z > 0$ for $\Lam \to 0$.
To leading order in a low-energy expansion, the propagator has the
form $G(p) = Z/(\xi-v_F^*p_r)$. 
This implies a discontinuity $Z$ in the momentum distribution function, 
and the existence of fermionic single-particle excitations, Landau's
{\em quasi-particles}, with a linear dispersion relation and velocity
$v_F^*$. 
For the single-particle properties residual interactions are important 
only at next-to-leading order in a low-energy expansion,
where they yield a damping of quasi-particle excitations and contribute
to a smooth background of incoherent excitations. 
By contrast, residual interactions contribute already in leading order 
to low-energy response functions in a Fermi liquid. 
In particular, the low-energy long-wavelength (small $\bq$) charge- or 
spin-density response is described exactly by an RPA summation with
the Landau function 
$f_{\bk_F\bk'_F} = \lim_{\Lam \to 0} \bar g_{\bk_F\bk'_F}^{\Lam}(\b0)$ 
as interaction (where $\bar g_{\bk\bk'}^{\Lam}(\bq)$ is the 
renormalized running coupling function).
These residual interactions renormalize the compressibility and the
spin susceptibility by a finite factor.
\par
\smallskip
\pp In a {\em one-dimensional Luttinger liquid}\/ (Sec.\ 6) residual 
interactions in the forward scattering channel do not only affect 
response functions, but also the leading low-energy behavior of the 
single-particle propagator. 
The wave function renormalization factor $Z^{\Lam}$ vanishes for 
$\Lam \to 0$. 
The propagator obeys an anomalous scaling law $G(sp) = s^{\eta-1}G(p)$ 
with a non-universal (i.e.\ system-dependent) anomalous exponent $\eta$.
As a consequence, the momentum distribution has a continuous power-law
behavior $n_{\bk} \!-\! n_{\bk_F} \propto |k_r|^{\eta}$ near the Fermi
surface, and the density of single-particle excitations $N(\xi)$ 
vanishes as $|\xi|^{\eta}$ at low energies. 
In addition to properties related to anomalous scaling, the strong 
coupling of fermions to collective modes in the Luttinger liquid leads 
also to "spin-charge separation": 
an extra fermion inserted into the system decays in collective charge 
and spin fluctuations which propagate with two different velocities and 
thus separate.
Although residual forward scattering destroys Landau quasi-particles,
the low-energy, long-wavelength response of a Luttinger liquid has the 
same structure as for a Fermi liquid: 
an RPA sum of bubble chains with fixed point couplings 
$f_{\bk_F\bk'_F} = \lim_{\Lam \to 0} \bar g_{\bk_F\bk'_F}^{\Lam}(\b0)$ 
as interactions.
In contrast to the situation in a Fermi liquid, the $\beta$-function
for $\bar g_{\bk_F\bk'_F}^{\Lam}(\b0)$ does not vanish simply as a
consequence of reduced phase space for $\Lam \to 0$, but due to
cancellations imposed by the special conservation laws obeyed by the
forward scattering processes.
Singular self-energy and vertex corrections cancel in the polarization
bubble.
The charge and spin density response functions have poles in 
$\omega = u_c|\bq|$ and $\omega = u_s|\bq|$, respectively, which 
describe gapless bosonic collective excitations with a linear 
energy-momentum relation.
\par
\smallskip
\pp In Sec.\ 7 we have presented a quantitative analysis of the effects
of residual interactions in d-dimensional systems with {\em short-range}\/
interactions, clarifying in particular how Luttinger liquid behavior 
in one dimension crosses over to Fermi liquid behavior in higher 
dimensions as a function of continuous dimensionality. 
It turned out that a normal Fermi system with short-range 
interactions has two distinct characteristic dimensions where the low
energy behavior undergoes significant changes.
Below two dimensions, forward scattering of single-particle excitations
by long-wavelength spin- and charge-density fluctuations yields the 
dominant contribution to the self-energy, and makes the quasi-particle 
decay-rate scale as $|k_r|^d$ instead of the square law valid for $d > 2$. 
These scattering corrections do not, however, destroy Fermi liquid 
quasi-particle behavior until the {\em critical 
dimension}\/ $d_c = 1$ is reached, where small-$\bq$ scattering 
completely destroys the quasi-particle pole in the propagator. 
In Fig.\ 10.1 we show a schematic phase diagram of Fermi systems
with short-range interactions in the dimensionality-coupling plane, 
where $g$ is a typical coupling in the forward scattering channel
(e.g.\ a certain harmonic of the Landau function).
For strong effective couplings $|g| > g_c$ there may be Pomerantchuk
instabilities, i.e.\ strong coupling instabilities such as phase 
separation, signalled by an infinite compressibility, or ferromagnetism, 
indicated by a divergent spin susceptibility.
At weak coupling in two dimensions, the scenario in Fig.\ 10.1 is 
compatible with a rigorous result by Feldman et al.\ \cite{FMRT}, who
proved that in the absence of Cooper processes the perturbation 
expansion with respect to short-range interactions has a finite radius 
of convergence.
Other instabilities (than Cooper or forward scattering driven), such
as antiferromagnetism or charge density wave instabilities, may only 
occur at sufficiently strong coupling or for very special Fermi 
surfaces (with nesting).
\par
\smallskip
\pp Fermi systems with a pair-potential $V(\br) \propto 1/|\br|^{d-\gam}$
as a prototype for {\em long-range}\/ density-density interactions 
have been reviewed in Sec.\ 8.
For such interactions there is a critical dimension $d_c = 1 + \gamma/2
> 1$ separating Fermi liquid from non-Fermi liquid behavior. 
In spite of screening effects, for $d \leq d_c$ the forward scattering
is so strong that quasi-particles cannot exist. 
In the marginal case $d = d_c$ the momentum distribution function obeys
a power-law behavior as in a Luttinger liquid.
\par
\smallskip
\pp In Sec.\ 9 we have reviewed the low-energy physics of coupled
fermion-gauge-field systems, which have become important in the theory 
of the half-filled Landau level and also in attempts to describe the
"strange" normal metallic phase of high-$T_c$ superconductors.
Transverse gauge-fields are not screened and generate forward 
scattering amplitudes in the Fermi system. 
As a consequence of loop cancellation, the gauge-field propagator
and also the low-energy long-wavelength response of the coupled
system can be calculated in random phase approximation.
Reliable calculations of the detailed behavior of the fermion propagator 
$G(\xi,\bp)$ seem to be difficult, because vertex corrections are not yet
fully under control.
A peculiar feature that distinguishes the gauge theories from other
forward scattering dominated systems is that the most important
processes are characterized by (small) momentum transfers which are
essentially parallel to the Fermi surface. 
This makes the linearization of fermionic energies used in resummation
techniques such as asymptotic Ward identities or bosonization 
problematic.
As a consequence, there is a broad consensus only on those properties 
of the fermion propagator which seem to be insensitive to the 
different assumptions made for the vertex corrections.
In particular, it is clear that
Landau quasi-particles are destroyed by the gauge-field fluctuations 
(in physically relevant dimensions), 
and are replaced by overdamped fermionic excitations with a non-linear 
dispersion relation.
While the long-wavelength response of the coupled fermion-gauge-field
system behaves as in a Fermi liquid at low energies, 
the $2k_F$-response seems to obey anomalous power-laws.
\par
\smallskip
\pp Let us conclude with a few remarks on the normal phase of high 
temperature superconductors. The research reviewed in this work
has shown that a generic normal Fermi system with short-range or Coulomb
interactions in dimensions $d>1$ obeys Fermi liquid behavior, with
no weak coupling instabilities besides the Cooper instability. 
Here "generic" means that the Fermi surface should not have a special 
shape, and the Fermi velocity should be finite.
Singular long-range interactions (stronger that Coulomb) such as in
the gauge-theories can destroy Fermi liquid behavior in two and three
dimensions, and have therefore been invoked to explain the observed 
non-Fermi liquid properties of high-$T_c$ superconductors \cite{BA,IL,LN}. 
Nested Fermi surfaces \cite{VR} and strong van Hove singularities in the
density of states due to peculiarities of the bandstructure \cite{PKN,NTH}
have also been argued to be responsible for anomalous low-energy behavior.
These suggestions take support from photoemission experiments which
indicate extended pieces of the Fermi surface with very small curvature
and regions with "flat" energy-bands in the Brillouin zone \cite{PKC}. 
Other proposals identify the proximity to an instability (quantum
critical point) of antiferromagnetic \cite{MTU,MP} and/or charge 
\cite{EK,CDG,VAR} origin as a source of singular scattering, leading
to a doping dependent disruption of Fermi liquid behavior. Also 
these ideas are (more or less) motivated by experimental evidences.
All the proposals (gauge-theory, nesting or Van-Hove scenario, and quantum 
criticality) involve complicated infrared singularities and strong
coupling problems, which have so far prevented the construction of a
complete and generally accepted theory.
\par

\vfill\eject

\centerline{\large Acknowledgements}

\bigskip

\pp We are very grateful for valuable discussions with 
P.W. Anderson,
N. Andrei,
G. Gallavotti,
M. Grilli,
H. Fukuyama,
D. Haldane,
R. Hlubina,
L.B. Ioffe,
D. Khveshchenko,
H. Kn\"orrer,
P. Kopietz,
B. Marston,
V. Meden,
E. M\"uller-Hartmann,
M. Randeria,
D. Rainer,
T.M. Rice,
M. Salmhofer,
K. Sch\"onhammer,
E. Trubowitz,
and D. Vollhardt.
During the course of this work, we have benefitted from the hospitality
of the International Center for Theoretical Physics (Trieste), the
Institute for Scientific Interchange (Torino) and the Aspen Center for
Physics.
W.M.\ was partially supported by the Sonderforschungsbereich No.\  341 
of the Deutsche Forschungsgemeinschaft.
C.C.\ and C.D.\ have been supported within the "progetto ricerca
avanzata istituto nazionale fisica della materia (INFM)". 
\par

\vfill\eject

\def\vm{\vskip -4mm}
\def\bk{{\bf k}}
\def\bQ{{\bf Q}}
\def\bq{{\bf q}}
\def\bP{{\bf P}}
\def\bp{{\bf p}}
\def\b0{{\bf 0}}
\def\bi{{\bf i}}
\def\bj{{\bf j}}
\def\bJ{{\bf J}}
\def\bn{{\bf n}}
\def\br{{\bf r}}
\def\bR{{\bf R}}
\def\bv{{\bf v}}
\def\binf{{\bf\infty}}
\def\eps{\epsilon}
\def\up{\uparrow}
\def\down{\downarrow}
\def\bra{\langle}
\def\ket{\rangle}
\def\FS{\partial{\cal F}}
\def\Re{{\rm Re}}
\def\Im{{\rm Im}}
\def\xik{\xi_{\bk}}
\def\cO{{\cal O}}
\def\cD{{\cal D}}
\def\cF{{\cal F}}
\def\cG{{\cal G}}
\def\cZ{{\cal Z}}
\def\Lam{\Lambda}
\def\lam{\lambda}
\def\dbm{\delta\bar\mu}
\def\para{\parallel}
\def\xip{\xi_{\bp}}
\def\xik{\xi_{\bk}}
\def\xikq{\xi_{\bk+\bq}}
\def\tilk{\tilde k}
\def\tilth{\tilde\theta}
\def\tilxi{\tilde\xi}
\def\dph{\Delta^{ph}}
\def\dpp{\Delta^{pp}}
\def\Dt{\tilde D}
\def\tht{\tilde\theta}
\def\omt{\tilde\omega}
\def\q0t{\tilde q_0}
\def\qrt{\tilde q_r}
\def\qtt{\tilde q_t}
\def\xit{\tilde \xi}
\def\tt{\tilde t}
\def\Lt{\tilde L}
\def\cdotr{\!\cdot\!}
\def\om{\omega}
\def\sg{\sigma}
\def\Sg{\Sigma}
\def\pba{{\rm \bar p}}
\def\qba{{\rm \bar q}}
\def\omb{\bar\omega}

\vspace*{1cm}
\centerline{\Large APPENDICES}
\vskip 1cm

{\bf Appendix A: Loop-cancellation for $\bf N=3$:} \par
\medskip
\pp Here we demonstrate the loop-cancellation explicitly for the
case $N=3$. There are two ways of adding an insertion with momentum
$q_3 = q$ to a loop with two insertions carrying momenta $q_1$ and 
$q_2$, as shown in Fig.\ A.1. Using the relation (5.18), the two Feynman 
diagrams yield
$$ \int_k \> {h(\bk,\bk,\bk;q_1,q_2,q) \over 
   iq_0 - \bv_{\bk}\cdot\bq} \>
   [G_0(k-q/2+q_1/2) - G_0(k+q/2+q_1/2)] G_0(k-q/2-q_1/2) \> +        $$
\vskip -7mm
$$ \int_k \> {h(\bk,\bk,\bk;q_1,q_2,q) \over 
   iq_0 - \bv_{\bk}\cdot\bq} \>
   [G_0(k-q/2-q_1/2) - G_0(k+q/2-q_1/2)] G_0(k+q/2+q_1/2) \hskip 4mm  $$
to leading order in $\bq_{\nu}$. Note that $q_1 + q_2 + q = 0$ 
(momentum conservation). The above expression yields contains four terms.
The third term obviously cancels the second one in the above expression.
Shifting the integration variable $k$ by $\pm q/2$ and using 
$\bv_{\bk\pm\bq/2} \approx \bv_{\bk}$ and $h(\bk\pm\bq/2,\dots;q_1,\dots) 
\approx h(\bk,\dots;q_1,\dots)$, one finds that the two remaining 
contributions to the integral cancel each other to leading order in 
$\bq$.
\par

\bigskip

{\bf Appendix B: Spectral representation of $\bar D$} \par
\medskip
\pp The effective interaction has a spectral representation of the form
$$ D_{\bp}(\om,\bq) = \int_{-\infty}^{\infty} d\om'
   {\Delta_{\bp}(q'_0,\bq) \over \om - \om' + i0^+s(\om)} \> + \>
   D_{\bp}(\bq)                                              \eqno(B.1) $$
where $\Delta_{\bp}(\om,\bq) = -\pi^{-1} s(\om) \Im D_{\bp}(\om,\bq)$
and $D_{\bp}(\bq)$ is a real function of $\bq$. 
Performing the angular average, one has a spectral representation 
for $\bar D$ (cutoff-dependence not written)
$$ \bar D(\om,q_r) = \int_{-\infty}^{\infty} d\om'
   {\bar\Delta(\om',q_r) \over \om - \om' + i0^+s(\om)} \> + \> 
   \bar D(q_r)                                               \eqno(B.2) $$
Inserting this in our expression for $L(\om,q_r)$ and performing the
energy-time Fourier transformation by doing a simple contour integral, 
one finds
$$ L(t,q_r) = s(t) \int_{-\infty}^{\infty} d\om \>
  {\bar\Delta(\om,q_r) \over (\om - v_Fq_r)^2}
  \{\Theta(\om t)e^{-i\om t} + 
    \Theta(q_rt)[it(\om-v_Fq_r) - 1] e^{-iv_F q_r t}\}                  $$
\vskip -4mm
$$ - i \bar D(q_r) \Theta(q_r t) |t| e^{-iv_F q_r t}         \eqno(B.3) $$
Note that the double pole in $\om = v_Fq_r$ has been eliminated in this
representation of $L$, since the curly bracket has a double zero in
$\om = v_F q_r$. 
Note also that $L(t,q_r)$ is continuous in $t=0$, although it doesn't 
look so at first sight, and
$$ L(0^+,r) = L(0^-,r) = {1 \over 2\pi} 
   \int_{-\infty}^{\infty} dq_r \int_{-\infty}^{\infty} d\om \>
   \Theta(-\om q_r) {|\bar\Delta(\om,q_r)| \over (\om-v_Fq_r)^2}
   e^{iq_rr}                                                 \eqno(B.4) $$
which is manifestly real and positive. In particular,
$$ L_0 = {1 \over 2\pi} 
   \int_{-\infty}^{\infty} dq_r \int_{-\infty}^{\infty} d\om \>
   \Theta(-\om q_r) {|\bar\Delta(\om,q_r)| \over 
   (\om-v_Fq_r)^2}                                           \eqno(B.5) $$
is a positive real number.
\par
\smallskip
\pp Using the above spectral representation it is easy to show that the 
momentum distribution function obtained from our solution for $G$ is real,
as it should be for the expectation value of a hermitian operator.
From $n_{\bp} = -i \int_{-\infty}^{\infty} dr \> G(0^-,r) \> e^{-ip_r r}$ 
it is easy to see that $n_{\bp}$ is real if and only if $G^*(0^-,-r) = 
- G(0^-,r)$, where the asterisk means complex conjugation. 
For $G_0$ this is obviously satisfied. Hence, we must require that 
$$ L^*(0^-,-r) = L(0^-,r)                                    \eqno(B.6) $$ 
This property follows immediately from the expression for $L(0^-,r)$ in
terms of $\bar\Delta$.
\par
\smallskip
\pp In systems with {\em short-range}\/ interactions in dimensions $d<2$ 
the scaling behavior (7.55) for $\bar D$ implies
$$ \bar\Delta(\om,q_r) \sim 
   s(\om) |\om|^{d-1} \tilde\Delta(q_r/\om) =
   s(q_r) |q_r|^{d-1} s(\omt) |\omt|^{d-1} \tilde\Delta(\omt^{-1})
                                                            \eqno(B.7a) $$
and
\vskip -5mm
$$ \bar D(q_r) = |q_r|^{d-1} \Dt                            \eqno(B.7b) $$
where $\Dt$ is a real $q_r$-independent number.
Note that $\tilde\Delta(\omt^{-1}) > 0$ by definition.
Inserting this in (B.3), and substituting $\om$ by $q_r \omt$, one can 
carry out the $q_r$-integration explicitly, and obtains
$$ L(t,r) = 
   {1 \over 2\pi} \int_{-\infty}^{\infty} d\omt \>
   {|\omt|^{d-1} \tilde\Delta(\omt^{-1}) \over (\omt-v_F)^2} \>
   \Big[ I_{d,\Lam}^{s(\omt t)}(r\!-\!\omt t)
    - s(\omt) I_{d,\Lam}^{s(t)}(r\!-\!v_Ft)  \hskip 2cm                 $$
\hskip -7mm
$$  + s(\omt) i|t|(\omt\!-\!v_F) 
    I_{d+1,\Lam}^{s(t)}(r\!-\!v_Ft) \Big] \>
    - {i \Dt \over 2\pi} \> |t| \> 
    I_{d+1,\Lam}^{s(t)}(r\!-\!v_Ft)                          \eqno(B.8) $$
where
$$ I_{d,\Lam}^{\alpha}(y) = 
   \int_{-\infty}^{\infty} dq_r \Theta(\alpha q_r) |q_r|^{d-2}
   e^{iq_r y - |q_r|/\Lam} =
   \Gamma(d\!-\!1) {e^{i\alpha (d-1) {\rm arctg}(y\Lam)} \over
   (y^2 + \Lam^{-2})^{(d-1)/2}}                              \eqno(B.9) $$
Due to the double zero of the curly bracket for $\omt = v_F$ in (B.8),
the $\omt$-integral is well-defined and finite. Replacing it by a
principal value integral around $v_F$ changes nothing. 
Exploiting the relation
$$ P\!\int_{-\infty}^{\infty} d\omt \> 
   {s(\omt) |\omt|^{d-1} \tilde\Delta(\omt^{-1}) \over v_F - \omt}
   + \Dt \> = \> \Re\Dt(v_F^{-1})                           \eqno(B.10) $$
the two terms proportional to $|t|$ in (B.8) can be combined to one
term proportional to $\Re\Dt(v_F^{-1})$:
$$ L(t,r) = 
   {1 \over 2\pi} P\!\int_{-\infty}^{\infty} d\omt \>
   {|\omt|^{d-1} \tilde\Delta(\omt^{-1}) \over (\omt-v_F)^2} \> 
   \big[ I_{d,\Lam}^{s(\omt t)}(r - \omt t)
    - s(\omt) I_{d,\Lam}^{s(t)}(r-v_Ft) \big]                           $$
\vskip -4mm
$$ \hskip - 4cm  - {i \Re\Dt(v_F^{-1}) \over 2\pi} \> |t| \> 
    I_{d+1,\Lam}^{s(t)}(r-v_Ft)                             \eqno(B.11) $$
Note that $\Dt(v_F^{-1}) = 0$ in the one-dimensional Luttinger model,
i.e.\ the $t$-linear term vanishes in that case.
\par
\pp The constant $\Lt_0$ defined in (7.61) can be written as
$$ \Lt_0 = {\Gamma(d\!-\!1) \over \pi} \int_{-\infty}^0
   d\omt \> |\omt|^{d-1} 
   {\tilde\Delta(\omt^{-1}) \over (\omt-v_F)^2}
                                                            \eqno(B.12) $$
Note that the logarithmic "ultraviolet" divergence of $\Lt_0$ for 
$d \to 2$ does of 
course {\em not}\/ imply that $Z \to 0$ for $d \to 2$, since the range
of applicability of the leading small-$\Lam$ behavior in $d<2$ shrinks 
for increasing dimensionality. The divergence of $\Lt_0$ probably signals
the expected substitution of leading terms proportional to $\Lam^{d-1}$
in $d<2$ by terms of order $\Lam \log\Lam$ in $d=2$. 
\par

\bigskip

{\bf Appendix C: Bubbles for a quadratic dispersion relation} \par
\medskip
\pp Here we list explicit formulae for particle-hole and particle-particle 
bubbles in d-dimensional systems with the usual {\em quadratic}\/
(i.e.\ not linearized) dispersion relation $\eps_{\bk} = \bk^2/2m$.
For this $\eps_{\bk}$ exact analytic expressions can be obtained for
the spectral functions associated with the bubbles for arbitrary 
excitation energies $\om$.
\par
\pp We introduce dimensionless momentum and energy variables
$\qba = |\bq|/k_F$, $\pba = |\bp|/k_F$ and $\omb = \om/v_F k_F = 
\om/2\eps_F$, where $v_F = k_F/m$ and $\eps_F = k_F^2/2m$.
\par
\smallskip
\pp Let us start with the {\em particle-hole}\/ bubble $\Pi_0(\om,\bq)$,
as defined in (7.8).
One has to distinguish several regimes in the $(\qba,\omb)$-plane,
which are separated by the functions $\pm \omb_+(\qba)$ and
$\pm \omb_-(\qba)$, where
$$ \omb_{\pm}(\qba) = \pm\qba + \qba^2/2                    \eqno(C.1) $$
For the spectral function associated with the particle-hole bubble 
one finds \cite{BW}
$$ \Delta^{ph}(\om,\bq) = 
   {S_{d-1} \over (2\pi)^d} {1 \over d\!-\!1}
   {k_F^{d-1} \over v_F} {1 \over \qba^d}
   \Big[ \big( \qba^2 - (\omb - \qba^2/2)^2 \big)^{d-1 \over 2} -
         \big( \qba^2 - (\omb + \qba^2/2)^2 \big)^{d-1 \over 2} 
   \Big]                                                  \eqno(C.2a) $$
for $\> 0 < \omb < - \omb_-(\qba)$ with $\qba \in [0,2]$,
$$ \Delta^{ph}(\om,\bq) = 
   {S_{d-1} \over (2\pi)^d} {1 \over d\!-\!1}
   {k_F^{d-1} \over v_F} {1 \over \qba^d} \>
   \Big[ \qba^2 - (\omb - \qba^2/2)^2 \Big]^{d-1 \over 2} \eqno(C.2b) $$
for $|\omb_-(\qba)| < \omb < \omb_+(\qba)$, and
$$ \Delta^{ph}(\om,\bq) = 0                               \eqno(C.2c) $$
for $\omb > \omb_+(\qba)$ with $\qba \in [0,2]$, 
or $0 < \omb < \omb_-(\qba)$ with $\qba > 2$. 
The corresponding results for $\omb < 0$ can be obtained from 
$\Delta^{ph}(\om,\bq) = - \Delta^{ph}(-\om,\bq)$.
\par
\smallskip
\pp We now turn to the {\em particle-particle}\/ bubble $K_0(\om,\bp)$,
defined in (7.20). 
One has to distinguish several regimes in the $(\pba,\omb)$-plane,
which are separated by the functions $\omb_+(\pba)$, $\omb_-(\pba)$ 
and $\omb_0(\qba)$, where
$$ \omb_{\pm}(\qba) = \pm\qba + \qba^2/2 \quad {\rm and} 
   \quad \omb_0(\qba) = (\qba/2)^2 - 1                     \eqno(C.3) $$

\pp For the spectral function associated with the {\em 
particle-particle}\/ bubble we have derived the expressions
$$ \Delta^{pp}(\om,\bp) = 
   {S_{d-1} \over (2\pi)^d} {k_F^{d-1} \over v_F} \>
   [\omb - \omb_0(\pba)]^{d-2 \over 2} \int_0^{x(\omb,\pba)} 
   (1-x^2)^{d-3 \over 2} \quad {\rm where} \quad
   x(\omb,\pba) = 
   {\omb/\pba \over \sqrt{\omb \!-\! \omb_0(\pba)}}       \eqno(C.4a) $$
for $\> \omb_-(\pba) < \omb < \omb_+(\pba)$,
$$ \Delta^{pp}(\om,\bp) = 
   {S_{d-1} \over (2\pi)^d} {k_F^{d-1} \over v_F} \>
    s(\omb) [\omb - \omb_0(\pba)]^{d-2 \over 2} \int_0^1 
   (1-x^2)^{d-3 \over 2} =
   {s(\omb) k_F^{d-1}/v_F \over 2^d \pi^{d/2} \Gamma(d/2)} \>
   [\omb - \omb_0(\pba)]^{d-2 \over 2}                    \eqno(C.4b) $$
for $\> \omb_0(\pba) < \omb < \omb_-(\pba)\> $ or 
$\> \omb > \omb_+(\pba)$, and
$$ \Delta^{pp}(\om,\bp) = 0                               \eqno(C.4c) $$
for $\> \omb < \omb_0(\pba)$.
\par
\smallskip
\pp The division of energy-momentum space in different regimes does
not depend on dimensionality. An illustration of these regimes and
explicit expressions for $\Pi_0$ and $K_0$ in $d=1,2,3$ dimensions
(including the real parts) can be found in a paper by Fukuyama et al.\ 
\cite{FHN}.
\par

\bigskip

\vfill\eject

{\bf Appendix D: Perturbative self-energy in one dimension} \par
\medskip
\pp In the one-dimensional g-ology model (see Sec.\ 6) without umklapp
processes and couplings $g_{i\perp} =: g_i \neq 0$, $g_{i\para} = 0$ 
for $i=1,2,4$, the Feynman diagram in Fig.\ 7.2 yields the contribution
$$ - {g_1^2 + g_2^2 \over 16\pi v_F^2} (\xi - \xip) \Theta(|\xi|-|\xip|) 
   - {g_4^2 \over 8\pi v_F^2} s(\xip) \xi^2 \delta(\xi - \xip)
                                                            \eqno(D.1) $$  
to $\Im\Sigma(\xi,\bp)$. For $\xip = 0$, this yields, via Kramers-Kronig,
a contribution
$$ {g_1^2 + g_2^2 \over 8\pi^2 v_F^2} \> 
   \xi \log|\xi/v_F\Lam|                                    \eqno(D.2) $$
to the real part $\Re\Sigma(\xi,\bp)$, where $\Lam$ is a band-width
cutoff. For interactions between opposite spins (only) these are the 
only second order contributions to the self-energy (except trivial
constant Hartree terms).
\par

\bigskip

{\bf Appendix E: Limit $\bf d \to 1$ for the propagator} \par
\medskip
\pp It is instructive to recover the one-dimensional solution for $G$
by taking the limit $d \to 1$ in the general d-dimensional result (B.11).
In $d = 1$, the effective interaction of two particles with parallel
spins near the same Fermi point has the form
$$ D(\om,q_r) = (\om-v_Fq_r) \sum_{\nu=c,s}
   \left[ {A_{\nu} \over \om - u_{\nu}q_r + i0^+s(\om)} +
          {B_{\nu} \over \om + u_{\nu}q_r + i0^+s(\om)} 
   \right]                                                  \eqno(E.1) $$
Since there is no transverse component $q_t$ in one dimension, $D$ and
$\bar D$ are identical. The spectral weight of the dynamical part of
$D$ is obviously given by 
$$ \Delta(\om,q_r) = q_r \sum_{\nu}
   [(u_{\nu}-v_F)A_{\nu} \delta(\om-u_{\nu}q_r) - 
    (u_{\nu}+v_F)B_{\nu} \delta(\om+u_{\nu}q_r)]            \eqno(E.2) $$
while the $\om$-independent part is a constant 
$D(q_r) = \sum_{\nu} (A_{\nu} + B_{\nu}) = g_4^{\sg\sg}$.
All these functions are scale-invariant.
The scaling function $\tilde\Delta$ is obtained as
$$ \tilde\Delta(\omt^{-1}) = \sum_{\nu}
   [(u_{\nu}-v_F)A_{\nu} \delta(\omt-u_{\nu}) + 
    (u_{\nu}+v_F)B_{\nu} \delta(\omt+u_{\nu})]              \eqno(E.3) $$
while $\Dt(v_F^{-1})$ is obviously zero.
To obtain $L(t,r)$, we must expand the function $I_{d,\Lam}^{\alpha}(y)$
around $d=1$:
$$ I_{d,\Lam}^{\alpha}(y) = {1 \over d-1} 
   - \log(y+i\alpha/\Lam) + i\alpha\pi/2 + {\cal O}(d\!-\!1) 
                                                            \eqno(E.4) $$
Inserting all this in our expression (B.11) for $L(t,r)$, performing 
the $\omt$-integrals (trivial due to $\delta$-functions), and collecting
all terms, one gets
$$ L(t,r) \> \to \> {\eta \over d-1} \> + \> 
   \log(r-v_Ft+is(t)/\Lam)                                  \hskip 7cm $$
\vskip -5mm 
$$  - \sum_{\nu} \Big[ (1/2+\eta_{\nu}/2) \log(r-u_{\nu}t+is(t)/\Lam)
    + (\eta_{\nu}/2) \log(r+u_{\nu}t-is(t)/\Lam) \Big]      \eqno(E.5) $$
where $\eta_{\nu} = B_{\nu}/\pi (u_{\nu}+v_F) = 
2 - A_{\nu}/\pi (u_{\nu}-v_F)$ and $\eta = \sum_{\nu} \eta_{\nu}$. 
Finally, $L_0$ is determined by
$$ L_0 = L(0,0) = {\eta \over d-1} + \eta \log\Lam          \eqno(E.6) $$
Thus the constant $\eta/(d-1)$ is cancelled in $L(r,t)-L_0$, and
we are left with the well-known result for the Luttinger model in 
one dimension \cite{SOL}, discussed already in Sec.\ 6.
\par

\vfill\eject


\vfill\eject


\def\pp{\hskip 5mm}
\def\vm{\vskip -4mm}
\def\ba{{\bf a}}
\def\bk{{\bf k}}
\def\bQ{{\bf Q}}
\def\bq{{\bf q}}
\def\bP{{\bf P}}
\def\bp{{\bf p}}
\def\b0{{\bf 0}}
\def\bi{{\bf i}}
\def\bj{{\bf j}}
\def\bJ{{\bf J}}
\def\bn{{\bf n}}
\def\br{{\bf r}}
\def\bR{{\bf R}}
\def\bv{{\bf v}}
\def\bx{{\bf x}}
\def\binf{{\bf\infty}}
\def\eps{\epsilon}
\def\up{\uparrow}
\def\down{\downarrow}
\def\bra{\langle}
\def\ket{\rangle}
\def\FS{\partial{\cal F}}
\def\Re{{\rm Re}}
\def\Im{{\rm Im}}
\def\xik{\xi_{\bk}}
\def\cO{{\cal O}}
\def\cD{{\cal D}}
\def\cF{{\cal F}}
\def\cG{{\cal G}}
\def\cZ{{\cal Z}}
\def\Gam{\Gamma}
\def\gam{\gamma}
\def\Lam{\Lambda}
\def\lam{\lambda}
\def\dbm{\delta\bar\mu}

\def\xip{\xi_{\bp}}
\def\xik{\xi_{\bk}}
\def\xikq{\xi_{\bk+\bq}}
\def\tilk{\tilde k}
\def\tilth{\tilde\theta}
\def\tilxi{\tilde\xi}
\def\dph{\Delta^{ph}}
\def\dpp{\Delta^{pp}}
\def\nsim{\sim \hskip -4truemm / \>}
\def\Dt{\tilde D}
\def\tht{\tilde\theta}
\def\omt{\tilde\omega}
\def\q0t{\tilde q_0}
\def\qrt{\tilde q_r}
\def\qtt{\tilde q_t}
\def\xit{\tilde \xi}
\def\tt{\tilde t}
\def\Lt{\tilde L}
\def\cdotr{\!\cdot\!}
\def\alf{\alpha}
\def\om{\omega}
\def\sg{\sigma}
\def\Sg{\Sigma}

\centerline{\bf\large Figure Captions} \par
\bigskip

\begin{enumerate}

\item[{\bf Fig.\ 2.1:}] Diagrammatic representation of the bare 
 interaction $g_{\bk\bk';\bq}$ with momentum labels according to the 
 notations used in this paper.

\item[{\bf Fig.\ 2.2:}] Mode elimination in momentum space. 
 Fermi fields $\psi_k$, $\psi^*_k$ with $d(\bk,\FS)$ $>$ $\Lam$ 
 are integrated out. The remaining effective theory is defined on a 
 shell of width $2\Lam$ around the Fermi surface in momentum space.

\item[{\bf Fig.\ 2.3:}] Examples for Feynman diagrams contributing to 
 the self-energy. Shaded boxes connected to $2n$ lines represent
 renormalized effective n-particle interactions. 

\item[{\bf Fig.\ 2.4:}] Contribution to the effective three-particle 
 interaction (in terms of bare interactions); the internal line must 
 carry a momentum outside the $\Lam$-shell.

\item[{\bf Fig.\ 2.5:}] Scattering processes on the Fermi surface in 
 two dimensions.

\item[{\bf Fig.\ 2.6:}] Diagrams contributing to effective 
 two-particle interactions in second order perturbation theory.
 The shaded squares represent (antisymmetrized) effective two-particle 
 interactions.

\item[{\bf Fig.\ 4.1:}] The two-particle vertex $\Gam$ in terms of the 
 particle-hole irreducible vertex $\Gam_{irr}$. 
 The internal lines correspond to dressed propagators $G$.

\item[{\bf Fig.\ 5.1:}] Dyson equation relating the vertex part
 $\Gam$ to irreducible components.

\item[{\bf Fig.\ 5.2:}] Fermionic loop with $N$ insertions.

\item[{\bf Fig.\ 5.3:}] Diagrams contributing to the RPA effective 
 interaction $D$.

\item[{\bf Fig.\ 5.4:}] The current-current response function 
 $J^{\mu\nu}$ in terms of $D$.

\item[{\bf Fig.\ 5.5:}] Perturbation expansion of the current vertex
 $\Lam^{\mu}(p;q)$.

\item[{\bf Fig.\ 5.6.}] Two cancelling contributions to 
 $\Lam^{\mu}(p;q)$.

\item[{\bf Fig.\ 5.7:}] Dyson equation relating the self-energy $\Sg$
 to the irreducible density vertex $\Lam^0$, the effective interaction
 $D$ and the propagator $G$.

\item[{\bf Fig.\ 6.1:}] Band structure of the g-ology model.

\item[{\bf Fig.\ 6.2:}] Interaction terms in the g-ology model;
 "-" and "+" indicate the left and right Fermi point, respectively.

\item[{\bf Fig.\ 7.1:}] Particle-hole bubble $\Pi_0(q)$ for small $q$
 as a function of $\tilde\om = \om/|\bq|$ in dimensions $d = 1.1$, 
 $1.5$, $2$ and $3$.

\item[{\bf Fig.\ 7.2:}] Feynman diagram contributing to the second
order self-energy.

\item[{\bf Fig.\ 7.3:}] $\Im\Sigma(\xi,\bp)$ from second order perturbation
 theory as a function of $\xi$ for fixed $p_r = 0.1k_F$ in dimensions
 $d = 1$, $1.5$, $2$, and $3$ ($g = k_F = v_F = 1$).

\item[{\bf Fig.\ 7.4:}] Spectral function $\rho(\xi,\bp)$ from second 
 order perturbation theory as a function of $\xi$ for fixed $p_r = 0.1k_F$
 in $d = 1.5$ ($g = 2$, $k_F = v_F = 1$).

\item[{\bf Fig.\ 7.5:}] Effective interaction $D^{\sg\sg}(\om,\bq)$ as a 
 function of $\tilde\om = \om/|\bq|$ for a constant coupling $g=2$ 
 between opposite spins in dimensions $d = 1.1$, $1.5$ and $2$.

\item[{\bf Fig.\ 7.6:}] Feynman diagram representing the RPA self-energy.

\item[{\bf Fig.\ 7.7:}] RPA-result for 
 $\Im\Sg(\xi,\bp)$ as a function of $\xi$ at fixed
 $p_r = 0.1k_F$ for a constant coupling $g=2$ between opposite spins in 
 $d = 1.5$ dimensions ($k_F = v_F = 1$). 

\item[{\bf Fig.\ 9.1:}] The gauge-field propagator $D$ in random
 phase approximation: 
 The polarization insertion $\Pi$ is approximated by its 1-loop
 result $\Pi_0$. 

\item[{\bf Fig.\ 9.2:}] The fermion self-energy within RPA,
 where $D$ is the RPA gauge-field propagator, and $\lam^{\mu}(k) =
 (1,\bv_{\bk})$ the bare fermion-gauge-field vertex.

\item[{\bf Fig.\ 9.3:}] Dyson equation relating the fermion
 self-energy $\Sg$ to the fermion propagator $G$, the gauge-field
 propagator $D^{\mu\nu}$ and the irreducible fermion-gauge-field
 vertex $\Lam^{\mu}$.

\item[{\bf Fig.\ 10.1:}] Phase-diagram of Fermi systems with
 short-range interactions in the dimens\-ionality-coupling plane, 
 where $g$ is a typical renormalized coupling in the forward scattering
 channel.

\item[{\bf Fig.\ A.1:}] Two loops with $N = 3$ insertions.

\end{enumerate}


\begin{thebibliography}{99}

\bibitem{LAN} L.D. Landau, Sov. Phys. JETP {\bf 3}, 920 (1956); 
 ibid {\bf 5}, 101 (1957).

\bibitem{NOZ} A particularly thorough discussion of Fermi liquid theory
 is given by P. Nozi\`eres, {\em Theory of Interacting Fermi Systems} 
 (Benjamin, Amsterdam 1964).

\bibitem{VW} D. Vollhardt and P. W\"olfle, {\em The superfluid phases of
 Helium 3} (Taylor and Francis, London 1990).

\bibitem{HFS} For reviews on heavy fermion systems, see 
 G. Czycholl, Phys. Rep. {\bf 143}, 277 (1986); 
 P. Fulde, J. Keller, and G. Zwicknagl, in {\em Solid State Physics}, 
 Vol. 41, ed. by H. Ehrenreich and D. Turnbull (Academic, San Diego 1988);
 P.A. Lee, T.M. Rice, J.W. Serene, L.J. Sham, and J.W. Wilkins,
 Comments Cond. Matt. Phys. {\bf 12}, 99 (1986).

\bibitem{WYD} See, for example, Proceedings of the International Conference
 on {\em Materials and Mechanisms of Superconductivity: High Temperature
 Superconductors IV}, ed.\ P. Wyder (North Holland 1994).

\bibitem{PKC} For a review on the Fermi surface structure 
 of high-$T_c$ superconductors see, for example,
 W.E. Pickett, H. Krakauer, R.E. Cohen and D.J. Singh, Science
 {\bf 255}, 46 (1992);
 Z.X. Shen and D.S. Dessau, Phys. Rep. {\bf 253}, 1 (1995).

\bibitem{AND90} P. W. Anderson, Phys. Rev. Lett. {\bf 64}, 1839 (1990); 
 Phys. Rev. Lett. {\bf 65}, 2306 (1990).

\bibitem{VLS} C. M. Varma, P. B. Littlewood, S. Schmitt-Rink, E. Abrahams
 and A. E. Ruckenstein, Phys. Rev. Lett. {\bf 63}, 1996 (1989).

\bibitem{AT} B. Andraka and A.M. Tsvelik, Phys. Rev. Lett. {\bf 67},
 2886 (1991).

\bibitem{AS} B. Andraka and G.R. Stewart, Phys. Rev. B {\bf 47}, 3208
 (1993).

\bibitem{SML} C.L. Seaman et al., Phys. Rev. Lett. {\bf 67}, 2882 (1991).

\bibitem{LPP} H. v. L\"ohneysen, T. Pietrus, G. Portisch, H.G. Schlager,
 A. Schr\"oder, M. Sieck, and T. Trappmann, Phys. Rev. Lett. {\bf 72},
 3262 (1994).

\bibitem{OTT} L. Degiorgi, H.R. Ott, and F. Hulliger, Phys. Rev. B 
 {\bf 52}, 42 (1995); L. Degiorgi and H.R. Ott, J. Phys.: Cond. Matt.
 (in press).

\bibitem{AGD} A.A. Abrikosov, L.P. Gorkov, and I.E. Dzyaloshinski,
 {\em Methods of Quantum Field Theory in Statistical Physics}
 (Dover, New York 1975).

\bibitem{FT} J. Feldman and E. Trubowitz, Helv. Phys. Act. {\bf 63},
 156 (1990); {\bf 64}, 213 (1991); {\bf 65}, 679 (1992).

\bibitem{BG} G. Benfatto and G. Gallavotti, Phys. Rev. {\bf B42}, 9967 
 (1990); J. Stat. Phys. {\bf 59}, 541 (1990).

\bibitem{FMRT} J. Feldman, J. Magnen, V. Rivasseau, and E. Trubowitz in
 {\em The State of Matter}, M. Aizenmann and H. Araki eds, Advanced Series
 in Mathematical Physics Vol. {\bf 20} (World Scientific, 1994).

\bibitem{SHA91} R. Shankar, Physica A {\bf 177}, 530 (1991). 

\bibitem{SHA94} R. Shankar, Rev. Mod. Phys. {\bf 66}, 129 (1994).

\bibitem{SOL} For an extensive early review on 1D and quasi-1D electronic 
 systems, see J. Solyom, Adv. Phys. {\bf 28}, 201 (1979).

\bibitem{TL} S. Tomonaga, Prog. Theor. Phys. {\bf 5}, 544 (1950);
 J.M. Luttinger, J. Math. Phys. {\bf 4}, 1154 (1963).

\bibitem{ML} D.C. Mattis and E.H. Lieb, J. Math. Phys. {\bf 6}, 304 (1965).

\bibitem{VOI95} For a comprehensive up-to-date review on one-dimensional
Fermi systems, see J. Voit, Rep. Prog. Phys. {\bf 58}, 977 (1995).

\bibitem{SCH95} For a recent review on one-dimensional Fermi systems
 (including a discussion of the one-dimensional Hubbard model), 
 see H.J. Schulz, preprint (1995).

\bibitem{HAL81} F.D.M. Haldane, J. Phys. C {\bf 14}, 2585 (1981);
 Phys. Rev. Lett. {\bf 45}, 1358 (1980); {\bf 47}, 1840 (1981);
 Phys. Lett. {\bf 81A}, 153 (1981). 

\bibitem{VR} A. Viroztek and J. Ruvalds, Phys. Rev. B {\bf 42}, 4064
 (1990).

\bibitem{PKN} P.C. Pattnaik, C.L. Kane, D.M. Newns, and C.C. Tsuei,
 Phys. Rev. B {\bf 45}, 5714 (1992).

\bibitem{NTH} D.M. Newns, C.C. Tsuei, R.P. Huebener, P.J.M. van Bentum,
 P.C. Pattnaik, and C.C. Chi, Phys. Rev. Lett. {\bf 73}, 1695 (1994).

\bibitem{HR} R. Hlubina and T.M. Rice, Phys. Rev. B {\bf 51}, 9253 (1995).

\bibitem{FHN} H. Fukuyama, Y. Hasegawa, and O. Narikiyo, J. Phys. Soc. Jpn.
 {\bf 60}, 2013 (1991).

\bibitem{ER92} J.R. Engelbrecht and M. Randeria, Phys. Rev. B {\bf 45},
 12419 (1992).

\bibitem{STA93} P.C.E. Stamp, J. Phys. I (France) {\bf 3}, 625 (1993).

\bibitem{MC} W. Metzner and C. Castellani, Int. J. Mod. Phys. B (1995).

\bibitem{FKL} J. Feldman, H. Kn\"orrer, D. Lehmann, and E. Trubowitz,
 in {\em Constructive Physics}, V. Rivasseau (ed.), Springer Lecture
 Notes in Physics (Springer, 1995).

\bibitem{CDM94} C. Castellani, C. Di Castro and W. Metzner, Phys. Rev. Lett.
{\bf 72}, 316 (1994).

\bibitem{STA92} P.C.E. Stamp, Phys. Rev. Lett. {\bf 68}, 2180 (1992).

\bibitem{BW} P. Bares and X.G. Wen, Phys. Rev. B {\bf 48}, 8636 (1993).

\bibitem{CD94} C. Castellani and C. Di Castro, Physica C {\bf 235-240},
 99 (1994);
 A. Maccarone, M.Sc. thesis at Universit\`a "La Sapienza", Roma (1994).

\bibitem{BA} G. Baskaran and P.W. Anderson, Phys. Rev. B {\bf 37}, 580
 (1988).

\bibitem{IL} L.B. Ioffe and A.I. Larkin, Phys. Rev. B {\bf 39}, 8988 (1989).

\bibitem{LN} P.A. Lee and N. Nagaosa, Phys. Rev. B {\bf 46}, 5621 (1992).

\bibitem{HNP} T. Holstein, R. Norton, and P. Pincus, Phys. Rev. B {\bf 8},
 2649 (1973).

\bibitem{REI} M. Reizer, Phys. Rev. B {\bf 39}, 1602 (1989).

\bibitem{HLR} B.I. Halperin, P.A. Lee and N. Read, Phys. Rev. B {\bf 47},
 7312 (1993).

\bibitem{KZ} V. Kalmeyer and S.C. Zhang, Phys. Rev. B {\bf 46}, 9889 
 (1992).

\bibitem{LOE} H. von L\"ohneysen, J. Phys.: Cond. Matt. (in press).

\bibitem{STE} F. Steglich et al., Proceedings of Euroconference on 
 Correlations and Unconventional Quantum Liquids (Evora 1996).

\bibitem{MTU} T. Moriya, Y. Takahashi, and K. Ueda, J. Phys. Soc. Jpn.
 {\bf 59}, 2905 (1990).

\bibitem{MP} P. Monthoux and D. Pines, Phys. Rev. B {\bf 49}, 4261 (1994)
 and references therein.

\bibitem{EK} V.J. Emery and S. Kivelson, Physica {\bf 209C}, 597 (1993).

\bibitem{CDG} C. Castellani, C. Di Castro and M. Grilli, Phys. Rev.
 Lett. {\bf 75}, 4650 (1995).;
 A. Perali, Castellani, C. Di Castro and M. Grilli, Phys. Rev. B, in
 press (1996).
 
\bibitem{VAR} C.M. Varma, preprint (1996).

\bibitem{NB} P. Nozi\`eres and A. Blandin, J. Phys. {\bf 41}, 193 (1980).

\bibitem{COX} D.L. Cox, Phys. Rev. Lett. {\bf 59}, 1240 (1987).

\bibitem{AD} N. Andrei and C. Destri, Phys. Rev. Lett. {\bf 52}, 364 (1984).

\bibitem{WT} P.B. Wiegmann and A.M. Tsvelick, Z. Phys. B {\bf 54}, 201
 (1985).

\bibitem{AL} I. Affleck and A.W.W. Ludwig, Nucl. Phys. {\bf B360}, 641 
 (1991); Phys. Rev. Lett. {\bf 67}, 161 (1991);
 A.W.W. Ludwig and I. Affleck, Phys. Rev. Lett. {\bf 67}, 3160 (1991).

\bibitem{SKG} Q. Si, G. Kotliar, and A. Georges, Phys. Rev. B {\bf 46},
 1261 (1992).

\bibitem{SK} Q. Si and G. Kotliar, Phys. Rev. Lett. {\bf 70}, 3143 (1993); 
 Phys. Rev. B {\bf 48}, 13881 (1993).

\bibitem{SRK} Q. Si, M.J. Rozenberg, G. Kotliar, and A.E. Ruckenstein,
 Phys. Rev. Lett. {\bf 72}, 2761 (1994).

\bibitem{DL} I.E. Dzyaloshinskii and A.I. Larkin, Sov. Phys. JETP {\bf 38},
 202 (1974).

\bibitem{KHS} P. Kopietz, J. Hermisson, and K. Sch\"onhammer, Phys. Rev. B 
 {\bf 52}, 10877 (1995).

\bibitem{ILA} L.B. Ioffe, D. Lidsky, and B.L. Altshuler, Phys. Rev. Lett.
 {\bf 73}, 472 (1994);
 B.L. Altshuler, L.B. Ioffe, and A.J. Millis, Phys. Rev. B {\bf 50}, 
 14048 (1994).

\bibitem{KSgau} D.V. Khveshchenko and P.C.E. Stamp, Phys. Rev. Lett. 
 {\bf 71}, 2118 (1994); Phys. Rev. B {\bf 49}, 5227 (1994).

\bibitem{LP} A. Luther and I. Peschel, Phys. Rev. B {\bf 9}, 2911 (1974).

\bibitem{MAT} D.C. Mattis, J. Math. Phys. {\bf 15}, 609 (1974).

\bibitem{LUT79} A. Luther, Phys. Rev. B {\bf 19}, 320 (1979).

\bibitem{HAL92} F.D.M. Haldane, Helv. Phys. Acta, {\bf 65}, 152 (1992);
 Proceedings of the International School of Physics "Enrico Fermi",
 Course CXXI, eds. R.A. Broglia and J.R. Schrieffer (North Holland,
 Amsterdam, 1994). 

\bibitem{HM} A. Houghton and J.B. Marston, Phys. Rev. B {\bf 48}, 7790 
 (1993).

\bibitem{HKM} A. Houghton, H.-J. Kwon, and J.B. Marston, Phys. Rev. B
 {\bf 50}, 1351 (1994); 
 H.-J. Kwon, A. Houghton, and J.B. Marston, Phys. Rev. B {\bf 52}, 8002
 (1995).

\bibitem{CF} A.H. Castro Neto and E. Fradkin, Phys. Rev. Lett. {\bf 72},
 1393 (1994); Phys. Rev. B {\bf 49}, 10877 (1994).

\bibitem{KHVbos} D.V. Khveshchenko, Phys. Rev. B {\bf 49}, 16893 (1994);
 Phys. Rev. B {\bf 52}, 4833 (1995).

\bibitem{KS} P. Kopietz and K. Sch\"onhammer, Z. Phys. B {\bf 100}, 561 
 (1996).

\bibitem{UR} K. Ueda and T.M. Rice, Phys. Rev. B {\bf 29}, 1514 (1984).

\bibitem{WIL} K.G. Wilson, Rev. Mod. Phys. {\bf 47}, 773 (1975);
 K.G. Wilson and J.B. Kogut, Phys. Rep. {\bf 12}, 7 (1974), 
 and references therein.

\bibitem{DIS} For a review, see P.A. Lee and T.V. Ramakrishnan, Rev. Mod.
 Phys. {\bf 57}, 287 (1985); 
 C. Castellani, C. Di Castro and G. Strinati in {\em Lecture Notes in
 Physics}\/ {\bf 268} (Springer 1986);
 A.M. Finkelstein, Sov. Sci. Rev. A. Phys. {\bf 14}, 1 (1990); 
 D. Belitz and T.R. Kirkpatrick, Rev. Mod. Phys. {\bf 66}, 261 (1994). 

\bibitem{SR} See, in particular, the review articles by 
 J.W. Serene and D. Rainer, Phys. Rep. {\bf 101}, 221 (1983);
 D. Rainer, Prog. Low Temp. Phys. {\bf X}, 371 (1986).

\bibitem{BG1d} G. Benfatto, G. Gallavotti, and V. Mastropietro,
 Phys. Rev. B {\bf 42}, 9967 (1990);
 G. Benfatto, G. Gallavotti, A. Procacci, and B. Scoppola, 
 Comm. Math. Phys. {\bf 160}, 93 (1994).

\bibitem{BGb} G. Benfatto and G. Gallavotti, {\em Renormalization
 Group}\/ (Princeton University Press 1995).

\bibitem{FKLT} J. Feldman, H. Kn\"orrer, D. Lehmann and E. Trubowitz
 in {\em Constructive Physics}, ed. V. Rivasseau, Springer Lecture
 Notes in Physics (Springer 1995).

\bibitem{CFS} T. Chen, J. Fr\"ohlich, and M. Seifert, in Proceedings
 of the Les Houches Summer School 1994. 

\bibitem{POL} J. Polchinski in Proceedings of 1992 Theoretical Advanced
 Studies Institute in Elementary Particle Physics, eds. J. Harvey and
 J. Polchinski (World Scientific, Singapore, 1993).

\bibitem{NO} J.W. Negele and H. Orland, {\em Quantum Many-Particle 
 Systems}\/ (Addison-Wesley 1988).
 
\bibitem{POP} V.N. Popov, {\em Functional Integrals and Collective
 Excitations}\/ (Cambridge University Press).

\bibitem{FST} J. Feldman, M. Salmhofer and E. Trubowitz, preprint 
 (1995).

\bibitem{MD} W. Metzner and C. Di Castro, Phys. Rev. B {\bf 47}, 16107 
 (1993).

\bibitem{KL65} W. Kohn and J.M. Luttinger, Phys. Rev. Lett. {\bf 15},
 524 (1965).

\bibitem{Tpriv} E. Trubowitz, private communication.

\bibitem{FL} D. Fay and A. Layzer, Phys. Rev. Lett. {\bf 20}, 187 (1968).

\bibitem{KW} S. K\"uchenhoff and P. W\"olfle, Phys. Rev. B {\bf 38},
 935 (1988).

\bibitem{KC} M.Yu. Kagan and A.V. Chubukov, JETP Lett. {\bf 47}, 614 (1988);
 M.A. Baranov, A.V. Chubukov and M.Yu. Kagan, Int. J. Mod. Phys. B {\bf 6},
 2471 (1992).

\bibitem{BOH} T. Bohr, Nordita preprint 81/4, {\em Lectures on the
 Luttinger Model} (1981).

\bibitem{DC} G.Y. Chitov and D. S\'en\'echal, Phys. Rev. B {\bf 52},
 13487 (1995).

\bibitem{BP} G. Baym and C. Pethick, {\em Landau Fermi Liquid Theory}
 (Wiley, New York 1991).

\bibitem{EHR} M. Eschrig, J. Heym, and D. Rainer, J. Low Temp. Phys.
 {\bf 95}, 323 (1994).

\bibitem{ZIM} See, for example, J.M. Ziman, {\em Electrons and Phonons}
 (Oxford University Press 1960).

\bibitem{POM} I.Ja. Pomerantchuk, Sov. Phys. JETP {\bf 8}, 361 (1958).

\bibitem{COP} L.N. Cooper, Phys. Rev. {\bf 104}, 1189 (1956).

\bibitem{MET95} W. Metzner, unpublished notes (1995).

\bibitem{KFWL} Y.B. Kim, A. Furusaki, X.G. Wen, and P.A. Lee,
 Phys. Rev. B {\bf 50}, 17917 (1994).

\bibitem{HK} J.A. Hertz and M.A. Klenin, Phys. Rev. B {\bf 10}, 1084 
 (1974).

\bibitem{FKST} J. Feldman, H. Kn\"orrer, R. Sinclair and E. Trubowitz,
 preprint (ETH Z\"urich, 1996).

\bibitem{FGMbos} J. Fr\"ohlich, R. G\"otschmann and P.A. Marchetti,
 J. Phys. A {\bf 28}, 1169 (1995).

\bibitem{FOG} H.C. Fogedby, J. Phys. C {\bf 9}, 3757 (1976).

\bibitem{LC} D.K.K. Lee and Y. Chen, J. Phys. A {\bf 21}, 4155 (1988).

\bibitem{KOPhs} P. Kopietz, Habilitationsschrift, Georg-August 
 University G\"ottingen (1995).

\bibitem{LW} E.H. Lieb and F.Y. Wu, Phys. Rev. Lett. {\bf 20}, 1445 (1968).

\bibitem{FK} H. Frahm and V.E. Korepin, Phys. Rev. B {\bf 42}, 10553 
 (1990); Phys. Rev. B {\bf 43}, 5653 (1991).

\bibitem{HT} F.D.M. Haldane and Y. Tu (unpublished).

\bibitem{KY} N. Kawakami and S.K. Yang, Phys. Lett. A {\bf 148}, 359 (1990).

\bibitem{SCH90} H.J. Schulz, Phys. Rev. Lett. {\bf 64}, 2831 (1990).

\bibitem{OS} M. Ogata and H. Shiba, Phys. Rev. B {\bf 41}, 2326 (1990).

\bibitem{PS90} A. Parola and S. Sorella, Phys. Rev. Lett. {\bf 64}, 1831 
 (1990); S. Sorella, A. Parola, M. Parrinello and E. Tosatti, Europhys. Lett.
 {\bf 12}(8), 721 (1990).

\bibitem{PMS} K. Penc, F. Mila and H. Shiba, Phys. Rev. Lett. {\bf 75},
 894 (1995).

\bibitem{CH} See, in particular, J.M.P. Carmelo and P. Horsch, 
 Phys. Rev. Lett. {\bf 68}, 871 (1992).

\bibitem{ES} H.U. Everts and H. Schulz, Sol. State Comm. {\bf 15}, 1413 
 (1974).

\bibitem{EME} V.J. Emery in {\em Highly Conducting One-Dimensional
 Solids}, edited by J.T. Devreese, R.P. Evrard, and V.E. van Doren
 (Plenum, New York 1979).

\bibitem{DJ} For a review, see C. Di Castro and G. Jona-Lasinio in 
 {\em Phase Transitions and Critical Phenomena}, edited by C. Domb and 
 M.S. Green (Academic, London, 1976), Vol.6.

\bibitem{AMI} For an introduction to the field theoretic renormalization 
 group, see D. Amit, {\em Field Theory, the Renormalization Group and 
 Critical Phenomena} (World Scientific, Singapore, 1984). 

\bibitem{DM} C. Di Castro and W. Metzner, Phys. Rev. Lett. {\bf 67}, 3852 
 (1991).

\bibitem{LE} A. Luther and V.J. Emery, Phys. Rev. Lett. {\bf 33}, 589 (1974).

\bibitem{JON} K. Johnson, Nuovo Cimento {\bf 20}, 773 (1961).

\bibitem{MS} V. Meden and K. Sch\"onhammer, Phys. Rev. B {\bf 46} (1992).

\bibitem{VOI93} J. Voit, Phys. Rev. B {\bf 47}, 6740 (1993).

\bibitem{MER} N.D. Mermin, Phys. Rev. {\bf 159}, 161 (1967). 

\bibitem{FO} H. Fukuyama and M. Ogata, J. Phys. Soc. Japan. {\bf 63}, 
 3923 (1995); 
 H. Fukuyama and M. Ogata in {\em Spectroscopy of Mott Insulator and
 Correlated Metals}, ed. A. Fujimori and Y. Tokura (Springer Verlag 1995),
 p.\ 34.

\bibitem{HSW} C. Hodges, H. Smith, and J.W. Wilkins, Phys. Rev. B {\bf 4},
 302 (1971).

\bibitem{BL} P. Bloom, Phys. Rev. B {\bf 12}, 125 (1975).

\bibitem{HKMS} A. Houghton, H.-J. Kwon, J.B. Marston, and R. Shankar,
 J. Phys. C {\bf 6}, 4909 (1994).

\bibitem{JAI} K. Jain, Phys. Rev. Lett. {\bf 63}, 199 (1989);
 Phys. Rev. B {\bf 41}, 7653 (1990); Adv. Phys. {\bf 41}, 105 (1992).

\bibitem{LF} A. Lopez and E. Fradkin, Phys. Rev. B {\bf 44}, 5246 (1991);
 Phys. Rev. Lett. {\bf 69}, 2126 (1992).

\bibitem{WILC} F. Wilczek, {\em Fractional Statistics and Anyon
 Superconductivity} (World Scientific, Singapore, 1990).

\bibitem{GW} J. Gan and E. Wong, Phys. Rev. Lett. {\bf 71}, 4226 (1993).

\bibitem{FGMgau} J. Fr\"ohlich, R. G\"otschmann, and P.A. Marchetti,
 Commun. Math. Phys. {\bf 173}, 417 (1995).

\bibitem{POL94} J. Polchinski, Nucl. Phys. B {\bf 422}, 617 (1994).

\bibitem{NW} C. Nayak and F. Wilczek, Nucl. Phys. B {\bf 417}, 359 (1994);
 {\bf 430}, 534 (1994).

\bibitem{AW} A.G. Aronov and P. W\"olfle, Phys. Rev. Lett. {\bf 72}, 2239
 (1994); Phys. Rev. B {\bf 50}, 16574 (1994).

\bibitem{LEE} P.A. Lee, Phys. Rev. Lett. {\bf 63}, 680 (1989).

\bibitem{BM} B. Blok and H. Monien, Phys. Rev. B {\bf 47}, 3454 (1993).

\bibitem{KHM} H.-J. Kwon, A. Houghton, and J.B. Marston, Phys. Rev. Lett.
 {\bf 73}, 284 (1994); preprint (1995).

\bibitem{KOP95} P. Kopietz, Phys. Rev. B {\bf 53}, 12761 (1996).

\bibitem{CNS} S. Chakravarty, R.E. Norton, and O.F. Sylju{\aa}sen,
 Phys. Rev. Lett. {\bf 74}, 1423 (1995).

\end{thebibliography}
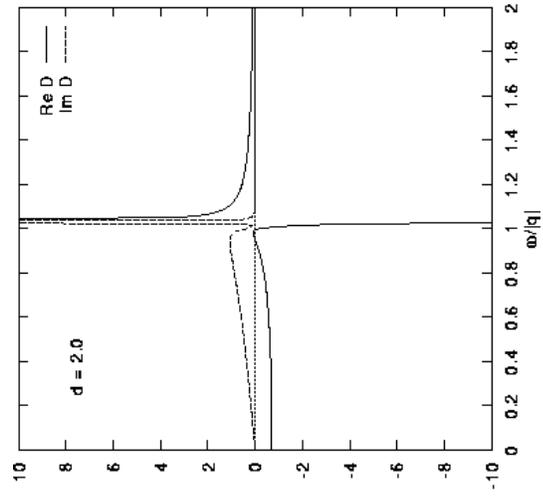
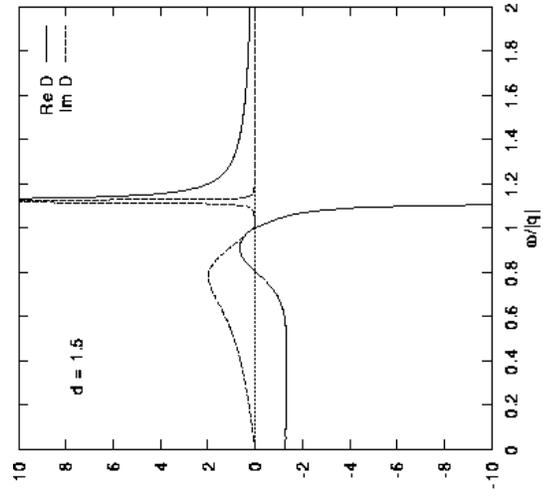
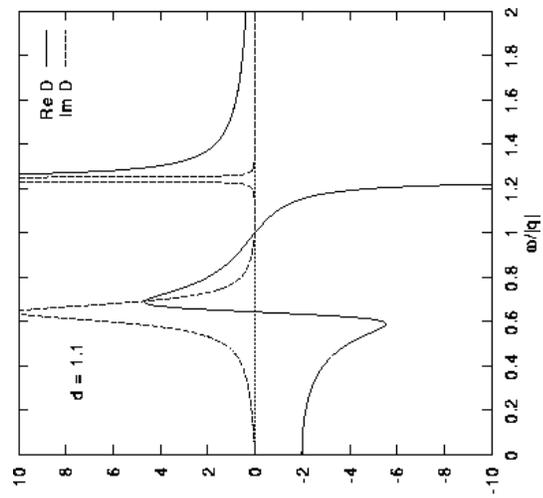
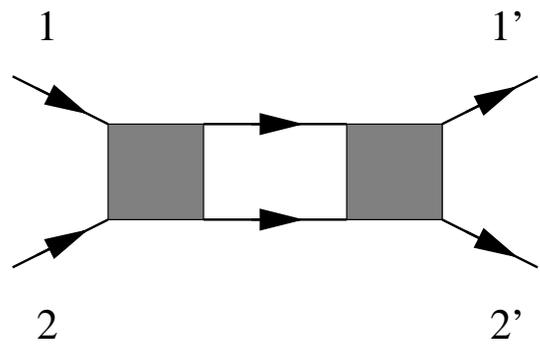
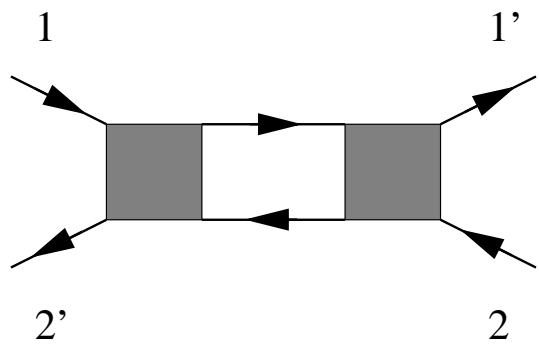
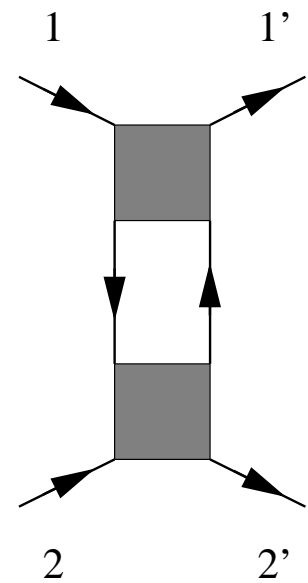
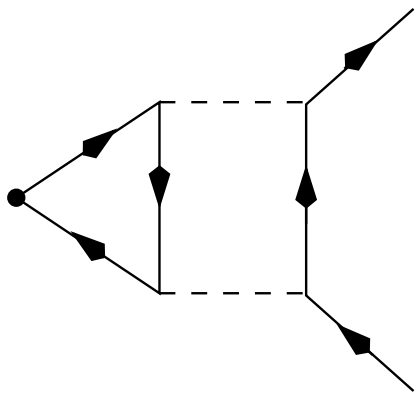
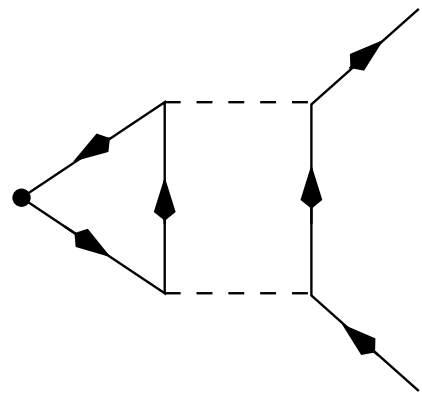
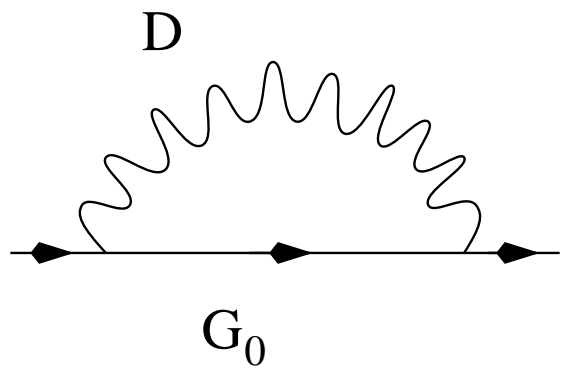
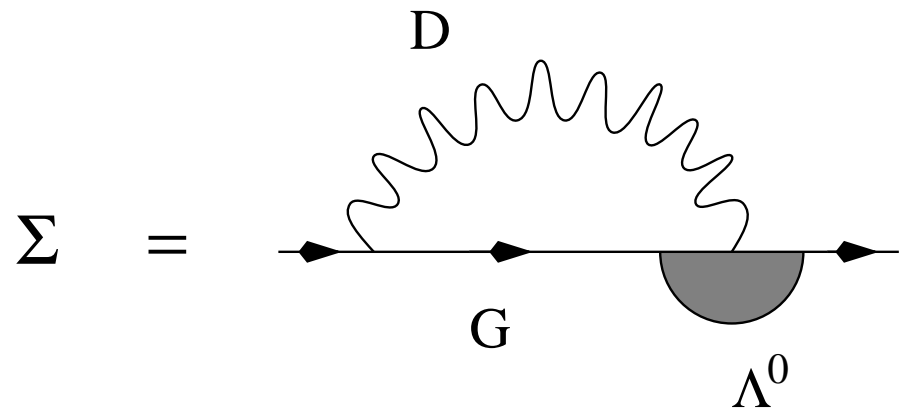
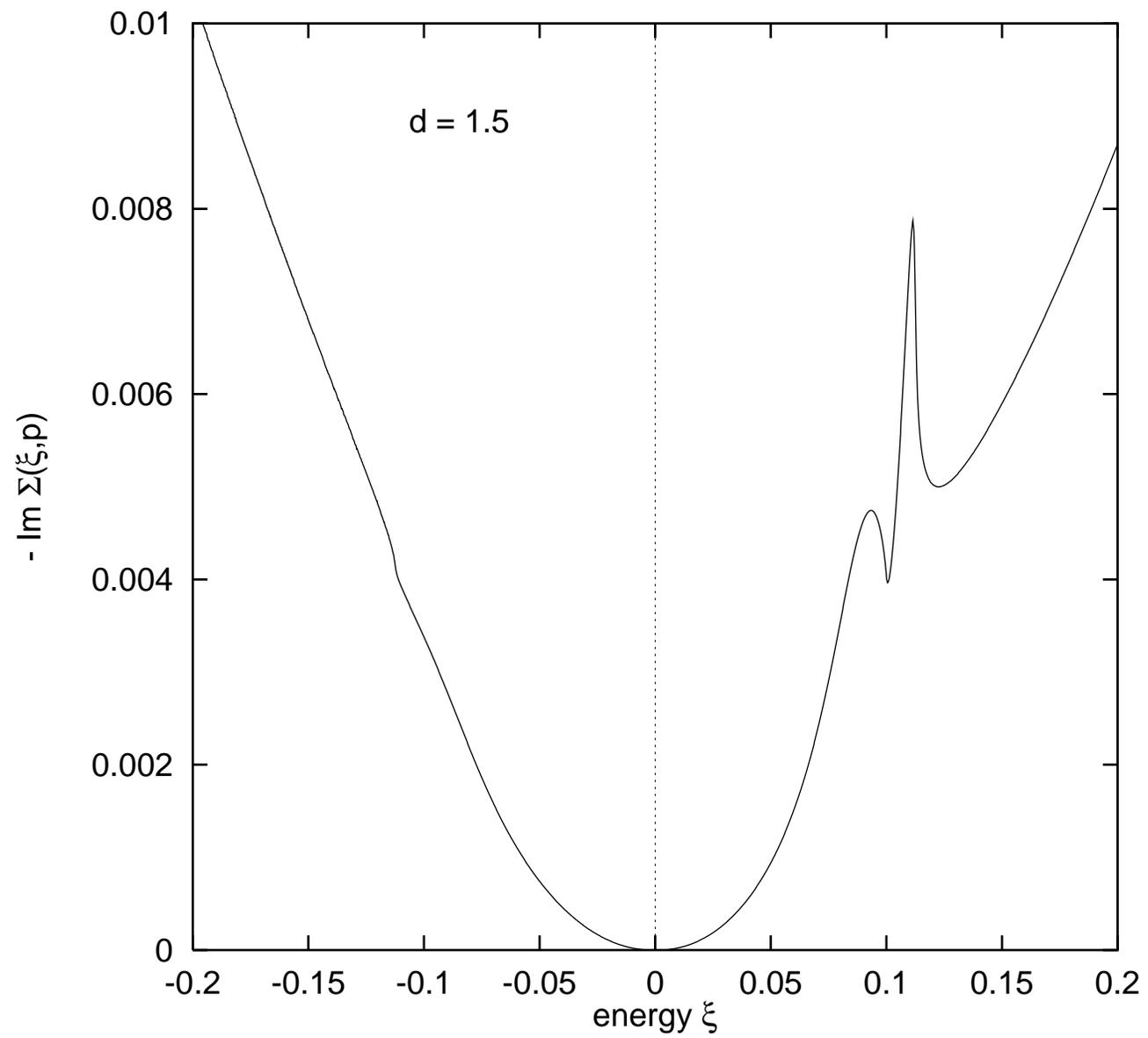
\end{document}